\pdfoutput=1
\documentclass[12pt,a4paper]{article}
\usepackage{ifthen} % for conditional statements
\usepackage{multirow}
\newboolean{pdflatex}
\setboolean{pdflatex}{true} % False for eps figures 
\usepackage{booktabs}

\newboolean{articletitles}
\setboolean{articletitles}{true} % False removes titles in references

\newboolean{uprightparticles}
\setboolean{uprightparticles}{false} %True for upright particle symbols

\def\paperauthors{LHCb collaboration} % Leave as is for PAPER, CONF and FIGURE
\def\paperasciititle{Amplitude analysis of B0 -> D0bar Ds+ pi- and B+ -> D- Ds+ pi+ decays} % Set ASCII title here !! MAKE sure it's only ASCII characters !! 
\def\papertitle{Amplitude analysis of $B^0 \rightarrow \offsetoverline{D}^0 D_s^+ \pi^-$ and $B^+ \rightarrow D^- D_s^+ \pi^+$ decays} % Latex formatted title
\def\paperkeywords{{High Energy Physics}, {LHCb}} % Comma separated list
\def\papercopyright{\the\year\ CERN for the benefit of the LHCb collaboration} % new since 9/Apr/2018
\def\paperlicence{CC BY 4.0 licence}
\def\paperlicenceurl{https://creativecommons.org/licenses/by/4.0/}

\usepackage[top=1in, bottom=1.25in, left=1in, right=1in]{geometry}

\usepackage{microtype}
\usepackage{lineno}  % for line numbering during review
\usepackage{xspace} % To avoid problems with missing or double spaces after
\usepackage{caption} %these three command get the figure and table captions automatically small

\usepackage{graphicx}  % to include figures (can also use other packages)
\usepackage{color}
\usepackage{colortbl}
\graphicspath{{./figs/}} % Make Latex search fig subdir for figures
\usepackage{amsmath} % Adds a large collection of math symbols
\usepackage{amssymb}
\usepackage{amsfonts}
\usepackage{upgreek} % Adds in support for greek letters in roman typeset

\newcommand*\patchAmsMathEnvironmentForLineno[1]{%
\expandafter\let\csname old#1\expandafter\endcsname\csname #1\endcsname
\expandafter\let\csname oldend#1\expandafter\endcsname\csname
end#1\endcsname
 \renewenvironment{#1}%
   {\linenomath\csname old#1\endcsname}%
   {\csname oldend#1\endcsname\endlinenomath}%
}
\newcommand*\patchBothAmsMathEnvironmentsForLineno[1]{%
  \patchAmsMathEnvironmentForLineno{#1}%
  \patchAmsMathEnvironmentForLineno{#1*}%
}
\AtBeginDocument{%
\patchBothAmsMathEnvironmentsForLineno{equation}%
\patchBothAmsMathEnvironmentsForLineno{align}%
\patchBothAmsMathEnvironmentsForLineno{flalign}%
\patchBothAmsMathEnvironmentsForLineno{alignat}%
\patchBothAmsMathEnvironmentsForLineno{gather}%
\patchBothAmsMathEnvironmentsForLineno{multline}%
\patchBothAmsMathEnvironmentsForLineno{eqnarray}%
}

\usepackage{hyperxmp}

\usepackage[pdftex,
            pdfauthor={\paperauthors},
            pdftitle={\paperasciititle},
            pdfkeywords={\paperkeywords},
            pdfcopyright={Copyright (C) \papercopyright},
            pdflicenseurl={\paperlicenceurl}]{hyperref}
\usepackage[colorinlistoftodos,textsize=scriptsize]{todonotes}

\usepackage[bottom,flushmargin,hang,multiple]{footmisc}

\usepackage[all]{hypcap} % Internal hyperlinks to floats.

\usepackage{xspace} 
\usepackage{upgreek}

\def\lhcb   {\mbox{LHCb}\xspace}

\def\MagUp {\mbox{\em Mag\kern -0.05em Up}\xspace}

\ifthenelse{\boolean{uprightparticles}}%
{

 \def\Pmu         {\ensuremath{\upmu}\xspace}

 \def\Ppi         {\ensuremath{\uppi}\xspace}

 \def\Ppsi        {\ensuremath{\uppsi}\xspace}

 \def\PDelta      {\ensuremath{\Delta}\xspace}                 
 \def\PXi         {\ensuremath{\Xi}\xspace}                 
 \def\PLambda     {\ensuremath{\Lambda}\xspace}                 
 \def\PSigma      {\ensuremath{\Sigma}\xspace}                 
 \def\POmega      {\ensuremath{\Omega}\xspace}                 
 \def\PUpsilon    {\ensuremath{\Upsilon}\xspace}
 \let\oldPi\Pi
 \def\PPi         {\ensuremath{\oldPi}\xspace}

 \def\PB      {\ensuremath{\mathrm{B}}\xspace}                 
                  
 \def\PD      {\ensuremath{\mathrm{D}}\xspace}

 \def\PJ      {\ensuremath{\mathrm{J}}\xspace}                 
 \def\PK      {\ensuremath{\mathrm{K}}\xspace}

 \def\Pb      {\ensuremath{\mathrm{b}}\xspace}                 
 \def\Pc      {\ensuremath{\mathrm{c}}\xspace}                 
                  
 \def\Pe      {\ensuremath{\mathrm{e}}\xspace}

 \def\Pi      {\ensuremath{\mathrm{i}}\xspace}

 \def\Pp      {\ensuremath{\mathrm{p}}\xspace}

 \def\Ps      {\ensuremath{\mathrm{s}}\xspace}

 \def\thebaroffset{0.0em}
}
{

 \def\Pmu         {\ensuremath{\mu}\xspace}

 \def\Ppi         {\ensuremath{\pi}\xspace}

 \def\Ppsi        {\ensuremath{\psi}\xspace}                 
                  
 \mathchardef\PDelta="7101
 \mathchardef\PXi="7104
 \mathchardef\PLambda="7103
 \mathchardef\PSigma="7106
 \mathchardef\POmega="710A
 \mathchardef\PUpsilon="7107
 \mathchardef\PPi="7105
                  
 \def\PB      {\ensuremath{B}\xspace}                 
                  
 \def\PD      {\ensuremath{D}\xspace}

 \def\PJ      {\ensuremath{J}\xspace}                 
 \def\PK      {\ensuremath{K}\xspace}

 \def\Pb      {\ensuremath{b}\xspace}                 
 \def\Pc      {\ensuremath{c}\xspace}                 
                  
 \def\Pe      {\ensuremath{e}\xspace}

 \def\Pi      {\ensuremath{i}\xspace}

 \def\Pp      {\ensuremath{p}\xspace}

 \def\Ps      {\ensuremath{s}\xspace}

 \def\thebaroffset{0.18em}
}
\newcommand{\offsetoverline}[2][\thebaroffset]{\kern #1\overline{\kern -#1 #2}}%

\makeatletter
\ifcase \@ptsize \relax% 10pt
  \newcommand{\miniscule}{\@setfontsize\miniscule{4}{5}}% \tiny: 5/6
\or% 11pt
  \newcommand{\miniscule}{\@setfontsize\miniscule{5}{6}}% \tiny: 6/7
\or% 12pt
  \newcommand{\miniscule}{\@setfontsize\miniscule{5}{6}}% \tiny: 6/7
\fi
\makeatother

\DeclareRobustCommand{\optbar}[1]{\shortstack{{\miniscule (\rule[.5ex]{1.25em}{.18mm})}
  \\ [-.7ex] $#1$}}

\def\ep         {{\ensuremath{\Pe^+}}\xspace}

\def\mumu       {{\ensuremath{\Pmu^+\Pmu^-}}\xspace}

\def\squark    {{\ensuremath{\Ps}}\xspace}

\def\cquark    {{\ensuremath{\Pc}}\xspace}

\def\bquark    {{\ensuremath{\Pb}}\xspace}

\def\pion   {{\ensuremath{\Ppi}}\xspace}
\def\piz    {{\ensuremath{\pion^0}}\xspace}
\def\pip    {{\ensuremath{\pion^+}}\xspace}
\def\pim    {{\ensuremath{\pion^-}}\xspace}
\def\pipm   {{\ensuremath{\pion^\pm}}\xspace}

\def\kaon    {{\ensuremath{\PK}}\xspace}
\def\KorKbar {\kern \thebaroffset\optbar{\kern -\thebaroffset \PK}{}\xspace}

\def\Kp      {{\ensuremath{\kaon^+}}\xspace}
\def\Km      {{\ensuremath{\kaon^-}}\xspace}

\def\Dbar    {{\ensuremath{\offsetoverline{\PD}}}\xspace}
\def\D       {{\ensuremath{\PD}}\xspace}

\def\DorDbar {\kern \thebaroffset\optbar{\kern -\thebaroffset \PD}\xspace}
\def\Dz      {{\ensuremath{\D^0}}\xspace}
\def\Dzb     {{\ensuremath{\Dbar{}^0}}\xspace}
\def\Dp      {{\ensuremath{\D^+}}\xspace}
\def\Dm      {{\ensuremath{\D^-}}\xspace}

\def\DpDm    {\ensuremath{\Dp {\kern -0.16em \Dm}}\xspace}
\def\Dstar   {{\ensuremath{\D^*}}\xspace}

\def\Dstarp  {{\ensuremath{\D^{*+}}}\xspace}
\def\Dstarm  {{\ensuremath{\D^{*-}}}\xspace}

\def\Ds      {{\ensuremath{\D^+_\squark}}\xspace}
\def\Dsp     {{\ensuremath{\D^+_\squark}}\xspace}

\def\B       {{\ensuremath{\PB}}\xspace}

\def\BorBbar {\kern \thebaroffset\optbar{\kern -\thebaroffset \PB}\xspace}
\def\Bz      {{\ensuremath{\B^0}}\xspace}

\def\Bd      {{\ensuremath{\B^0}}\xspace}

\def\BdorBdbar {\kern \thebaroffset\optbar{\kern -\thebaroffset \Bd}\xspace}
\def\Bu      {{\ensuremath{\B^+}}\xspace}
\def\Bub     {{\ensuremath{\B^-}}\xspace}
\def\Bp      {{\ensuremath{\Bu}}\xspace}
\def\Bm      {{\ensuremath{\Bub}}\xspace}

\def\Bs      {{\ensuremath{\B^0_\squark}}\xspace}

\def\BsorBsbar {\kern \thebaroffset\optbar{\kern -\thebaroffset \Bs}\xspace}

\def\jpsi     {{\ensuremath{{\PJ\mskip -3mu/\mskip -2mu\Ppsi}}}\xspace}

\def\Y#1S{\ensuremath{\PUpsilon{(#1S)}}\xspace}

\def\proton      {{\ensuremath{\Pp}}\xspace}
\def\antiproton  {{\ensuremath{\overline \proton}}\xspace}

\def\Lz          {{\ensuremath{\PLambda}}\xspace}
\def\Lbar        {{\ensuremath{\offsetoverline{\PLambda}}}\xspace}
\def\LorLbar     {\kern \thebaroffset\optbar{\kern -\thebaroffset \PLambda}\xspace}

\def\Xires       {{\ensuremath{\PXi}}\xspace}

\def\Omegares    {{\ensuremath{\POmega}}\xspace}

\def\Lb           {{\ensuremath{\Lz^0_\bquark}}\xspace}
\def\Lbbar        {{\ensuremath{\Lbar{}^0_\bquark}}\xspace}

\def\Xibm         {{\ensuremath{\Xires^-_\bquark}}\xspace}

\def\Omegab       {{\ensuremath{\Omegares^-_\bquark}}\xspace}

\newcommand{\decay}[2]{\ensuremath{#1\!\to #2}\xspace} 

\def\to                 {\ensuremath{\rightarrow}\xspace}

\def\AT#1     {\ensuremath{A_{\mathrm{T}}^{#1}}\xspace}           % 2

\def\C#1      {\ensuremath{\mathcal{C}_{#1}}\xspace}                       % 9
\def\Cp#1     {\ensuremath{\mathcal{C}_{#1}^{'}}\xspace}                    % 7
\def\Ceff#1   {\ensuremath{\mathcal{C}_{#1}^{\mathrm{(eff)}}}\xspace}        % 9  
\def\Cpeff#1  {\ensuremath{\mathcal{C}_{#1}^{'\mathrm{(eff)}}}\xspace}       % 7
\def\Ope#1    {\ensuremath{\mathcal{O}_{#1}}\xspace}                       % 2
\def\Opep#1   {\ensuremath{\mathcal{O}_{#1}^{'}}\xspace}                    % 7

\newcommand{\nospaceunit}[1]{\ensuremath{\text{#1}}}       
\newcommand{\aunit}[1]{\ensuremath{\text{\,#1}}}       
\newcommand{\tev}{\aunit{Te\kern -0.1em V}\xspace}
\newcommand{\gev}{\aunit{Ge\kern -0.1em V}\xspace}
\newcommand{\mev}{\aunit{Me\kern -0.1em V}\xspace}
\newcommand{\kev}{\aunit{ke\kern -0.1em V}\xspace}
\newcommand{\ev}{\aunit{e\kern -0.1em V}\xspace}
 
\newcommand{\mevc}{\ensuremath{\aunit{Me\kern -0.1em V\!/}c}\xspace}
\newcommand{\gevc}{\ensuremath{\aunit{Ge\kern -0.1em V\!/}c}\xspace}
\newcommand{\mevcc}{\ensuremath{\aunit{Me\kern -0.1em V\!/}c^2}\xspace}
\newcommand{\gevcc}{\ensuremath{\aunit{Ge\kern -0.1em V\!/}c^2}\xspace}
\def\mum  {\ensuremath{\,\upmu\nospaceunit{m}}\xspace}

\newcommand{\chisq}{\ensuremath{\chi^2}\xspace}

\newcommand{\chisqip}{\ensuremath{\chi^2_{\text{IP}}}\xspace}

\def\deriv {\ensuremath{\mathrm{d}}}

\def\gsim{{~\raise.15em\hbox{$>$}\kern-.85em
          \lower.35em\hbox{$\sim$}~}\xspace}
\def\lsim{{~\raise.15em\hbox{$<$}\kern-.85em
          \lower.35em\hbox{$\sim$}~}\xspace}

\def\pt         {\ensuremath{p_{\mathrm{T}}}\xspace}

\def\ptot       {\ensuremath{p}\xspace}

\def\evtgen     {\mbox{\textsc{EvtGen}}\xspace}

\def\geant      {\mbox{\textsc{Geant4}}\xspace}

\def\photos     {\mbox{\textsc{Photos}}\xspace}

\def\pythia     {\mbox{\textsc{Pythia}}\xspace}

\def\tell1  {TELL1\xspace}
\def\ukl1   {UKL1\xspace}

\newcommand{\lhcborcid}[1]{\href{https://orcid.org/#1}{\hspace*{0.1em}\raisebox{-0.45ex}{\includegraphics[width=1em]{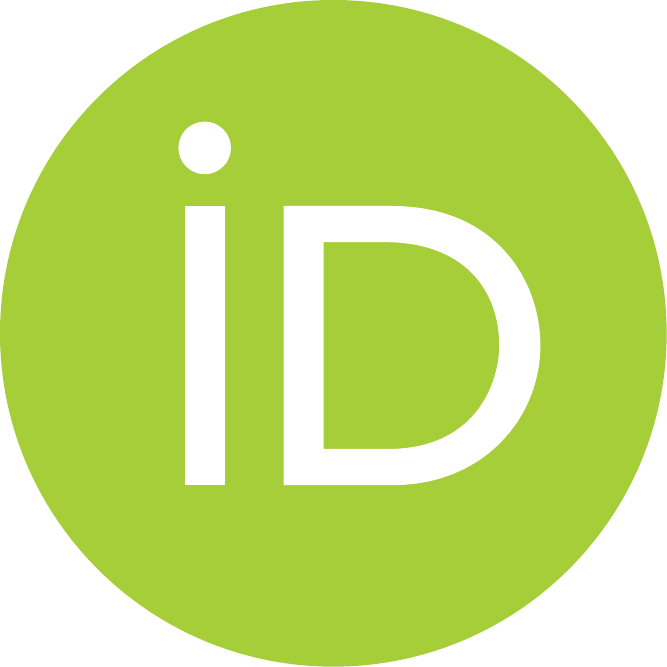}}}}

\usepackage{cite} % Allows for ranges in citations
\usepackage{mciteplus}

\usepackage{longtable} % only for template; not usually to be used in PAPERs

\begin{document}

%%%%%%%%%%%%%%%%%%%%%%%%%
%%%%% Title     %%%%%%%%%
%%%%%%%%%%%%%%%%%%%%%%%%%
\renewcommand{\thefootnote}{\fnsymbol{footnote}}
\setcounter{footnote}{1}

% %%%%%%% CHOOSE TITLE PAGE--------
%\onecolumn
%\input{title-LHCb-PAPER}
%%%%%%%%%%%%%%%%%%%%%%%%%
%%%%%  TITLE PAGE  %%%%%%
%%%%%%%%%%%%%%%%%%%%%%%%%
\begin{titlepage}
\pagenumbering{roman}

% Header ---------------------------------------------------
\vspace*{-1.5cm}
\centerline{\large EUROPEAN ORGANIZATION FOR NUCLEAR RESEARCH (CERN)}
\vspace*{1.5cm}
\noindent
\begin{tabular*}{\linewidth}{lc@{\extracolsep{\fill}}r@{\extracolsep{0pt}}}
\ifthenelse{\boolean{pdflatex}}% Logo format choice
{\vspace*{-1.5cm}\mbox{\!\!\!\includegraphics[width=.14\textwidth]{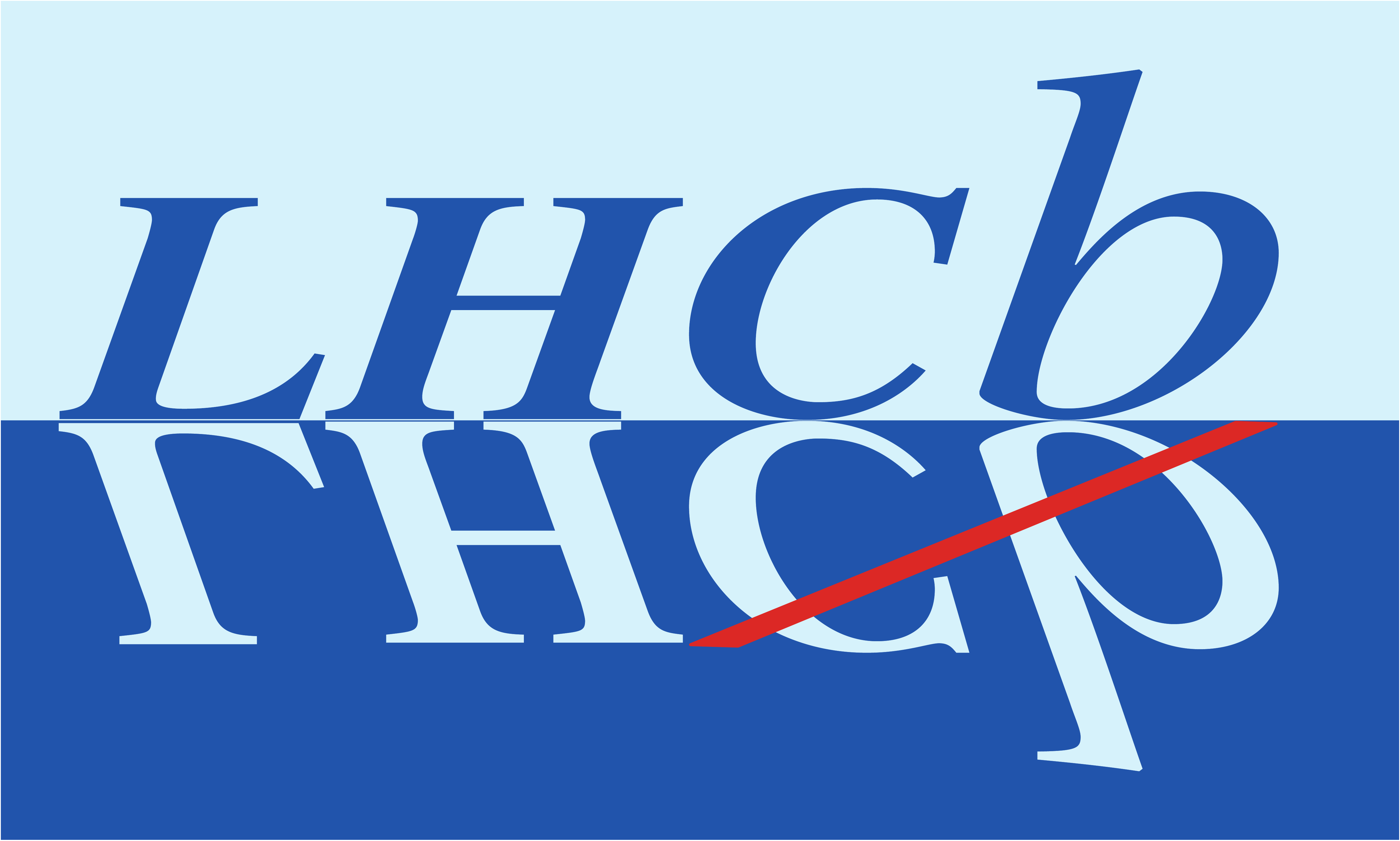}} & &}%
{\vspace*{-1.2cm}\mbox{\!\!\!\includegraphics[width=.12\textwidth]{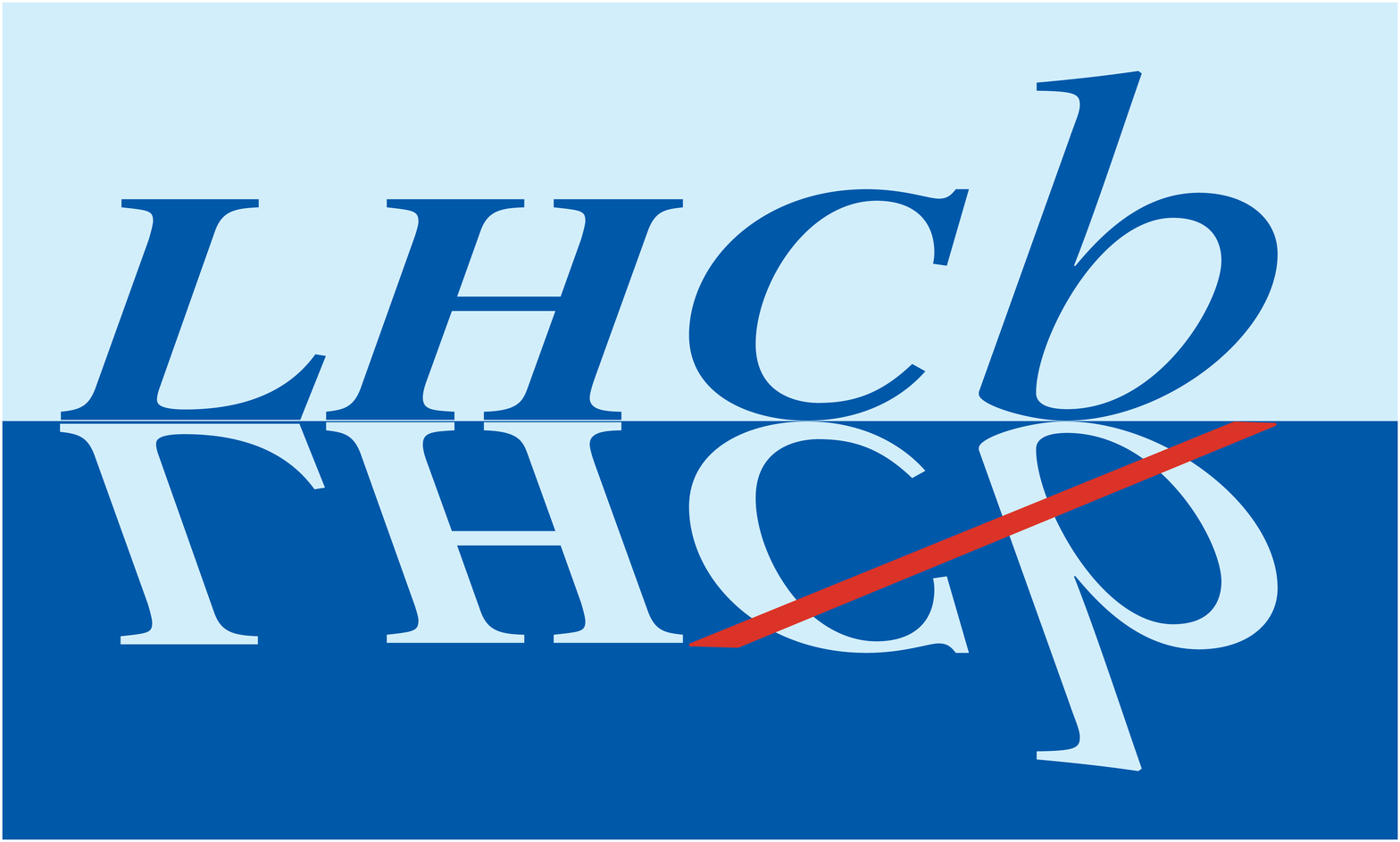}} & &}%
\\
 & & CERN-EP-2022-246 \\  % ID 
 & & LHCb-PAPER-2022-027 \\  % ID 
 & & July 27, 2023 \\ % Date - Can also hardwire e.g.: 23 March 2010
 & & \\
% not in paper \hline
\end{tabular*}

\vspace*{4.0cm}

% Title --------------------------------------------------
{\normalfont\bfseries\boldmath\huge
\begin{center}
% DO NOT EDIT HERE. Instead edit macro in main.tex to keep metadata correct
  \papertitle 
\end{center}
}

\vspace*{2.0cm}

% Authors -------------------------------------------------
\begin{center}
%In the footnote, replace 'paper' by 'Letter' in case of submission to PRL or PLB 
% Edit macro in main.tex to keep metadata correct
\paperauthors\footnote{Authors are listed at the end of this paper.}
\end{center}

\vspace{\fill}

% Abstract -----------------------------------------------
\begin{abstract}
  \noindent
Resonant contributions in $B^0 \rightarrow \offsetoverline{D}{}^0 D^+_s\pi^-$ and $B^+\rightarrow D^- D^+_s\pi^+$ decays are determined with an amplitude analysis, which is performed both separately and simultaneously, where in the latter case isospin symmetry between the decays is assumed. The analysis is based on data collected
by the \mbox{LHCb} detector in proton-proton collisions at center-of-mass energies of 7, 8 and 13 TeV. The full data sample corresponds to an integrated luminosity of 9 $\rm fb^{-1}$. A doubly charged spin-0 open-charm tetraquark candidate together with a neutral partner, both with masses near $2.9\,\rm{GeV}$, are observed in the $D_s\pi$ decay channel. 
  
\end{abstract}

\vspace*{2.0cm}

\begin{center}
  Published in \href{https://journals.aps.org/prd/abstract/10.1103/PhysRevD.108.012017}{Phys. Rev. D 108 (2023) 012017}
\end{center}

\vspace{\fill}

{\footnotesize 
% Edit macro in main.tex to keep metadata correct
\centerline{\copyright~\papercopyright. \href{\paperlicenceurl}{\paperlicence}.}}
\vspace*{2mm}

\end{titlepage}

%%%%%%%%%%%%%%%%%%%%%%%%%%%%%%%%
%%%%%  EOD OF TITLE PAGE  %%%%%%
%%%%%%%%%%%%%%%%%%%%%%%%%%%%%%%%

%  empty page follows the title page ----
\newpage
\setcounter{page}{2}
\mbox{~}
%\twocolumn
% %%%%%%%%%%%%% ---------

\renewcommand{\thefootnote}{\arabic{footnote}}
\setcounter{footnote}{0}

%%%%%%%%%%%%%%%%%%%%%%%%%%%%%%%%
%%%%%  Table of Content   %%%%%%
%%%%%%%%%%%%%%%%%%%%%%%%%%%%%%%%
%%%% Uncomment if desired
%\tableofcontents
\cleardoublepage

%%%%%%%%%%%%%%%%%%%%%%%%%
%%%%% Main text %%%%%%%%%
%%%%%%%%%%%%%%%%%%%%%%%%%

\pagestyle{plain} % restore page numbers for the main text
\setcounter{page}{1}
\pagenumbering{arabic}

%% Uncomment during review phase. 
%% Comment before a final submission.
%\linenumbers

%% This is the main body
%\input{1_IntroductionAndSelection}
%\input{2_DalitzPlotFit}
%%%%%%%%%% symbol define
\def\Zz         {{\ensuremath{T^a_{c\bar{s}0}(2900)}}\xspace}
\def\Zzz        {{\ensuremath{T^a_{c\bar{s}0}(2900)^0}}\xspace}
\def\Zzpp       {{\ensuremath{T^a_{c\bar{s}0}(2900)^{++}}}\xspace}

\def\Zzzo       {{\ensuremath{T^a_{c\bar{s}1}(2900)^0}}\xspace}
\def\Zzppo      {{\ensuremath{T^a_{c\bar{s}1}(2900)^{++}}}\xspace}

\def\DzKpi      {{\ensuremath{\D^0_{K\pi}}}\xspace}
\def\DzbKpi     {{\ensuremath{\Dbar{}^0_{K\pi}}}\xspace}
\def\DzKtpi      {{\ensuremath{\D^0_{K3\pi}}}\xspace}
\def\DzbKtpi     {{\ensuremath{\Dbar{}^0_{K3\pi}}}\xspace}

\def\BztoDzbarDsppim        {{\ensuremath{\Bz \rightarrow \Dzb \Dsp \pim}}\xspace}
\def\BztoDzbarKpiDsppim     {{\ensuremath{\Bz \rightarrow \DzbKpi \Dsp \pim}}\xspace}
\def\BztoDzbarKtpiDsppim    {{\ensuremath{\Bz \rightarrow \DzbKtpi \Dsp \pim}}\xspace}
\def\BptoDmDsppip           {{\ensuremath{\Bp \rightarrow \Dm \Dsp \pip}}\xspace}

%%%%%%%%%%%%%%%%%%%%%%%%%%%%%%%%%%%%%%%%%%%%%%%%%
%%%%%           Introduction                %%%%%
%%%%%%%%%%%%%%%%%%%%%%%%%%%%%%%%%%%%%%%%%%%%%%%%%
\section{Introduction}
\label{sec:Introduction}

The decays of $b$ hadrons into final states involving two open-charm hadrons form a large family of topologically similar processes that 
include many intermediate states such as charmonia, highly excited $D_{\left(s\right)}$ states, and possible exotic hadrons. The Dalitz plot distributions of $\Bz\to\Dz\Dm\Kp$, $\Bp\to\Dz\Dzb\Kp$ and $\Bp\to\Dp\Dm\Kp$ decays\footnote{Charge conjugation is implied throughout this paper.} have already been explored by the Belle~\cite{Belle:2007hht}, BaBar~\cite{BaBar:2014jjr} and LHCb collaborations~\cite{LHCb-PAPER-2020-024,LHCb-PAPER-2020-025}. In these studies, the discovery of the charm-strange meson $D_{s1}(2700)^+$, the charmonium-like state $\chi_{c0}(3930)$, and the open-charm tetraquark state $X_{0,1}(2900)$, were reported, prompting many theoretical investigations into the internal structure of these states~\cite{Zhu:2022arXiv}.

\begin{figure}[!b]
  \begin{center}
    \includegraphics[width=0.45\linewidth]{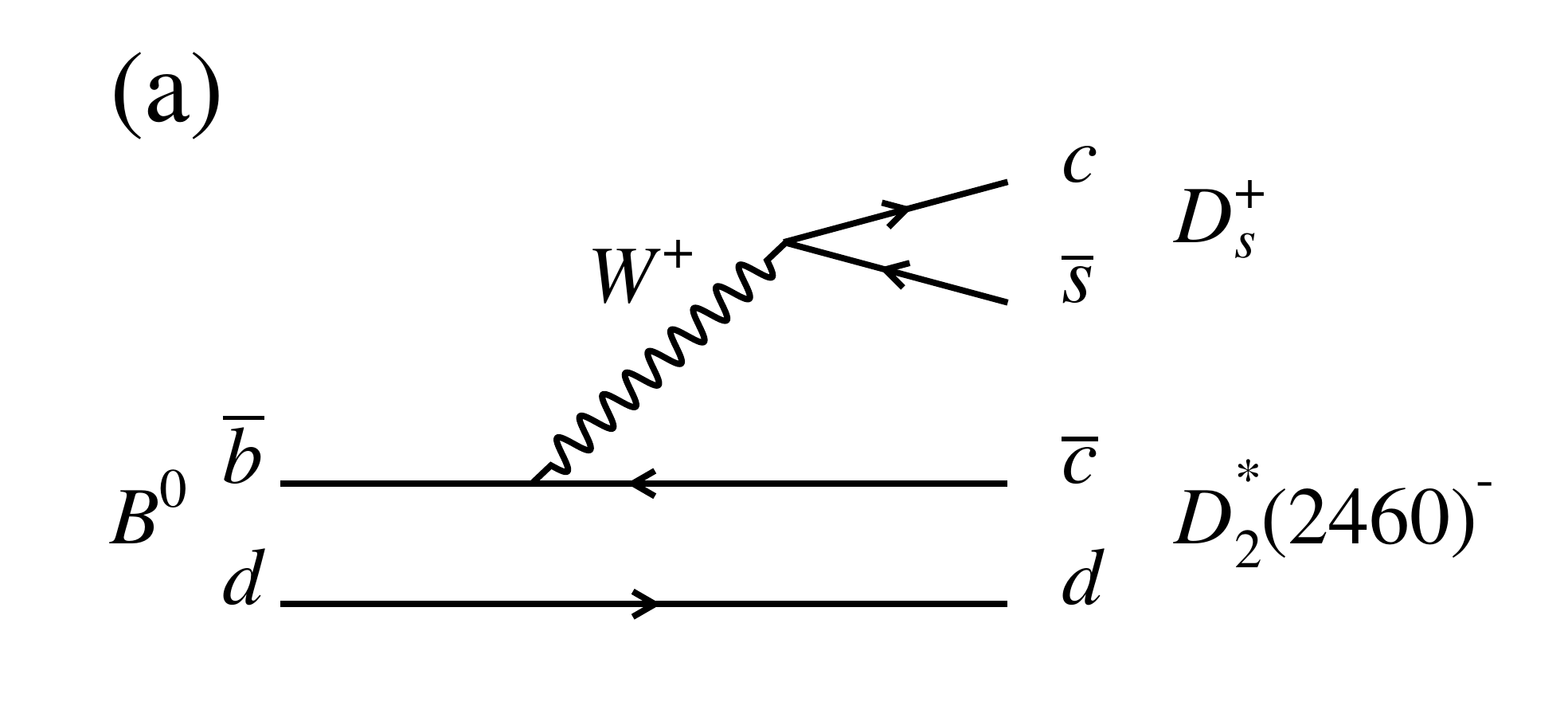}
    \includegraphics[width=0.45\linewidth]{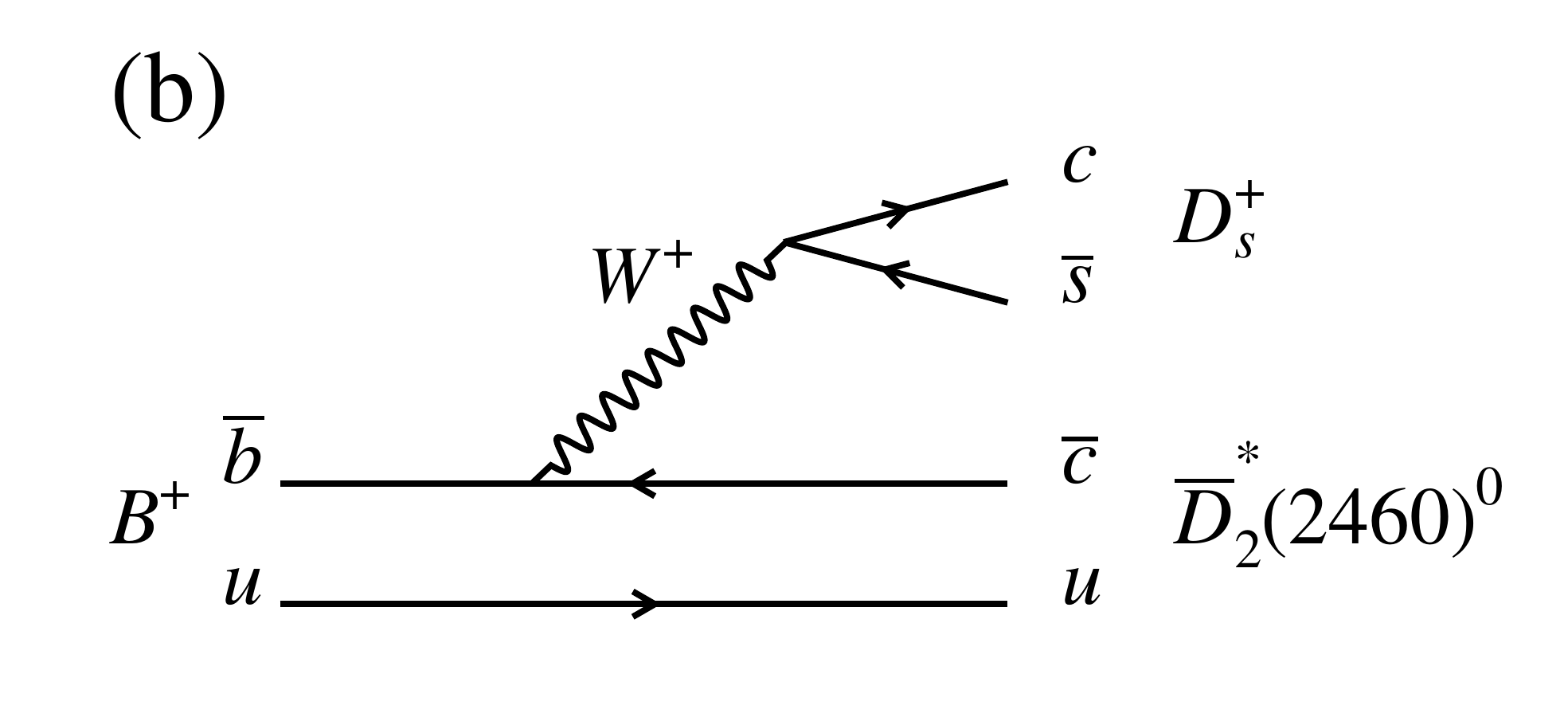}\\
    \includegraphics[width=0.45\linewidth]{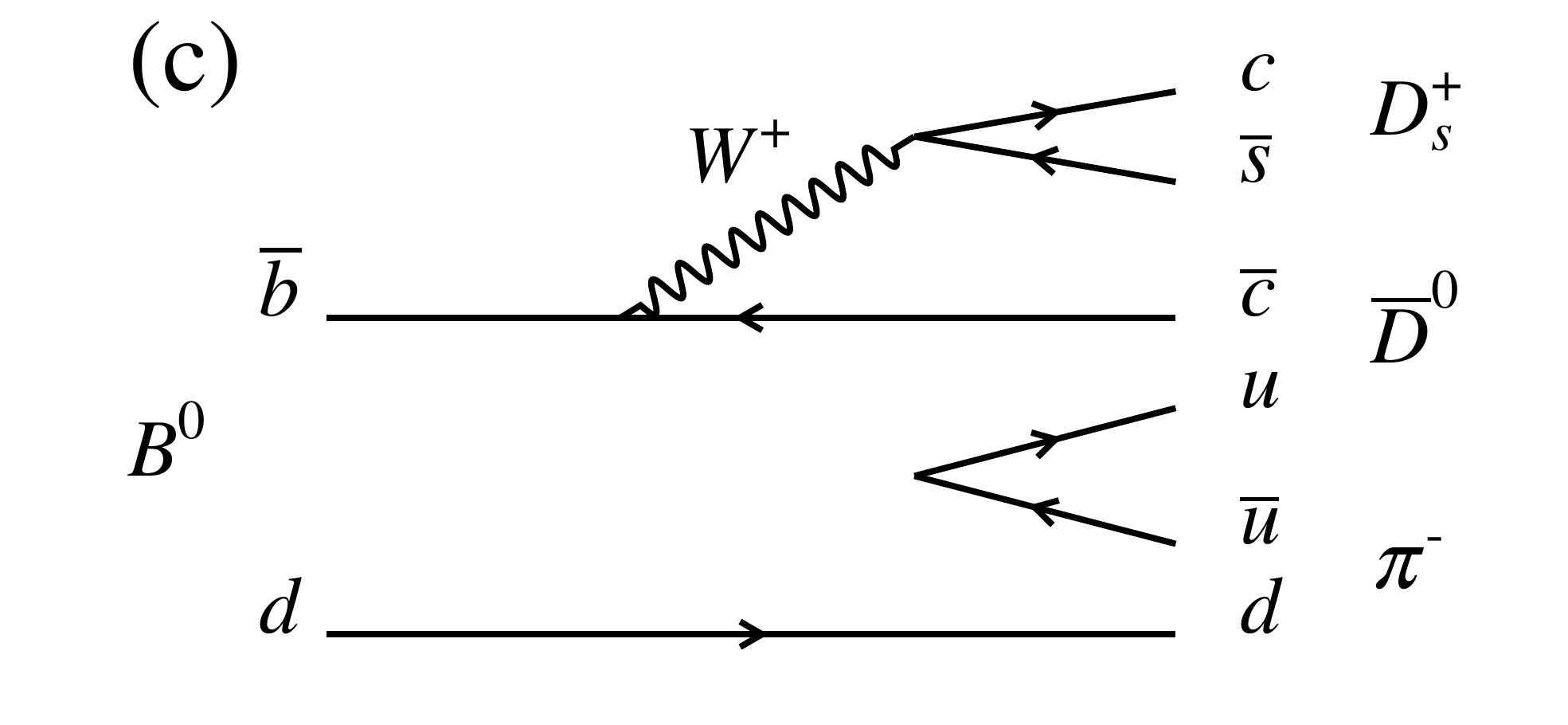}
    \includegraphics[width=0.45\linewidth]{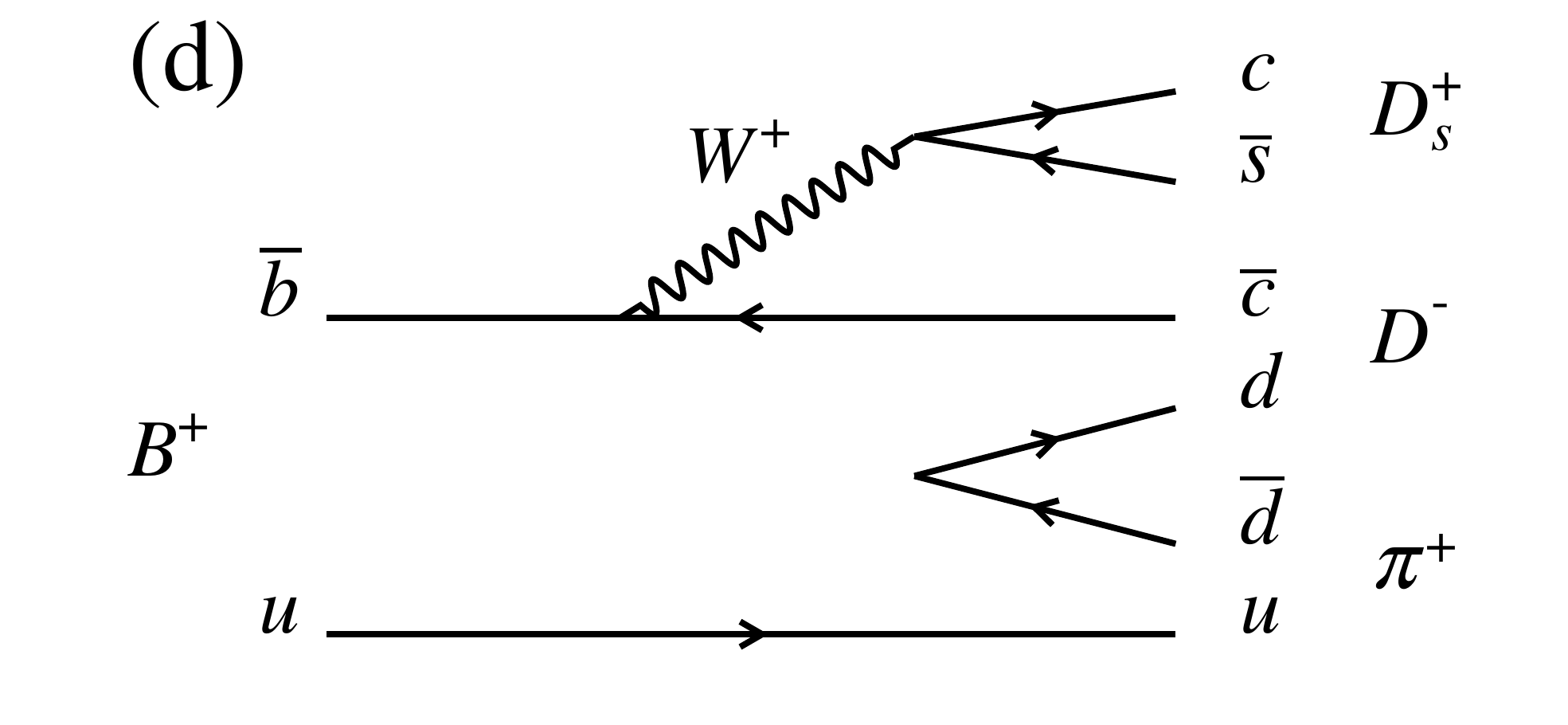}

  \end{center}
  \caption{Feynman diagrams for the dominant tree-level amplitudes contributing to (a) \mbox{\BztoDzbarDsppim} and (b) \BptoDmDsppip decays with intermediate $D\pi$ resonances; and nonresonant three body decays of (c) \BztoDzbarDsppim and (d) \BptoDmDsppip decays.}
  \label{fig:Feynman}
\end{figure}

The decays $\BptoDmDsppip$ and $\BztoDzbarDsppim$ are yet to be explored. They are ideal to study excited $D$ mesons ($D^{**}$) with natural spin-parity, to test isospin symmetry in the charged and neutral $D\pi$ resonances, and to test quantum chromodynamics (QCD) predictions~\cite{Chen:2016spr}. The $D^*(2007)^0$, $D^*(2010)^+$, $D_0^*(2300)$, and $D_2^*(2460)$ mesons are already well-established. The $D_1^*(2600)^0$ and $D_J^*(3000)^0$ mesons were recently discovered in the inclusive proton-proton ($pp$) collisions and in $B$ decays~\cite{PDG2022}, while their charged isospin partners have not been observed, although some measurements suggest their existence~\cite{LHCb-PAPER-2013-026}. These states could also be explored in $B \to \Dbar \Ds \pi$ decays.
Figure~\ref{fig:Feynman} shows the Feynman diagrams of the dominant tree-level amplitudes contributing to the two decays.

Studies of $B \to \Dbar \Ds \pi$ decays also provide an excellent opportunity to search for exotic hadrons decaying into the $\Ds\pi$ and $\Dbar \Ds$ final states. The discoveries of the $D^{*}_{s0}(2317)^+$~\cite{BaBar:2003oey} and $D_{s1}(2460)^+$~\cite{CLEO:2003ggt} states prompted speculation that they may have a tetraquark component~\cite{PDG2022,Chen:2016spr}. No evidence for isospin partners has been found in explicit searches~\cite{BaBar:2006eep,Belle:2015glz}, but if they exist they should contribute to the $B \to \Dbar \Ds \pi$ decays. The D0 collaboration claimed evidence for an $X(5568)$ state~\cite{D0:2016mwd,D0:2017qqm}, which however was not confirmed by other experiments~\cite{LHCb-PAPER-2016-029,CMS:2017hfy,ATLAS:2018udc,CDF:2017dwr}. An open-charm tetraquark state with four different quark flavors, analogous to the $X(5568)$ state, is predicted by the diquark-antidiquark model ~\cite{Agaev:2016lkl,Chen:2017rhl}, and can be investigated in $\Dsp\pi^\pm$ final states.
In particular, the $\Dsp\pi^+$ channel presents an attractive potential according to lattice QCD calculation~\cite{Baeza-Ballesteros:2022azb}, which can form a tetraquark resonance. 
Some theoretical studies on the $X_{0,1}(2900)$ state suggest searching for a potential doubly charged charm-strange tetraquark candidate, together with its neutral isospin partner, in the $\Dsp\pip[c\bar{s}u\bar{d}]$ and $\Dsp\pim[c\bar{s}\bar{u}d]$ final states~\cite{He:2020jna,Lu:2020qmp,Burns:2020xne,Agaev:2021knl,Agaev:2021jsz,Azizi:2021aib}.
In addition, searches for possible $\Dbar \Ds$ resonances are well-motivated by the recent observations of the open-strange hidden-charm tetraquark state $Z_{cs}(3985)$ decaying into $\overline{D}{}^* D_s^+ + \Dbar D_s^{*+}$ at BESIII~\cite{BESIII:2020qkh,BESIII:2022qzr}, as well as the $Z_{cs}(4000)$ and $Z_{cs}(4220)$ states decaying into $K J/\psi$ at LHCb~\cite{LHCb-PAPER-2020-044}.

In this paper, an amplitude analysis of $\BptoDmDsppip$ and $\BztoDzbarDsppim$ decays is presented for the first time, revealing the contributions of $\Dbar\pi$ resonances in the two decays, and allowing searches for possible exotic states. As the two channels are closely related by isospin, three different fit scenarios are performed: a fit performed independently on the two decay channels is called the separate fit; a simultaneous fit of the \BztoDzbarDsppim and \BptoDmDsppip decays, by assuming that all the $\Dbar\pi$ resonances in the two decays are isospin-related, is denoted the simultaneous $\Dbar\pi$ fit; a simultaneous fit, in which all the parameters of $\Dbar\pi$ states and potential $\Ds\pi$ or $D\Ds$ resonances are shared between the two decays, is called the full simultaneous fit. The separate fit and simultaneous $\Dbar\pi$ fit are discussed in this paper, while the full simultaneous fit, which is considered as the default result of this analysis, is described in Ref.~\cite{PRLsister}. The analysis is based on $pp$ collision data collected using the \lhcb detector, corresponding to a total integrated luminosity of $3~\rm fb^{-1}$ at center-of-mass energies $\sqrt{s}=7,~8\tev$, referred to as Run 1, and $6~\rm fb^{-1}$ at $\sqrt{s}=13\tev$, referred to as Run 2.

The paper is organized as follows. A brief introduction of \lhcb detector and reconstruction and simulation procedures is provided in Sec.~\ref{sec:LHCbDetector}. The event selection criteria are shown in Sec.~\ref{sec:Selection}, and signal and background yields are determined in Sec.~\ref{sec:InvMassFit}.  The formalism of amplitude analysis is summarized in Sec.~\ref{sec:Analysis formalism}. The signal efficiency and background models in amplitude analysis are studied in Sec.~\ref{sec:EffAndBG}. The separate fit result is presented in Sec.~\ref{sec:Model description}, while the result from simultaneous $\Dbar\pi$ fit is provided in Sec.~\ref{sec:simuDpi}. The systematic uncertainties are evaluated in Sec.~\ref{sec:Systematic uncertainties}. All the results are summarized in Sec.~\ref{sec:Summary}.

%%%%%%%%%%%%%%%%%%%%%%%%%%%%%%%%%%%%%%%%%%%%%%%%%%%%%%%%%%%%%%%%
%%%%%          LHCb detector and Simulation                %%%%%
%%%%%%%%%%%%%%%%%%%%%%%%%%%%%%%%%%%%%%%%%%%%%%%%%%%%%%%%%%%%%%%%
\section{LHCb detector and simulation}
\label{sec:LHCbDetector}

The \lhcb detector~\cite{LHCb-DP-2008-001,LHCb-DP-2014-002} is a single-arm forward
spectrometer covering the \mbox{pseudorapidity} range $2<\eta <5$,
designed for the study of particles containing \bquark or \cquark
quarks. The detector includes a high-precision tracking system
consisting of a silicon-strip vertex detector surrounding the $pp$
interaction region, a large-area silicon-strip detector located
upstream of a dipole magnet with a bending power of about
$4{\mathrm{\,Tm}}$, and three stations of silicon-strip detectors and straw
drift tubes
placed downstream of the magnet.
The tracking system provides a measurement of the momentum, \ptot, of charged particles with
a relative uncertainty that varies from 0.5\% at low momentum to 1.0\% at 200\gev.\footnote{Natural units with $\hbar=c=1$ are used throughout this paper.}
The minimum distance of a track to a primary $pp$ collision vertex (PV), the impact parameter (IP), 
is measured with a resolution of $(15+29/\pt)\mum$,
where \pt is the component of the momentum transverse to the beam, in\,\gev.
Different types of charged hadrons are distinguished using information
from two ring-imaging Cherenkov detectors. 
Photons, electrons and hadrons are identified by a calorimeter system consisting of
scintillating-pad and preshower detectors, an electromagnetic
% calorimeter
and a hadronic calorimeter. Muons are identified by a
system composed of alternating layers of iron and multiwire
proportional chambers.
The online event selection is performed by a trigger, 
which consists of a hardware stage, based on information from the calorimeter and muon
systems, followed by a software stage, which applies a full event
reconstruction.

At the hardware trigger stage, events are required to have a muon with high \pt or a
hadron, photon or electron with high transverse energy in the calorimeters. For hadrons,
the transverse energy threshold is 3.5\gev.
The software trigger requires a two-, three- or four-track
secondary vertex with a significant displacement from any primary
$pp$ interaction vertex. At least one charged particle
must have a transverse momentum $\pt > 1.6\gev$ and be
inconsistent with originating from a PV.
A multivariate algorithm~\cite{BBDT,LHCb-PROC-2015-018} is used for
the identification of secondary vertices consistent with a decay
of a \bquark hadron.

Simulation is used to model the effects of the detector acceptance and the
imposed selection requirements.
In the simulation, $pp$ collisions are generated using
\pythia~8~\cite{Sjostrand:2007gs,Sjostrand:2006za} 
with a specific \lhcb configuration~\cite{LHCb-PROC-2010-056}.
Decays of unstable particles
are described by \evtgen~\cite{Lange:2001uf}, in which final-state
radiation is generated using \photos~\cite{davidson2015photos}.
The interaction of the generated particles with the detector, and its response,
are implemented using the \geant
toolkit~\cite{Allison:2006ve, *Agostinelli:2002hh} as described in
Ref.~\cite{LHCb-PROC-2011-006}. 
The underlying $pp$ interaction is reused multiple times, with an independently generated signal decay for each~\cite{LHCb-DP-2018-004}.

The particle identification (PID) response for charged tracks in the simulated samples is corrected based on special samples of $D^{*+}\to\Dz\pip$, $\Dz\to\Km\pip$ decays. For each PID response of a track, the unbinned four-dimensional probability density functions (PDF) for the data, $p_{\textrm{data}}(x|\pt, \eta, N_{\textrm{tr}})$, and for the simulated samples $p_{\textrm{sim}}(x|\pt, \eta, N_{\textrm{tr}})$ are extracted based on a kernel density estimation~\cite{Poluektov:2014rxa}, where $x$ is the PID response, $\pt$ and $\eta$ are the transverse momentum and pseudorapidity of the track, and $N_{\textrm{tr}}$ is the number of tracks in the event. The cumulative distribution functions for the data $P_{\textrm{data}}(x|\pt, \eta, N_{\textrm{tr}})$ and for the simulated samples $P_{\textrm{sim}}(x|\pt, \eta, N_{\textrm{tr}})$ are determined, and the corrected PID response in the simulated samples is evaluated by transforming the $x_{\textrm{sim}}$ into $x_{\textrm{corr}}$ with
\begin{equation}
    x_{\textrm{corr}} = P^{-1}_{\textrm{data}}\left(P_{\textrm{sim}}\left(x_{\textrm{sim}}|\pt, \eta, N_{\textrm{tr}}\right)|\pt, \eta, N_{\textrm{tr}}\right).
    \label{eq:PIDCorr}
\end{equation}
During the transformation, the $N_{\textrm{tr}}$ distribution in the simulated samples is scaled by a factor to match the same distribution in the corresponding datasets. The PID response in the simulated samples shows good agreement with that in the data samples after the correction.

The momentum scale is calibrated using control samples of $\decay{\jpsi}{\mumu}$ 
and \mbox{$\decay{\Bu}{\jpsi\Kp}$}~decays collected concurrently
with the~data samples used for this analysis~\cite{LHCb-PAPER-2012-048,LHCb-PAPER-2013-011}.
The~relative uncertainty on the momentum scale is $3 \times 10^{-4}$.

%%%%%%%%%%%%%%%%%%%%%%%%%%%%%%%%%%%%%%%%%%%%%%%%%%%%%%%%%%%%%%
%%%%%                     Selection                      %%%%%
%%%%%%%%%%%%%%%%%%%%%%%%%%%%%%%%%%%%%%%%%%%%%%%%%%%%%%%%%%%%%%
\section{Selection}
\label{sec:Selection}

In \lhcb, trigger decisions are associated with reconstructed particles.
Selection requirements can therefore be made on the trigger selection itself
and on whether the decision was due to the signal candidate, other particles produced in the $pp$ collision, or a combination of both.

In the analysis, the \BztoDzbarDsppim and \BptoDmDsppip candidates are formed using charged kaon and pion candidates, in which the $\Dzb$ candidates is reconstructed through the $\Dzb\to\Kp\pim$ and $\Dzb\to\Kp\pim\pim\pip$ decays, $\Dp$ through the $\Dp\to\Km\pip\pip$ decays, and $\Dsp$ through the $\Dsp\to\Kp\Km\pip$ decays.
The invariant mass of $\Dzb\pim$ is required to be larger than $2.05\gev$ in order to veto the contribution from the $\Bz\to\Dstarm\Dsp$ decay, which is not the focus of this analysis. To consider potential variations of signal efficiencies and background distributions, selections are designed and optimized separately for the six datasets, namely Run 1 and Run 2 datasets with three reconstruction channels, $\Bz \rightarrow \Dzb_{\Kp\pim} \Dsp \pim$, $\Bz \rightarrow \Dzb_{\Kp\pim\pim\pip} \Dsp \pim$, and \BptoDmDsppip decay.

%To begin with obtaining visible $B$ signals, 
Loose requirements on the $p$, $p_{\rm T}$, PID response, and  minimum \chisqip of the charged tracks are first applied to improve track quality and remove tracks originating directly from the $pp$ collision. Here \chisqip\ is defined as the difference in the vertex-fit \chisq of a given PV reconstructed with and without the considered track.
To separate $B$ and $D_{(s)}$ candidates from random combinations of tracks directly produced in $pp$ collisions, loose requirements on the invariant mass and on the $\chisq$ of the flight distance of $D$ candidates with respect to the associated PV are imposed, where the associated PV is defined as the PV yielding the smallest \chisqip\ for the considered $B$ candidate.
The quantity $\cos{\theta_{\textrm{dir}}}$ for a $B$ candidate, where $\theta_{\textrm{dir}}$ is the angle between the momentum direction and the vector from the associated PV to the vertex of the candidate, is required to be close to unity.

A boosted decision tree~(BDT) classifier~\cite{Breiman,AdaBoost} implemented in the TMVA toolkit~\cite{Hocker:2007ht,*TMVA4} is used to further suppress the combinatorial background. The BDT classifier depends on the $\chisqip$ of the $B$, $D$, $D_s$ candidates and the final tracks, the $\chisq$ of the $B$ decay vertex and its $\cos{\theta_{\textrm{dir}}}$, the PID response of all the final tracks, and the signed significance of the separation of $D_{(s)}$ and $B$ vertices parallel to the beam pipe ($s_{z-\rm{FD}}^{D_{(s)}}$). For the BDT classifier training, the signal samples are the simulated signal candidates, % from \lhcb simulation framework,
and the backgrounds are $B$ sideband candidates in data with $B$ invariant mass within $[5500, 6950]\mev$. The requirement on the BDT response is determined by maximising $S^2/\left(S+B\right)^\frac{3}{2}$, where the $S$ and $B$ are the expected signal and background yields in the signal mass window $|M(B)-m(B)|<20\mev$, which is $2.5$ to $3$ times wider than the mass resolution for the different channels, where the $M(B)$ and $m(B)$ are the reconstructed and known masses of the corresponding $B$ meson~\cite{PDG2022}, respectively.

Two kinds of misidentified (misID) backgrounds are vetoed through additional requirements. A background to the \BptoDmDsppip signal occurs at the mass threshold of the $\Dsp\pip$ spectrum due to $\ep-\pip$ misidentification. It is removed by a stringent requirement on the PID response of the companion pion.
For \BztoDzbarDsppim, $\Bs\to\Dzb\Dsp\Km$ and $\Lbbar\to\Dzb\Dsp\antiproton$ decays, misID candidates are vetoed by tightening PID requirements of the companion \pim for candidates with invariant mass within $\pm 30\mev$ of the $\Bs$ or $\Lbbar$ mass~\cite{PDG2022}, after replacing the mass hypothesis of the companion $\pim$ to $\Km$ or $\antiproton$. 

A track-swapped background is found to peak in the $B$ signal region where the $\pim$ from $\Dstarm$ in the  $\Bz\to\Dstarm\Dsp$, $\Dstarm\to\Dzb\pim$, $\Dzb\to\Kp\pim\pim\pip$ candidates is swapped with a $\pim$ in the $\Dzb$ decays. As the momentum of the $\pim$ in $\Dstarm$ decays is almost zero in the $\Dstarm$ rest frame, this type of background is removed by requiring $M(K^+\pi^-\pi^-\pi^+)-M(K^+\pi^-\pi^+)>160\mev$ in the $\Dzb$ decays, to veto those $\pim$ with negligible momentum.

Backgrounds with the same final-state tracks but with one or even zero intermediate charmed mesons, called non-double-charm (NDC) backgrounds, are suppressed by the requirements on $s_{z-\rm{FD}}^{D_{(s)}}$. After the selection, the invariant mass distributions of all possible two-body and three-body combinations of the final charged tracks in $D_{(s)}$ sideband samples are further checked, and all the visible narrow structures, namely $K^*(892)^0$, $\phi$, $\Dz$, $\Dp$ and $\Dsp$ particles, are vetoed, to further suppress the NDC backgrounds.

After applying the full offline selection, the $D$ and $D_s$ masses are required to lie within $\pm15\mev$ around their known mass~\cite{PDG2022}, corresponding to two to three times the detector resolution.
Roughly 1\% to 2\% of events contain more than one $B$ candidate; one is retained at random in these cases.

To improve the resolution of $B$ candidate invariant mass distributions, a fit based on the Kalman filter method~\cite{Hulsbergen:2005pu}, which contains topological and kinematic information of the decay chain, is applied. By updating the four-momentum of all the final-state tracks, the invariant mass of $D$ and $D_s$ candidates is constrained to the their known masses~\cite{PDG2022}, and the updated invariant mass distributions of $B$ candidates are used to determine the signal and background yields. The four-momenta of charged tracks from another fit, which additionally constrains $B$ candidate mass to the known $B$ mass, are used to calculate kinematic variables used in the amplitude fit.

%%%%%%%%%%%%%%%%%%%%%%%%%%%%%%%%%%%%%%%%%%%%%%%%%%%%%%%%%%%%%
%%%%%           B meson invariant mass fit              %%%%%
%%%%%%%%%%%%%%%%%%%%%%%%%%%%%%%%%%%%%%%%%%%%%%%%%%%%%%%%%%%%%
\section{Signal yield determination}
\label{sec:InvMassFit}

\begin{figure}[!tb]
  \begin{center}
    \includegraphics[width=0.45\linewidth]{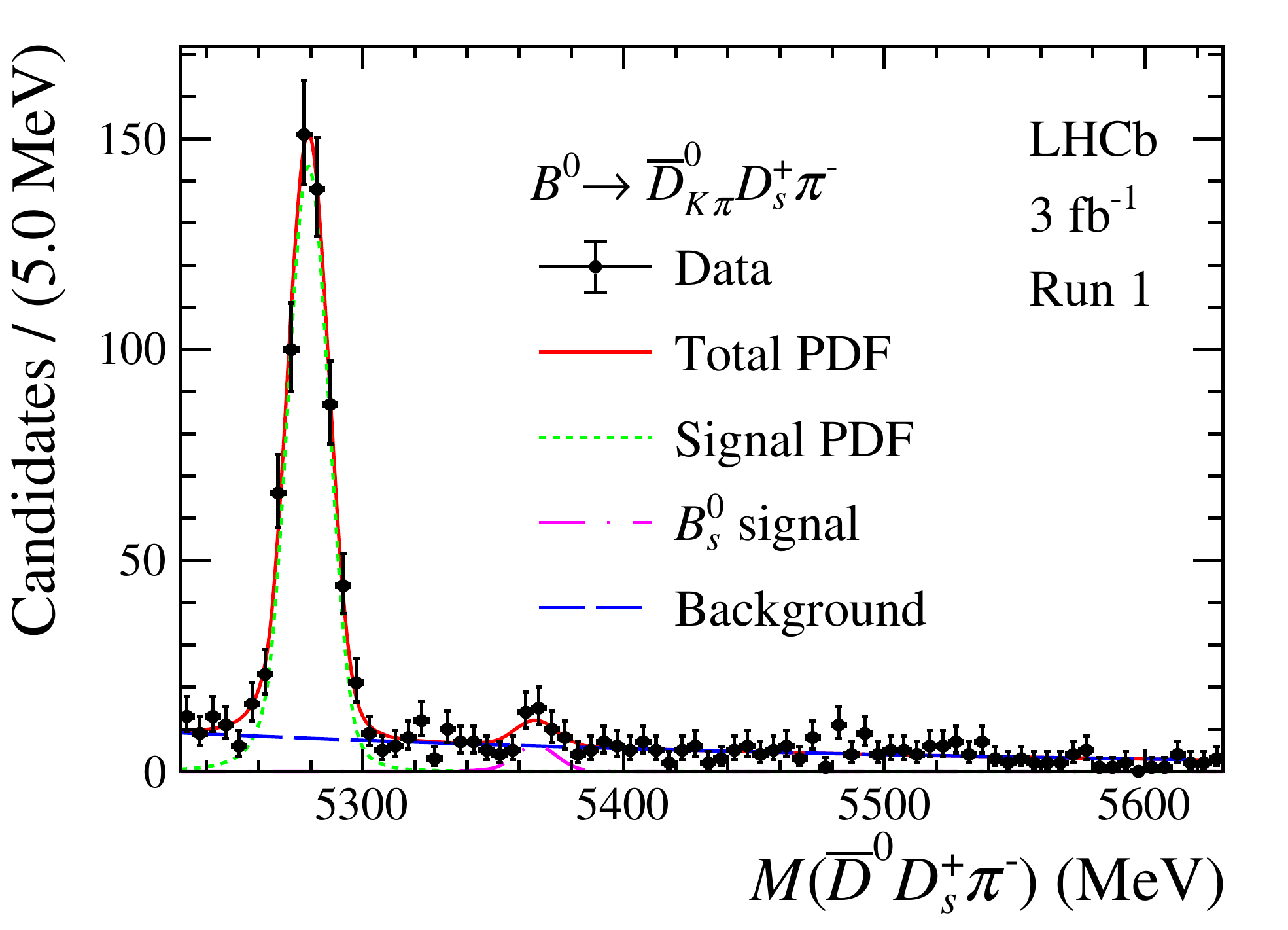}
    \includegraphics[width=0.45\linewidth]{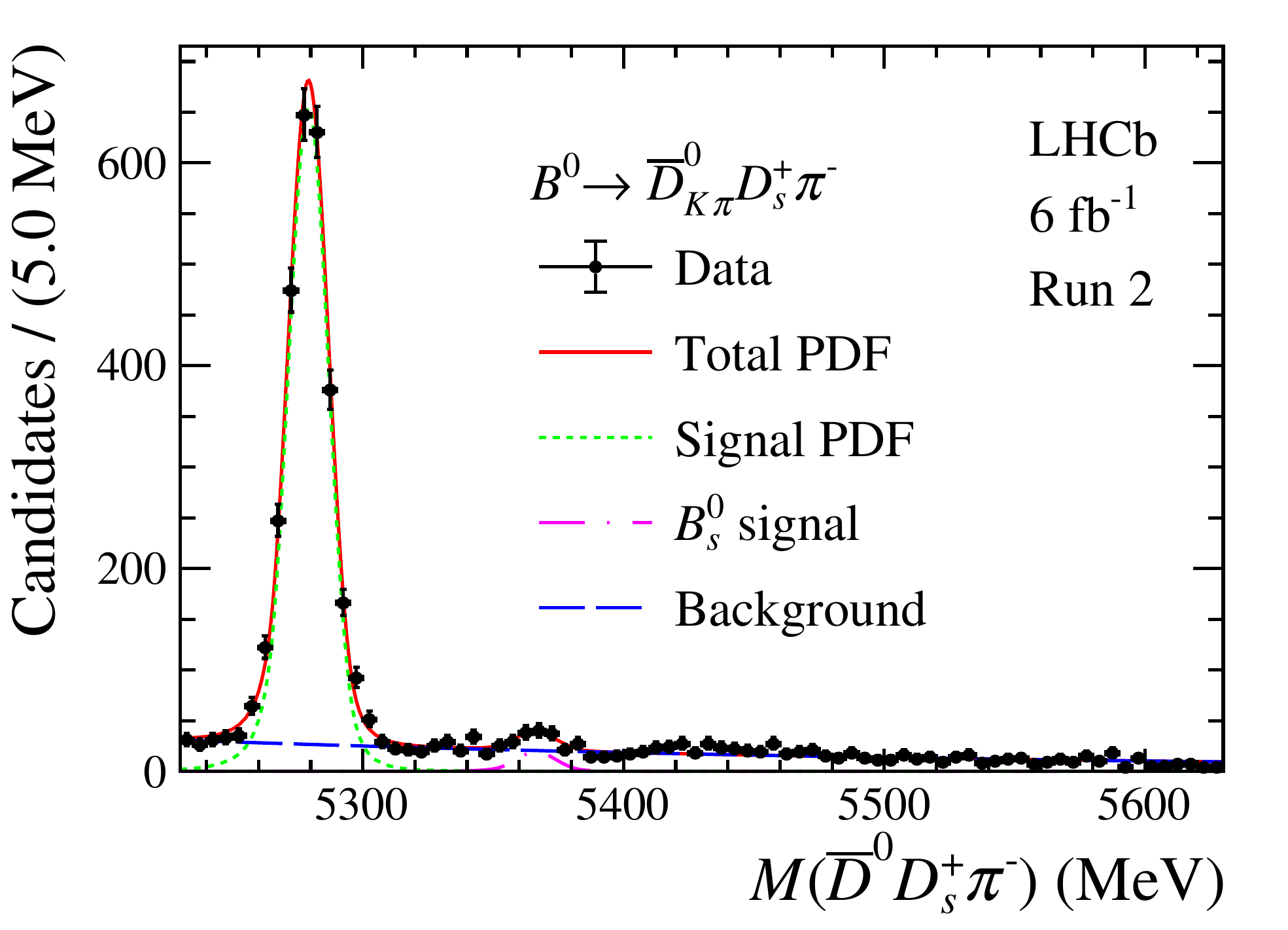}\\
    \includegraphics[width=0.45\linewidth]{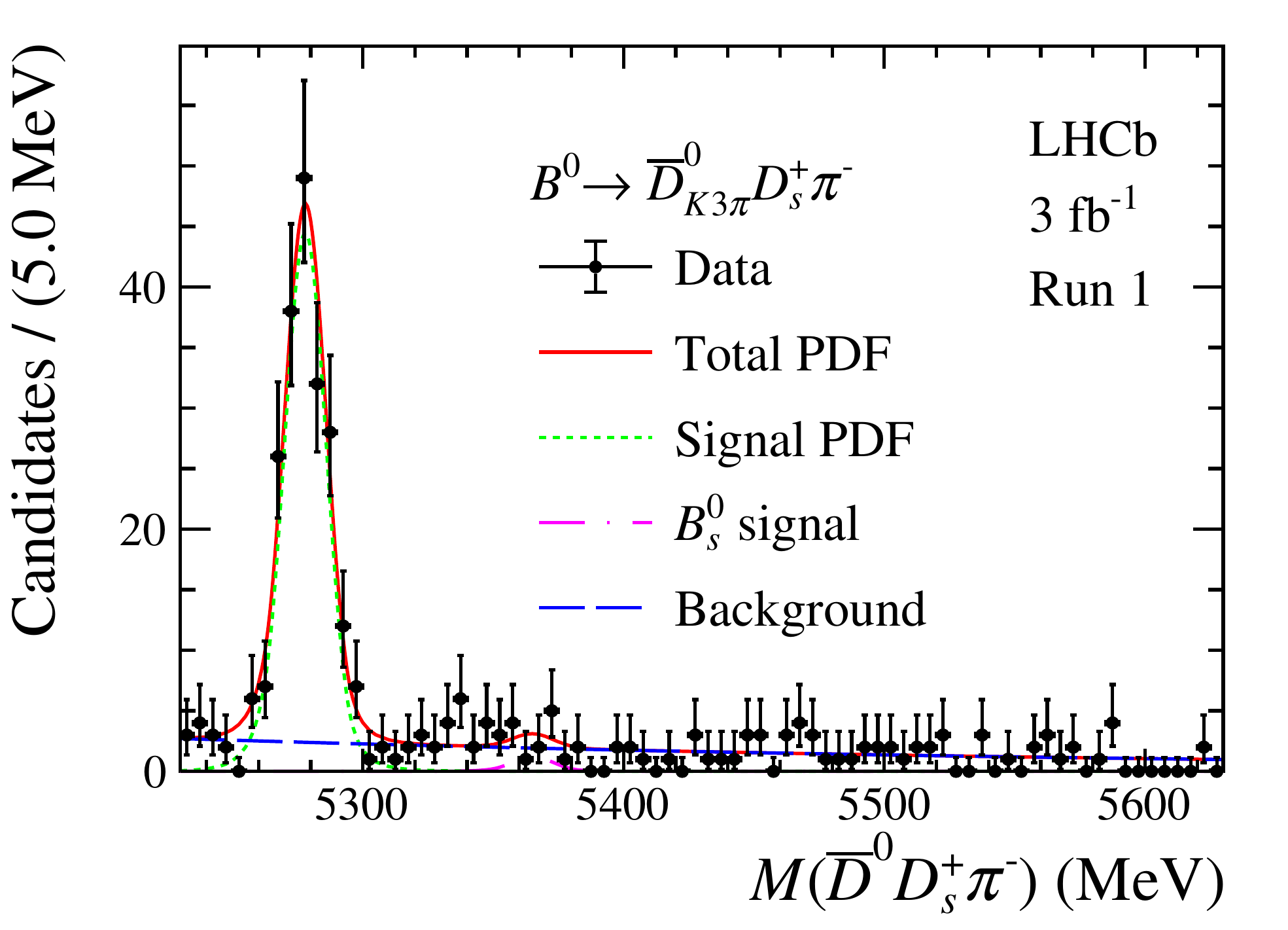}
    \includegraphics[width=0.45\linewidth]{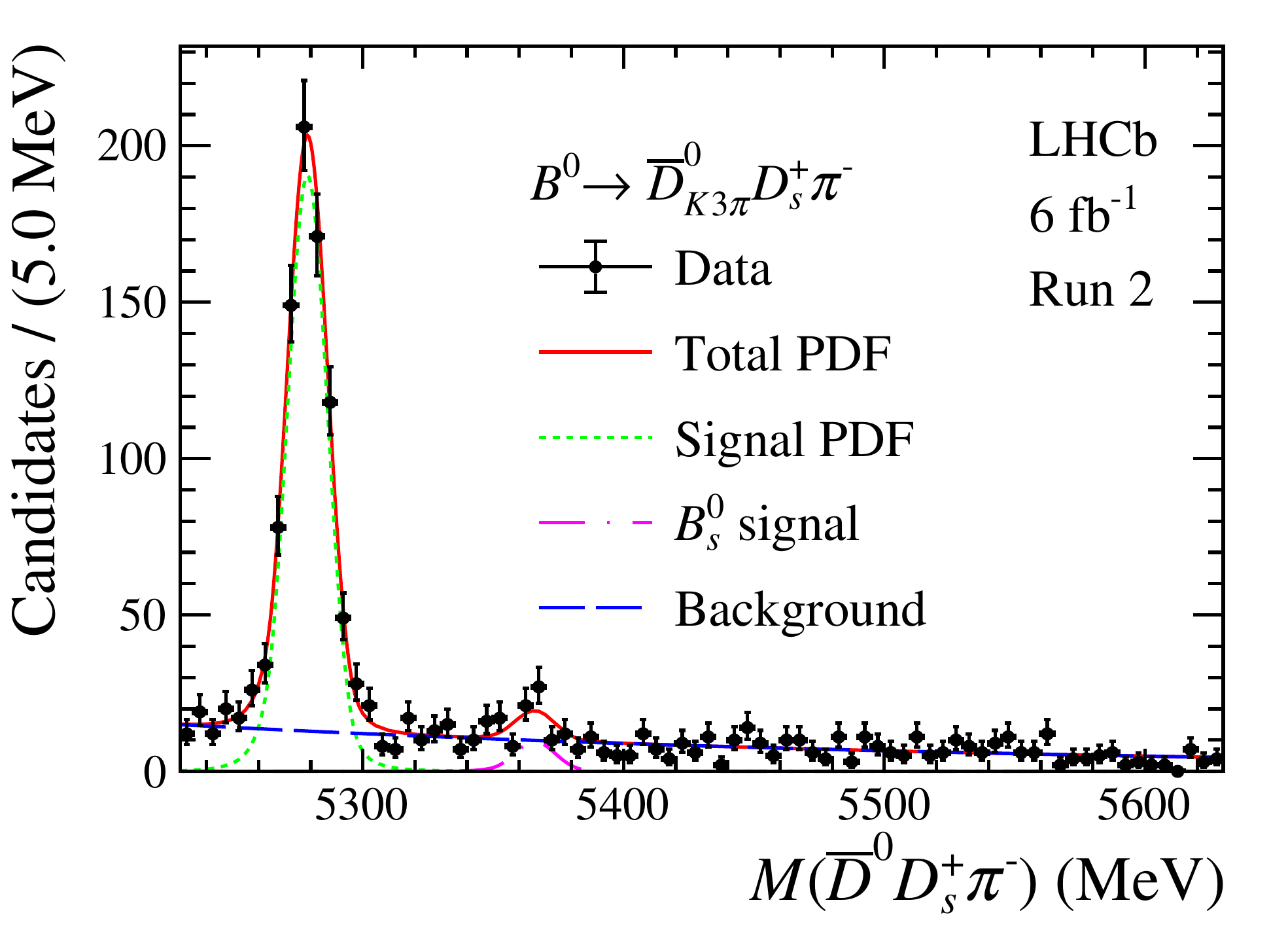}\\
    \includegraphics[width=0.45\linewidth]{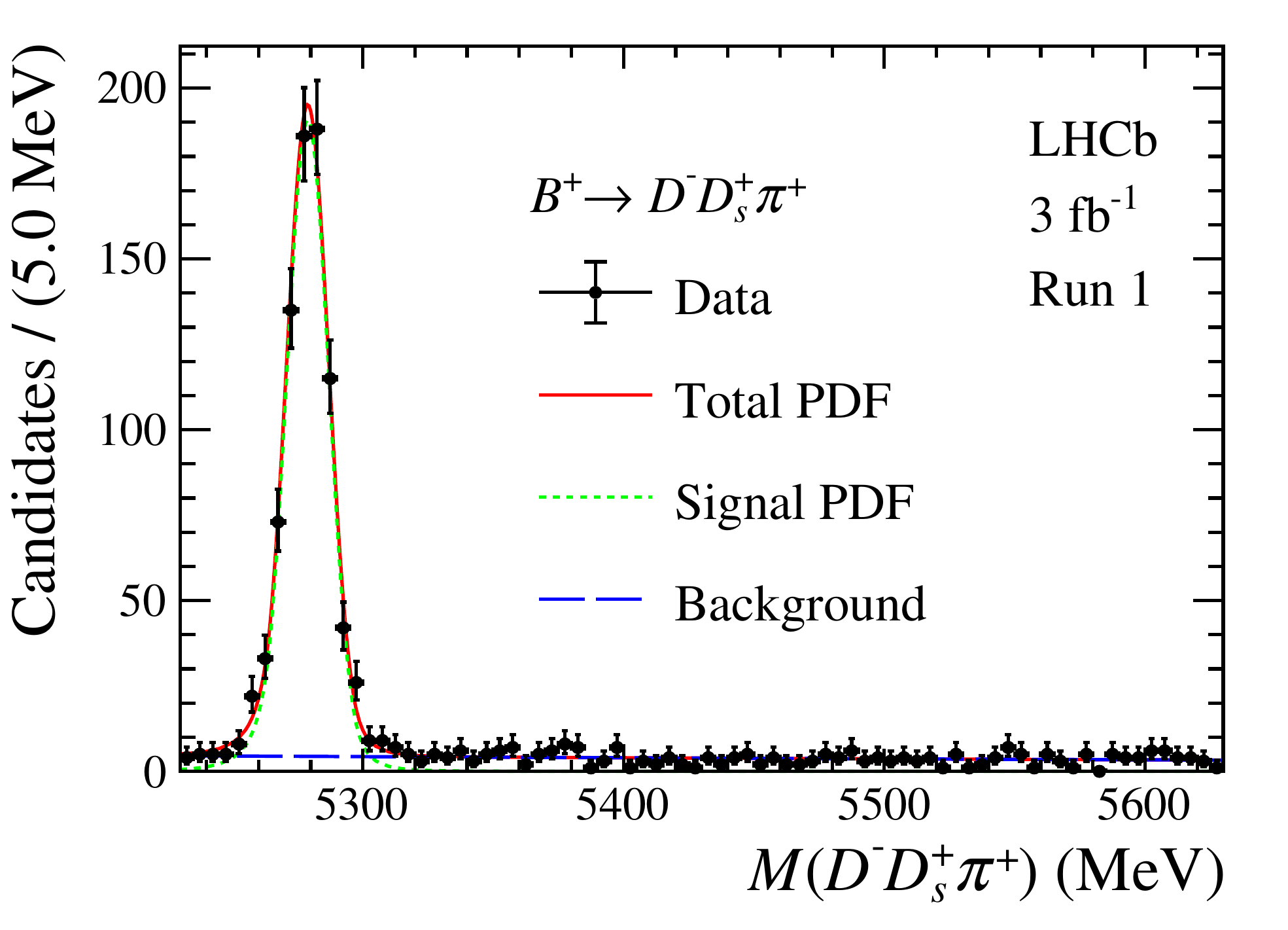}
    \includegraphics[width=0.45\linewidth]{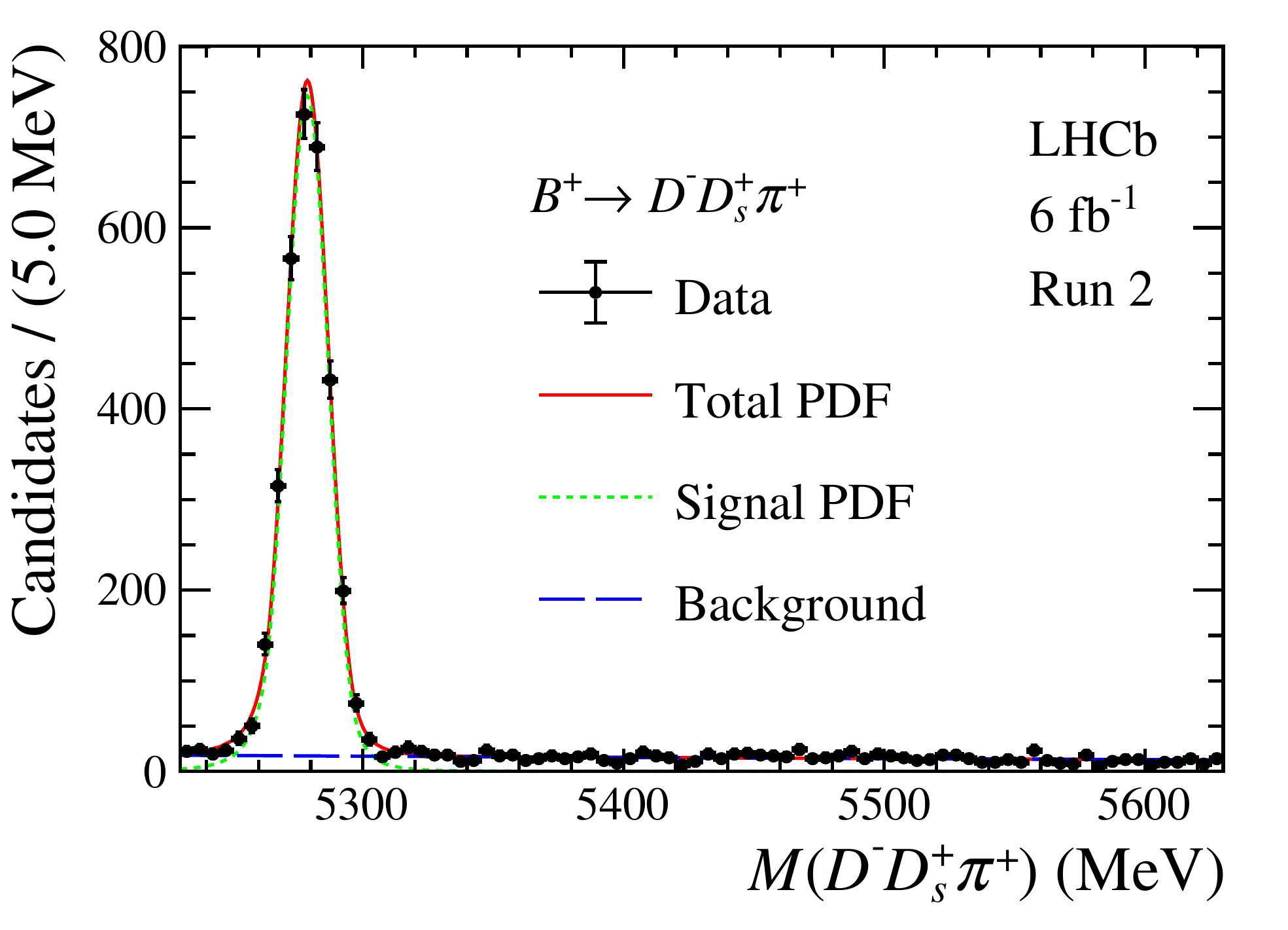}
  \end{center}
  \caption{Invariant mass spectrum of the signal candidates, split by decay mode and run period. The data are overlaid with the results of the fit.}
  \label{fig:FitToData}
\end{figure}

Figure~\ref{fig:FitToData} shows the invariant mass distributions of the $\Bz$ and $\Bp$ candidates in each dataset after the application of the selection requirements. An extended unbinned maximum-likelihood fit to the data in the mass range $[5230, 5630]\mev$ is used to extract the signal and background yields, which are used later in the amplitude fit. 

The total PDF comprises a signal PDF and an exponential function to describe the distribution of combinatorial background. 
The signal PDF is a double-sided Crystal Ball (DSCB) function~\cite{Skwarnicki:1986xj} which consists of a Gaussian kernel and independent tail parameters on both the left and right sides to model effects such as the detector resolution and final-state radiation. In the fit, the signal and background yields, the parameters of the exponential function, and those of the Gaussian kernel in the DSCB function are allowed to vary independently in each dataset, while the tail parameters are fixed to the values obtained from a fit to the corresponding simulated sample. 

In the $\Dzb\Dsp\pim$ mode, an additional DSCB function is included in the fit model to describe the singly Cabibbo-suppressed $\Bs\to\Dzb\Dsp\pim$ contribution, whose yield is determined from data. The width of the Gaussian kernel is shared with the $\Bz$ signal and the mean value $\mu_{\Bs}$ is defined as $\mu_{\Bz} + (m(\Bs)-m(\Bz))$, where the $\mu_{\Bz}$ is the mean value of the Gaussian kernel of the $\Bz$ signal PDF, and $m(\Bz)$ and $m(\Bs)$ are the known masses of the $\Bz$ and $\Bs$ mesons~\cite{PDG2022}. The tail parameters are fixed to the same values as the $\Bz$ signal.

The results of the fit to the data samples are shown in Fig.~\ref{fig:FitToData}. The values of the fitted parameters are listed in Table~\ref{tab:FitResults}, and Table~\ref{tab:ResultsForDalitzFit} summarises the signal yields, the number of candidates, and the purity inside the signal mass window used for the amplitude analysis.

\begin{table}[!tb]
  \caption{Results of the fit parameters of invariant mass fit to the data samples. The uncertainties shown are statistical.}
\centering
\resizebox{\textwidth}{!}{
\begin{tabular}{l|l|ll}
    \hline
    Decay                   & Parameter             & Run 1                         & Run 2                     \\
    \hline   
                            & Signal yield          & $587\pm27$                    & $2641\pm57$               \\
    \BztoDzbarKpiDsppim     & $\Bs$ signal          & $25.3\pm8.3$                  & $77\pm15$                 \\
                            & Background yield      & $421\pm26$                    & $1440\pm49$               \\
                            & Mean (MeV)            & $5279.12\pm0.38$              & $5279.16\pm0.18$          \\
                            & Width (MeV)           & $7.89\pm0.35$                 & $7.73\pm0.17$             \\
                            & Exponential slope     & $-(3.08\pm0.52)\times10^{-3}$ & $-(2.98\pm0.29)\times10^{-3}$ \\
    \hline
                            & Signal yield          & $185\pm15$                    & $759\pm32$                \\
    \BztoDzbarKtpiDsppim    & $\Bs$ signal          & $4.9\pm4.6$                   & $38\pm11$                 \\
                            & Background yield      & $136\pm14$                    & $692\pm33$                \\
                            & Mean (MeV)            & $5277.98\pm0.70$              & $5278.79\pm0.34$          \\
                            & Width (MeV)           & $8.01\pm0.59$                 & $7.72\pm0.33$             \\
                            & Exponential slope     & $-(2.56\pm0.90)\times10^{-3}$ & $-(3.03\pm0.41)\times10^{-3}$ \\
    \hline
                            & Signal yield          & $798\pm30$                    & $3123\pm59$               \\
    \BptoDmDsppip           & Background yield      & $311\pm21$                    & $1201\pm40$               \\
                            & Mean (MeV)            & $5278.88\pm0.33$              & $5278.74\pm0.16$          \\
                            & Width (MeV)           & $8.08\pm0.30$                 & $8.05\pm0.14$             \\
                            & Exponential slope     & $-(0.82\pm0.61)\times10^{-3}$ & $-(0.90\pm0.31)\times10^{-3}$ \\
    \hline
    \end{tabular}
    }
\label{tab:FitResults}
\end{table}

\begin{table}[!tb]
  \caption{
    Signal and background yields inside the $B$ mass signal window, together with the signal purity, split by run period and decay mode. The uncertainties shown are statistical.
  }
\begin{center}\begin{tabular}{l|l|ll}
    \hline
    Decay                   & Parameter         & Run 1             & Run 2         \\
    \hline   
                            & Signal yield      & $564\pm26$        & $2534\pm55$   \\
    \BztoDzbarKpiDsppim     & Total candidates  & 633               & 2753          \\
                            & Purity            & 89.1\%            & 92.1\%        \\
    \hline
                            & Signal yields     & $177\pm14$        & $734\pm31$   \\
    \BztoDzbarKtpiDsppim    & Total candidates  & 199               & 835           \\
                            & Purity            & 88.9\%            & 87.9\%        \\
    \hline
                            & Signal yield      & $766\pm29$        & $2984\pm57$   \\
    \BptoDmDsppip           & Total candidates  & 797               & 3143          \\
                            & Purity            & 96.1\%            & 94.9\%        \\
    \hline
    \end{tabular}\end{center}
\label{tab:ResultsForDalitzFit}
\end{table}

%%%%%%%%%%%%%%%%%%%%%%%%%%%%%%%%%%%%%%%%%%%%%%%%%%%%%
%%%%%           Analysis formalism              %%%%%
%%%%%%%%%%%%%%%%%%%%%%%%%%%%%%%%%%%%%%%%%%%%%%%%%%%%%
\section{Analysis formalism}
\label{sec:Analysis formalism}
The amplitude formalism and fit method of three-body $B\to abc$ decays, where $abc$ denotes any sequence of $\Dbar$, $\Ds$ and $\pi$ states, is established in this Section.
\subsection{Amplitude model}
\label{sec:Amplitude model}
The amplitude of the three-body $\B\to \Dbar D_s\pi$ decays is constructed following the isobar formalism~\cite{PhysRev.135.B551,PhysRev.166.1731,PhysRevD.11.3165}, which is a coherent sum of quasi two-body amplitudes, either resonant or nonresonant, 
\begin{equation}
    \mathcal{A}(x;\Theta)=\sum c_i\cdot\mathcal{A}_i(x;\Theta_i),
\end{equation}
where $c_i$ is a complex parameter for the $i$-th contribution that is determined from data, $x$ denotes variables calculated from the four-momenta of the final-state particles and $\Theta_i$ is a set of parameters used to describe the $i$-th lineshape. 
The amplitude of the $i$-th quasi two-body decay to $a$ and $b$ ($a$, $b$ represent any pair of $D$, $D_s$, $\pi$ mesons) is 
\begin{equation}
    \mathcal{A}_i(x;\Theta_i)=T(\theta_{ab})\cdot f(m_{ab}^2;\Theta_i),
\end{equation}
where $T(\theta_{ab})$ describes the angular distribution which depends on the spin $J$ of the intermediate resonant state $R(ab)$. The helicity angle, $\theta_{ab}$, is defined as the angle between the $R(ab)$ momentum direction in the $\B$ rest frame, and the momentum direction of $a$ as determined in the $R(ab)$ rest frame. The definitions of $T(\theta_{ab})$ up to $J = 4$ are
\begin{equation}
\label{eq:Angular}
    T(\theta_{ab}) = \begin{cases}
        1 & J = 0, \\
        \cos{\theta_{ab}} & J = 1,\\
        \cos^2{\theta_{ab}} - \frac{1}{3} & J = 2,\\
        \cos^3{\theta_{ab}} - \frac{3}{5} \cos{\theta_{ab}} & J = 3,\\
        \cos^4{\theta_{ab}} - \frac{30}{35} \cos^2{\theta_{ab}}+\frac{3}{35} & J = 4.
    \end{cases}
\end{equation}
The function $f(M_{ab})$ 
is the lineshape of the $R(ab)$ resonance where $M_{ab}$ is the invariant mass of the pair. The complex relativistic Breit--Wigner (RBW) function is used as the default lineshape,
\begin{align}
    f_{RBW}(M) = q(M)^{L_{1}}F(M,L_{1})\cdot p(M)^{L_{2}}F(M,L_{2})\cdot \frac{1}{m_{0}^{2}-M^{2}-i m_{0}\Gamma(M)},
\label{eqn:Breit-Wigner}
\end{align}
where $p$ is the momentum of particle $a$ in the rest frame of the resonance $R(ab)$, and $q$ denotes the momentum of $R(ab)$ in the $\B$ rest frame.
The mass-dependent running width is
\begin{align}
    \Gamma(M) = \Gamma_{0}\left(\frac{q(M)}{q_{0}}\right)^{2L_{2}+1}\frac{m_{0}}{M}F^{2}(M,L_{2}),
\label{eqn:running width}
\end{align}
where $m_{0}$ and $\Gamma_{0}$ are the mass and width of the resonance, respectively. The quantities $p_{0}$ and $q_{0}$ are these momenta evaluated when $M = m_0$. The orbital angular momentum between  $R(ab)$ and $c$ is denoted by $L_{1}$, while $L_{2}$ refers to the orbital angular momentum between particles $a$ and $b$. Conservation of angular momentum implies that $L_{1} = L_{2} = J$.
The Blatt-Weisskopf form-factor~\cite{VonHippel:1972fg} $F(M,L)$ is parameterized as
\begin{equation}
        F\left(M,L\right) = \begin{cases}
                1 & L = 0, \\
                \sqrt{\frac{1+z^{2}(M)}{1+z_{0}^{2}}} & L = 1,\\
                \sqrt{\frac{9+3z^{2}(M)+z^{4}(M)}{9+3z^{2}_{0}+z^{4}_{0}}} & L = 2,\\
                \sqrt{\frac{225+45z^{2}(M)+6z^{4}(M)+z^{6}}{225+45z^{2}_{0}+6z^{4}_{0}+z_{0}^{6}}} & L = 3,
        \end{cases}
   \label{eq:BWFactor}
\end{equation}
where $z(M) = pd$, $z_{0}=p_{0}d$, and $d$ stands for the radial parameter, which is taken to be $3.0\gev^{-1}$ by default for all resonances. 

Nonresonant (NR) contributions are parameterized using an exponential function, 
\begin{align}
    f_{NR} = & \exp{\left[-\alpha\left(M^2- m^2_{\rm min}(D\pi)\right)\right]},
\label{eqn:exponential}
\end{align}
where $\alpha$ is the slope parameter that is allowed to vary in the fit, and $m^2_{\rm min}(D\pi) = 4 \gev^2$ is an approximation of the $M^2(D\pi)$ lower threshold.

The RBW functions do not provide an adequate description of overlapping resonant states.
Furthermore, the latest experimental~\cite{LHCb-PAPER-2016-026} and theoretical~\cite{Du:2020pui} studies show that a simple BW lineshape is not sufficient for the $\overline{D}{}^*_0(2300)$ resonance.
A quasi-model-independent (qMI) parameterization~\cite{LHCb-PAPER-2016-026} is used for the $D\pi$ S-wave, where the $M(D\pi)$ range is divided into $k$ slices, and the line shape is replaced by a set of complex coefficients assigned to each slice, each free to vary in the fit. The real and imaginary parts are independently interpolated using cubic splines. Details are given in Sec.~\ref{sec:Model description}. 

\subsection{Maximum likelihood fit}
The normalized PDF for the signal is expressed as
\begin{align}\label{eq:sigpdf}
    P^{\rm norm}_{\rm sig}\left(x;\Theta\right) & = \frac{1}{I_{\rm sig}\left(\Theta\right)}\epsilon(x)\vert\mathcal{A}\left(x;\Theta\right)\vert^{2}.
\end{align}
The normalization factor, $I_{\rm sig}\left(\Theta\right)$, which is obtained by integrating over the phase space using simulated samples after full selection and thus including $\epsilon(x)$ implicitly, can be expressed as
\begin{align}\label{eq:sigint}
    I_{\rm sig}\left(\Theta\right) & = \frac{\sum_{j}w_{j}\vert\mathcal{A}\left(x_j;\Theta\right)\vert^{2}}{\sum_{j}w_{j}}.
\end{align}
The signal efficiency, $\epsilon(x)$, is obtained as described in Sec.~\ref{sec:EffAndBG}. The $w_{j}$, also described in Sec.~\ref{sec:EffAndBG}, are applied to the simulated events to correct for discrepancies between simulated samples and data, where the subscript $j$ runs over all the events of the sample. The resolutions of the Dalitz plot coordinates are much smaller than the widths of the narrowest structures and thus related effects can be neglected. The total PDF is given by
\begin{align}
    {\rm  PDF}\left(x;\Theta\right) &= f_{\rm sig}P^{\rm norm}_{\rm sig}\left(x;\Theta\right) + f_{\rm bkg}P^{\rm norm}_{\rm bkg}(x),
\end{align}
where $f_{\rm sig}$ and $f_{\rm bkg}$ are the fractions of signal and background contributions, determined from the fit to the $m(D\D_s\pi)$ invariant mass distributions, and $P^{\rm norm}_{\rm bkg}(x)$ is the normalized background PDF described in Sec.~\ref{sec:EffAndBG}.

An amplitude fit is performed, minimising the unbinned negative log-likelihood
\begin{align}
\begin{split}
    \mathrm{NLL} \equiv -\ln\mathcal{L} = - \sum_{j}\ln {\rm  PDF}(x_{j}; \Theta).
\end{split}
\end{align}
The signal efficiency map and background map are obtained separately for different samples. For the \mbox{\BztoDzbarDsppim} decay, the different LHC Run (1 or 2) and reconstruction channels (\mbox{$\Bz \rightarrow \Dzb_{\Kp\pim} \Dsp \pim$}, \mbox{$\Bz \rightarrow \Dzb_{\Kp\pim\pim\pip} \Dsp \pim$}) are fitted simultaneously. For the \BptoDmDsppip decay, the Run 1 and Run 2 datasets are simultaneously fitted. Where isospin symmetry is imposed, 
the fit is performed simultaneously on all datasets.

The fit fraction $\mathcal{F}_i$ for a given contribution $i$ is calculated from the fitted parameters, $\Theta_{0}$, and is defined as
\begin{align}
    \mathcal{F}_{i} = \frac{\int\vert c_i\mathcal{A}_{i}(x;\Theta_{0,i})\vert^2 \deriv\boldsymbol{x}}{\int\vert\mathcal{A}(x;\Theta_{0})\vert^2 \deriv\boldsymbol{x}}.
\end{align}
The fit fractions do not necessarily add up to 1 due to interference effects between components. The interference term between any pair of components is defined as
\begin{align}
    \mathcal{F}_{ij} = \frac{\int 2\mathcal{R}(c_i c_j^*\mathcal{A}_{i}\mathcal{A}^{*}_{j})\deriv\boldsymbol{x}}{\int\vert\mathcal{A}(x;\Theta_{0})\vert^2 \deriv\boldsymbol{x}},
\end{align}
and thus we have $\sum_{i}\mathcal{F}_{i}+\sum_{i< j}\mathcal{F}_{ij}=1$.

%%%%%%%%%%%%%%%%%%%%%%%%%%%%%%%%%%%%%%%%%%%%%%%%%%%%%%%%%%%%%%%%%%%%
%%%%%           Signal efficiency and background models        %%%%%
%%%%%%%%%%%%%%%%%%%%%%%%%%%%%%%%%%%%%%%%%%%%%%%%%%%%%%%%%%%%%%%%%%%%
\section{Signal efficiency and background models}
\label{sec:EffAndBG}

The amplitude analysis is only sensitive to signal efficiency variations across the Dalitz plot, not the absolute efficiency. These are extracted for each dataset as a function of position in the square Dalitz plot (SDP), whose coordinates are defined by
\begin{eqnarray}
    & m'(D\pi)=\frac{1}{\pi}{\rm arccos}\left(2\times\frac{m\left(D\pi\right)-m^{\rm min}_{D\pi}}{m^{\rm max}_{D\pi}-m^{\rm min}_{D\pi}}-1\right),\\
    & \theta'(D\pi)=\frac{1}{\pi}\theta\left(D\pi\right).
    \label{eq:mPAndThetaP}
\end{eqnarray}
Here $m(D\pi)$ is the invariant mass of the $D\pi$ combination, $m^{\rm min}_{D\pi}$ and $m^{\rm max}_{D\pi}$ are the kinematic limits of $m(D\pi)$ in $B\to D D_s \pi$ decays, and $\theta(D\pi)$ is the $D\pi$ helicity angle. 

\begin{figure}[!tb]
  \begin{center}
    \includegraphics[width=0.45\linewidth]{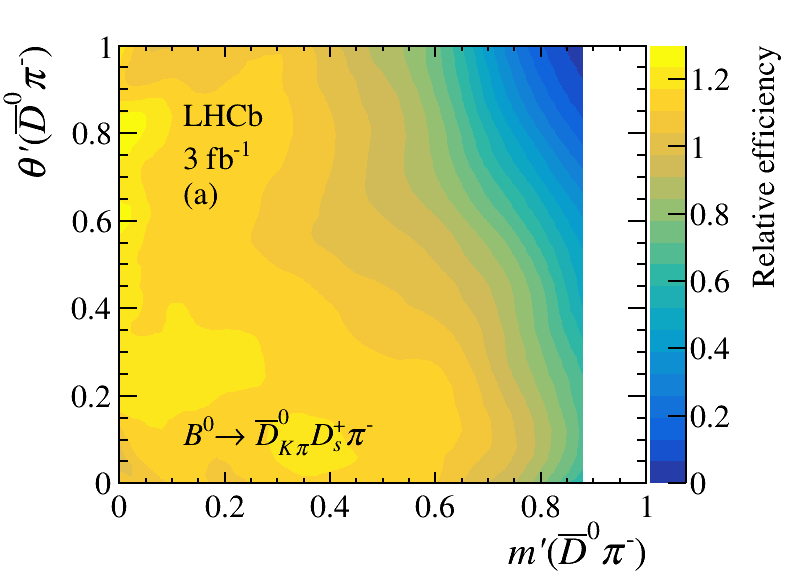}
    \includegraphics[width=0.45\linewidth]{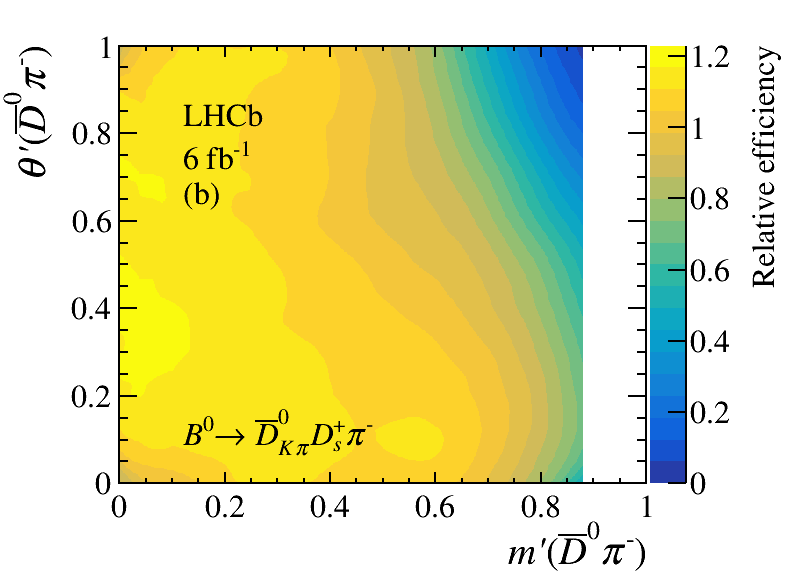}\\
    \includegraphics[width=0.45\linewidth]{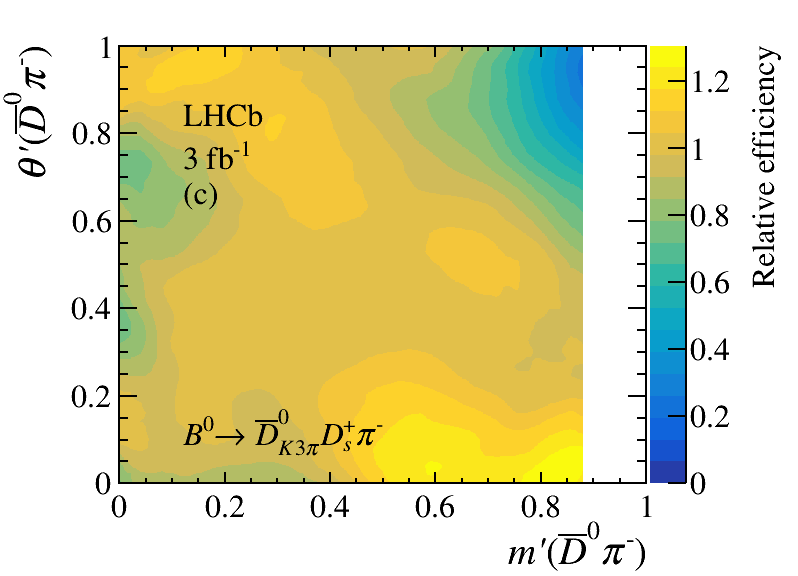}
    \includegraphics[width=0.45\linewidth]{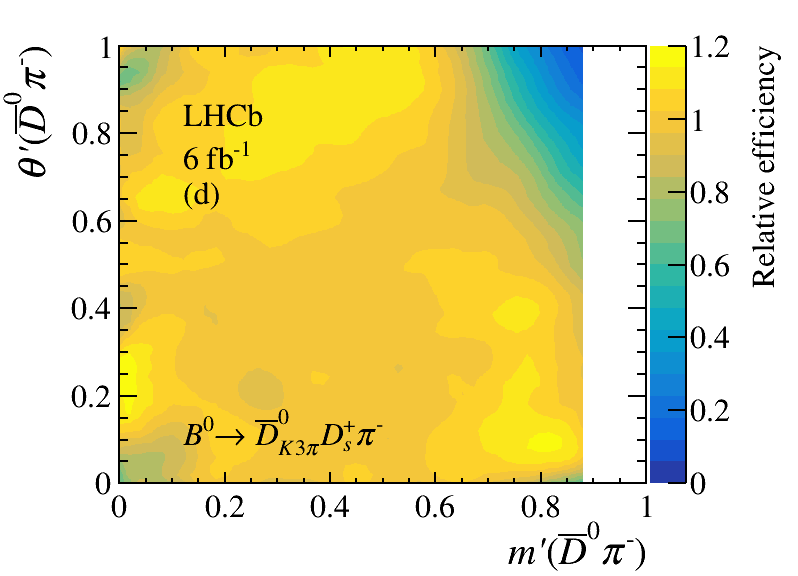}\\
    \includegraphics[width=0.45\linewidth]{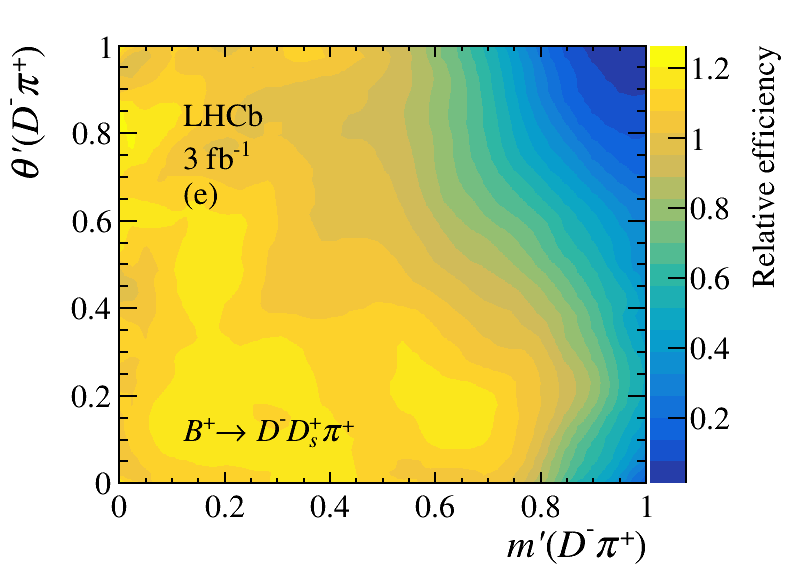}
    \includegraphics[width=0.45\linewidth]{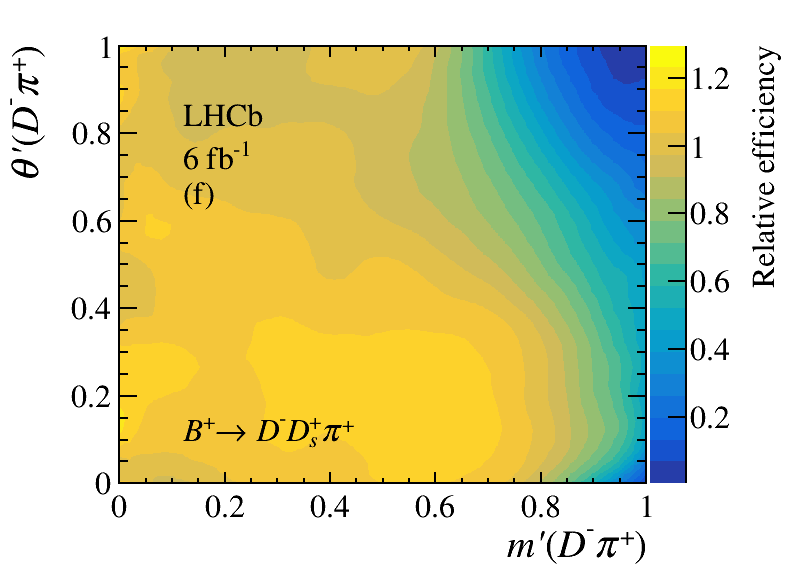}
  \end{center}
  \caption{Signal efficiency maps for the $\BztoDzbarDsppim$, $\Dzb\to\Kp\pim$ decays in (a) Run 1 and (b) Run 2; $\BztoDzbarDsppim$, $\Dzb\to\Kp\pim\pim\pip$ decays in (c) Run 1 and (d) Run 2; \mbox{$\BptoDmDsppip$} decays in (e) Run 1 and (f) Run 2. White regions are caused by \mbox{$\Bz\to\Dstarm\Dsp$} veto.}
  \label{fig:EffMap}
\end{figure}

The efficiency maps across the SDP are evaluated from the simulated samples by kernel density estimation~\cite{Poluektov:2014rxa} and are shown in Fig.~\ref{fig:EffMap}. The tracking efficiency and the efficiency of the trigger requirements have been corrected using control samples in data.

\begin{figure}[!tb]
  \begin{center}
    \includegraphics[width=0.45\linewidth]{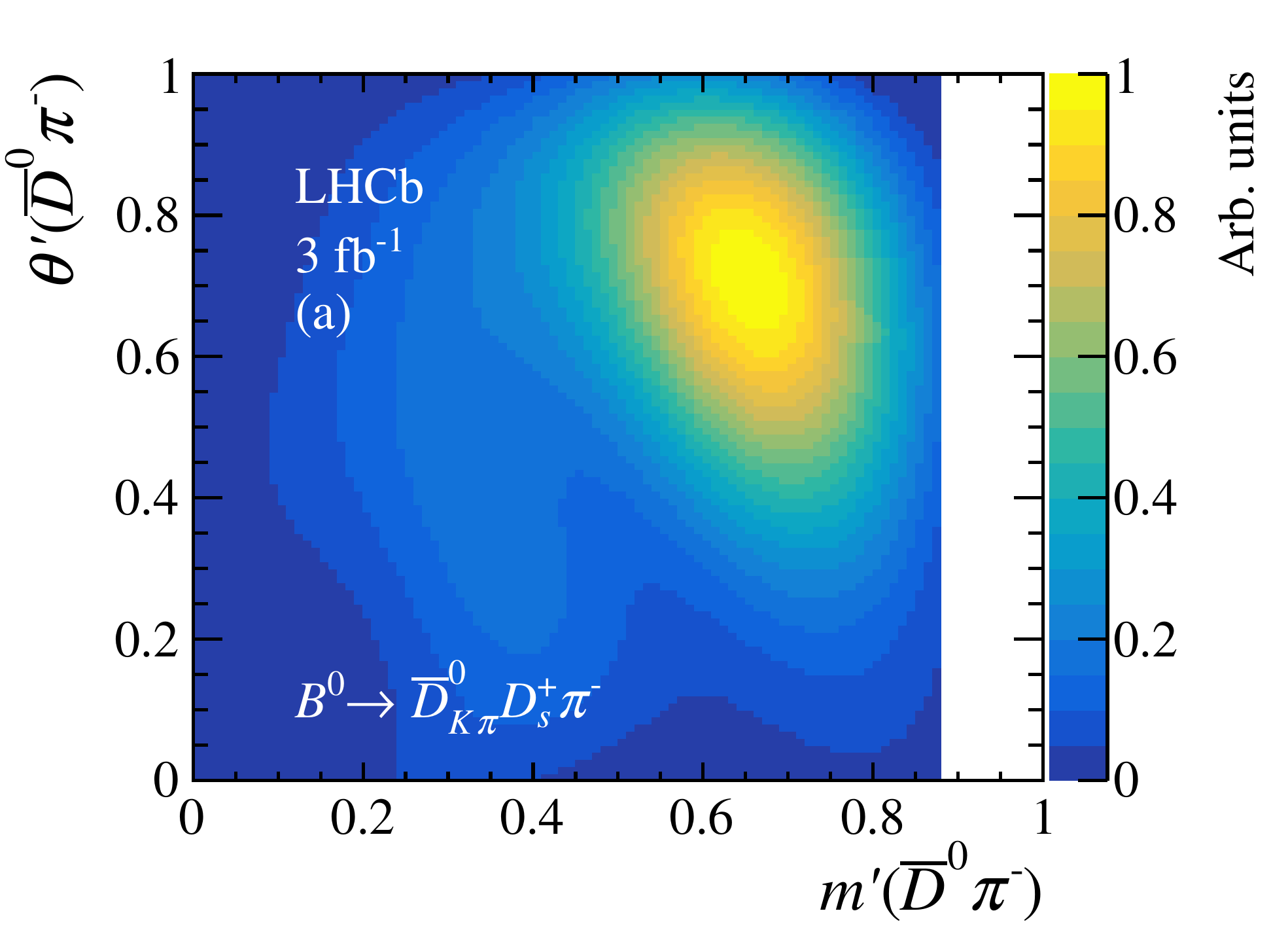}
    \includegraphics[width=0.45\linewidth]{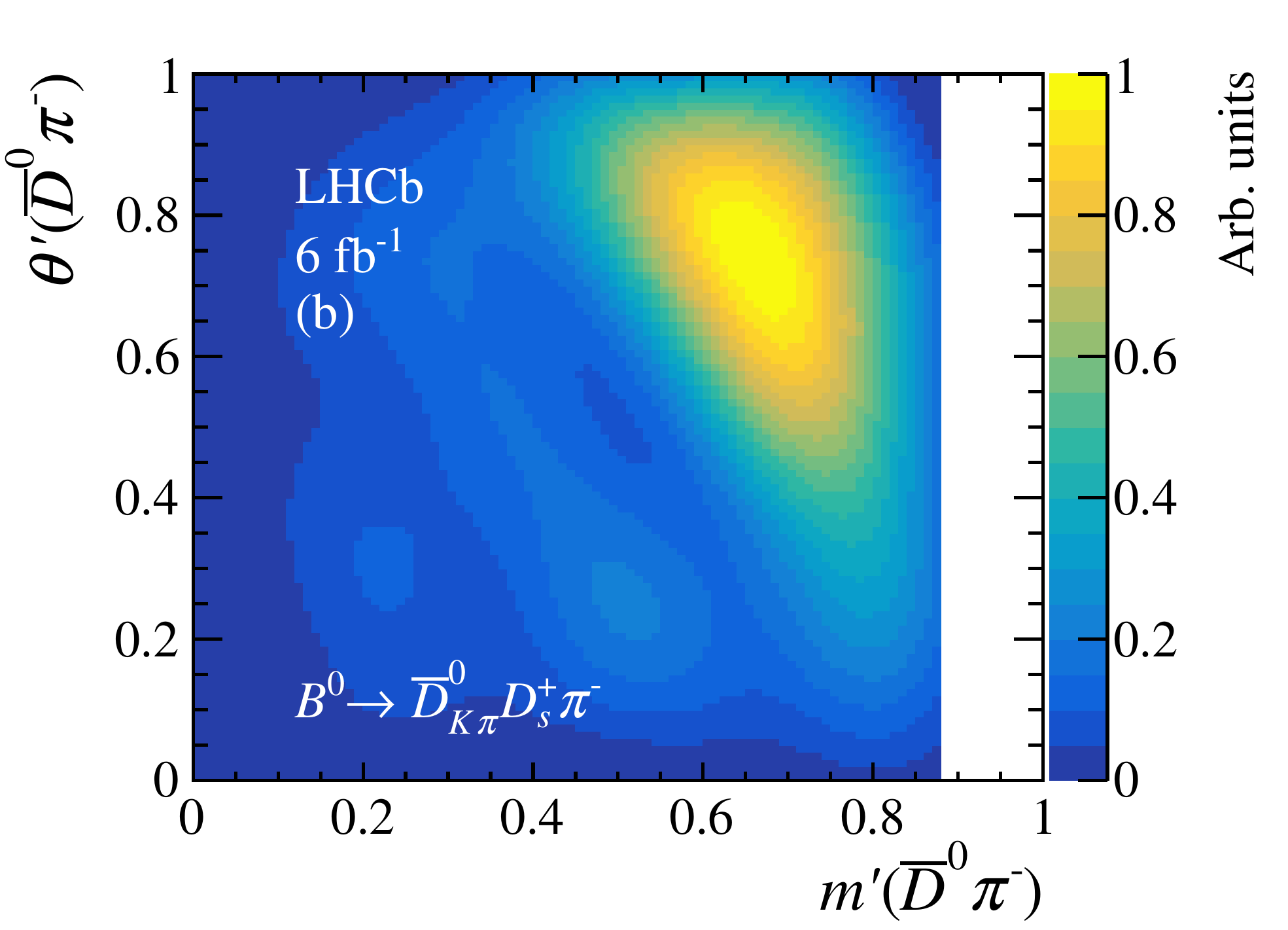}\\
    \includegraphics[width=0.45\linewidth]{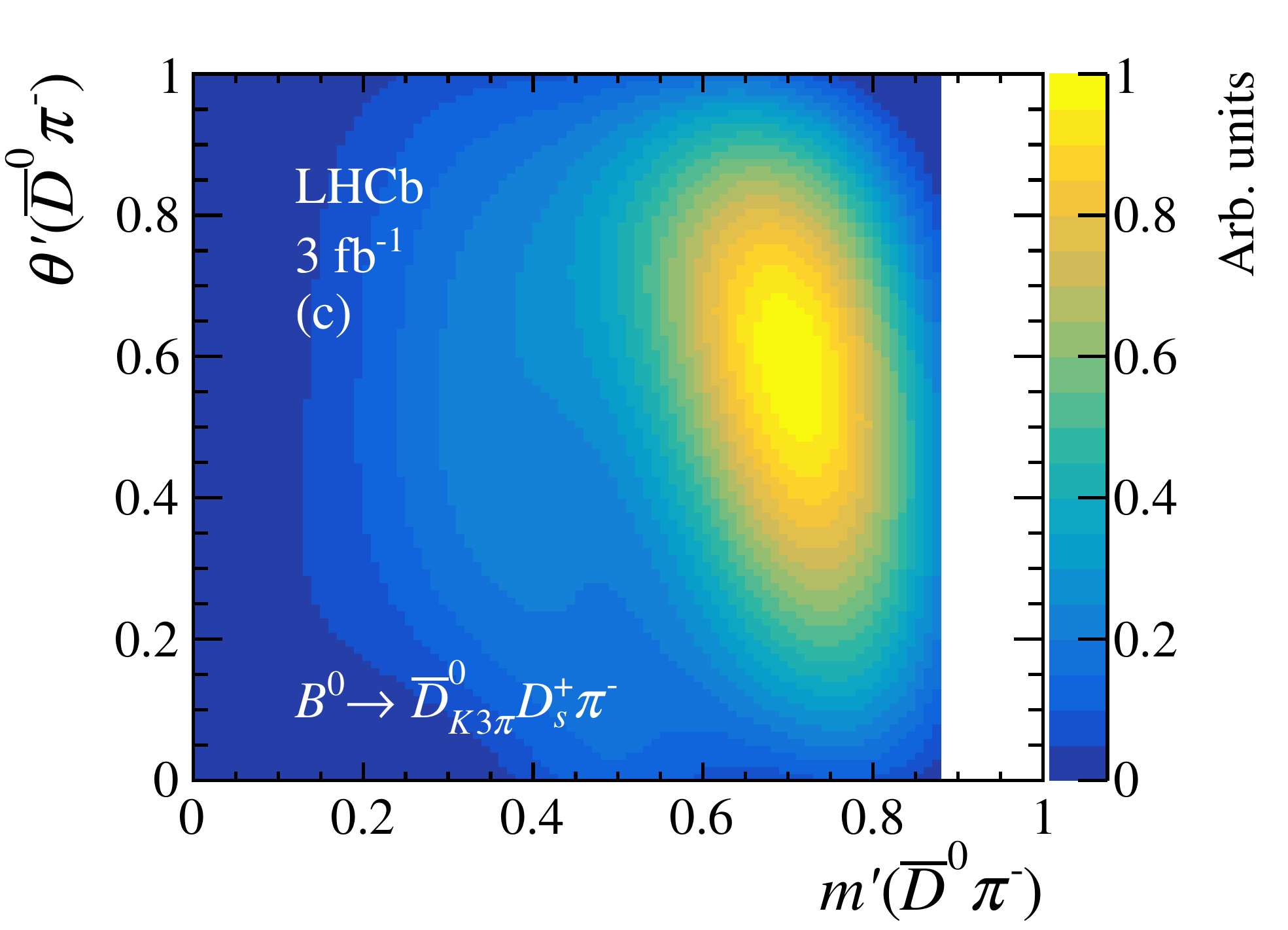}
    \includegraphics[width=0.45\linewidth]{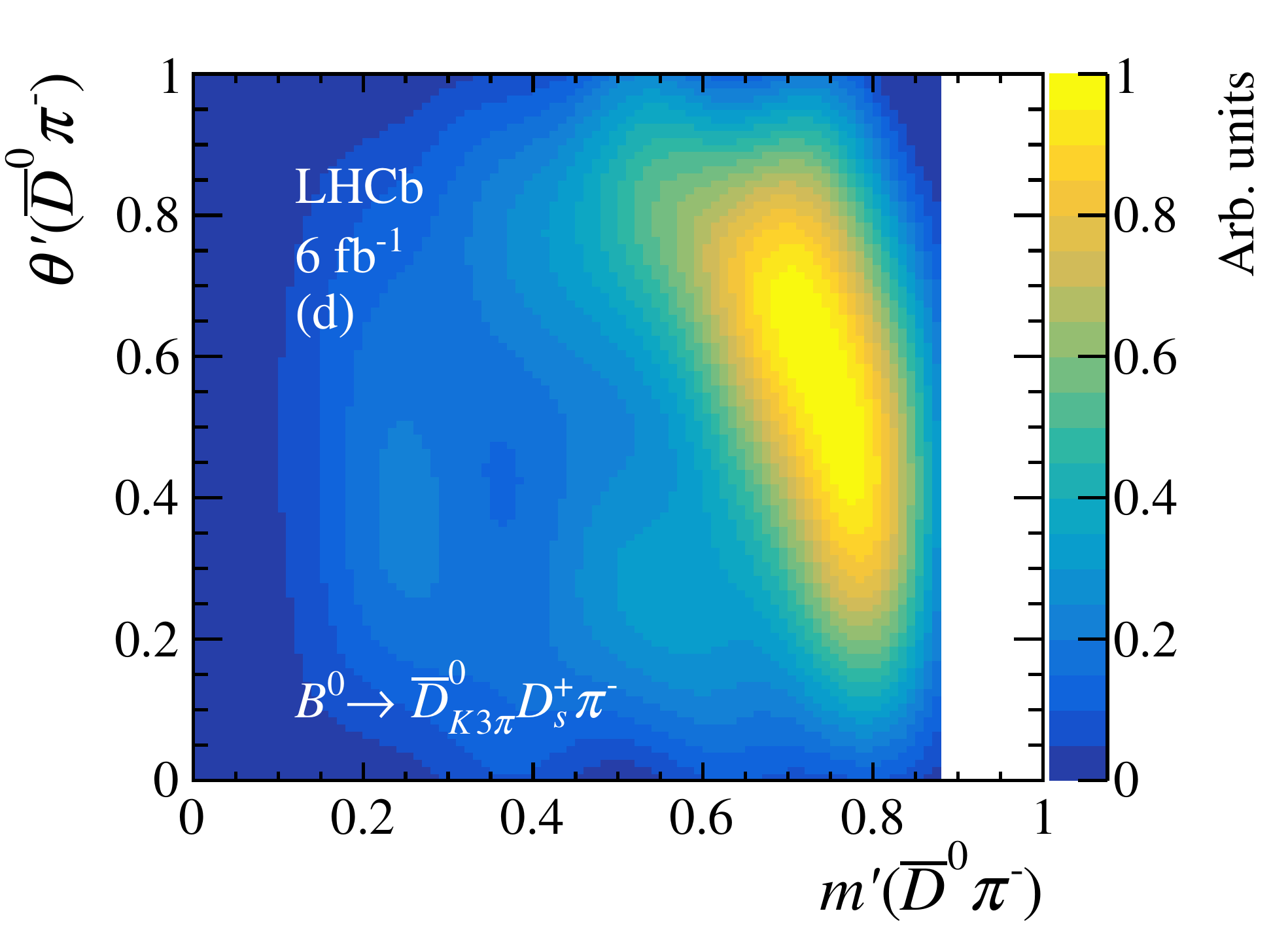}\\
    \includegraphics[width=0.45\linewidth]{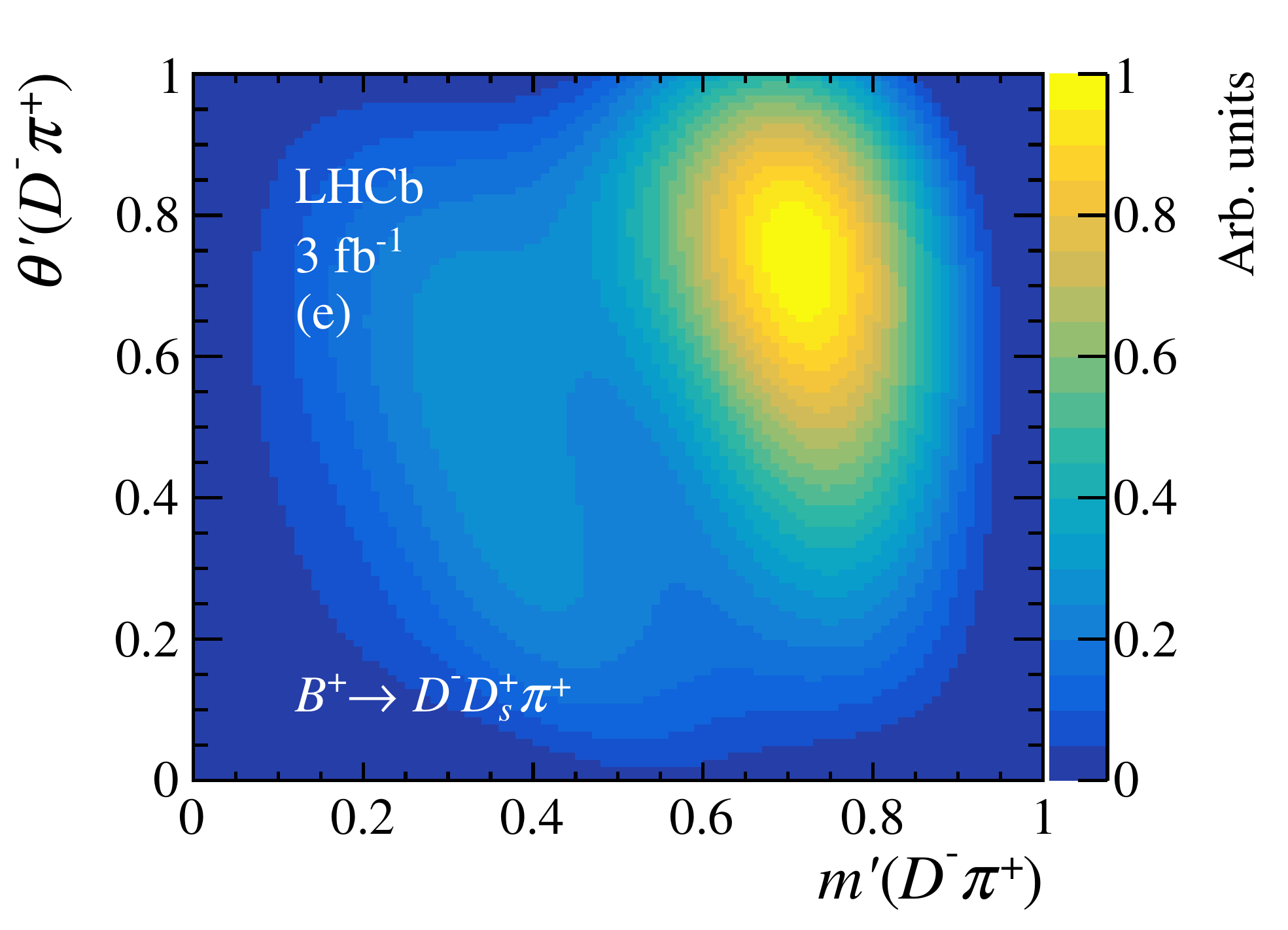}
    \includegraphics[width=0.45\linewidth]{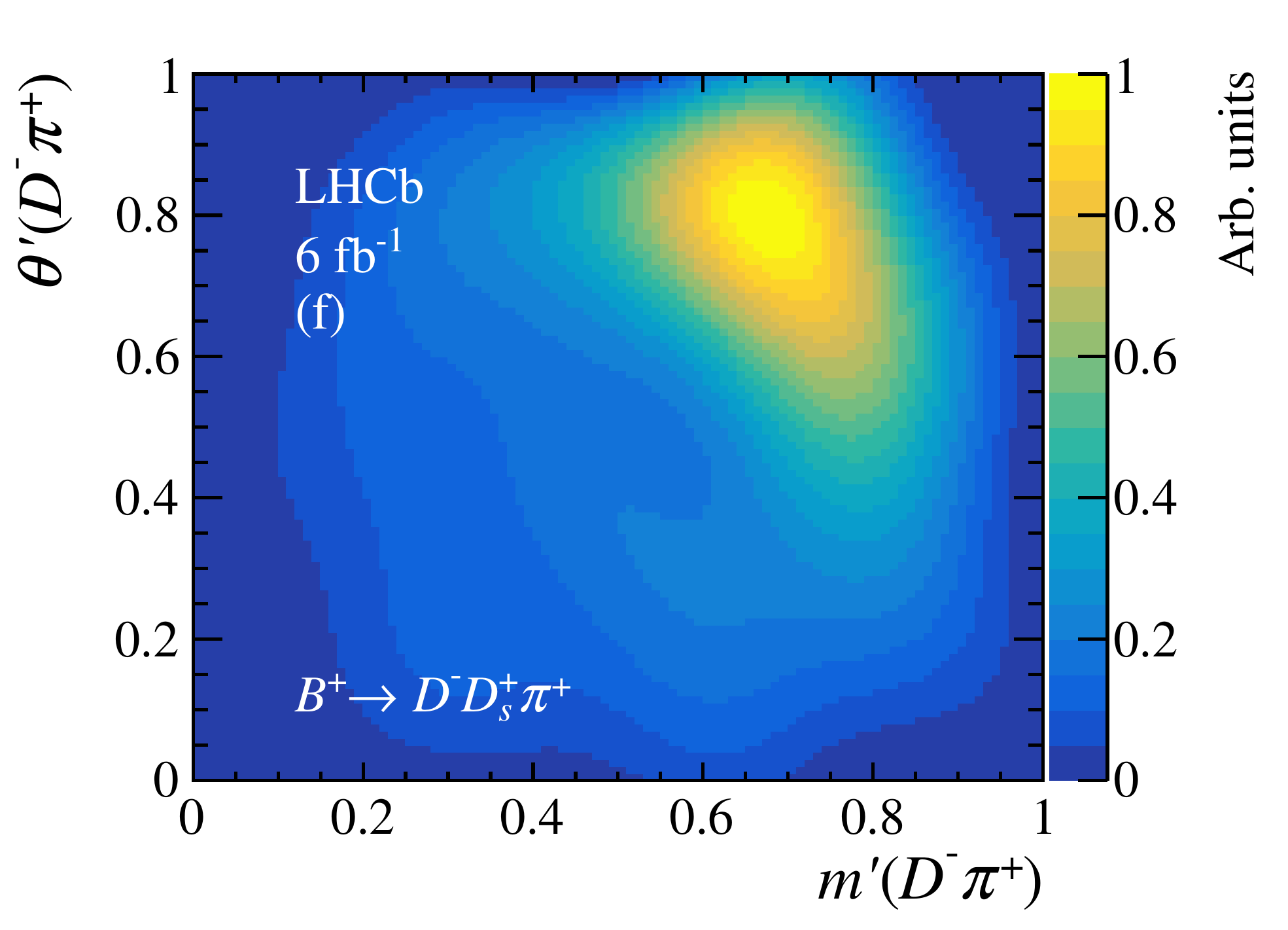}
  \end{center}
  \caption{Distributions of the SPD background shapes for the $\BztoDzbarDsppim$, $\Dzb\to\Kp\pim$ decays in (a) Run 1 and (b) Run 2; $\BztoDzbarDsppim$, $\Dzb\to\Kp\pim\pim\pip$ decays in (c) Run 1 and (d) Run 2; $\BptoDmDsppip$ decays in (e) Run 1 and (f) Run 2. White regions are caused by \mbox{$\Bz\to\Dstarm\Dsp$} veto.}
  \label{fig:BGMap}
\end{figure}

The background distributions over the phase space are estimated using candidates in the $B$ mass sidebands between $[5450, 6000]\mev$ for the \BztoDzbarDsppim decays and $[5400, 6000]\mev$ for the \BptoDmDsppip decays. The requirement on the BDT response is relaxed to increase the number of events in these regions as no significant change in the shapes of the distributions of sideband samples is observed. A Gaussian process extrapolation method~\cite{mathad2021efficient} is applied to extrapolate the background Dalitz plot distribution into the $B$-meson signal region, in order to account for the correlations between the Dalitz variables and the invariant mass of the $B$ candidates. The SDP distributions of the background shape for each dataset are shown in Fig.~\ref{fig:BGMap}. The broad structures in the SDP distributions are related to the accumulation of combinatorial background with low pion momentum.

%%%%%%%%%%%%%%%%%%%%%%%%%%%%%%%%%%%%%%%%%%%%%%%
%%%%%           Amplitude analysis        %%%%%
%%%%%%%%%%%%%%%%%%%%%%%%%%%%%%%%%%%%%%%%%%%%%%%
\section{Amplitude analysis}
\label{sec:Model description}
Conventional resonances are only expected to decay to $\Dzb\pim$ and $\Dm\pip$ final states in the \BztoDzbarDsppim and the \BptoDmDsppip decays, respectively.
The straightforward way to perform the amplitude analysis is by including all the $D\pi$ resonances with natural spin-parity, as listed in Table~\ref{tab:candidates}, in the fit model. It is called a model-dependent (MD) description. The result with the MD description is shown in Sec.~\ref{sec:MDresult}. 
The description of the $\Dbar\pi$ S-wave distributions is improved by introducing a $0^+$ $\Dbar\pi$ qMI description, which accounts for both the broad spin-0 $\Dbar\pi$ states and nonresonant components.
The result with the qMI description is considered as the default, which is further discussed in this section. 

\subsection{Model including only \texorpdfstring{\boldmath$D\pi$}{Dpi} resonances}
\label{sec:onlyDpi}
The basic fit model is defined by considering all known $D^{**}$ states with natural spin-parity~\cite{PDG2022}, as listed in Table~\ref{tab:candidates}, except for the broad $D^*_0(2300)$ state. Their masses and widths are fixed to their default values.
As described in Sec.~\ref{sec:Amplitude model}, a qMI description of the $D\pi$ S-wave is used~\cite{LHCb-PAPER-2016-026}, with 11 spline points.\footnote{[1.9, 2.0, 2.1, 2.2, 2.3, 2.4, 2.5, 2.6, 2.7, 2.9, 3.4]$\gev$} The first and last points are outside of the invariant mass range, and their amplitudes are fixed to zero. The other points are each assigned a complex coefficient that varies freely in the fit. Moreover, as the $\overline{D}{}^*(2007)^0$ mass is lower than the $\Dm\pip$ mass threshold, the $q_0$ value in Eq.~\ref{eqn:running width} would be imaginary. The $q_0$ value in this case is taken as the value calculated from $\overline{D}{}^*(2007)^0\to\Dzb\piz$ rather than $\overline{D}{}^*(2007)^0\to\Dm\pip$ in the default model.
The $\overline{D}{}^*_J(3000)$ state was first observed in the $D^{*-}\pip$ decay mode, and its spin has not been determined yet~\cite{LHCb-PAPER-2013-026}. A similar structure has been seen in $\Bm\to\Dp\pim\pim$ decays~\cite{LHCb-PAPER-2016-026}, with $J^P=2^+$. In this analysis different $J^P$ hypotheses for the $\Dbar_J^*(3000)$ state are tested, either $1^-$, $2^+$, $3^-$, or $4^+$. In each case its mass and width are fixed to the corresponding default values~\cite{PDG2022}. The test results favor $J^P=4^+$, which is used as the default.

\begin{table}[!tb]
  \caption{Resonances expected in \BztoDzbarDsppim and \BptoDmDsppip decays~\cite{PDG2022}.
  The masses and widths of resonances marked with \# are shared for both the charged and neutral isospin partners.}
  \begin{center}
      \resizebox{\textwidth}{!}{
      \begin{tabular}{ c  c  c  c  c }
    \hline
    Resonance & $J^P$ & Mass (GeV) & Width (GeV) & Comments \\ 
    \hline
    \specialrule{0em}{1pt}{1pt}
    $\overline{D}{}^*(2007)^0$ & $1^-$ & $2.00685\pm0.00005$ & $<2.1\times10^{-3}$ & Width set to be 0.1\mev  \\ 
    $D^*(2010)^-$ & $1^-$ & $2.01026\pm0.00005$ &  $(8.34\pm0.18)\times10^{-5}$ &  \\
    $\overline{D}{}^*_0(2300)$ & $0^+$ & $2.343\pm0.010$ & $0.229\pm0.016$ & \# \\
    $\overline{D}{}^*_2(2460)$ & $2^+$ & $2.4611\pm0.0007$ & $0.0473\pm0.0008$ & \# \\
    $\overline{D}{}^*_1(2600)^0$ & $1^-$ & $2.627\pm0.010$& $0.141\pm0.023$& \# \\
    $\overline{D}{}^*_3(2750)$ & $3^-$ & $2.7631\pm0.0032$& $0.066\pm0.005$& \# \\
    $\overline{D}{}^*_1(2760)^0$ & $1^-$ & $2.781\pm0.022$& $0.177\pm0.040$& \# \\
    $\overline{D}{}^*_J(3000)^0$ & $?^?$ & $3.214\pm0.060$& $0.186\pm0.080$& \# $J^P = 4^+$ is assumed \\
    \hline
  \end{tabular}}\end{center}
  \label{tab:candidates}
\end{table}

\begin{figure}[!tb]
  \begin{center}
    \includegraphics[width=0.45\linewidth]{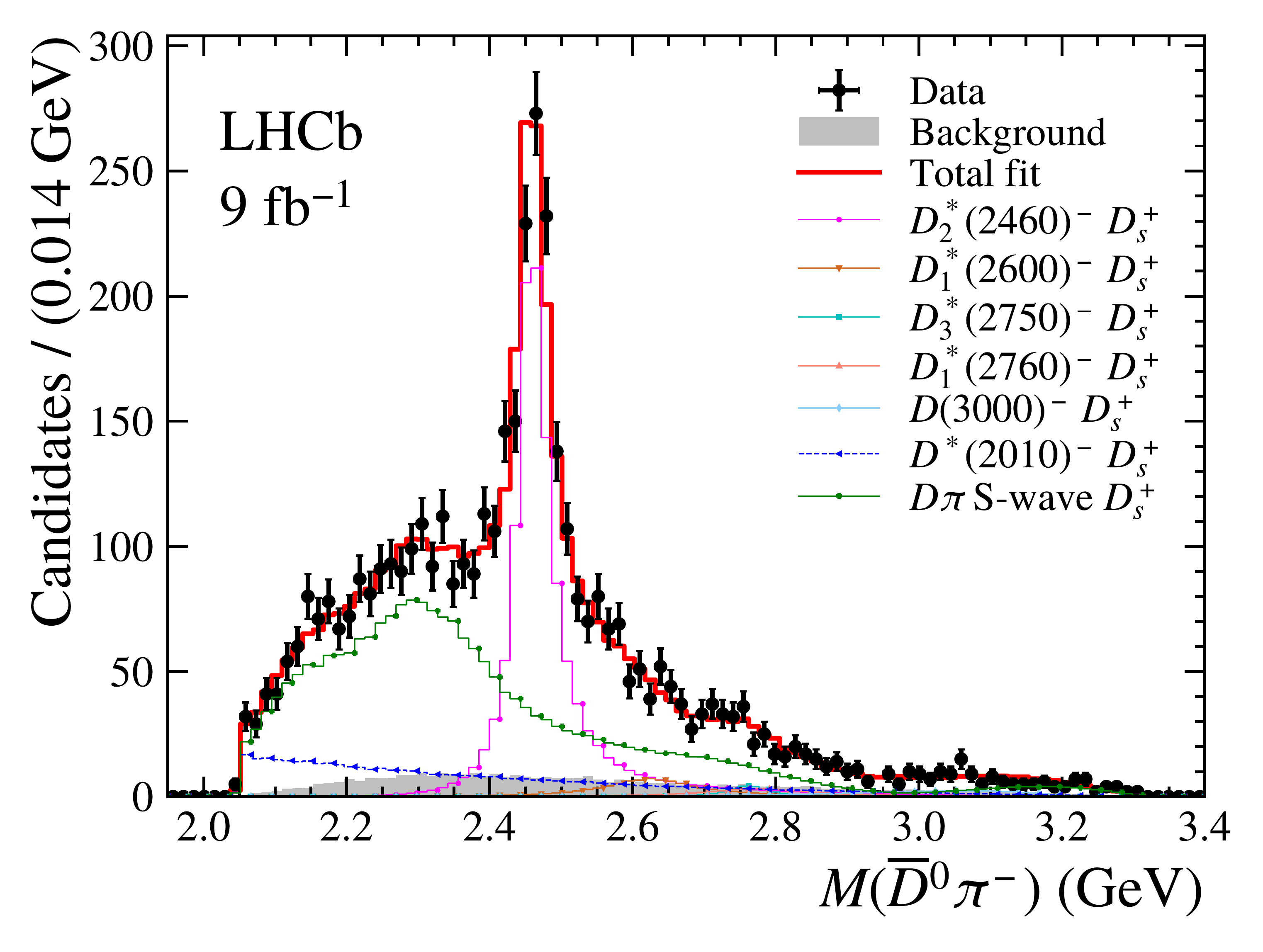}\put(-155,100){(a)}
    \includegraphics[width=0.45\linewidth]{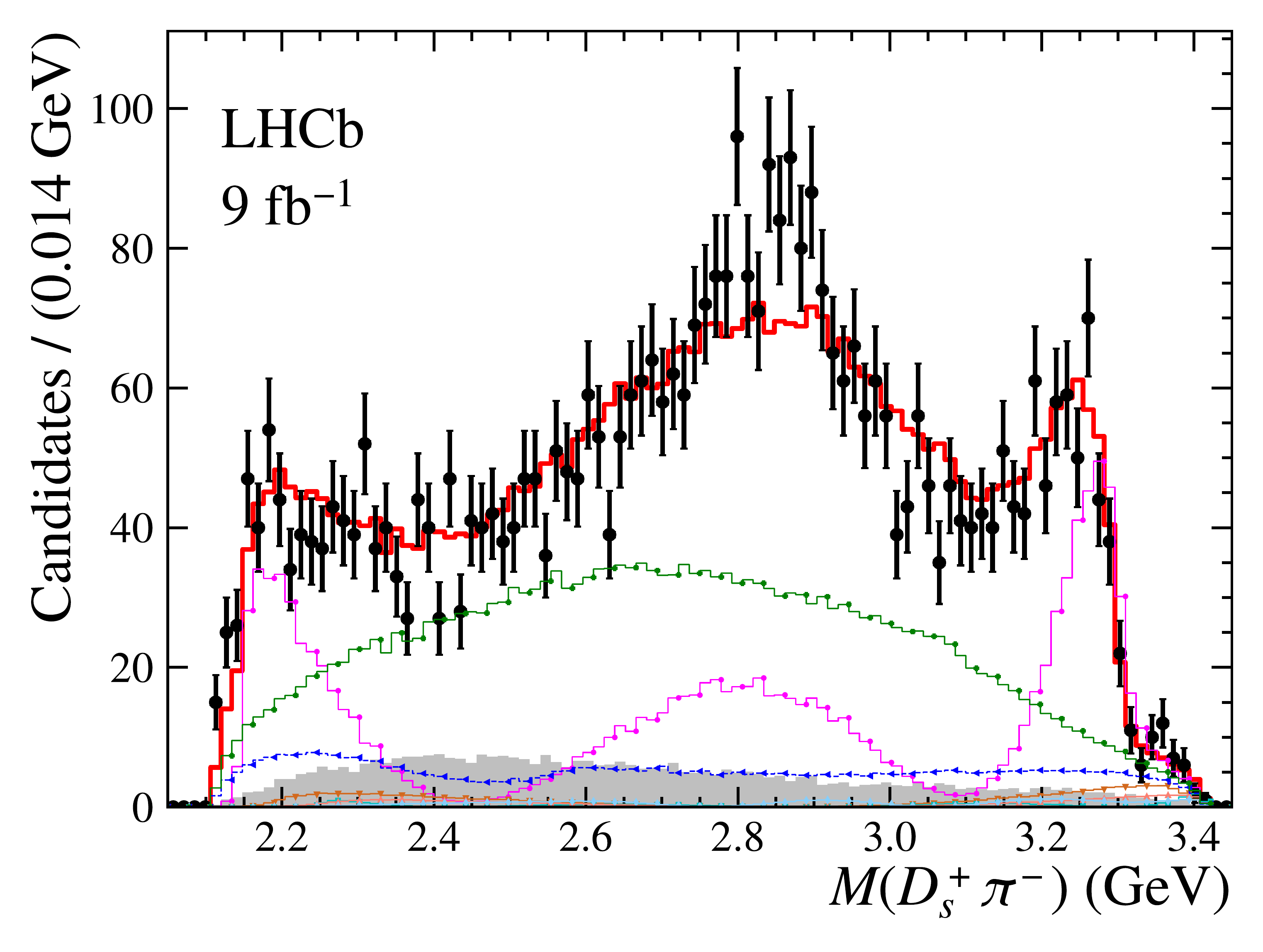}\put(-155,100){(b)}\\
    \includegraphics[width=0.45\linewidth]{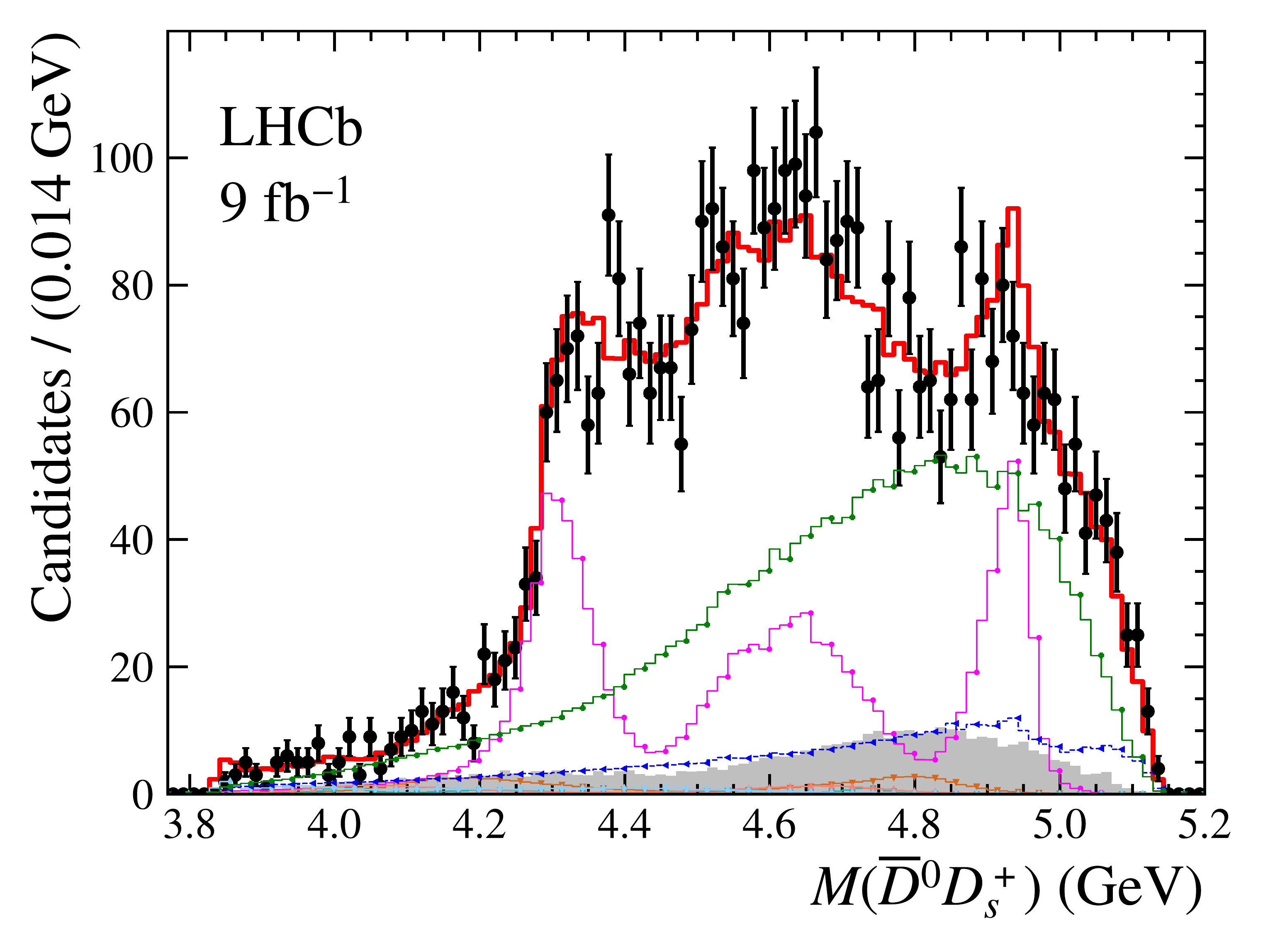}\put(-155,100){(c)}
    \end{center}
  \caption{Invariant mass distributions (a) $M(\Dzb\pim)$, (b) $M(\Dsp\pim)$ and (c) $M(\Dzb\Dsp)$ for the \BztoDzbarDsppim candidates compared with the fit results with only $D\pi$ resonances.}
  \label{fig:base_fit_Bz}
\end{figure}

\begin{figure}[!tb]
  \begin{center}
    \includegraphics[width=0.45\linewidth]{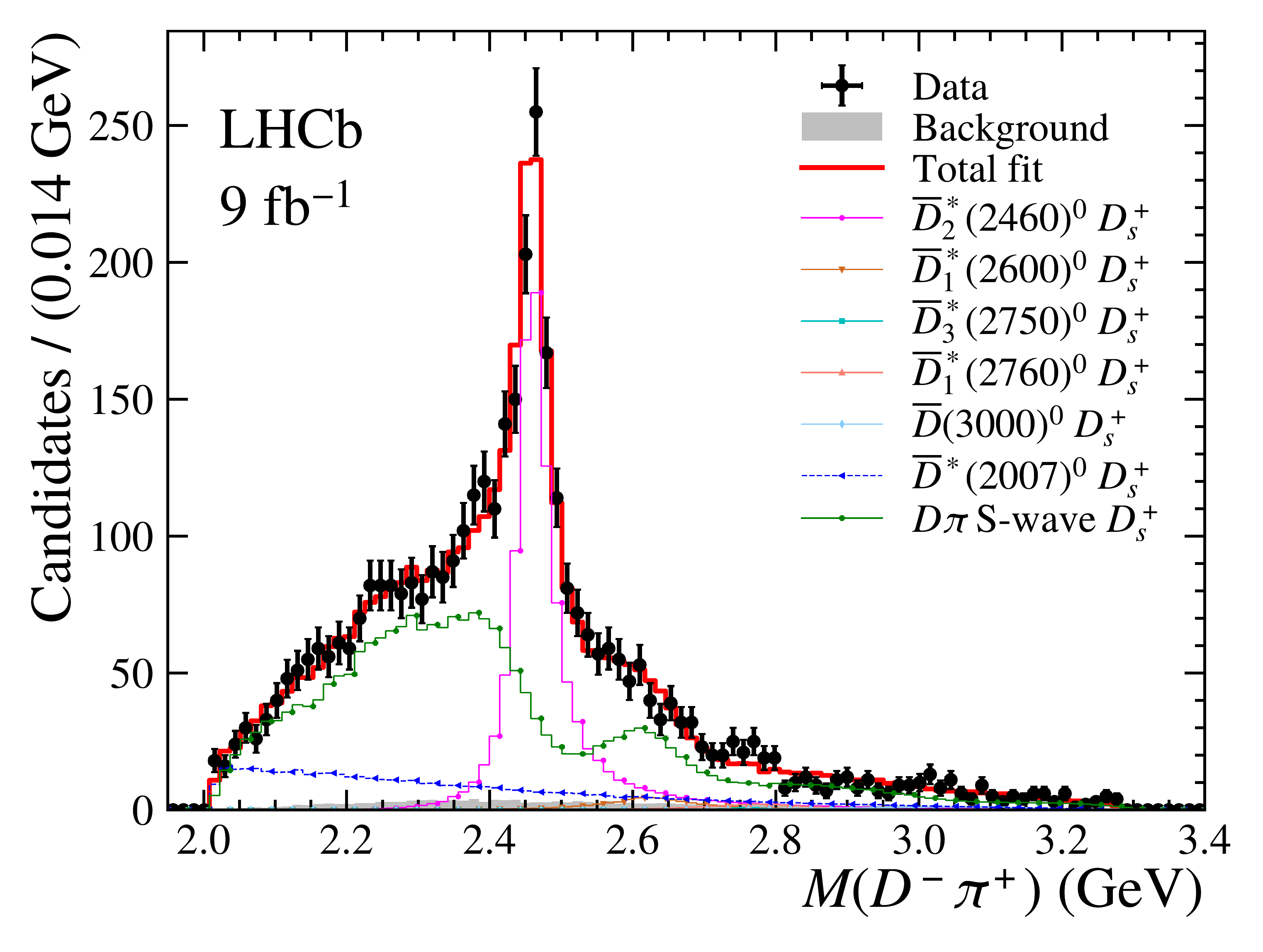}\put(-155,100){(a)}
    \includegraphics[width=0.45\linewidth]{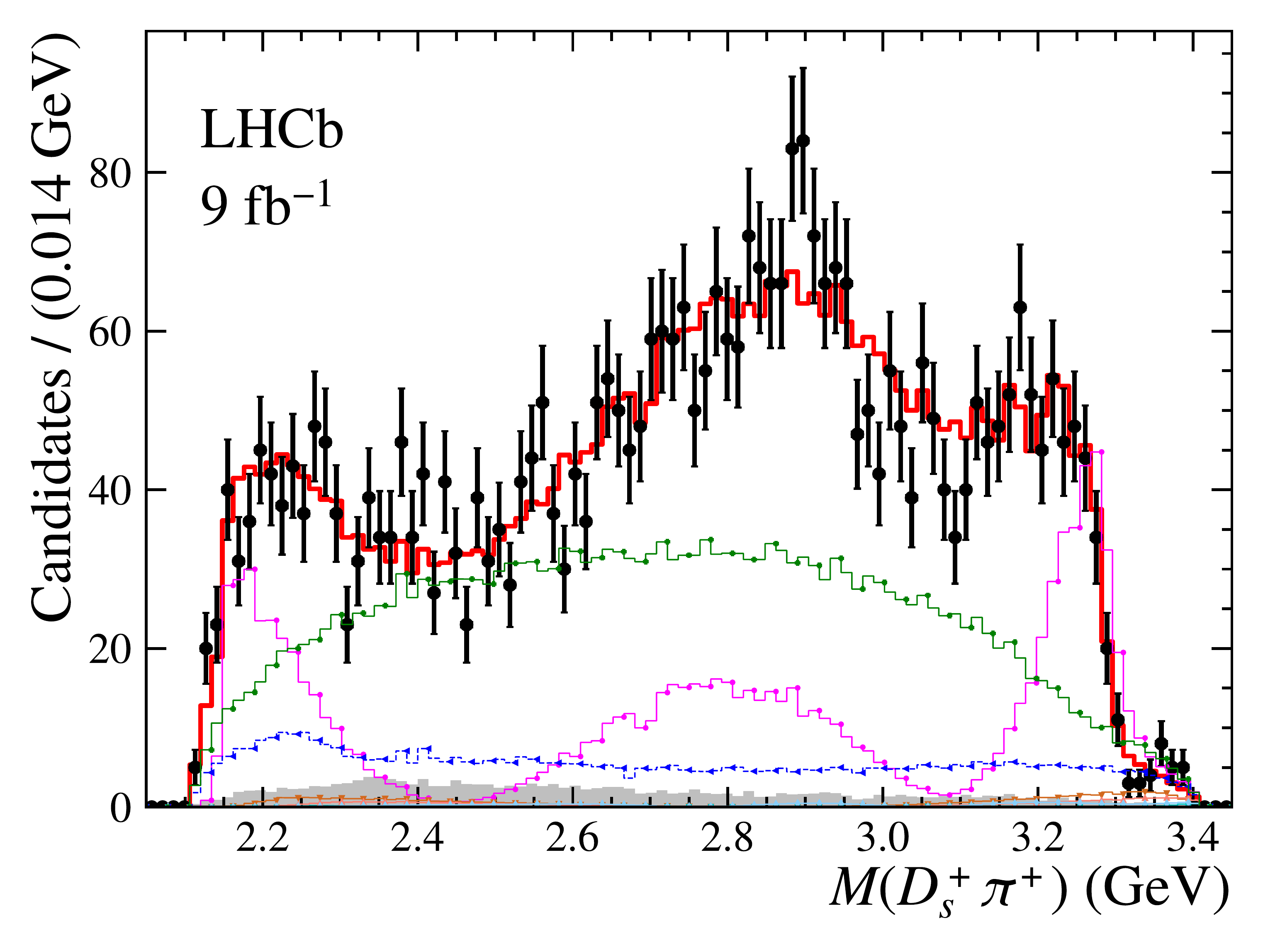}\put(-155,100){(b)}\\
    \includegraphics[width=0.45\linewidth]{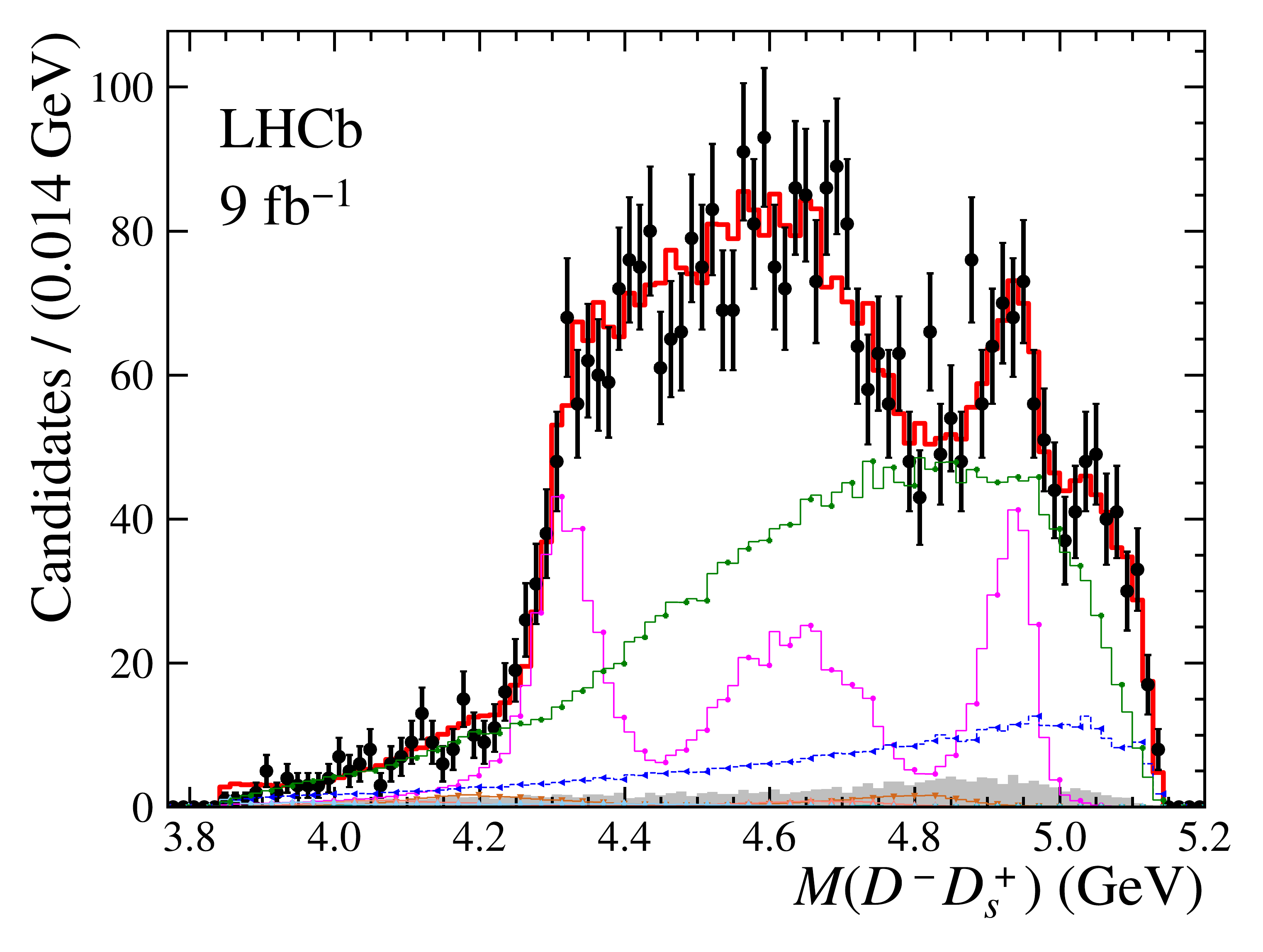}\put(-155,100){(c)}
  \end{center}
  \caption{Invariant mass distributions of the (a) $M(\Dm\pip)$, (b) $M(\Dsp\pip)$ and (c) $M(\Dm\Dsp)$ for the \BptoDmDsppip candidates compared with the fit results with only $D\pi$ resonances.}
  \label{fig:base_fit_Bp}
\end{figure}

\begin{figure}[!tb]
  \begin{center}
    \includegraphics[width=0.45\linewidth]{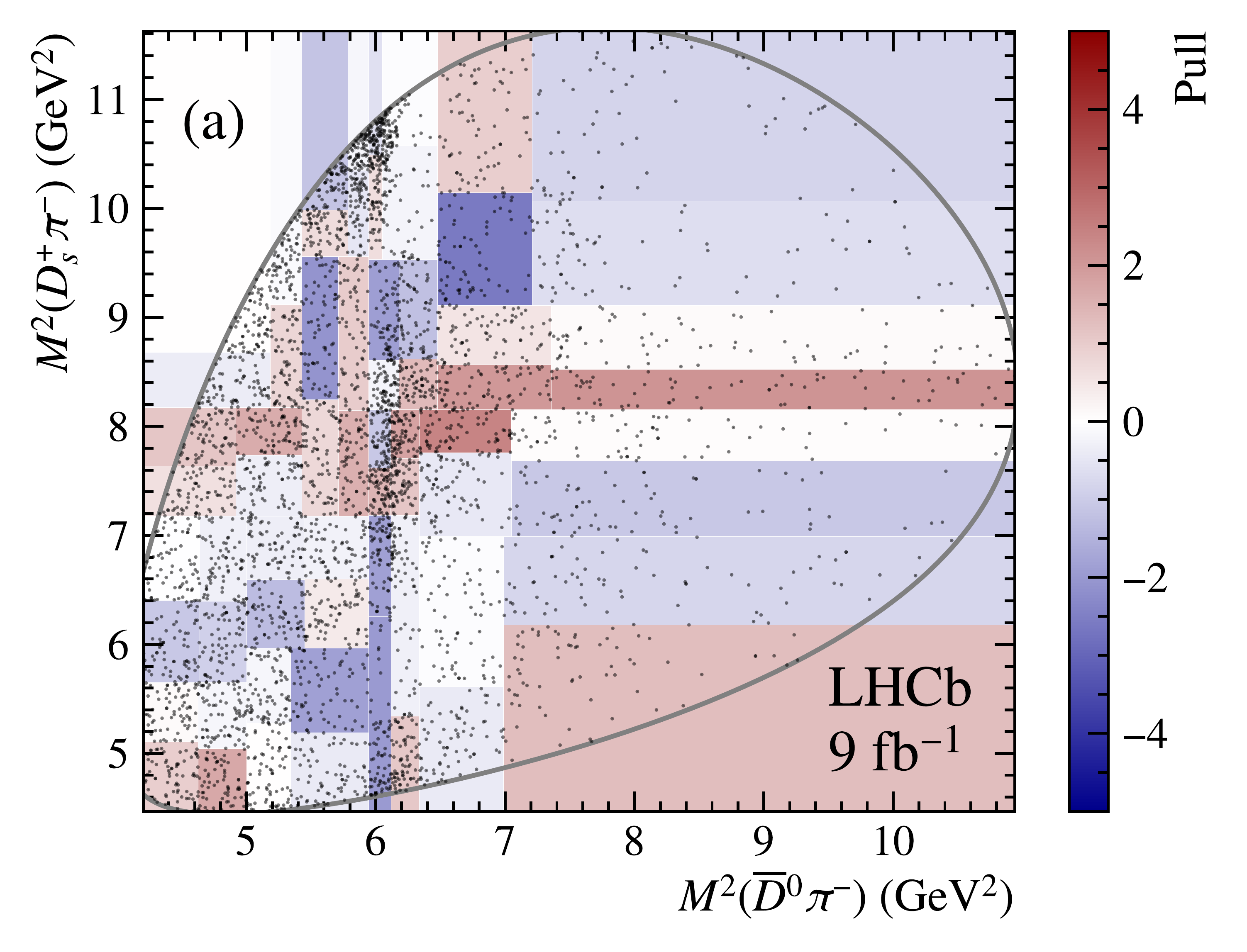}
    \includegraphics[width=0.45\linewidth]{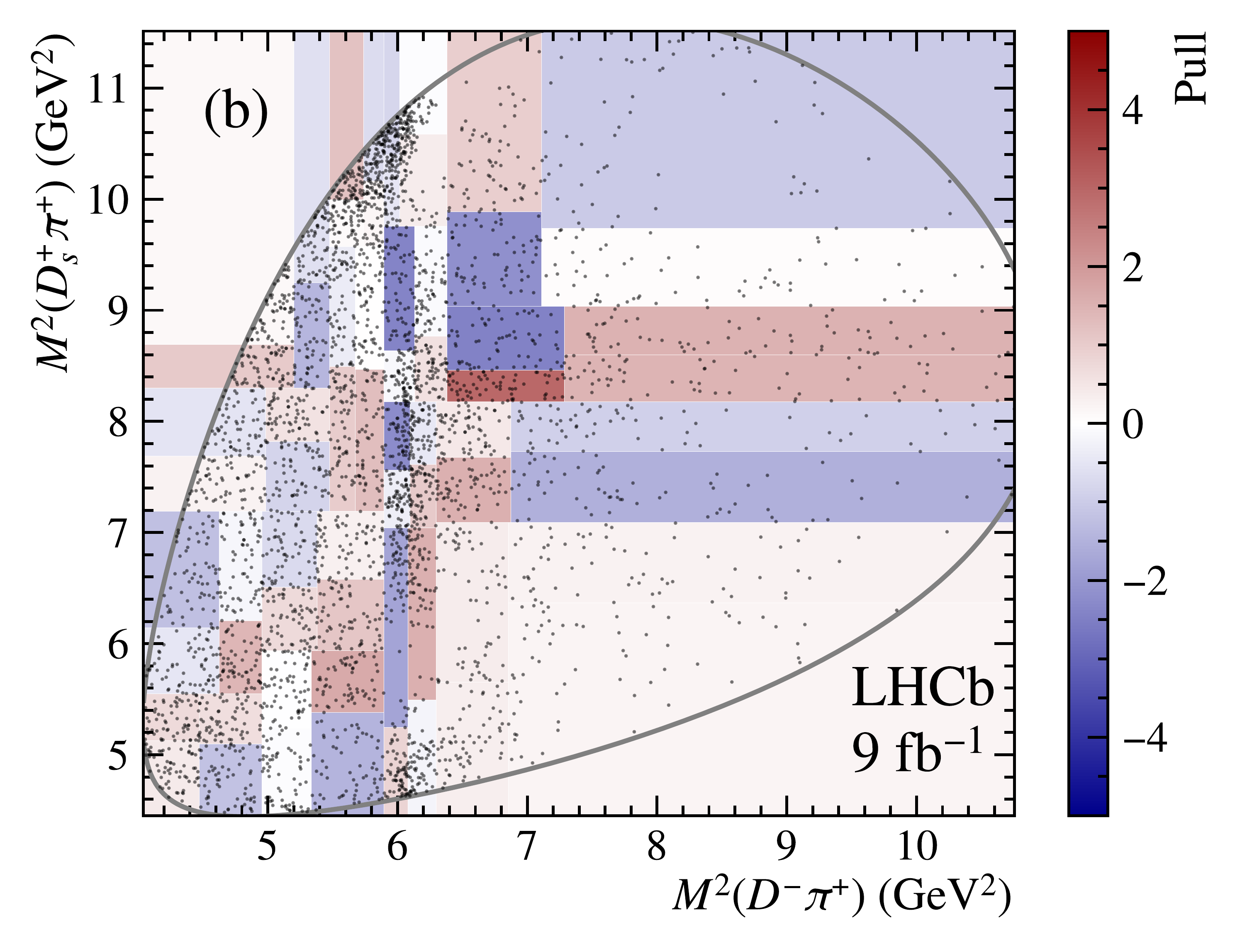}
  \end{center}
  \caption{Two-dimensional pull plots of the fits to the (a) \BztoDzbarDsppim and (b) \BptoDmDsppip samples.}
  \label{fig:2Dpull_onlyDpi}
\end{figure}

The fit results, where only $D\pi$ resonances are included and the two decays are considered independently, are given in Figs.~\ref{fig:base_fit_Bz} and \ref{fig:base_fit_Bp} for \BztoDzbarDsppim and \BptoDmDsppip decays, respectively. The $\Bz \rightarrow \Dzb_{\Kp\pim} \Dsp \pim$, $\Bz \rightarrow \Dzb_{\Kp\pim\pim\pip} \Dsp \pim$ decays are combined when plotting here and subsequently. A peaking structure at about $2.9\gev$ is visible in the $M(D_s\pi)$ distribution of each decay, and is not well described by the included contributions. Furthermore, the addition of further $\Dstar$ states up to $J^P=4^+$ does not resolve the discrepancy. The normalized residuals of the fits are shown in Fig.~\ref{fig:2Dpull_onlyDpi}. In the plot, the pull value in bin $i$ is defined as $(N^{i}_{\rm sig}-N^{i}_{\rm fit})/\sqrt{N^i_{\rm sig}}$, where the $N^i_{\rm sig}$ and $N^i_{\rm fit}$ are the number of signal candidates and number of expected candidates from the fit result. The $\chi^2/ndf$, where $ndf$ is the number of degrees of freedom, is $78.2/35$ for \BztoDzbarDsppim and $75.2/35$ for \BptoDmDsppip, which also indicates the existence of a new resonance.

\subsection{Model including \texorpdfstring{\boldmath$\Ds\pi$}{Dspi} resonances}\label{sec:sepDpi}
\begin{figure}[!tb]
  \begin{center}
    \includegraphics[width=0.45\linewidth]{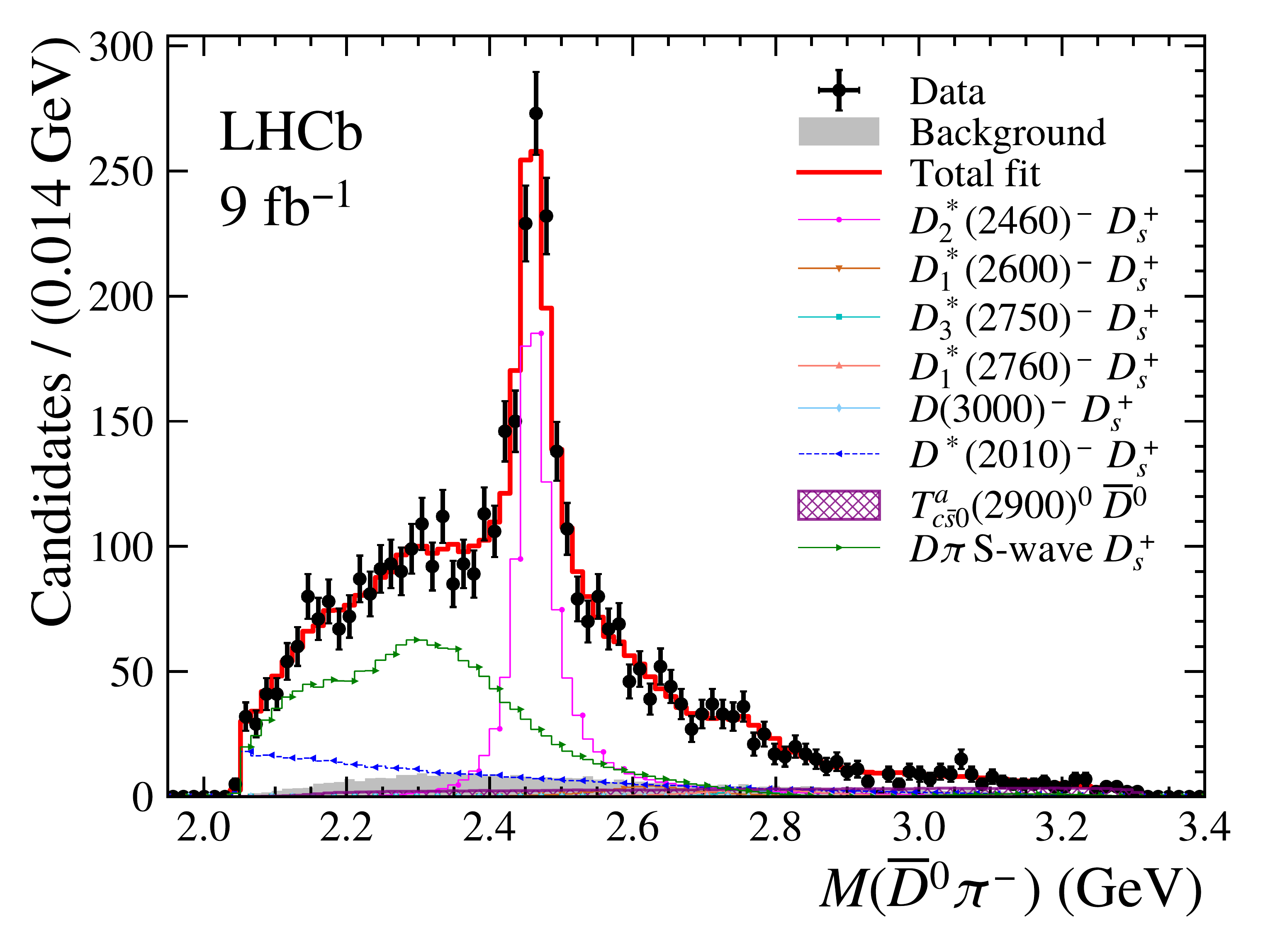}\put(-155,100){(a)}
    \includegraphics[width=0.45\linewidth]{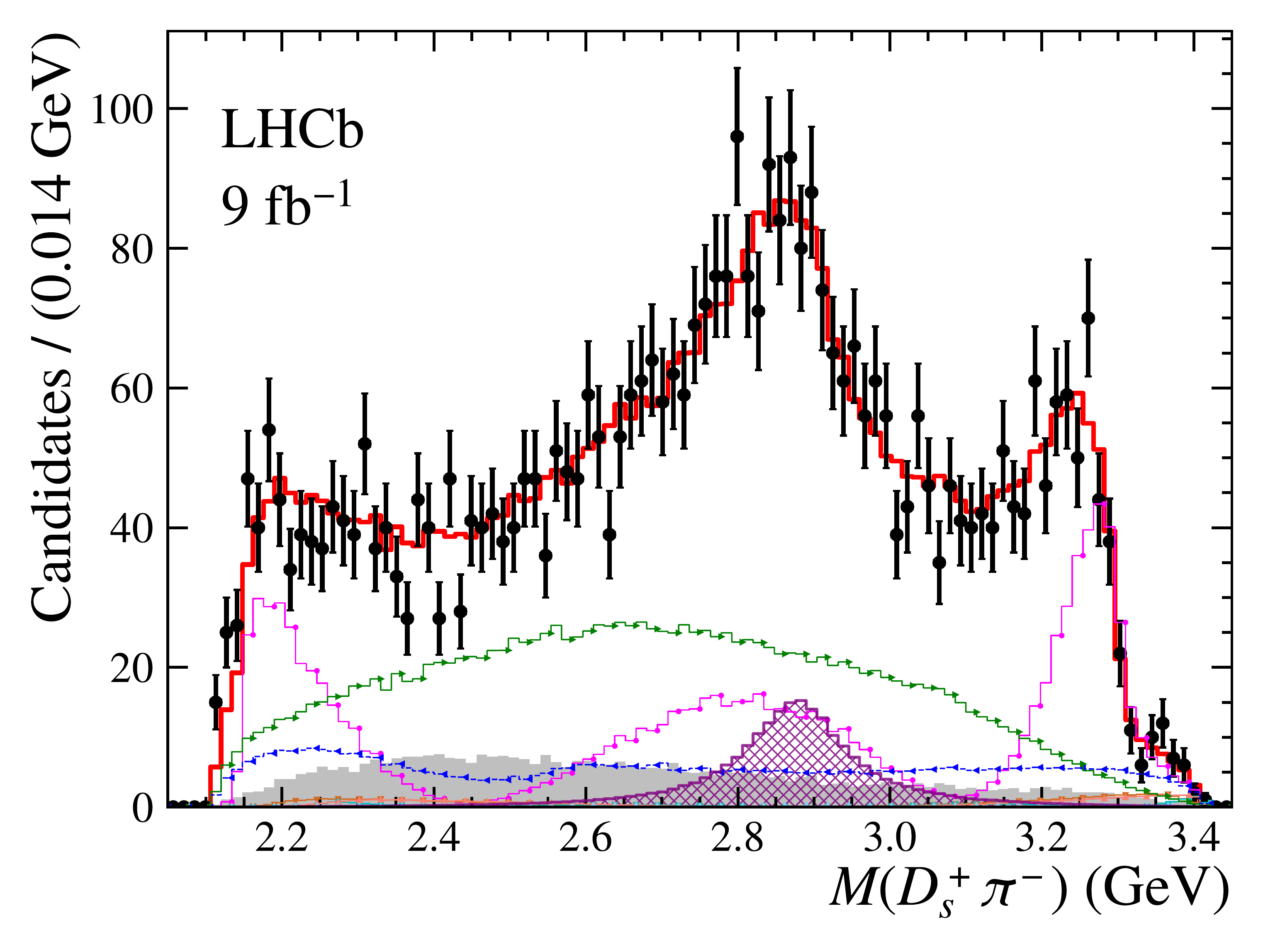}\put(-155,100){(b)}\\
    \includegraphics[width=0.45\linewidth]{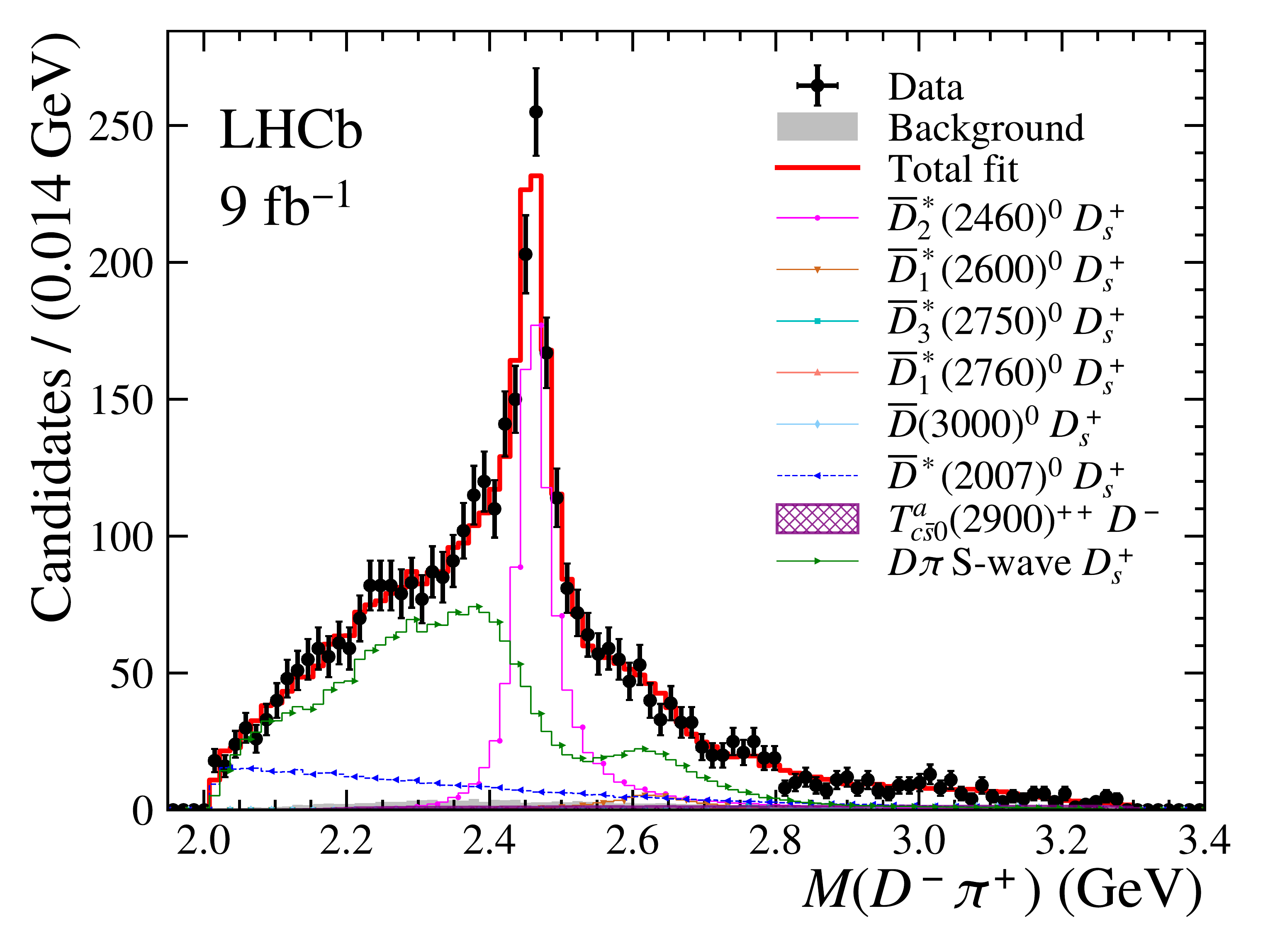}\put(-155,100){(c)}
    \includegraphics[width=0.45\linewidth]{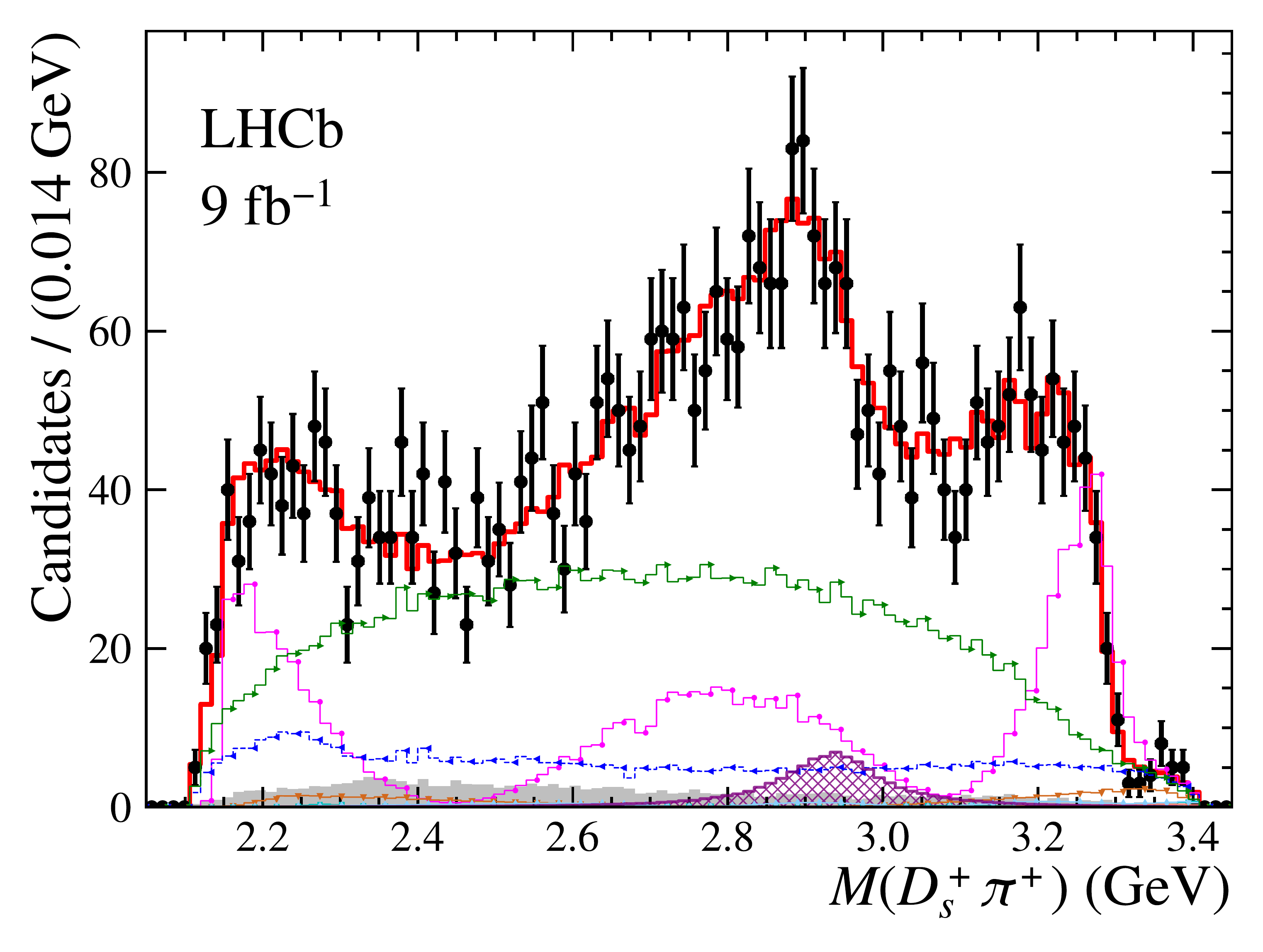}\put(-155,100){(d)}
  \end{center}
  \caption{Projection of the fit result on (a) $M(D\pi)$ and (b) $M(\Ds\pi)$ of $\BztoDzbarDsppim$ decays after including the $\Zzz$ state, and on (c) $M(D\pi)$ and (d) $M(\Ds\pi)$ of $\BptoDmDsppip$ decays after including the $\Zzpp$ state.}
  \label{fig:Z0_fit}
\end{figure}

To improve the description of the $M(\Ds\pi)$ distributions for the two decays, an additional $\Ds\pi$ state is added to each decay, whose mass and width are free parameters, and different $J^P$ assignments are tested.
No relationship is assumed for the two $\Ds\pi$ states.
Both states with $J^P = 0^+$ give the best description of the data, while the $\Ds\pi$ states with the other spin-parity are disfavored compared to the $0^+$ hypothesis (see Sec.~\ref{sec:OtherTest}).
The distributions of $M(\Dsp\pim)$ in \BztoDzbarDsppim and $M(\Dsp\pip)$ in \BptoDmDsppip are shown in Fig.~\ref{fig:Z0_fit}, where the two new $\Ds\pi$ resonances, which are named as $\Zzz$ and $\Zzpp$ following the convention in Ref.~\cite{Gershon:2022xnn}, are evident. 

\begin{figure}[!tb]
  \begin{center}
    \includegraphics[width=0.45\linewidth]{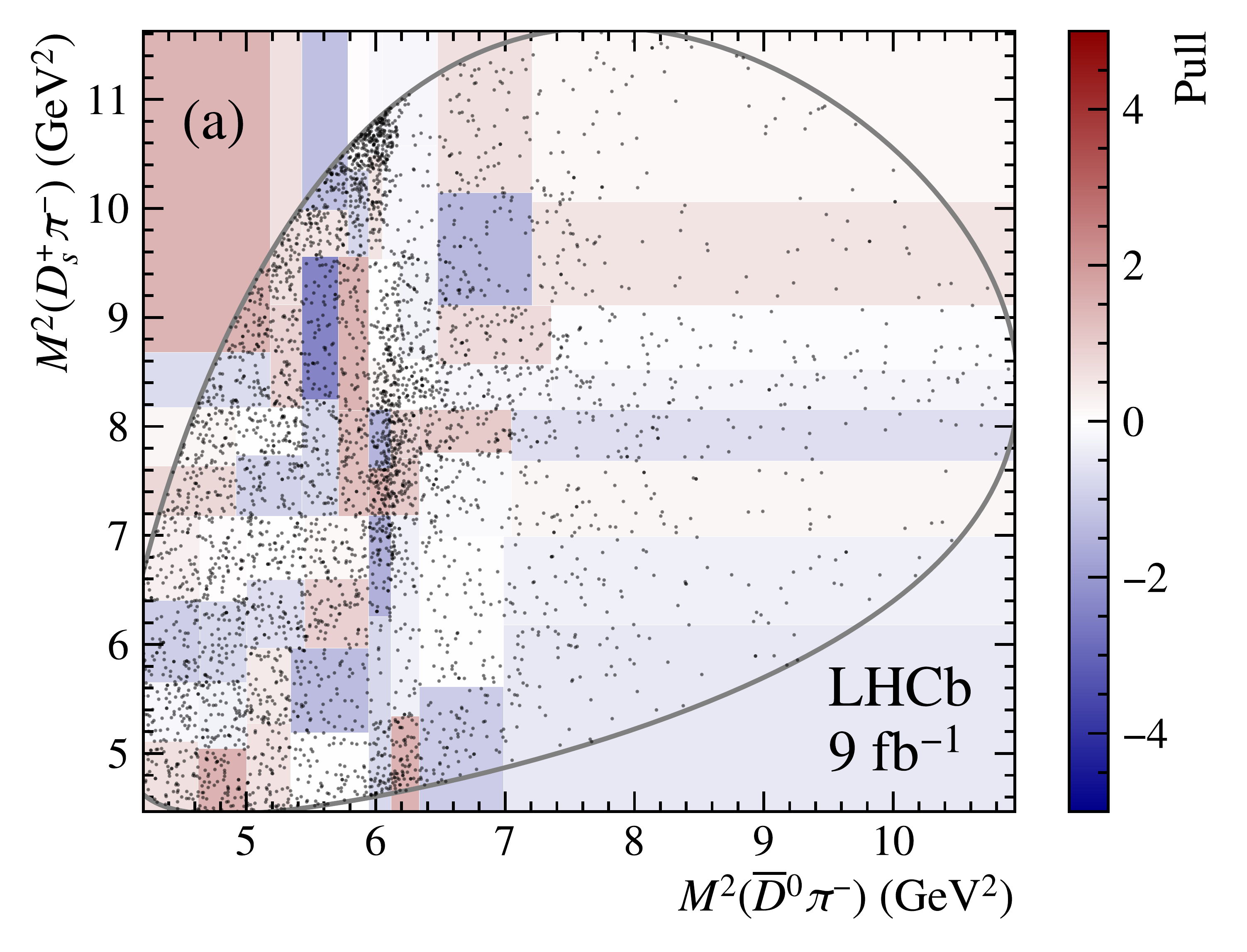}
    \includegraphics[width=0.45\linewidth]{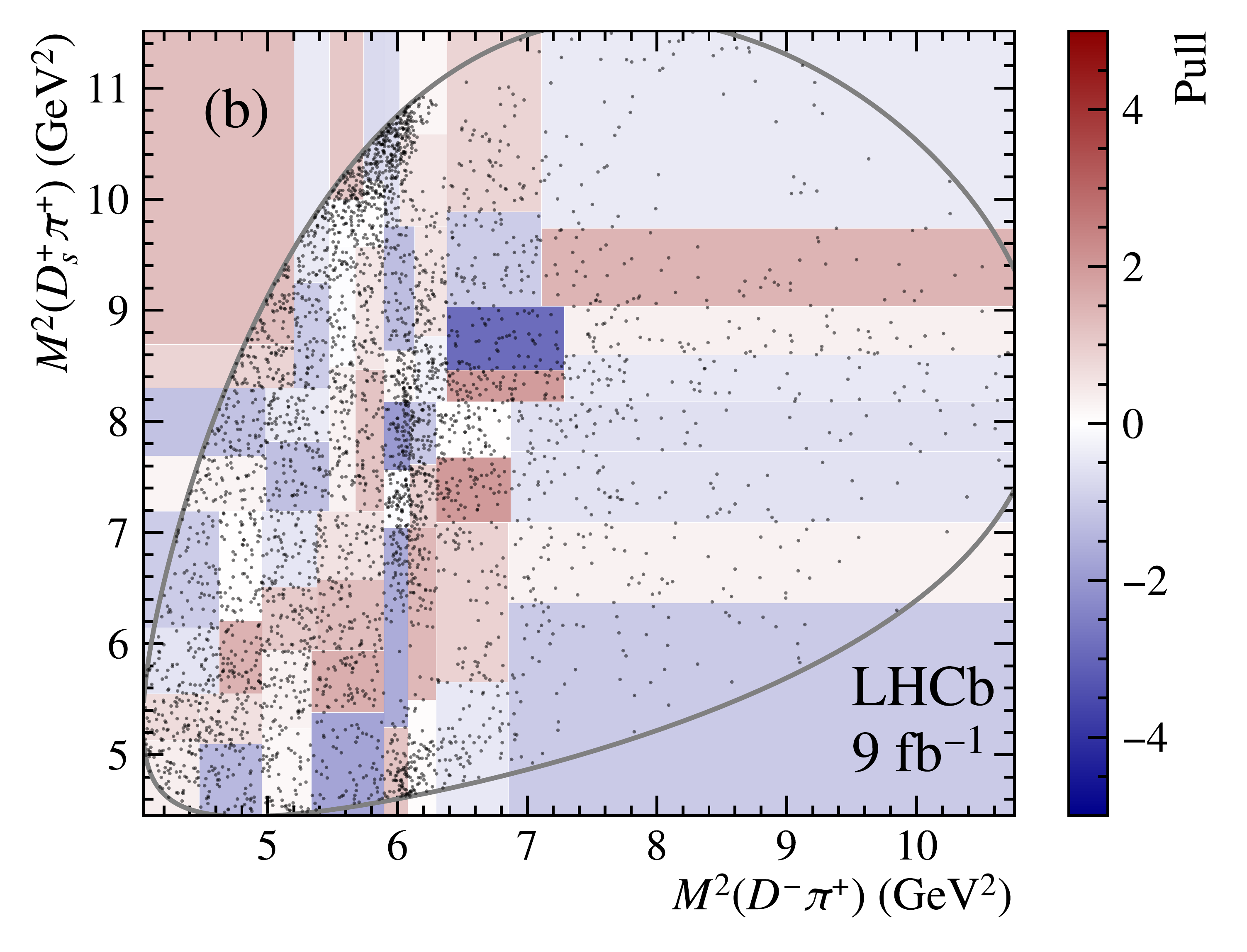}
  \end{center}
  \caption{Two-dimensional pull plots of the fits to (a) \BztoDzbarDsppim and (b) \BptoDmDsppip samples after including the \Zzz and \Zzpp states.}
  \label{fig:2Dpull}
\end{figure}

In the $M(\Dsp\pim)$ and $M(\Dsp\pip)$ distributions, both the peaks near $2.9\gev$ and the dips near $3.0\gev$ are better described by the presence of the new states and their interference with the existing $\Dstar$ states. The masses and widths of the $\Zzz$ and $\Zzpp$ states are listed in Table~\ref{tab:FitZMass}. 
Fit fractions are given in Tables~\ref{tab:FitResult_Bz} and~\ref{tab:FitResult_Bp} for \mbox{\BztoDzbarDsppim} and \mbox{\BptoDmDsppip} decays, respectively.
These results include the systematic uncertainties and corrections of fit bias, which are described in Sec.~\ref{sec:Systematic uncertainties}. The amplitudes and phases of the complex coefficients of the resonant contributions, relative to those of $\overline{D}{}^*_2(2460)$, are also displayed in Tables~\ref{tab:FitResult_Bz} and~\ref{tab:FitResult_Bp}. The two-dimensional pull plots are given in Fig.~\ref{fig:2Dpull}. 
The $\chi^2/ndf$ is $43.2/31$ and $63.0/31$ for \mbox{\BztoDzbarDsppim} and \mbox{\BptoDmDsppip} decays, respectively.
The distributions of Legendre polynomial weighted moments, together with the fit results with and without \Zzz and \Zzpp states, are shown in Appendix~\ref{app:Moments analysis}; these also suggest the existence of the new exotic states. The above model with a new $0^+$ \Zz is set as the default fit model. 

\begin{table}[!tb]
\centering
  \caption{
    Masses and widths of the $\Zzz$ and $\Zzpp$ states. The values are corrected for biases. The first and second uncertainties are statistical and systematic, respectively.
  }
    \begin{tabular}{lcc}
    \hline
    Particle        & Mass ($\gev$)                       & Width ($\gev$)                           \\
    \hline
$\Zzz$	            & 2.879$\,\pm\,$0.017$\,\pm\,$0.018	    & 0.153$\,\pm\,$0.028$\,\pm\,$0.020	\\
$\Zzpp$	            & 2.935$\,\pm\,$0.021$\,\pm\,$0.013	    & 0.143$\,\pm\,$0.038$\,\pm\,$0.025	\\
 
    \hline
    \end{tabular}
\label{tab:FitZMass}
\end{table}

\begin{table}[!tb]
\centering
  \caption{
    Amplitude, phase, and fit fraction of each component in the $\BztoDzbarDsppim$ fit result. The values are corrected for fit biases. The first and second uncertainties are statistical and systematic, respectively.
  }
    \begin{tabular}{lr@{$\,\pm\,$}c@{$\,\pm\,$}lr@{$\,\pm\,$}c@{$\,\pm\,$}lr@{$\,\pm\,$}c@{$\,\pm$\,}l}
    \hline
    Particle        & \multicolumn{3}{c}{Amplitude}                             & \multicolumn{3}{c}{Phase (rad)}                           & \multicolumn{3}{c}{Fraction (\%)}\\
    \hline
$\Zzz$	            & 0.223 & 0.044 & 0.048	    & $-$1.63 & 0.31 & 0.26	    & 4.8 & 1.1 & 1.4	\\
$D^{*}(2010)^{-}$	& 2.78 & 0.16 & 0.49	    & $-$2.90 & 0.11 & 0.10	    & 14.0 & 1.5 & 1.8	\\
$D_{2}^{*}(2460)$	& \multicolumn{3}{c}{1}                                     & \multicolumn{3}{c}{0}                                     & 22.1 & 1.1 & 0.6	\\
$D_{1}^{*}(2600)$	& 0.207 & 0.046 & 0.040	    & 0.36 & 0.24 & 0.19	    & 1.12 & 0.55 & 0.49	\\
$D_{3}^{*}(2750)$	& 0.174 & 0.046 & 0.062	    & $-$2.67 & 0.27 & 0.15	    & 0.40 & 0.25 & 0.31	\\
$D_{1}^{*}(2760)$	& 0.209 & 0.066 & 0.072	    & 0.22 & 0.29 & 0.27	    & 0.83 & 0.67 & 0.64	\\
$D_{J}^{*}(3000)$	& 0.72 & 0.32 & 0.61	    & 1.24 & 0.59 & 0.72	    & 0.09 & 0.13 & 0.16	\\
$D\pi$ S-wave	    & 0.995 & 0.067 & 0.081	    & $-$0.983 & 0.069 & 0.077	& 39.5 & 2.6 & 2.9	\\

    \hline
    \end{tabular}
\label{tab:FitResult_Bz}
\end{table}

\begin{table}[!tb]
\centering
  \caption{
    Amplitude, phase, and fit fraction of each component in the $\BptoDmDsppip$ fit result. The values are corrected for fit biases. The first and second uncertainties are statistical and systematic, respectively.
  }
    \begin{tabular}{lr@{$\,\pm\,$}c@{$\,\pm\,$}lr@{$\,\pm\,$}c@{$\,\pm\,$}lr@{$\,\pm\,$}c@{$\,\pm$\,}l}
    \hline
    Particle        & \multicolumn{3}{c}{Amplitude}                             & \multicolumn{3}{c}{Phase (rad)}                           & \multicolumn{3}{c}{Fraction (\%)}\\
    \hline
$\Zzpp$	            & 0.139 & 0.046 & 0.037	    & $-$0.79 & 0.37 & 0.25	& 1.96 & 0.87 & 0.88	\\
$D^{*}(2007)^{0}$	& 2.76 & 0.15 & 1.11	    & $-$3.03 & 0.10 & 0.43	& 15.7 & 1.5 & 2.0	\\
$D_{2}^{*}(2460)$	& \multicolumn{3}{c}{1}                                     & \multicolumn{3}{c}{0}                                 & 22.2 & 1.1 & 0.7	\\
$D_{1}^{*}(2600)$	& 0.228 & 0.050 & 0.086	    & $-$0.01 & 0.21 & 0.42	& 1.37 & 0.70 & 1.29	\\
$D_{3}^{*}(2750)$	& 0.110 & 0.043 & 0.042	    & 3.17 & 0.34 & 1.64	& 0.14 & 0.18 & 0.17	\\
$D_{1}^{*}(2760)$	& 0.089 & 0.056 & 0.207	    & $-$0.98 & 0.78 & 2.23	& 0.10 & 0.33 & 1.57	\\
$D_{J}^{*}(3000)$	& 1.74 & 0.34 & 1.87	    & 1.31 & 0.32 & 2.07	& 0.65 & 0.27 & 0.82	\\
$D\pi$ S-wave	    & 1.276 & 0.066 & 0.093	    & $-$0.926 & 0.063 & 0.113	& 52.6 & 3.1 & 2.5	\\

    \hline
    \end{tabular}
\label{tab:FitResult_Bp}
\end{table}

\subsection{Other models}
\label{sec:OtherTest}
Numerous additional resonances are tested to verify the stability of the default fit result and to better understand the system. First, the exotic $\Ds\pi$ states with other spin-parity hypotheses are tested. The $\Delta \rm LL$ values, with definition $\Delta \rm LL = NLL_{Default}-NLL_{Other}$, are summarized in Table~\ref{tab:othermodels}.
It is clear that at least one $\Ds\pi$ exotic state is needed to improve the value of NLL of the fit to each decay channel. 
The default model has the best fit quality, while the model with a spin-1 $\Ds\pi$ state also provides reasonably good description. The data are tested against the spin 0 and spin 1 hypotheses for the $\Zzz$ and $\Zzpp$ states. The results are shown in Sec.~\ref{sec:Exotics test} and $0^+$ is favored.

Secondly, the existence of extra $D\pi$, $\Ds\pi$ and $D\Ds$ states with natural spin-parity up to $3^-$  is explored when including the $0^+$ $\Zzz$ and $\Zzpp$ states. The masses and widths of the additional resonances are allowed to vary freely in the fit. When considering an additional spin-0 or spin-1 $\Ds\pi$ state for each decay, the mass of the new state converges near the $\Ds\pi$ mass threshold. 
The resulting changes in the value of NLL are insignificant, corresponding to statistical significance of less than $2\,\sigma$, and these states are not included in the default model.

\begin{table}[!tb]
  \small
  \caption{Tested fit models and the corresponding $\Delta \rm LL$ value.}
  \begin{center}\begin{tabular}{ l  c  c }
    \hline
    \multirow{2}{*}{Model}              & \multicolumn{2}{c}{$\Delta \rm LL$}\\
                                        & \BztoDzbarDsppim      & \BptoDmDsppip \\

    \hline
    Default model                       & \multirow{2}*{--}     & \multirow{2}*{--}     \\
    (One $\Ds\pi$ state ($0^+$))        &                               & \\
    \hline
    Variation of $\Ds\pi$ states & &\\
    No $\Ds\pi$ state                   & $-35$         & $-23$    \\
    One $\Ds\pi$ state ($1^-$)          & $-8$          & $-15$     \\
    One $\Ds\pi$ state ($2^+$)          & $-27$         & $-15$     \\
    Two $\Ds\pi$ states ($0^+$+$0^+$)   & $16$        & $11$     \\
    Two $\Ds\pi$ states ($0^+$+$1^-$)   & $11$        & $7$     \\
    \hline
    Additional $D\pi$ states & &\\
    $1^-$ $D\pi$ state                  & $5$         & $6$     \\
    $2^+$ $D\pi$ state                  & $15$        & $8$     \\
    $3^-$ $D\pi$ state                  & $6$         & $9$     \\
    \hline
    Additional $D\Ds$ states & &\\
    $0^+$ $D\Ds$ state                  & $15$        & $6$     \\
    $1^-$ $D\Ds$ state                  & $12$        & $9$     \\
    $2^+$ $D\Ds$ state                  & $6$         & $9$     \\

    \hline
  \end{tabular}\end{center}
  \label{tab:othermodels}
\end{table}

The $D\pi$ states with natural spin-parities have been well investigated by the amplitude analyses of $\Bz\to\Dzb\pip\pim$~\cite{LHCb-PAPER-2014-070}, $\Bp\to\Dm\pip\pip$~\cite{LHCb-PAPER-2016-026} and other topologically similar $B$-meson decays in \lhcb~\cite{LHCb-PAPER-2015-007,LHCb-PAPER-2015-017,LHCb-PAPER-2014-036} with large yields. No extra $D\pi$ state is expected to be observed in this analysis, which is consistent with the $D\pi$ results in Table~\ref{tab:othermodels}.
Additional $D\Ds$ exotic states with natural spin-parities are also found to be disfavored, which is consistent with the previous results~\cite{BESIII:2020qkh,BESIII:2022qzr,LHCb-PAPER-2020-044}, where only $1^{+-}$ $Z_{cs}$ states are observed.

\subsection{Results related to excited \texorpdfstring{\boldmath$D$}{D} states}
\label{sec:Dstar results}
With the $\Zzz$ and $\Zzpp$ states in place, the excited $D$ states are investigated. 
Some tension in the measured mass of the $D_2^*(2460)$ state between inclusive results~\cite{LHCb-PAPER-2013-026} and those obtained in amplitude fits~\cite{LHCb-PAPER-2014-070,LHCb-PAPER-2015-017,LHCb-PAPER-2015-007,LHCb-PAPER-2016-026} was
seen previously.
The mass and width of the $D_2^*(2460)$ states in the two decays are also investigated. The fit results in values of \mbox{$m_0=(2465.2\pm1.0)\mev$}, \mbox{$\Gamma_0=(38.7\pm2.5)\mev$} for the $D_2^*(2460)^-$ state and \mbox{$m_0=(2464.4\pm1.2)\mev$}, \mbox{$\Gamma_0=(44.6\pm2.8)\mev$} for the $D_2^*(2460)^0$ state. 
These results are in better agreement with earlier measurements in amplitude analyses. However they are also consistent with those obtained from inclusive results within $3\,\sigma$ when only considering statistical uncertainties\cite{PDG2022}.

The charged isospin partners of the $D_1^*(2600)^0$ and $D_J^*(3000)^0$ states have not yet been observed, their significances are estimated in the $\BztoDzbarDsppim$ decays based on the default fit model by fixing their masses and widths to the known values~\cite{PDG2022} and assuming the $D_J^*(3000)^0$ spin-parity to be $4^+$. The statistical significance of the $D_1^*(2600)^-$, and $D_J^*(3000)^-$ resonances is estimated to be $4.8\,\sigma$ and $2.2\,\sigma$, respectively. When the mass and width of the $D_1^*(2600)^-$ resonance are allowed to vary in the fit, they are determined to be $m_0=(2640\pm51)\mev$, $\Gamma_0=(122\pm35)\mev$ which are consistent with the default values. The masses and widths of the $D_J^*(3000)^-$ state cannot be determined due to the limited sample size. While the significance of these states is small they are still included in the default fit for a conservative evaluation of the exotic contributions.

\subsection{Search for \texorpdfstring{\boldmath$D_{s0}^*(2317)^{0}$}{Ds0*(2317)0} and \texorpdfstring{\boldmath$D_{s0}^*(2317)^{++}$}{Ds0*(2317)++}  states}
The nature of the $D_{s0}^*(2317)^+$ state is still in debate. Some theoretical models interpret the $D_{s0}^*(2317)^+$ state as an isoscalar $[cq\bar{s}\bar{q}]$ tetraquark state, and suggest searching for the isotriplet partners in the $\Dsp\pip$ and $\Dsp\pim$ final states~\cite{Cheng:2003kg,Browder:2003fk,Dmitrasinovic:2004cu,Dmitrasinovic:2012zz,Dmitrasinovic:2005gc,Maiani:2004vq}. The \BztoDzbarDsppim and \BptoDmDsppip decays are ideal for such studies.

However, in Sec.~\ref{sec:OtherTest}, additional $\Ds\pi$ exotic states with freely varying mass and width values, under different spin-parity hypotheses, are found to be insignificant. By assuming that the masses of the neutral and doubly charged partners are the same as that of the $D_{s0}^*(2317)^+$ state, the upper limit on fit fractions with three different scenarios is evaluated. The first scenario is that the natural width of the new $D_{s0}^*(2317)$ state is ignored. A Gaussian function is used to describe the lineshape of the new state, of which the width represents the detector resolution. 
The second is that a Breit-Wigner function is added to model the $J^P=0^+$, $\Ds\pi$ state, of which the width is set to $3.8\mev$, the current upper limit on the $D_{s0}^*(2317)^+$ width~\cite{PDG2022}. No resolution effect is considered. The third is that 
an additional $0^+$ $\Ds\pi$ Breit--Wigner function, the width of which is set to be the same as the $\Zzz$ and $\Zzpp$ states, is included in the fit model. The upper limits on the fit fractions of neutral and doubly charged $D_{s0}^*(2317)$ with different hypotheses at 90\% confidence level (C.L.), which are all less than 1\%, are summarized in Table~\ref{tab:Ds2317}.

\begin{table}[!tb]
  \small
  \caption{Upper limit on the fit fractions of neutral and doubly charged $D_{s0}^*(2317)$ with different hypotheses at 90\% C.L.}
  \begin{center}\begin{tabular}{ c  c  c }
    \hline
    Hypothesis  & \BztoDzbarDsppim      & \BptoDmDsppip \\
    \hline
    (1)         & 0.063\%     & 0.025\%     \\
    (2)         & 0.053\%     & 0.137\%     \\
    (3)         & 0.861\%     & 0.595\%     \\

    \hline
  \end{tabular}\end{center}
  \label{tab:Ds2317}
\end{table}

\subsection{Significances, spin analysis and Argand plot}
\label{sec:Exotics test} 
\begin{figure}[!tb]
  \begin{center}
    \includegraphics[width=0.48\linewidth]{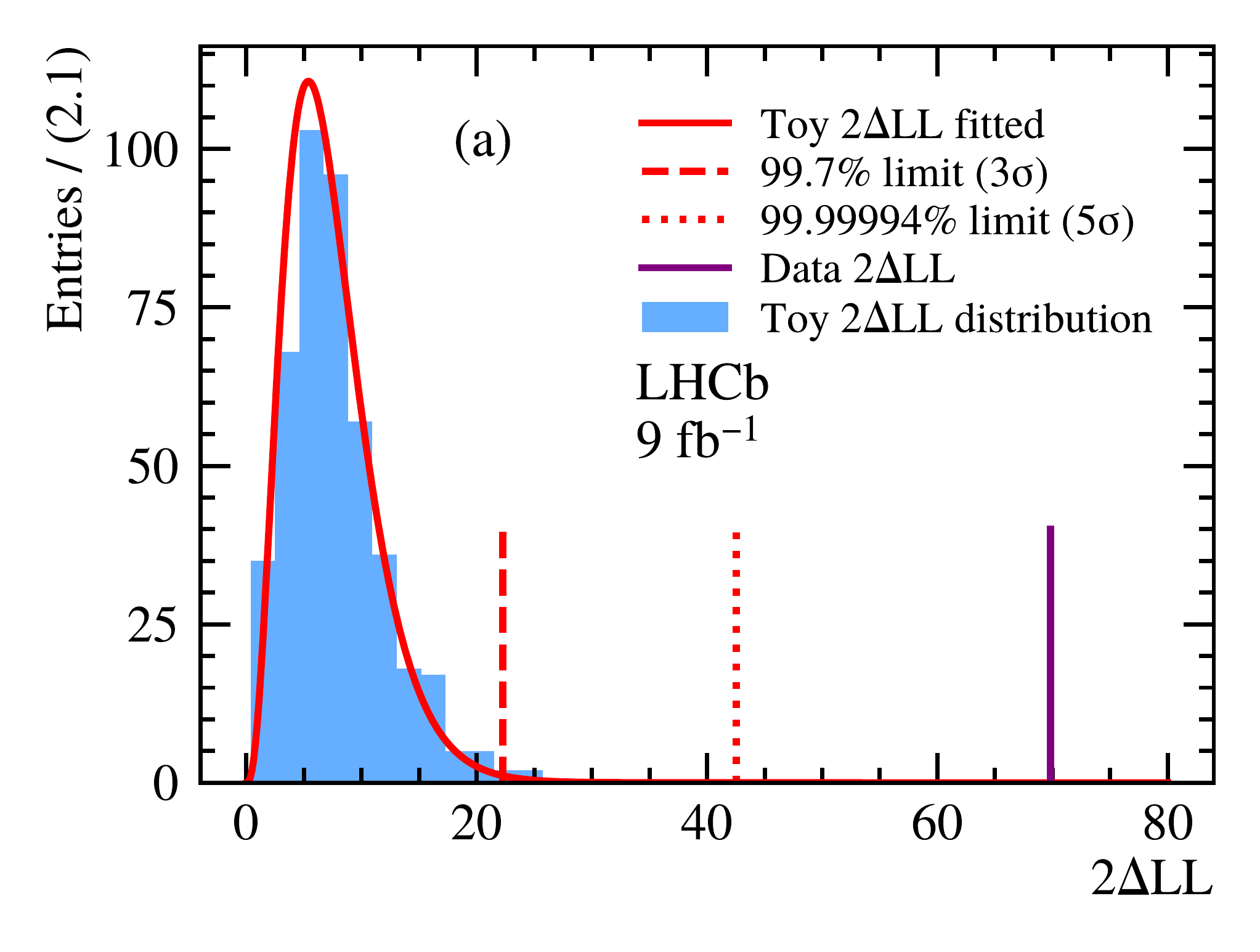}
    \includegraphics[width=0.48\linewidth]{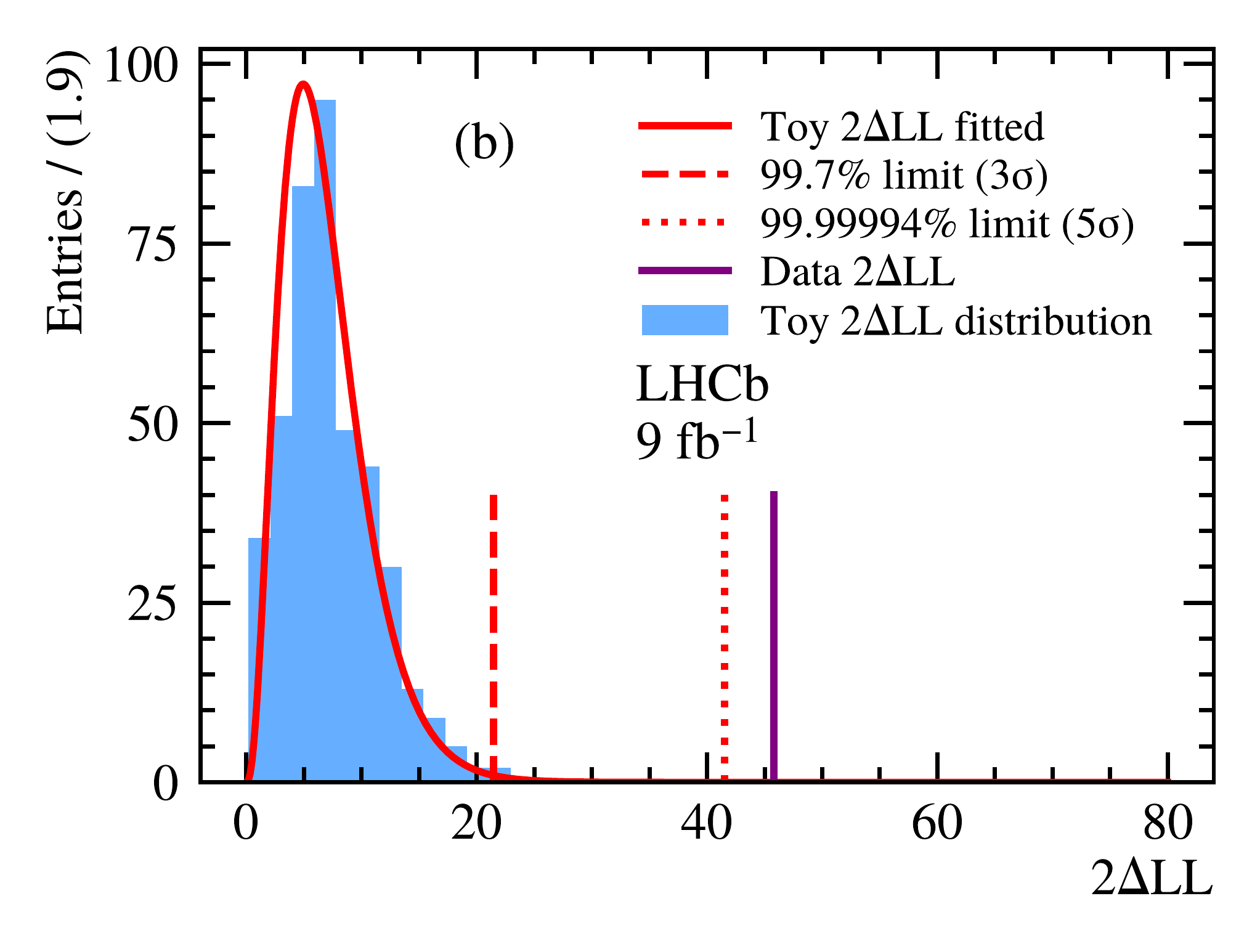}
  \end{center}
  \caption{Significance tests for the new $D_s\pi$ states in the (a) \BztoDzbarDsppim and (b) \BptoDmDsppip samples. The blue histogram is the distribution of $2\Delta{\rm LL}$ and the red solid curve shows the fitted $\chi^2$ PDF. The red dashed and dotted lines are the $2\Delta{\rm LL}$ values corresponding to $3 \sigma$ and $5 \sigma$, and the purple solid line is the $2\Delta{\rm LL}$ measured in the data.}
  \label{fig:sigtest}
\end{figure}

Pseudoexperiments are carried out to determine the significance of the $\Zzz$ and $\Zzpp$ states, accounting for the look-elsewhere effect~\cite{LHCb-PAPER-2020-044,Lyons:2008hdc}. The pseudoexperiments use data generated according to the fit results without the new $\Ds\pi$ resonances, and the yield of each generated sample follows a Poisson distribution whose mean is the yield in the corresponding dataset. The pseudodatasets are fitted with the model with and without the $\Ds\pi$ contributions. The difference in the value of $2\Delta{\rm LL}$ between these two fits is obtained, and fitted with a $\chi^2$ PDF. The distributions and fit results are shown in Fig.~\ref{fig:sigtest} for the two decays. The numbers of degrees of freedom after considering the look-elsewhere effect are found to be $7.39\pm0.17$ and $6.93\pm0.17$ for the $\Zzz$ and $\Zzpp$ states, respectively, with the corresponding significances estimated to be $7.3\,\sigma$ and $5.3\,\sigma$. After accounting for the systematic uncertainties discussed in Sec.~\ref{sec:Systematic uncertainties}, these are reduced to $6.6\,\sigma$ and $4.8\,\sigma$.

\begin{figure}[!tb]
  \begin{center}
    \includegraphics[width=0.48\linewidth]{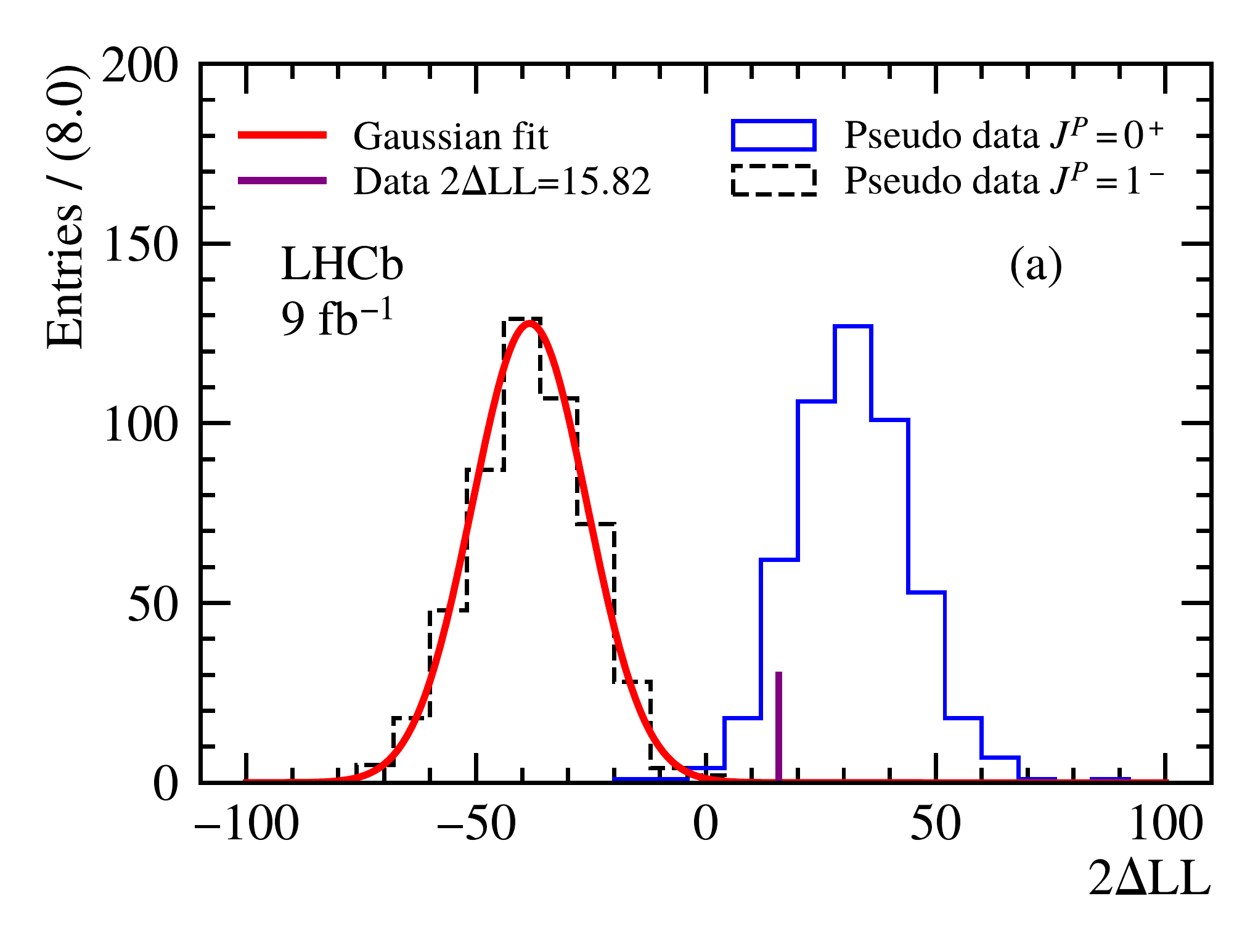}
    \includegraphics[width=0.48\linewidth]{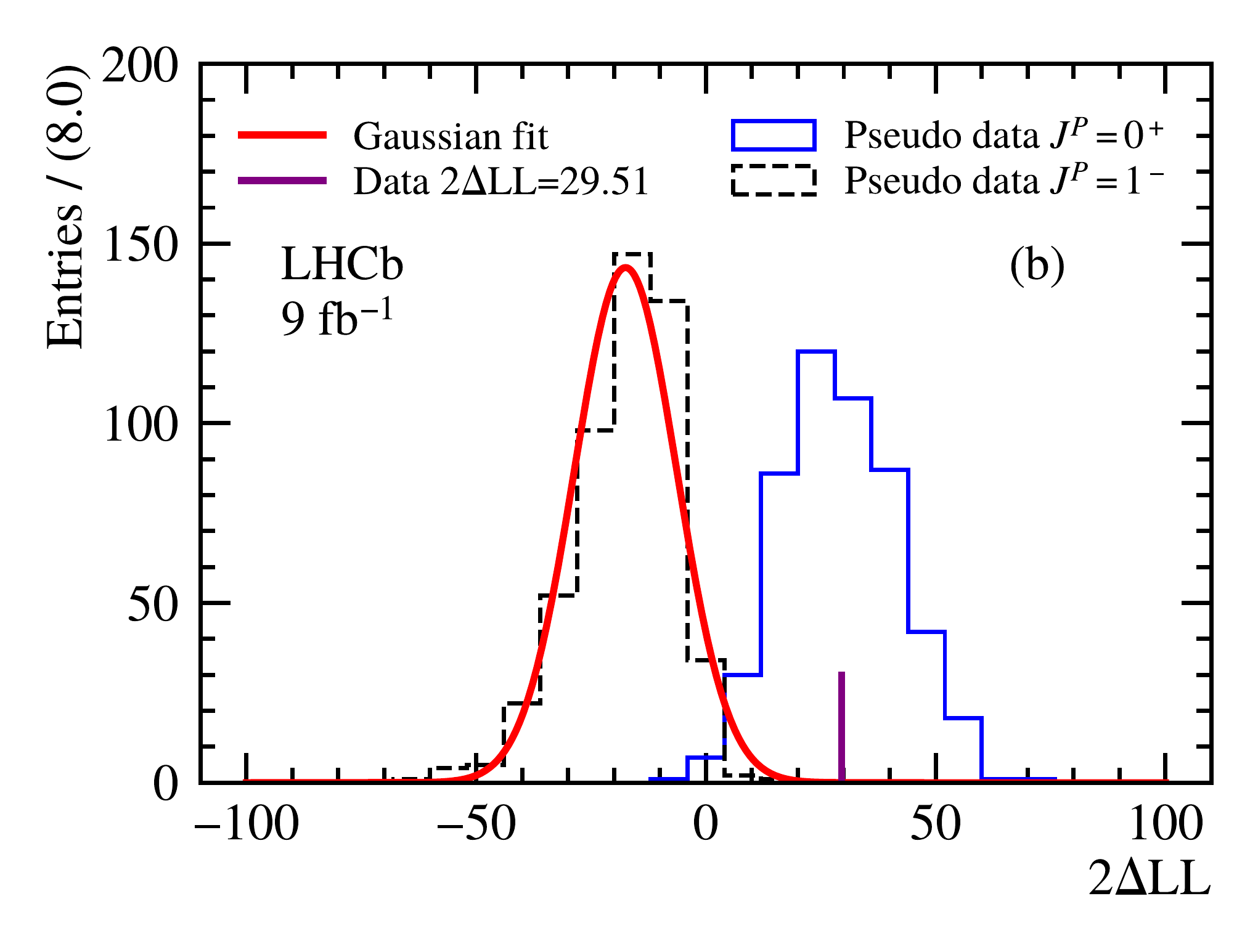}
  \end{center}
  \caption{Spin analysis of (a) \BztoDzbarDsppim and (b) \BptoDmDsppip decays. The blue solid and black dashed histograms are the distributions of the $2\Delta{\rm LL}$ for the pseudoexperiments generated based on the fit results with $0^+$ or $1^-$ $\Ds\pi$ exotic state, respectively. The purple vertical line shows the $2\Delta{\rm LL}$ value for the data fitted with the new $\Ds\pi$ exotic state under the $J^P=0^+$ and $J^P=1^-$ hypotheses. The red curve is the result of a fit to the black dashed histogram with a Gaussian function.}
  \label{fig:spinana}
\end{figure}

The spin-parity values for the $\Zzz$ and $\Zzpp$ states are also determined using pseudoexperiments. For each decay, 500 pseudoexperiments are generated based on the fit results with the $0^+$ $\Ds\pi$ state included, while another 500 pseudoexperiments are generated according to the fit results using a model assuming the $1^-$ spin hypothesis. Each pseudodataset is fitted in the same way as for data. The $2\Delta{\rm LL}$ between the two fits is calculated, and the distributions of $2\Delta{\rm LL}$ for the two sets of pseudoexperiments are shown in Fig.~\ref{fig:spinana}. The $2\Delta{\rm LL}$ distribution of the pseudoexperiments with the $1^-$ $\Ds\pi$ hypothesis is fitted with a Gaussian function, and the significance of the data disfavoring the $1^-$ hypothesis is evaluated to be $4.3\,\sigma$ for the $\Zzz$ state and $4.2\,\sigma$ for the $\Zzpp$ state when no isospin relationship is imposed on the $D\pi$ or $D_s\pi$ components.

\begin{figure}[!tb]
  \begin{center}
    \includegraphics[width=0.48\linewidth]{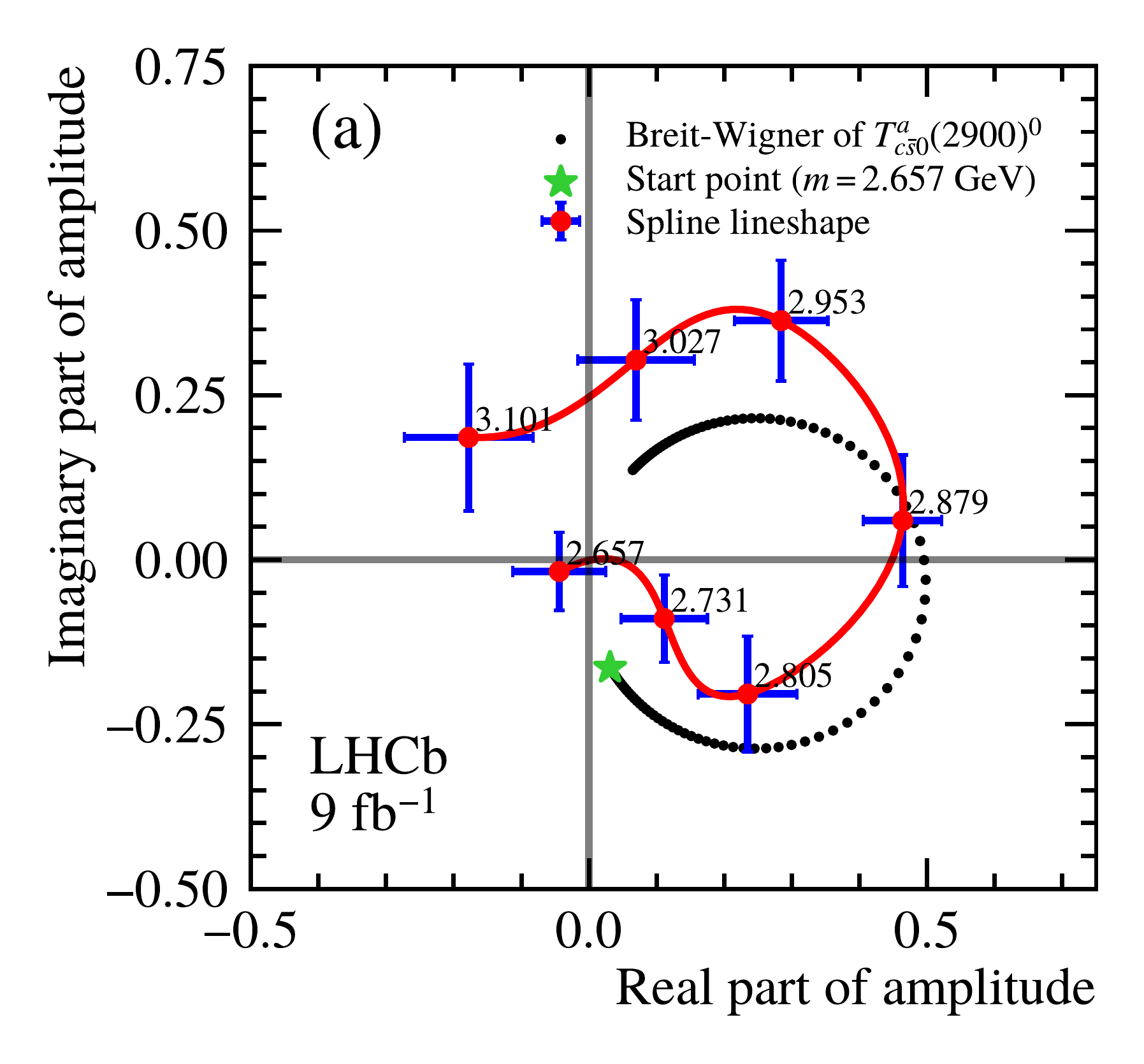}
    \includegraphics[width=0.48\linewidth]{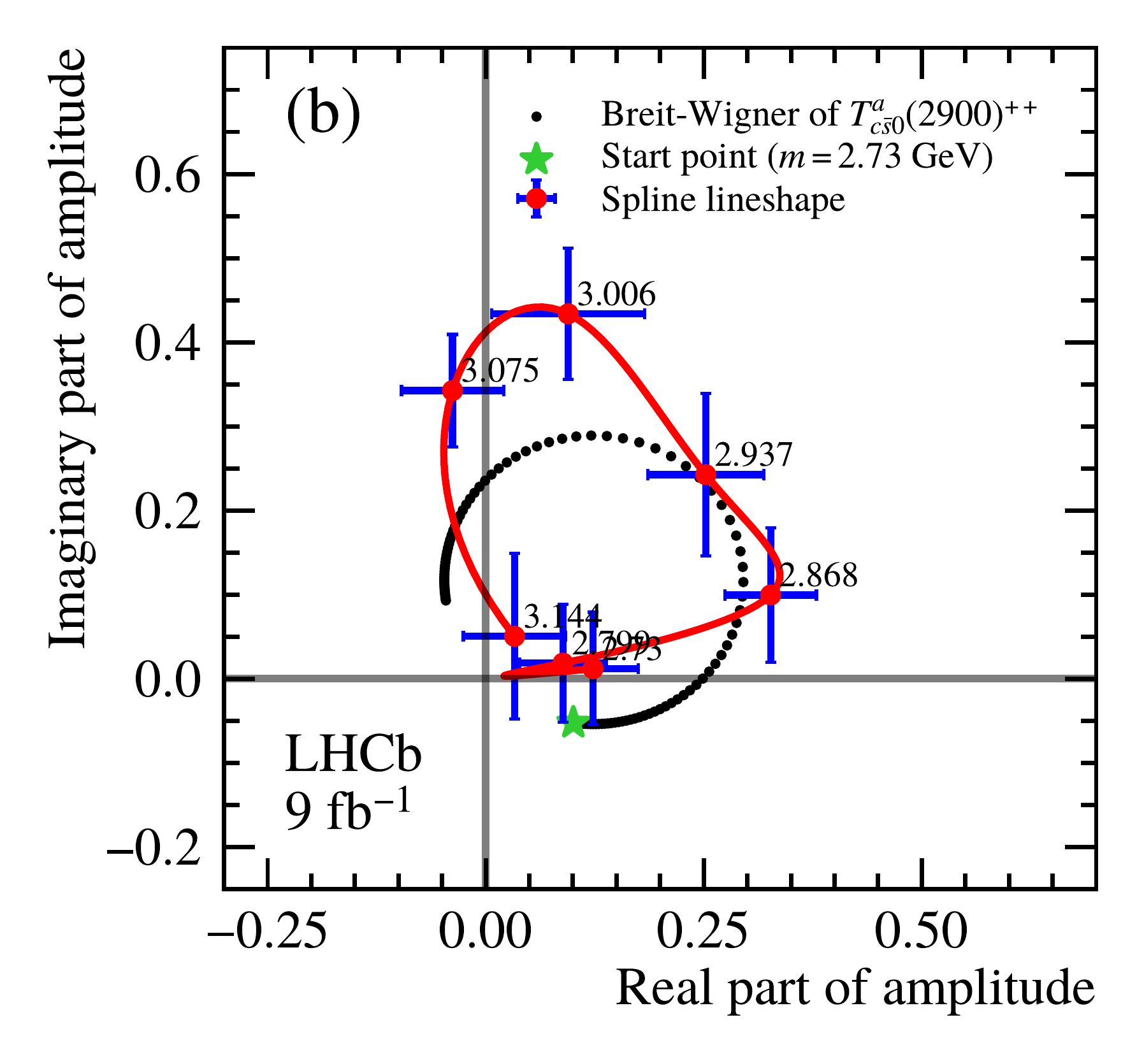}
  \end{center}
  \caption{Argand diagrams of the (a) \BztoDzbarDsppim and (b) \BptoDmDsppip decays. Black dots show the lineshape of the $\Zzz$ and $\Zzpp$ Breit--Wigner functions. The red solid line and blue error bars show the lineshape and fit results of spline $0^+$ $\Ds\pi$ model in $\Zzz$ and $\Zzpp$ mass region.}
  \label{fig:argand}
\end{figure}

The Argand diagrams~\cite{PDG2022} of the $\Zzz$ and $\Zzpp$ states are shown in Fig.~\ref{fig:argand}. The Breit--Wigner function follows a counter-clockwise circular path on the complex plane while contributions which are not genuine resonances are expected to have a different shape. Seven spline points on $M_{\Ds\pi}$ near the measured mass of the $\Zzz$ and $\Zzpp$ states ($m\pm1.5\Gamma$) are used to model these regions instead of Breit--Wigner functions. The complex parameters of all points are allowed to vary in the fit. 
%For the real resonances, the spline lineshape on the Argand diagrams should have similar behavior as the Breit--Wigner distribution, otherwise it would be messed.
The fitted parameters of the spline points, together with the $\Ds\pi$ lineshape are shown in Fig.~\ref{fig:argand}. The spline lineshape shows similar behavior as the Breit--Wigner distribution, which confirms the resonant character of the two new exotic states.

\subsection{Model-dependent results}
\label{sec:MDresult}
\begin{figure}[!tb]
  \begin{center}
    \includegraphics[width=0.45\linewidth]{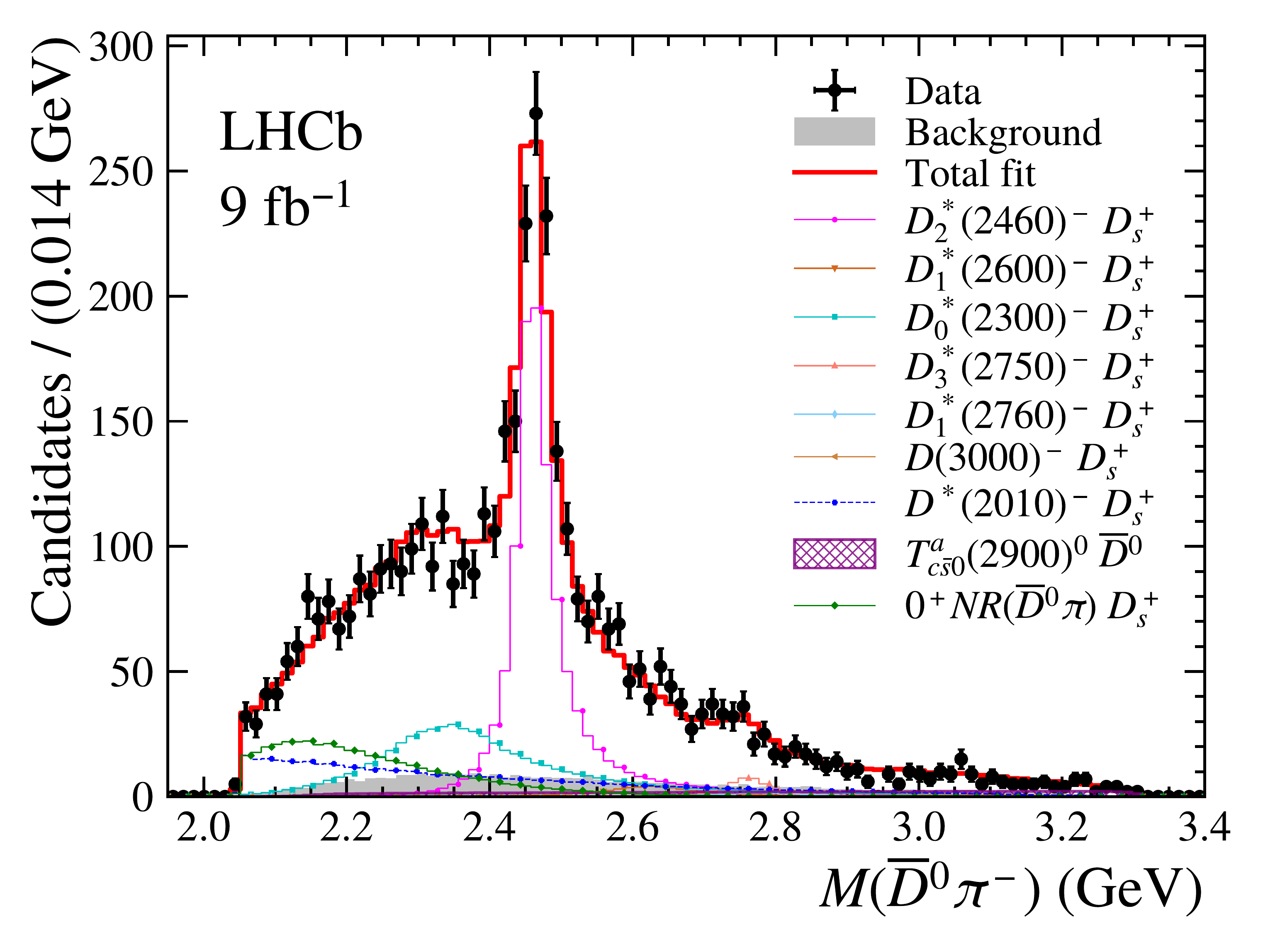}\put(-155,100){(a)}
    \includegraphics[width=0.45\linewidth]{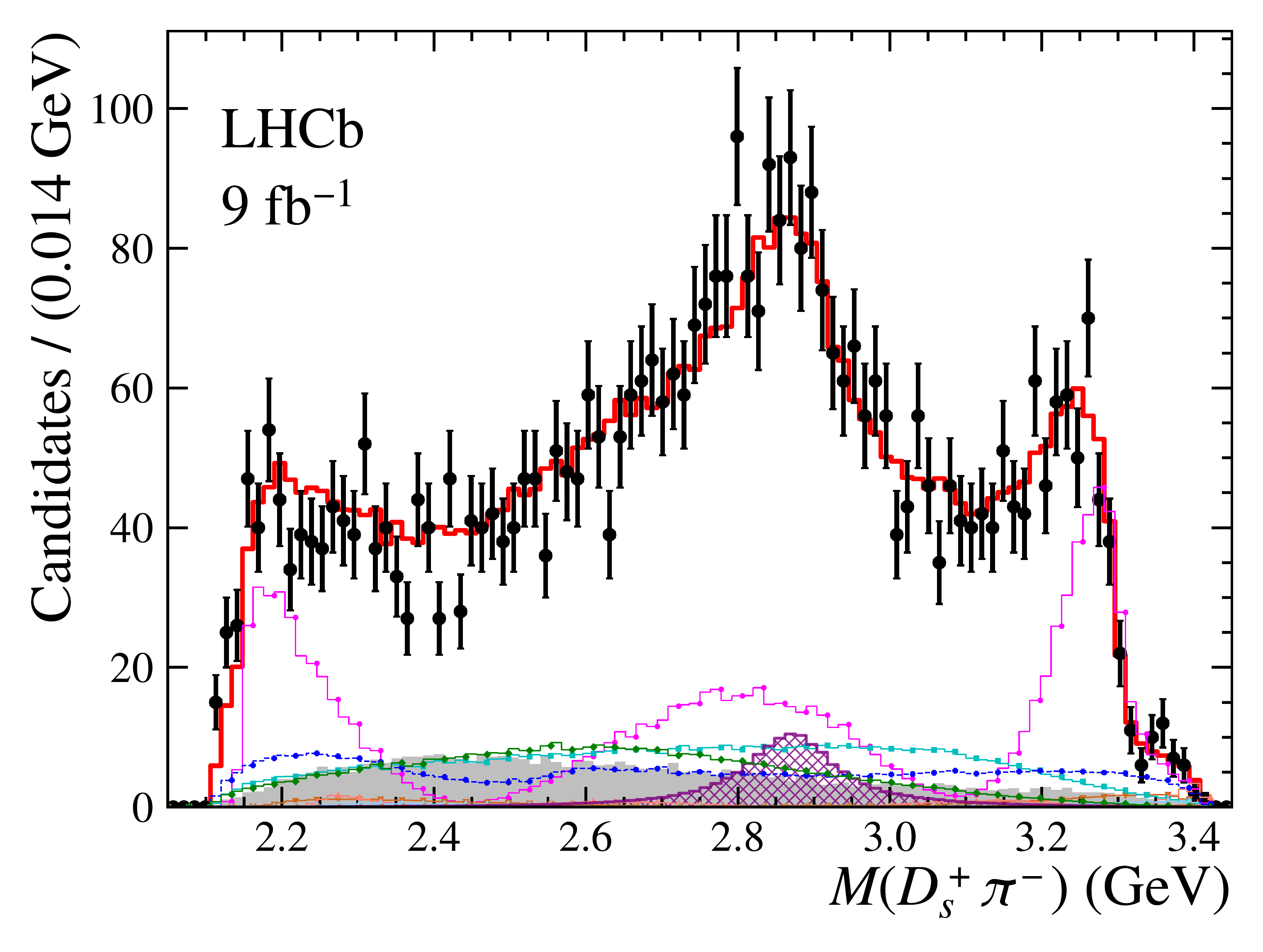}\put(-155,100){(b)}
    
    \end{center}
  \caption{The MD description of the (a) $M(\Dzb\pim)$ and (b) $M(\Dsp\pim)$ distributions for the \BztoDzbarDsppim decays.}
  \label{fig:MD_fit_Bz}
\end{figure}

\begin{figure}[!tb]
  \begin{center}
    \includegraphics[width=0.45\linewidth]{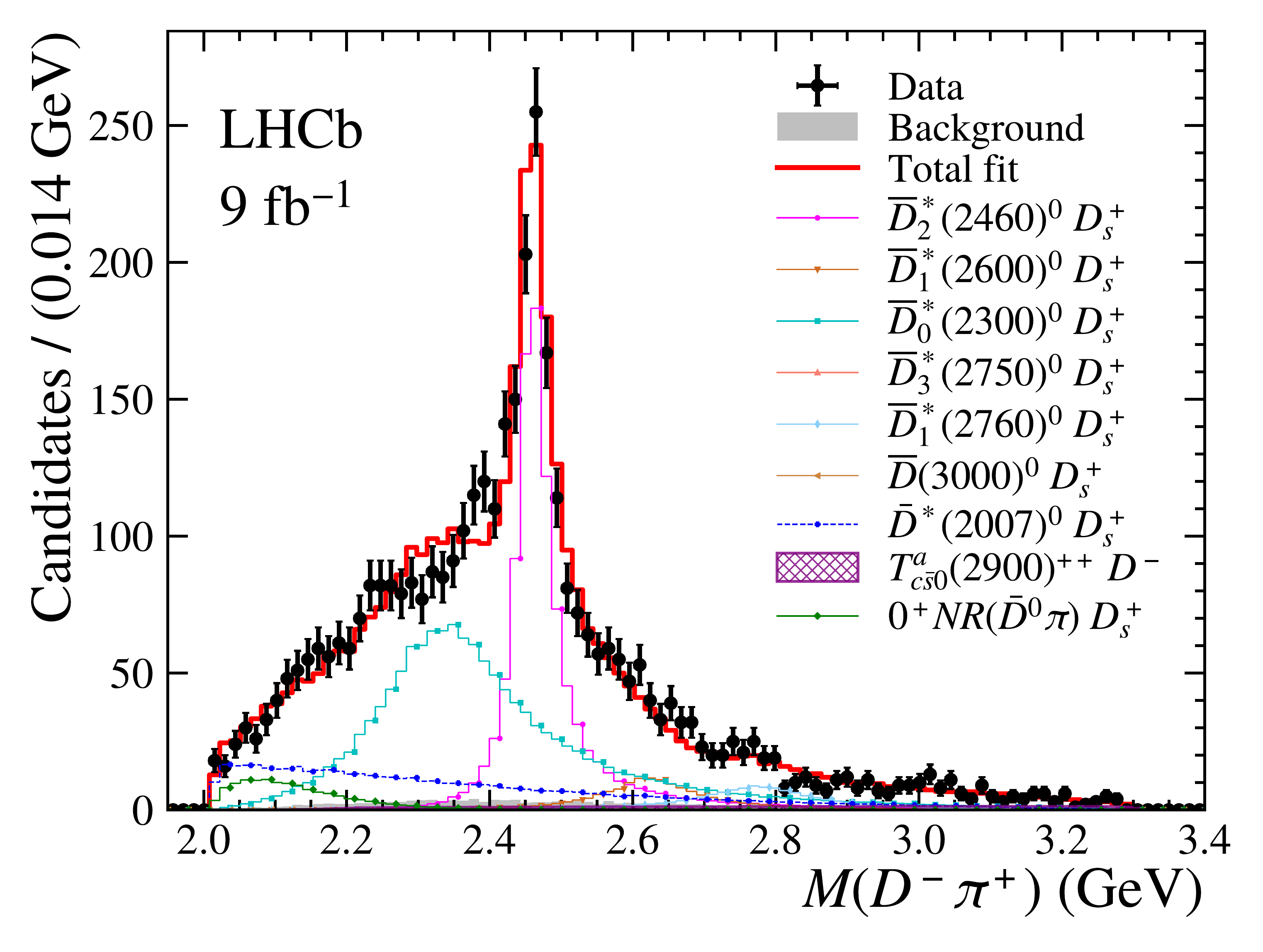}\put(-155,100){(a)}
    \includegraphics[width=0.45\linewidth]{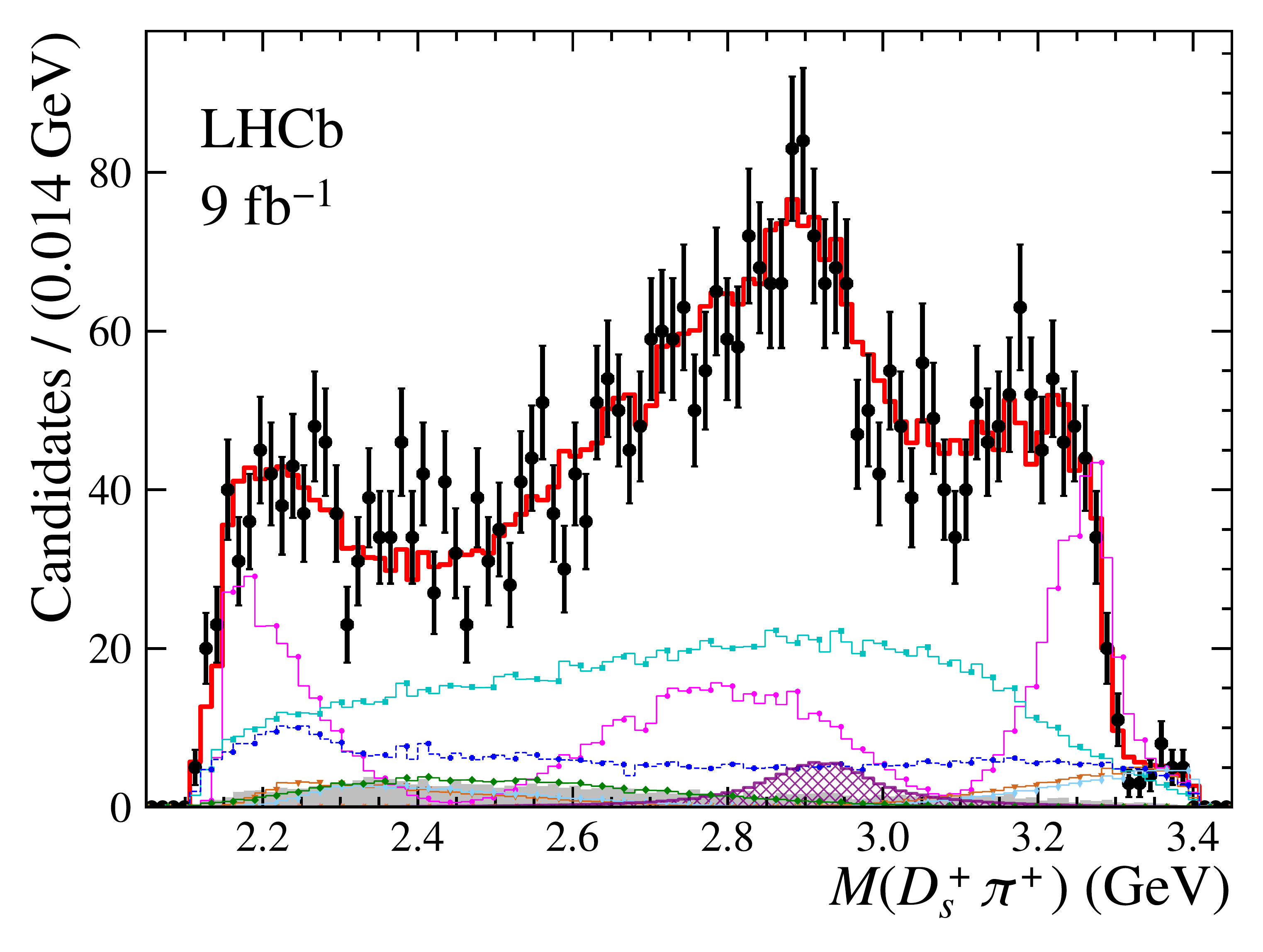}\put(-155,100){(b)}
    
  \end{center}
  \caption{The MD description of the (a) $M(\Dm\pip)$ and (b) $M(\Dsp\pip)$ distributions for the \BptoDmDsppip decays.}
  \label{fig:MD_fit_Bp}
\end{figure}

\begin{figure}[!tb]
  \begin{center}
    \includegraphics[width=0.48\linewidth]{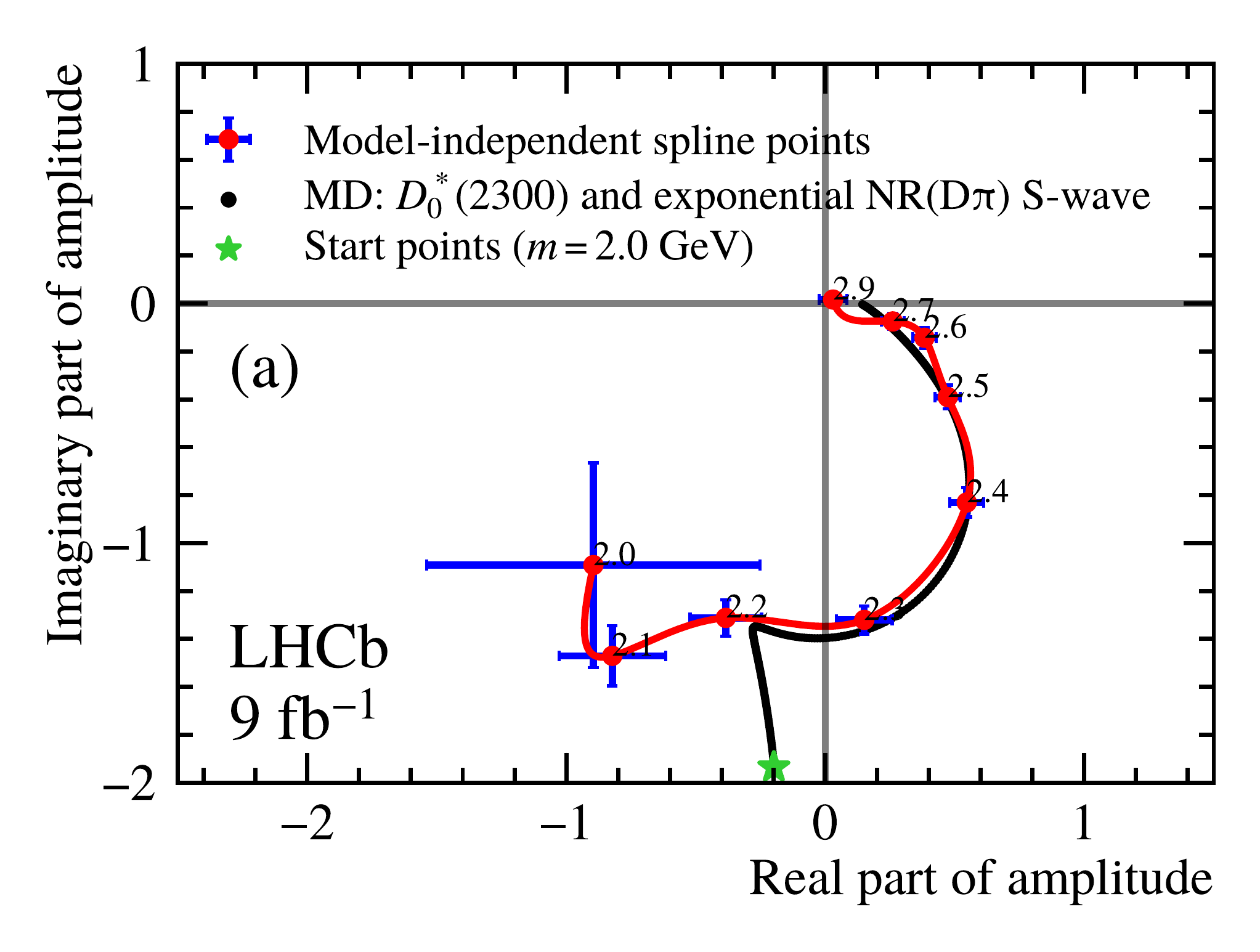}
    \includegraphics[width=0.48\linewidth]{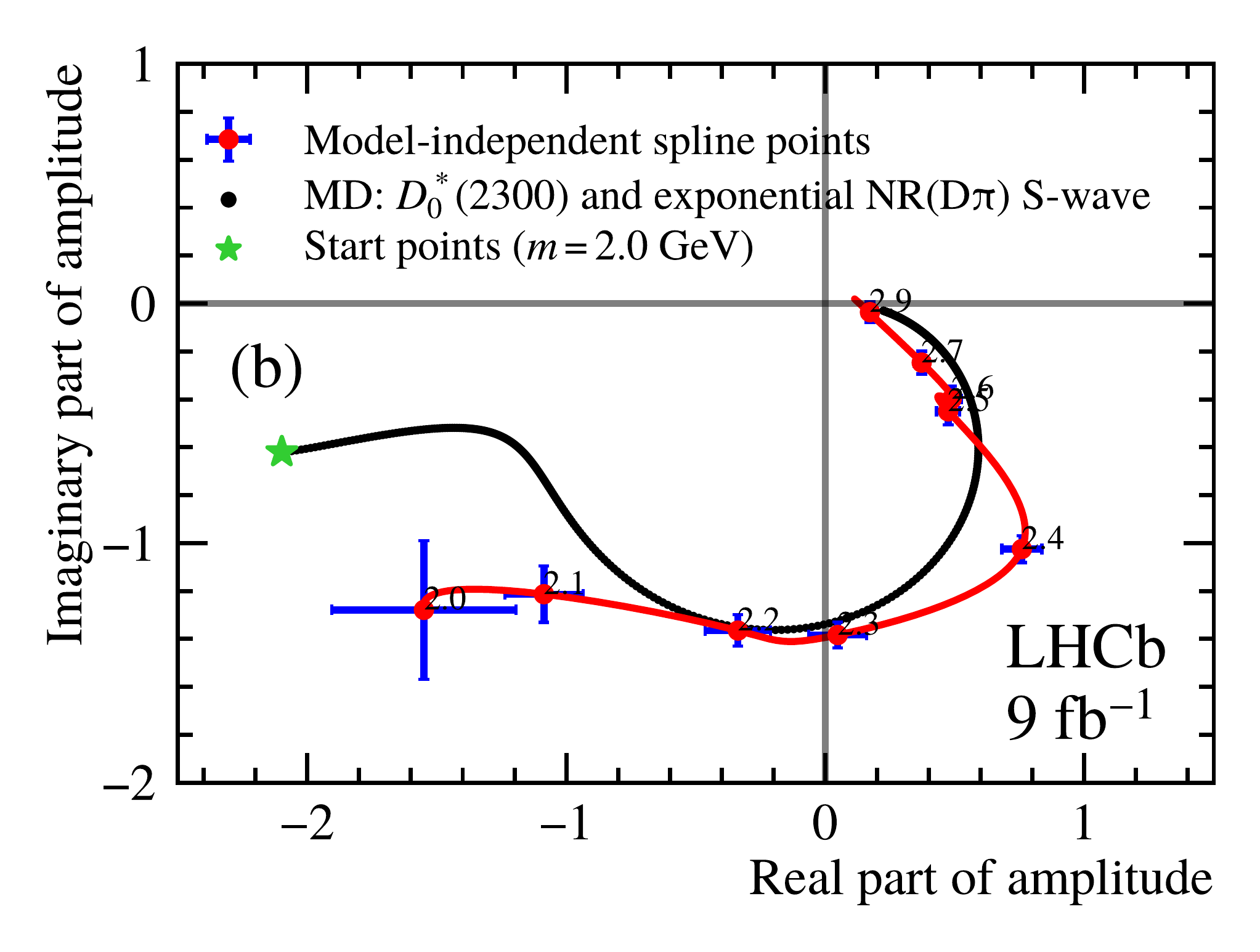}
  \end{center}
  \caption{Argand diagrams of the MD and qMI $0^+$ $D\pi$ components of the (a) \BztoDzbarDsppim and (b) \BptoDmDsppip decays. The red solid line and blue error bars show the lineshape and fit results of qMI $0^+$ $D\pi$ spline model, while the black line is the lineshape of the MD $0^+$ $D\pi$ component.}
  \label{fig:argand_DPi}
\end{figure}

Instead of modeling the $D\pi$ S-wave with the qMI description, 
the amplitude fit with a fully model-dependent (MD) description is carried out. In the MD description, the $0^+$ qMI model is replaced by the RBW of the $D_0^*(2300)$ component together with a $0^+$ nonresonant component, while the parameters of all the other resonances are set to be the same. The obtained parameters of the $\Zzz$ and $\Zzpp$ states are summarized in Table~\ref{tab:fittedZ_MD} and are consistent with those determined using the qMI model, within statistical uncertainties. Figures~\ref{fig:MD_fit_Bz} and \ref{fig:MD_fit_Bp} show the fit results of the MD description of the \BztoDzbarDsppim and \BptoDmDsppip decays, respectively. The $\chi^2/ndf$ is $79.1/44$ for the \BztoDzbarDsppim decay and $114.7/44$ for the \BptoDmDsppip decay. As a comparison, the Argand diagrams of the MD and qMI $0^+$ $D\pi$ components of the two decays are shown in Fig.~\ref{fig:argand_DPi}. The MD and qMI spline points after $2.2\gev$ are in good agreement. These results serve to validate the default fit.

\begin{table}[!tb]
\centering
  \caption{Masses, widths and fit fractions of the $\Zzz$ and $\Zzpp$ states obtained from the MD fit.}
\begin{tabular}{ l c c c c c c}
\hline
Model           & $\Delta{\rm LL}$        & Mass (GeV)   & Width (GeV)    & Fraction (\%) \\
\hline

$\Zzz$     & 70.1      & $2.871\pm0.012$   & $0.135\pm0.025$   & $3.0\pm0.5$   \\
$\Zzpp$    & 33.2      & $2.922\pm0.014$   & $0.161\pm0.033$   & $2.0\pm0.5$   \\
\hline
  \end{tabular}
  \label{tab:fittedZ_MD}
\end{table}

%%%%%%%%%%%%%%%%%%%%%%%%%%%%%%%%%%%%%%%%%%%%%%%%%%%%%%%
%%%%%           Simultaneous Dpi fit model        %%%%%
%%%%%%%%%%%%%%%%%%%%%%%%%%%%%%%%%%%%%%%%%%%%%%%%%%%%%%%
\section{Simultaneous \texorpdfstring{\boldmath$D\pi$}{Dpi} fit model}
\label{sec:simuDpi}
In the default fit model, all known $D^{**}$ states with natural spin-parity~\cite{PDG2022} are included.
However, their fit fractions, except those of the $D_2^*(2460)$ and $D^*_1(2600)$ states, are consistent with 0, as shown in Tables~\ref{tab:FitResult_Bz} and~\ref{tab:FitResult_Bp}. Moreover, the parameters of the qMI $0^+$ $D\pi$ spline points in the higher $D\pi$ mass region have large uncertainties due to the smaller sample size.
To improve the precision and stability of the fit results, a simultaneous fit of the \BztoDzbarDsppim and \BptoDmDsppip decays is performed, as the two decays are related by isospin symmetry. In the simultaneous fit, all complex parameters of the $D^{**}$ states are shared, 
except for the $\overline{D}{}^*(2007)^0$ and $D^*(2010)^-$ states allowing for small isospin symmetry breaking effects near the $D\pi$ mass threshold.
The simultaneous fit results with only $D\pi$ states are shown in Ref.~\cite{PRLsister}. The description of the mass spectra is consistent with the separate fit as shown in Sec.~\ref{sec:onlyDpi}, which supports the feasibility to perform simultaneous fit for the two decays. 

\begin{figure}[!tb]
  \begin{center}
    \includegraphics[width=0.45\linewidth]{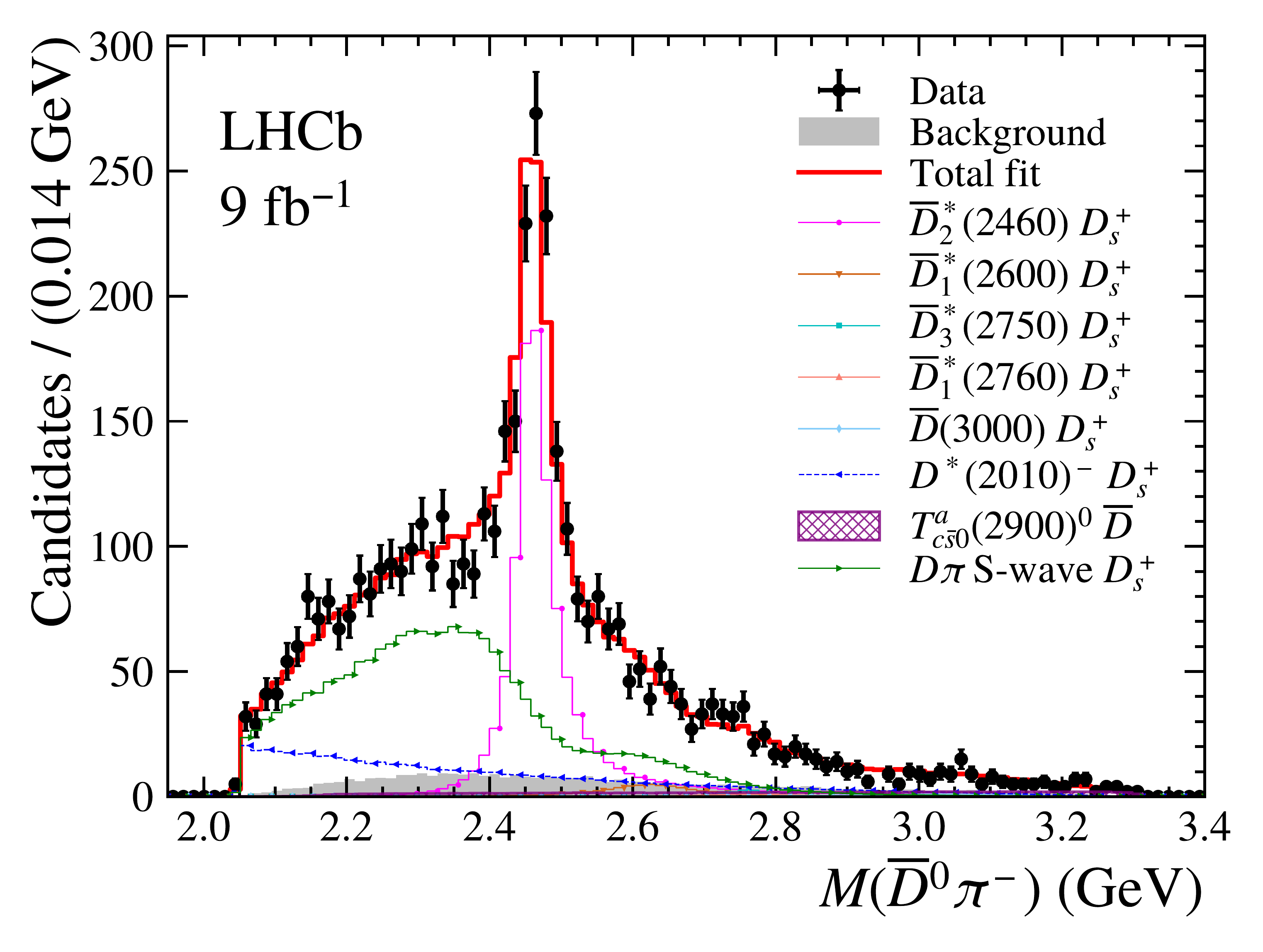}\put(-155,100){(a)}
    \includegraphics[width=0.45\linewidth]{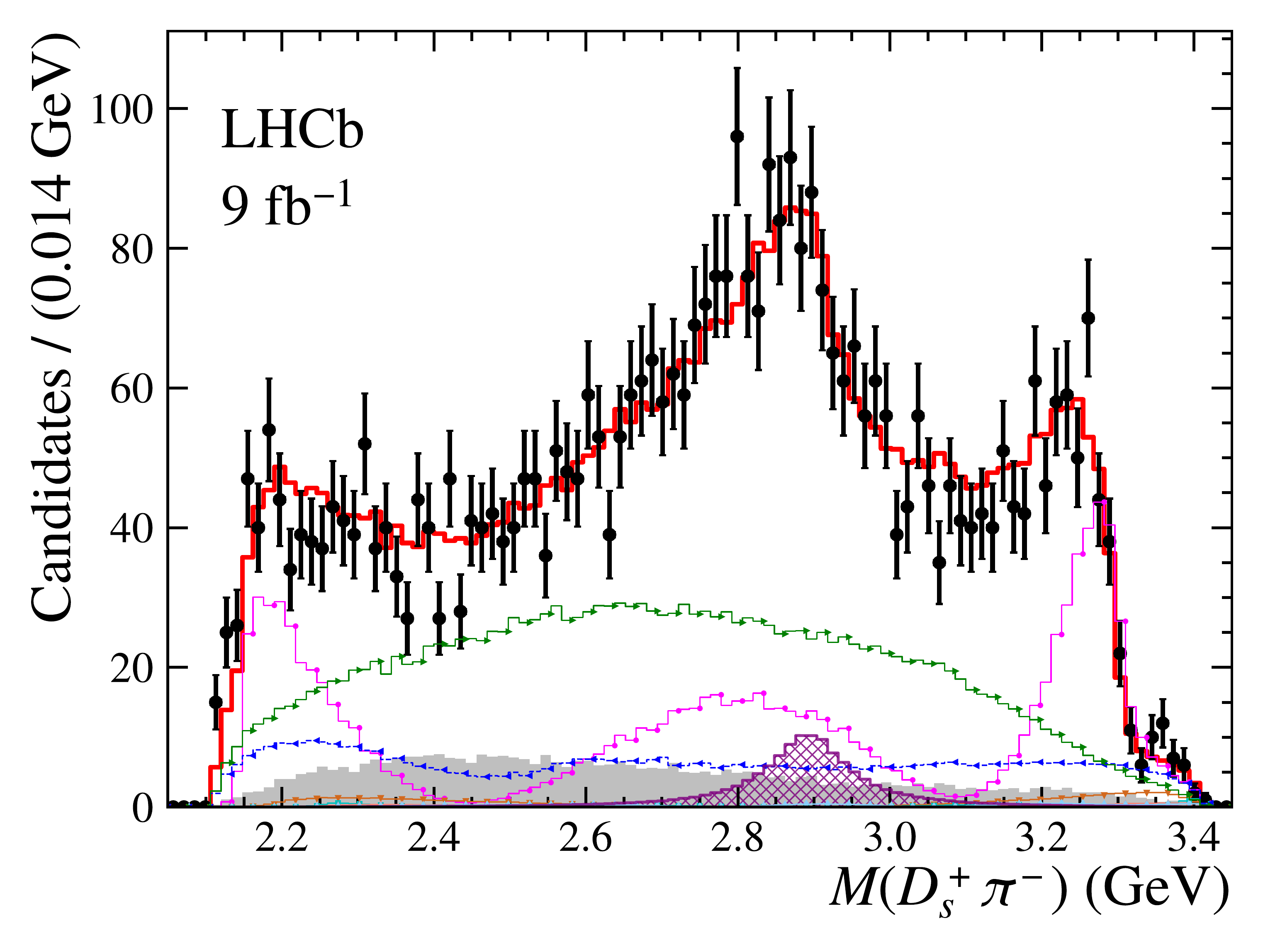}\put(-155,100){(b)}\\
    \includegraphics[width=0.45\linewidth]{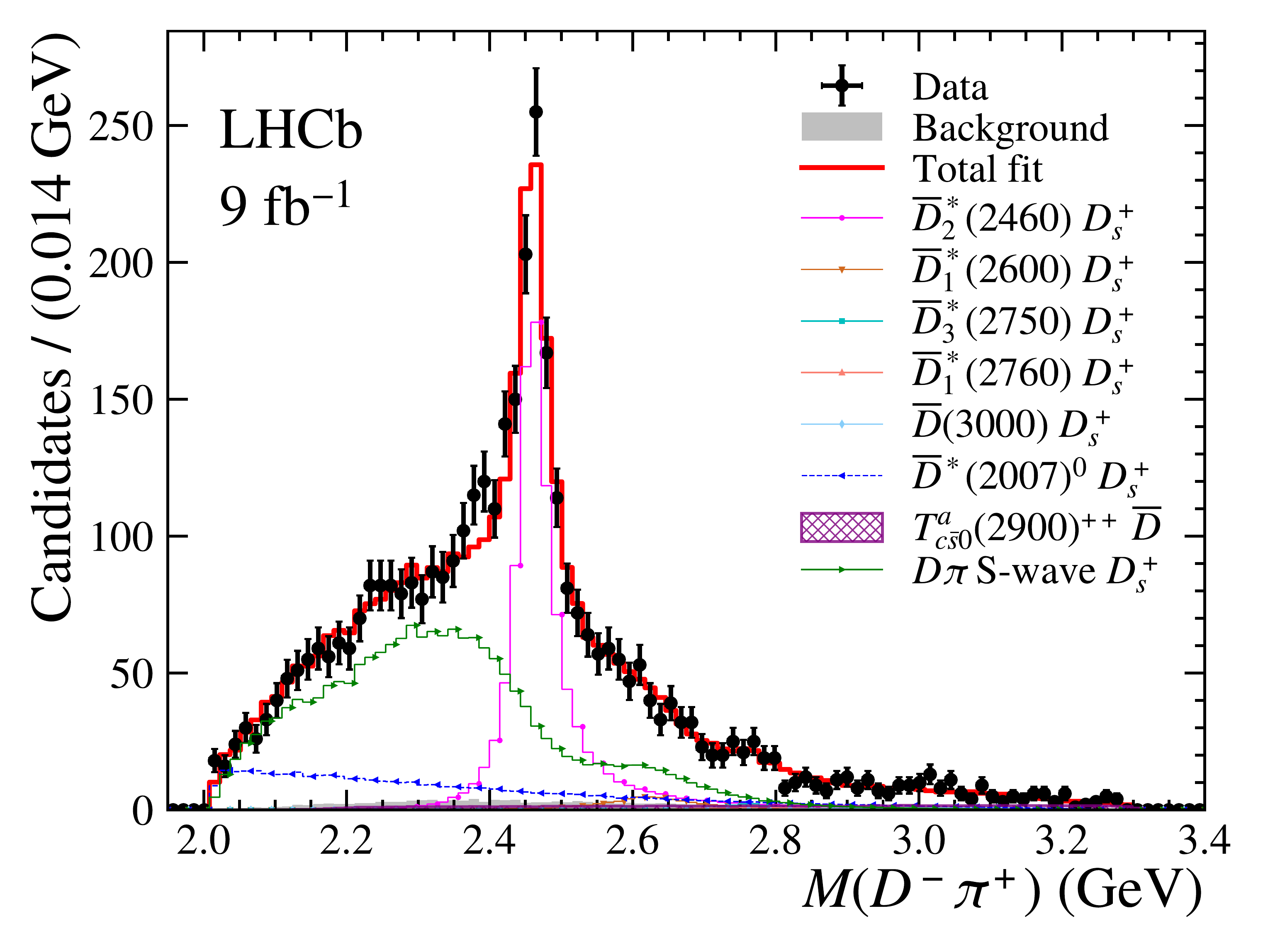}\put(-155,100){(c)}
    \includegraphics[width=0.45\linewidth]{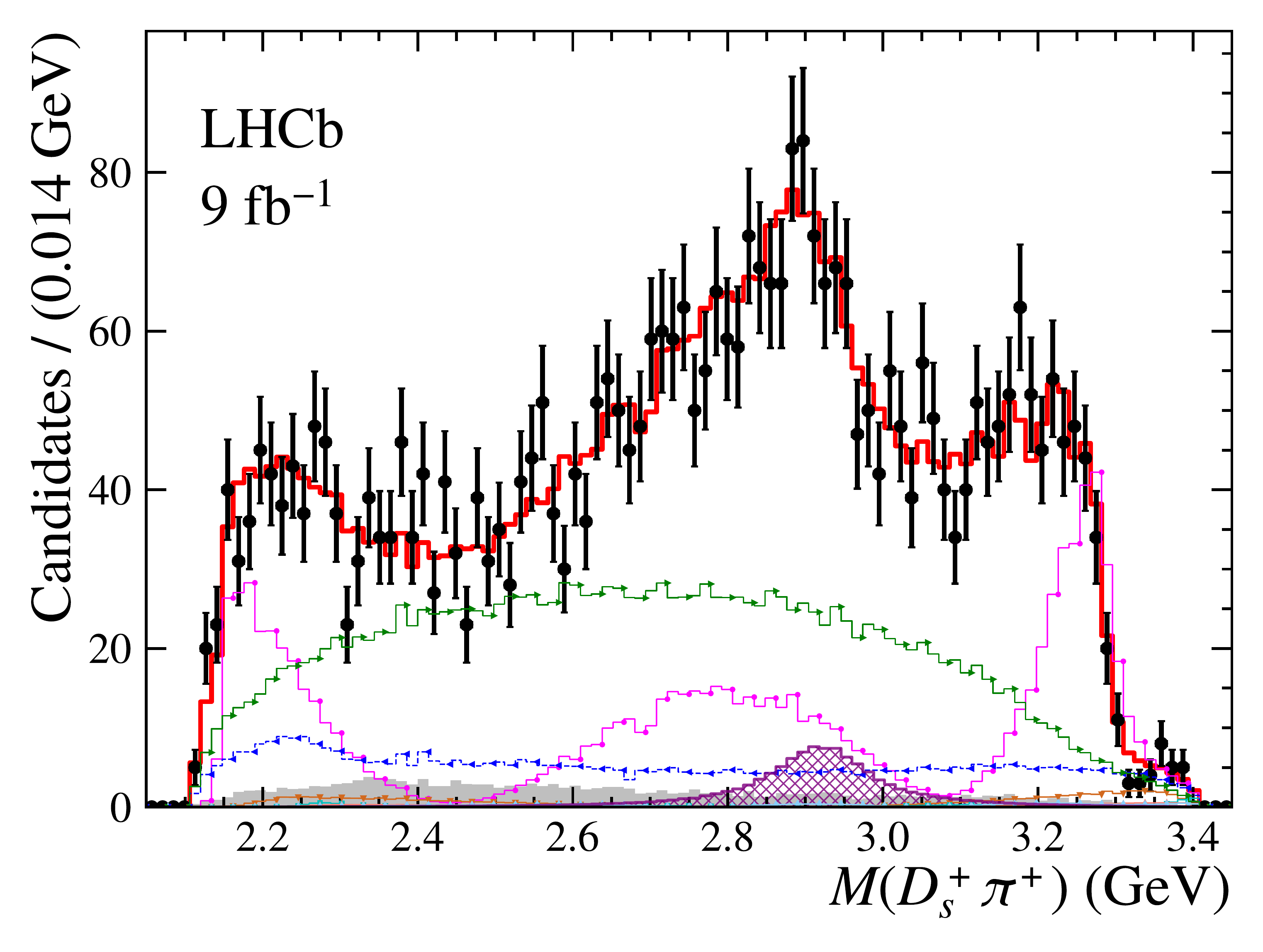}\put(-155,100){(d)}
  \end{center}
  \caption{Fit result of simultaneous $D\pi$ fit model, the (a) $M(D\pi)$ and (b) $M(\Ds\pi)$ distributions of $\BztoDzbarDsppim$ decays after including the $\Zzz$ state; the (c) $M(D\pi)$ and (d) $M(\Ds\pi)$ distributions of $\BptoDmDsppip$ decays after including the $\Zzpp$ state.}
  \label{fig:Z0_fit_Simu}
\end{figure}

The fit results are shown in Fig.~\ref{fig:Z0_fit_Simu}. As separate fit, neutral and doubly charged $\Ds\pi$ states are needed to describe the data well, where their parameters are set to be different. The masses, widths and fit fractions of the $\Zzz$ and $\Zzpp$ states after considering the systematic uncertainties and possible fit bias, which are summarized in Tables~\ref{tab:FitZMass_Simu} and~\ref{tab:FitResult_Simu}, show good agreement with the separate fit result, with significant improvement on the relative statistical uncertainties. The total $\chi^2/ndf$ is evaluated to be $140.3/89$.

The $ndf$ of the $\Zzz$ and $\Zzpp$ states in the simultaneous $D\pi$ fit model are evaluated to be $7.29\pm0.18$ and $8.57\pm0.17$ in the same way as described in Sec.~\ref{sec:Exotics test}, with the corresponding significances estimated to be $9.0\,\sigma$ and $7.4\,\sigma$, respectively. After accounting for the systematic uncertainties discussed in Sec.~\ref{sec:Systematic uncertainties}, these are reduced to $8.0\,\sigma$ and $6.5\,\sigma$. The constraints on the $D\pi$ contributions using isospin symmetry lead to higher significance, as expected. 

To estimate the isospin-breaking effects between the $\Zzz$ and $\Zzpp$ states, the mass difference, $\Delta M=M(\Zzpp)-M(\Zzz)$, and width difference, $\Delta\Gamma=\Gamma(\Zzpp)-\Gamma(\Zzz)$, are evaluated to be $28\pm20\pm12\mev$ and $15\pm39\pm16\mev$, respectively, where the first and second uncertainties are statistical and systematic. The statistical uncertainties in $\Delta M$ and $\Delta\Gamma$ are evaluated using pseudoexperiments to account for the correlations, and some of the systematic uncertainties are canceled. The masses and widths of the two exotic states are consistent with each other within $1\,\sigma$, which confirms that they are related by isospin symmetry.

The full simultaneous fit, where the parameters of the $\Zzz$ and $\Zzpp$ states are shared in the fit, is also performed, and described in a separate Letter~\cite{PRLsister}.

\begin{table}[!tb]
\centering
  \caption{
    Masses and widths of the $\Zzz$ and $\Zzpp$ states. The values are corrected for biases as described in the text. The first and second uncertainties are statistical and systematic, respectively.
  }
    \begin{tabular}{lcc}
    \hline
    Particle        & Mass ($\gev$)                       & Width ($\gev$)                           \\
    \hline
$\Zzz$	            & 2.892$\,\pm\,$0.014$\,\pm\,$0.015	    & 0.119$\,\pm\,$0.026$\,\pm\,$0.013	\\
$\Zzpp$	            & 2.921$\,\pm\,$0.017$\,\pm\,$0.020	    & 0.137$\,\pm\,$0.032$\,\pm\,$0.017	\\
 
    \hline
    \end{tabular}
\label{tab:FitZMass_Simu}
\end{table}

\begin{table}[!htb]
\centering
  \caption{
    Amplitude, phase and fit fraction of each component in the simultaneous $D\pi$ fit model. The values are corrected based on the results of pseudoexperiments. The first and second uncertainties are the statistical and systematic, respectively.
  }
  \resizebox{\textwidth}{!}{
    \begin{tabular}{l r@{$\,\pm\,$}c@{$\,\pm\,$}l r@{$\,\pm\,$}c@{$\,\pm\,$}l r@{$\,\pm\,$}c@{$\,\pm$\,}l r@{$\,\pm\,$}c@{$\,\pm$\,}l}
    \hline
    Particle        & \multicolumn{3}{c}{Amplitude}                             & \multicolumn{3}{c}{Phase (rad)}                             & \multicolumn{3}{c}{$B^{0}$ Fraction (\%)}             & \multicolumn{3}{c}{$B^{+}$ Fraction (\%)}    \\
\hline
\Zzz	            & 0.139 & 0.032 & 0.028	& $-$1.39 & 0.26 & 0.29	& 2.48 & 0.67 & 0.77	& \multicolumn{3}{c}{--}	\\
\Zzpp	            & 0.143 & 0.036 & 0.031	& $-$1.19 & 0.29 & 0.38	& \multicolumn{3}{c}{--}	                            & 2.25 & 0.67 & 0.77	\\
$D^{*}(2007)^{0}$	& 2.65 & 0.14 & 1.06	& $-$2.98 & 0.07 & 0.32	& \multicolumn{3}{c}{--}	                            & 14.7 & 1.3 & 2.7	\\
$D^{*}(2010)^{-}$	& 2.94 & 0.15 & 0.49	& $-$2.92 & 0.07 & 0.28	& 15.8 & 1.3 & 2.4	& \multicolumn{3}{c}{--}	\\
$D_{2}^{*}(2460)$	& \multicolumn{3}{c}{1}	& \multicolumn{3}{c}{0}	& 22.38 & 0.88 & 0.60	& 22.35 & 0.91 & 0.71	\\
$D_{1}^{*}(2600)$	& 0.223 & 0.033 & 0.052	& 0.15 & 0.16 & 0.26	& 1.35 & 0.40 & 0.59	& 1.37 & 0.42 & 0.62	\\
$D_{3}^{*}(2750)$	& 0.151 & 0.031 & 0.035	& $-$2.81 & 0.20 & 0.57	& 0.31 & 0.14 & 0.17	& 0.31 & 0.15 & 0.17	\\
$D_{1}^{*}(2760)$	& 0.121 & 0.043 & 0.164	& $-$0.19 & 0.35 & 0.98	& 0.28 & 0.25 & 1.48	& 0.28 & 0.26 & 1.53	\\
$D_{J}^{*}(3000)$	& 1.44 & 0.24 & 1.20	& 1.41 & 0.23 & 1.29	& 0.45 & 0.16 & 0.38	& 0.45 & 0.16 & 0.37	\\
$D\pi$ S-wave	    & 1.141 & 0.044 & 0.081	& $-$0.966 & 0.044 & 0.083	& 45.5 & 2.1 & 3.3	& 48.3 & 2.2 & 3.5	\\

    \hline
    \end{tabular}}
\label{tab:FitResult_Simu}
\end{table}

%%%%%%%%%%%%%%%%%%%%%%%%%%%%%%%%%%%%%%%%%%%%%%%%%%%%%
%%%%%           Systematic uncertainties        %%%%%
%%%%%%%%%%%%%%%%%%%%%%%%%%%%%%%%%%%%%%%%%%%%%%%%%%%%%
\section{Systematic uncertainties}
\label{sec:Systematic uncertainties}
The sources of systematic uncertainty fall into two categories: experimental and those related to the amplitude model. In the first category there are effects related to the fixed signal yields of $B$ candidates, the models of the background distributions, and the signal efficiency computation. Those arising from the amplitude model are mainly due to the fixed parameters of the model. The total systematic uncertainty is found by summing these in quadrature.

The signal yields in the amplitude analysis are taken from the results of the fits to the invariant mass distributions of $B$ candidates. 
To determine the systematic uncertainty, the signal yield of each dataset is varied according to a Gaussian distribution whose width corresponds to a signal-yield uncertainty that includes uncertainties due to the modelling of the invariant mass distribution. 
The amplitude fit is repeated with the new signal yields and 
the RMS value of each fit parameter is taken as the systematic uncertainty.

Backgrounds are modelled using a Gaussian process extrapolation method~\cite{mathad2021efficient} according to sideband distributions. To evaluate the associated systematic uncertainty, the background model is replaced by the result of a kernel density estimation~\cite{Poluektov:2014rxa} applied to the Dalitz-plot distributions of the sideband samples. The deviations of the fit parameters from the default result are taken as the associated systematic uncertainties. 

Knowledge of the signal efficiency variation over the Dalitz plot is limited by four effects: uncertainty in the PID response, trigger efficiency calibration uncertainties, signal efficiency determination, and simulation sample size. 
The systematic uncertainty due to the PID calibration is estimated by regenerating the PID responses with a perturbed kernel density estimation~\cite{Poluektov:2014rxa}, extracting the efficiencies, and repeating all the fitting procedures. For the trigger efficiency calibration effect, a conservative systematic uncertainty is determined by repeating all the fitting procedures with the signal efficiency maps without any trigger efficiency correction. The systematic uncertainty related to the signal efficiency determination is estimated by performing the amplitude analysis with the efficiency maps obtained using an alternative kernel density estimation. The deviations of the fit parameters in each fit are taken to be the systematic uncertainties.

The systematic uncertainty due to the limited size of the simulated samples is determined by generating 200 samples following the bootstrap method~\cite{Efron:1979bxm}, where the simulated candidates after all the selection criteria are allowed to be picked multiple times. The efficiencies are then extracted from each bootstrapped sample and applied in the amplitude analysis. The RMS of each fit parameter in these fits is taken as the systematic uncertainty.

The fixed parameters in the amplitude analysis include the radius $d$ in the Blatt--Weisskopf form factor, and the parameters of the $D\pi$ lineshapes.
The systematic uncertainty associated to $d$ is evaluated by setting $d$ to $1.5\gev^{-1}$ and $4.5\gev^{-1}$. 
The largest difference compared to the default results is assigned as the systematic uncertainty. The fixed $D\pi$ parameters consist of the masses and widths of all the $D^*$ states in Table~\ref{tab:candidates}, the choice of the qMI spline points, the constant $q_0$ in the $D^*(2007)^0$ state lineshape, and the spin hypothesis of $D_J(3000)$. The amplitude fits are performed several times, with one or more parameters changed. The fixed masses and widths of all the $D^*$ states are allowed to vary one at a time in the fit but are constrained by a Gaussian function within uncertainties in their default values~\cite{PDG2022}. The positions of the qMI spline points are chosen empirically in the default fit, and shifted by $\pm10\mev$ in the amplitude analysis.
The systematic uncertainty from the number of the qMI spline points is also explored, by adding a new point at $M(D\pi)=2.35\gev$ to try to improve the model description near the $\overline{D}{}^*_0(2300)$ mass region, or removing the point at $M(D\pi)=2.6\gev$ as there is no $0^+$ $D\pi$ state observed in this region.
The $q_0$ of the $D^*(2007)^0$ state is taken as the $q_0$ of the $D^*(2007)^0\to\Dz\piz$ decay in the default fit, and replaced by a value calculated from the $D^*(2007)^0$ effective mass~\cite{LHCb-PAPER-2016-026}. The spin of the $D_J(3000)$ state is found to be $4^+$ in the default fit, and altered to $2^+$, the result measured in Ref.~\cite{LHCb-PAPER-2016-026}. The deviations of the results between each fit and the default fit, are summed in quadrature, and taken as the systematic uncertainty.

Possible fit biases are investigated using pseudoexperiments, and used to correct the results. For each dataset, 500 samples are generated according to the default fit results, where the yield of each is sampled from a Poisson distribution whose mean value is the number of $B$ candidates in the corresponding dataset. The fit parameters of the pseudodata are then extracted using the default fit model. 
The residual distributions $(\mu_{\rm pseudo}-\mu_{\rm default})$ and pull distributions $(\mu_{\rm pseudo}-\mu_{\rm default})/\sigma_{\rm pseudo}$ are extracted from the pseudoexperiments and default fit result, and used to correct the fit results. Here the $\mu_{\rm pseudo}$ and $\sigma_{\rm pseudo}$ are the mean values and uncertainties in the pseudoexperiments.
Both the residual distributions and pull distributions of each parameter are fitted with a Gaussian function. The mean value of the residual distribution from the Gaussian fit is used to correct the mean value of the parameter, while the width of the pull distribution is used to scale the statistical uncertainty.
The results are summarized in Tables~\ref{tab:FitZMass}, \ref{tab:FitResult_Bz}, and \ref{tab:FitResult_Bp}.
The systematic uncertainties of the simultaneous $D\pi$ fit are also evaluated in the same way, and summarized in Tables~\ref{tab:FitZMass_Simu} and~\ref{tab:FitResult_Simu}.

%%%%%%%%%%%%%%%%%%%%%%%%%%%%%%%%%%%%%%%
%%%%%           Conclusion        %%%%%
%%%%%%%%%%%%%%%%%%%%%%%%%%%%%%%%%%%%%%%
\section{Conclusion}
\label{sec:Summary}

Amplitude analyses of $\BztoDzbarDsppim$ and $\BptoDmDsppip$ decays are performed for the first time, using \lhcb $pp$ collision data taken at center-of-mass energies of $7,~8$ and 13 $\tev$, corresponding to a total integrated luminosity of $9~\rm{fb^{-1}}$. In total, signal yields of $4009\pm70$ and $3750\pm64$ candidates are obtained from the $\BztoDzbarDsppim$ and $\BptoDmDsppip$ decays, respectively.

When all known $D\pi$ resonances with spin-parities of $1^-, 2^+, 3^-$ and $4^+$~\cite{PDG2022} are included, along with a qMI spline model to describe the $0^+$ $D\pi$ distributions, the results show that the $\Ds\pi$ invariant-mass distributions are not well described. 
To improve the model description, a $0^+$ $\Ds\pi$ resonance is added to each decay mode. The masses and widths of the two resonances are determined to be
\begin{equation}
    \begin{split}
    \Zzz:   M=&~(2.879\pm0.017\pm0.018)\gev,\\
            \Gamma=&~(0.153\pm0.028\pm0.020)\gev,\\
    \Zzpp:  M=&~(2.935\pm0.021\pm0.013)\gev,\\
            \Gamma=&~(0.143\pm0.038\pm0.025)\gev,\nonumber
    \end{split}
\end{equation}
where the first uncertainty is statistical and the second systematic. The significances, accounting for the look-elsewhere effect and systematic uncertainties, of the exotic $\Zzz$ and $\Zzpp$ states are $6.6\,\sigma$ and $4.8\,\sigma$, respectively. 

A simultaneous $D\pi$ amplitude fit assuming isospin symmetry in the $\BztoDzbarDsppim$ and $\BptoDmDsppip$ decays is also performed to provide better control on the contributions from $D\pi$ resonances, especially the $0^+$ $D\pi$ spline model, and to improve the precision of the measured parameters of exotic states. The masses and widths of the two resonances in the simultaneous $D\pi$ fit are measured to be
\begin{equation}
    \begin{split}
    \Zzz:   M=&~(2.892\pm0.014\pm0.015)\gev,\\
            \Gamma=&~(0.119\pm0.026\pm0.013)\gev,\\
    \Zzpp:  M=&~(2.921\pm0.017\pm0.020)\gev,\\
            \Gamma=&~(0.137\pm0.032\pm0.017)\gev,\nonumber
    \end{split}
\end{equation}
with the significances evaluated to be $8.0\,\sigma$ and $6.5\,\sigma$ for the $\Zzz$ and $\Zzpp$ states, including systematic uncertainties.
The mass and width differences between $\Zzpp$ and $\Zzz$ are evaluated to be
\begin{equation}
    \begin{split}
    \Delta M& =~(28\pm20\pm12)\mev,\\
    \Delta\Gamma & =~(15\pm39\pm16)\mev,\nonumber
    \end{split}
\end{equation}
based on simultaneous $D\pi$ amplitude fit, and consistent with zero.
A simultaneous fit with the parameters of the $\Ds\pi$ exotic states shared is also performed, and described in a separate Letter~\cite{PRLsister}. All the results of the different fit scenarios show good agreement.

This is the first observation of an isospin triplet of manifestly exotic mesons with four different quark flavors. The masses and widths of the two states are consistent with the $X_0(2900)$ and $X_1(2900)$ states~\cite{LHCb-PAPER-2020-024,LHCb-PAPER-2020-025}, but have an opposite strangeness number. No hint of a $D\Ds$ structure is observed in the analysis. With the significantly larger data samples that will be collected by the upgraded \lhcb detector in the coming years, the nature of the isospin triplet of exotic mesons, and the existence of the possible $\Ds\pi$ exotic states with $J^P = 1^-$ in the same region, will be further explored.

\section*{Acknowledgements}
%
% These Acknowledgements valid from 3-May-2019
%
\noindent We express our gratitude to our colleagues in the CERN
accelerator departments for the excellent performance of the LHC. We
thank the technical and administrative staff at the LHCb
institutes.
We acknowledge support from CERN and from the national agencies:
CAPES, CNPq, FAPERJ and FINEP (Brazil); 
MOST and NSFC (China); 
CNRS/IN2P3 (France); 
BMBF, DFG and MPG (Germany); 
INFN (Italy); 
NWO (Netherlands); 
MNiSW and NCN (Poland); 
MEN/IFA (Romania); 
%MSHE (Russia); 
MICINN (Spain); 
SNSF and SER (Switzerland); 
NASU (Ukraine); 
STFC (United Kingdom); 
DOE NP and NSF (USA).
We acknowledge the computing resources that are provided by CERN, IN2P3
(France), KIT and DESY (Germany), INFN (Italy), SURF (Netherlands),
PIC (Spain), GridPP (United Kingdom), 
%RRCKI and Yandex LLC (Russia), 
CSCS (Switzerland), IFIN-HH (Romania), CBPF (Brazil),
Polish WLCG  (Poland) and NERSC (USA).
We are indebted to the communities behind the multiple open-source
software packages on which we depend.
Individual groups or members have received support from
ARC and ARDC (Australia);
Minciencias (Colombia);
AvH Foundation (Germany);
EPLANET, Marie Sk\l{}odowska-Curie Actions and ERC (European Union);
A*MIDEX, ANR, IPhU and Labex P2IO, and R\'{e}gion Auvergne-Rh\^{o}ne-Alpes (France);
Key Research Program of Frontier Sciences of CAS, CAS PIFI, CAS CCEPP, 
Fundamental Research Funds for the Central Universities, 
and Sci. \& Tech. Program of Guangzhou (China);
%Key Research Program of Frontier Sciences of CAS, CAS PIFI,
%Thousand Talents Program, and Sci. \& Tech. Program of Guangzhou (China);
%RFBR, RSF and Yandex LLC (Russia);
GVA, XuntaGal, GENCAT and Prog.~Atracci\'on Talento, CM (Spain);
SRC (Sweden);
the Leverhulme Trust, the Royal Society
 and UKRI (United Kingdom).

\newpage
%%%%%%%%%%%%%%%%%%%%%%%%%%%%%%%%%%%%%
%%%%%           Appendix        %%%%%
%%%%%%%%%%%%%%%%%%%%%%%%%%%%%%%%%%%%%
\section*{Appendix}
\appendix
\section{Moments analysis results}
\label{app:Moments analysis}
Moments analysis is useful to suggest possible resonant structures in the decay. The formulation and the results are provided in this section. 

The Legendre polynomial of a certain order $k$, as expressed in Eq.~\ref{eqn:legendre}, is used to weight the data, 
\begin{equation}
    P_k(x)=\sqrt{\frac{2k+1}{2}}2^k\sum_{j=0}^k x^j\binom{k}{j}\binom{\frac{k+j-1}{2}}{k}.
\label{eqn:legendre}
\end{equation}
For example, when focusing on the resonant structures of the $D\pi$ channel, the variable $x$ in this case would be the helicity variable in that decay chain, namely $\cos\theta_{D}^{D\pi}$.
Therefore, the total amplitude modified by the order-$k$ Legendre polynomial can be expressed as
\begin{equation}
    \left<Y_k\right>=\sum_{i=1}^{N}w_iP_k(\cos\theta_{D}^{D\pi}),
\end{equation}
where $w_i$ is the original weight for the data point $i$. 

\begin{figure}[htb]
  \begin{center}
    \includegraphics[width=0.32\linewidth]{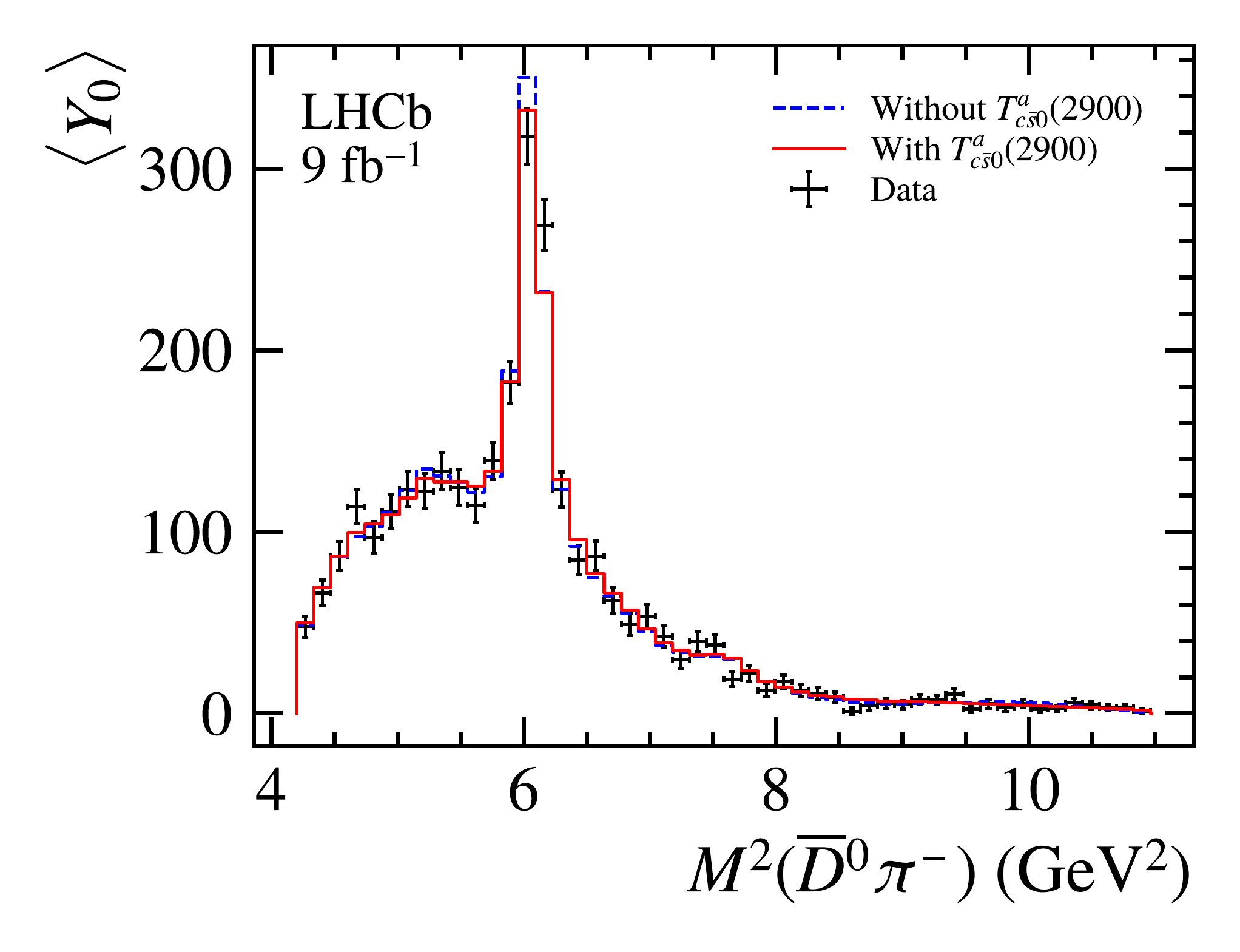}
    \includegraphics[width=0.32\linewidth]{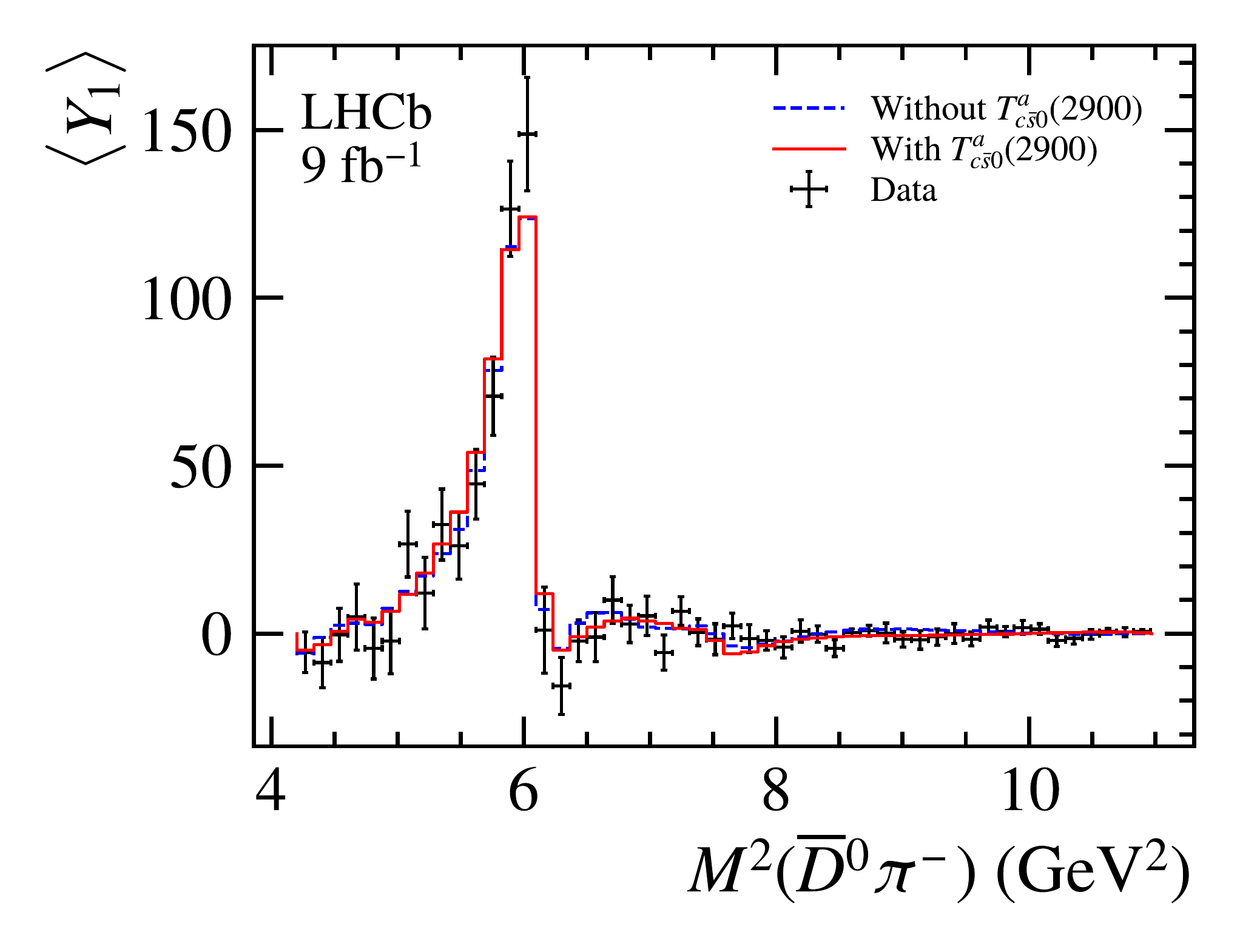}
    \includegraphics[width=0.32\linewidth]{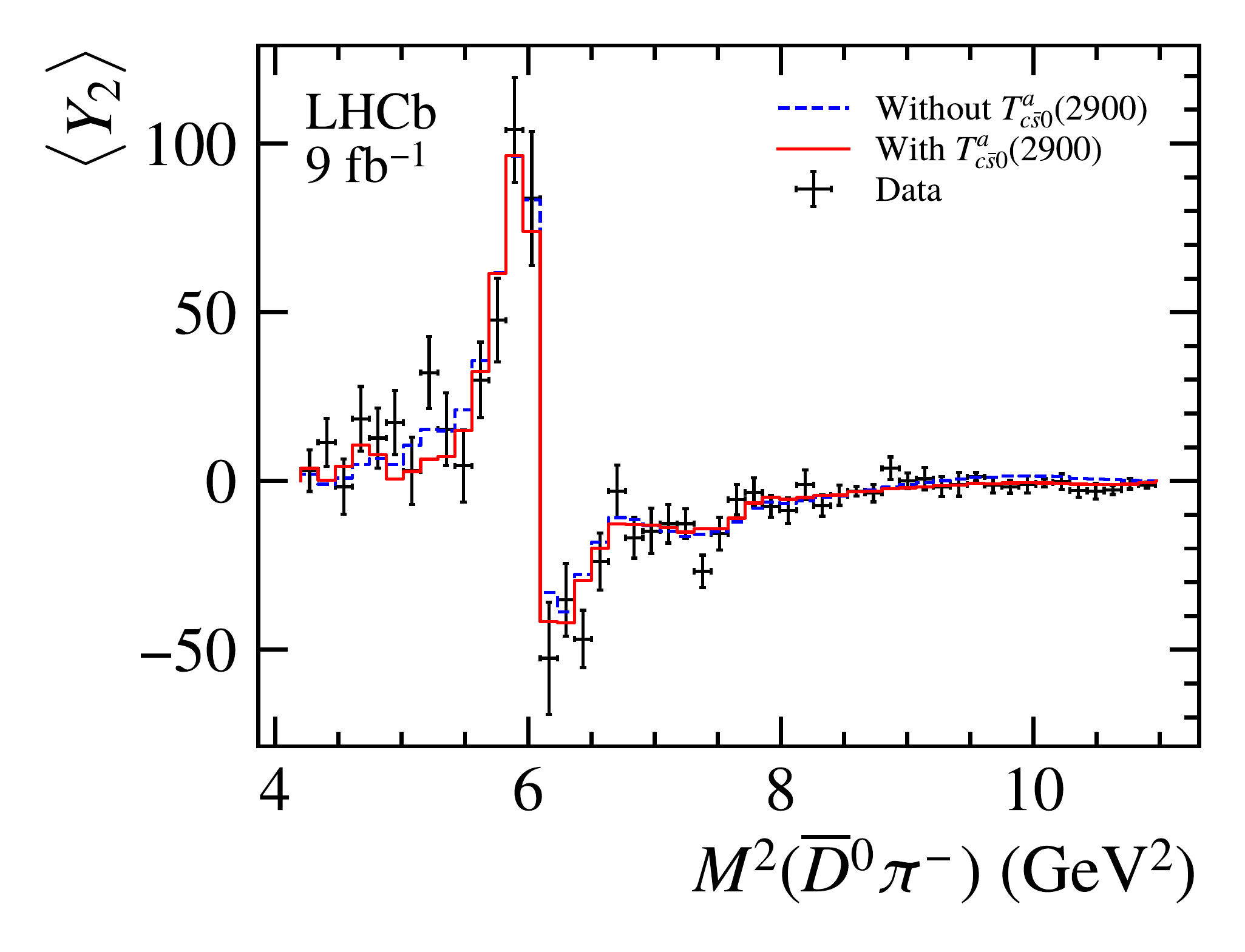}
    
    \includegraphics[width=0.32\linewidth]{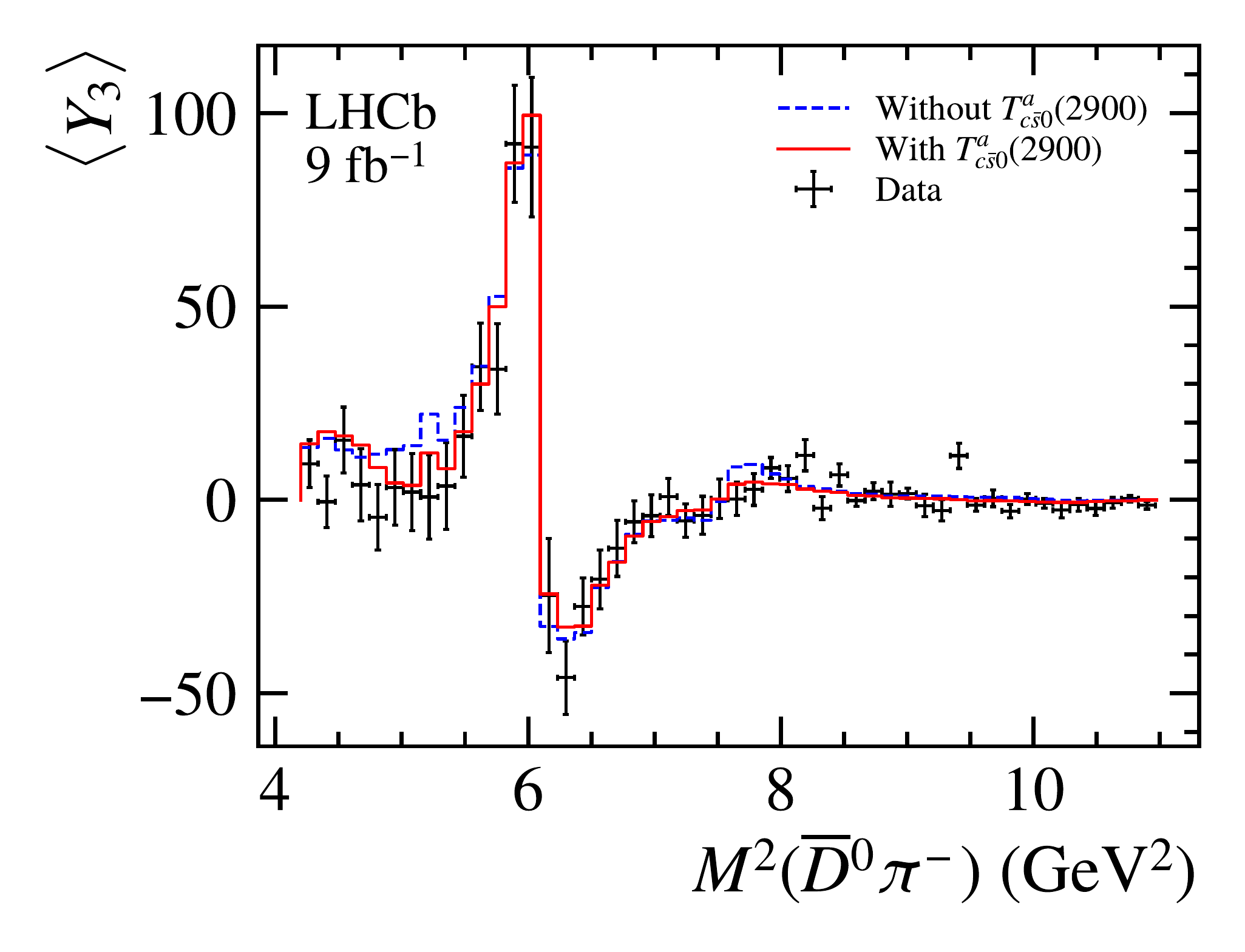}
    \includegraphics[width=0.32\linewidth]{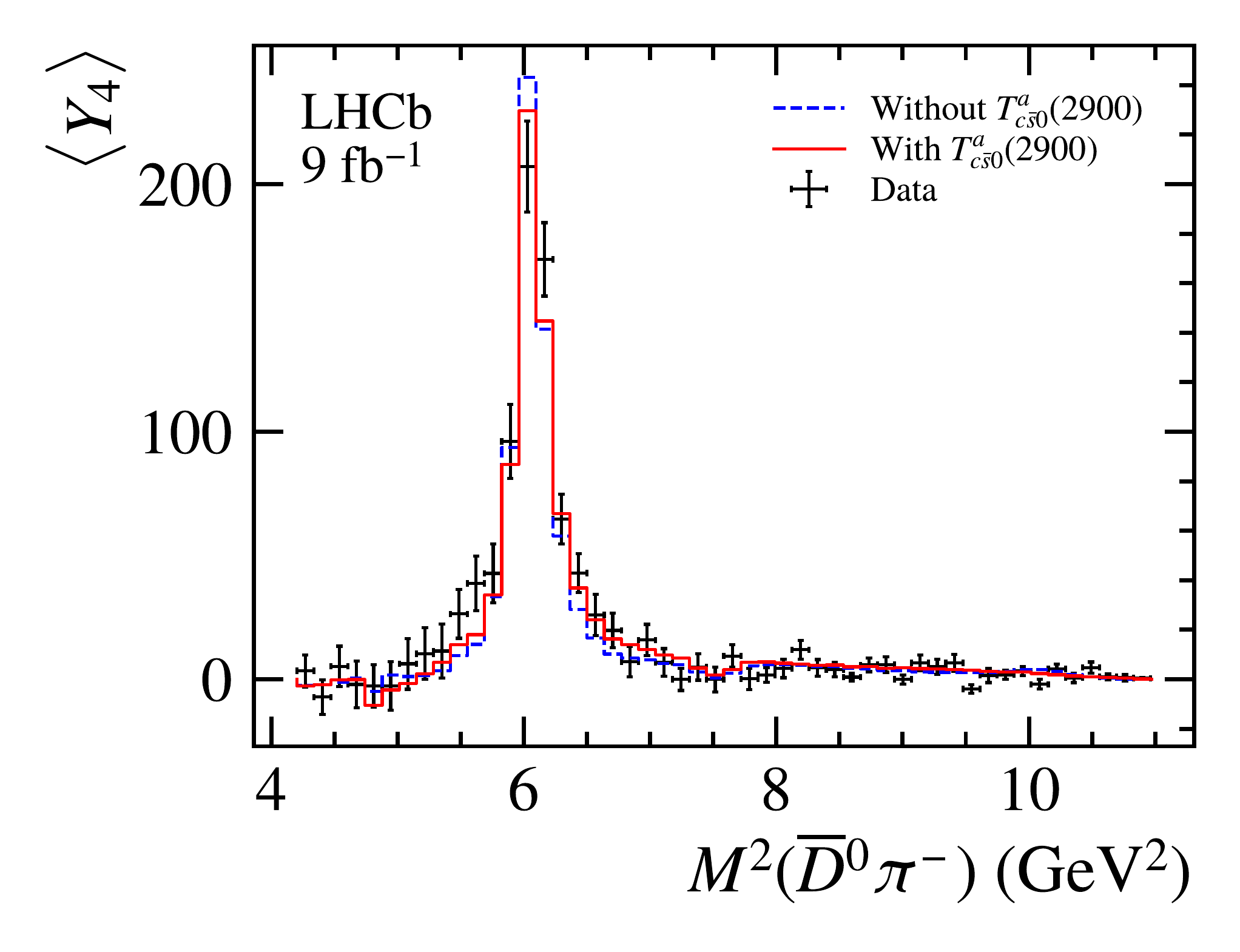}
    \includegraphics[width=0.32\linewidth]{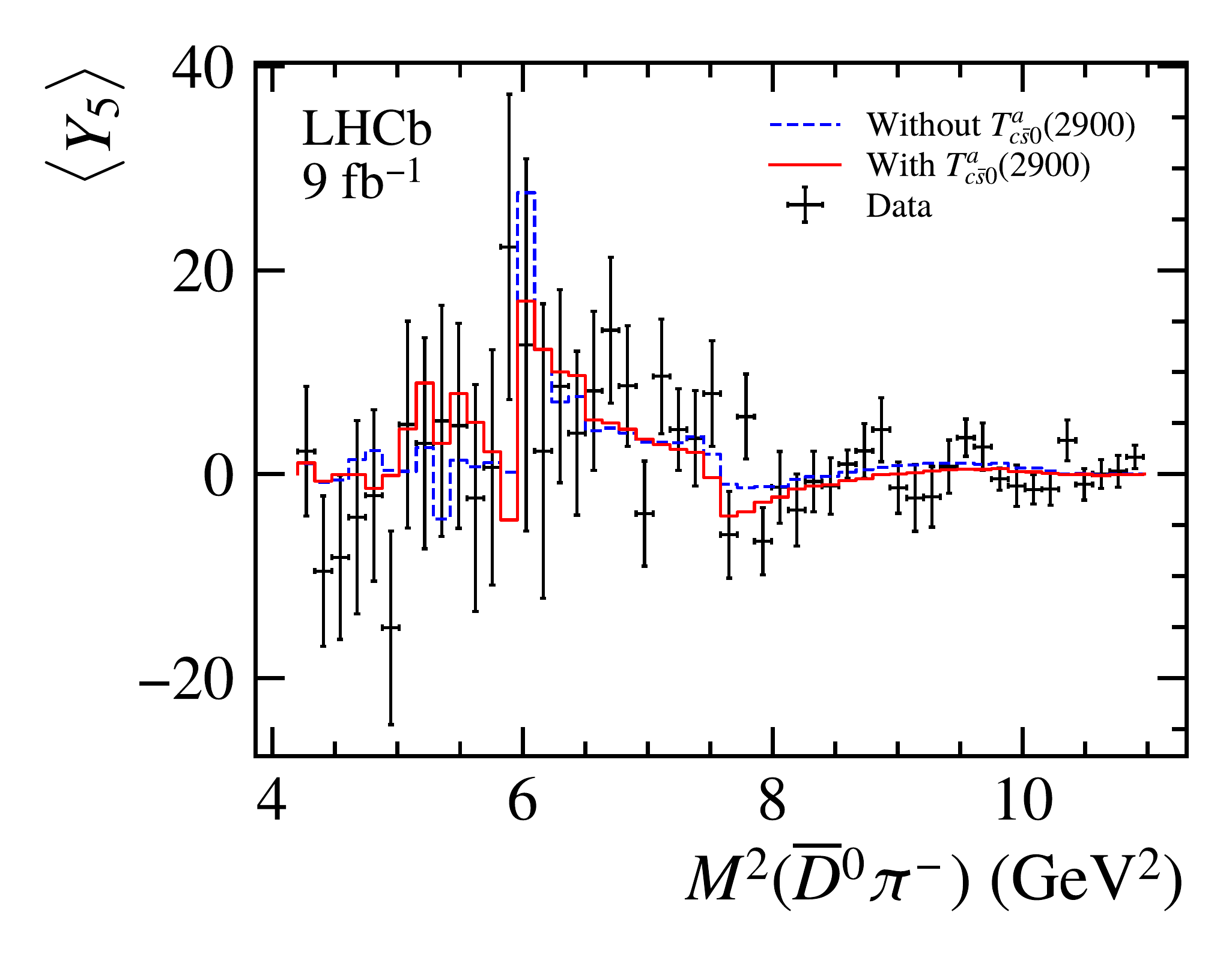}
    
    \includegraphics[width=0.32\linewidth]{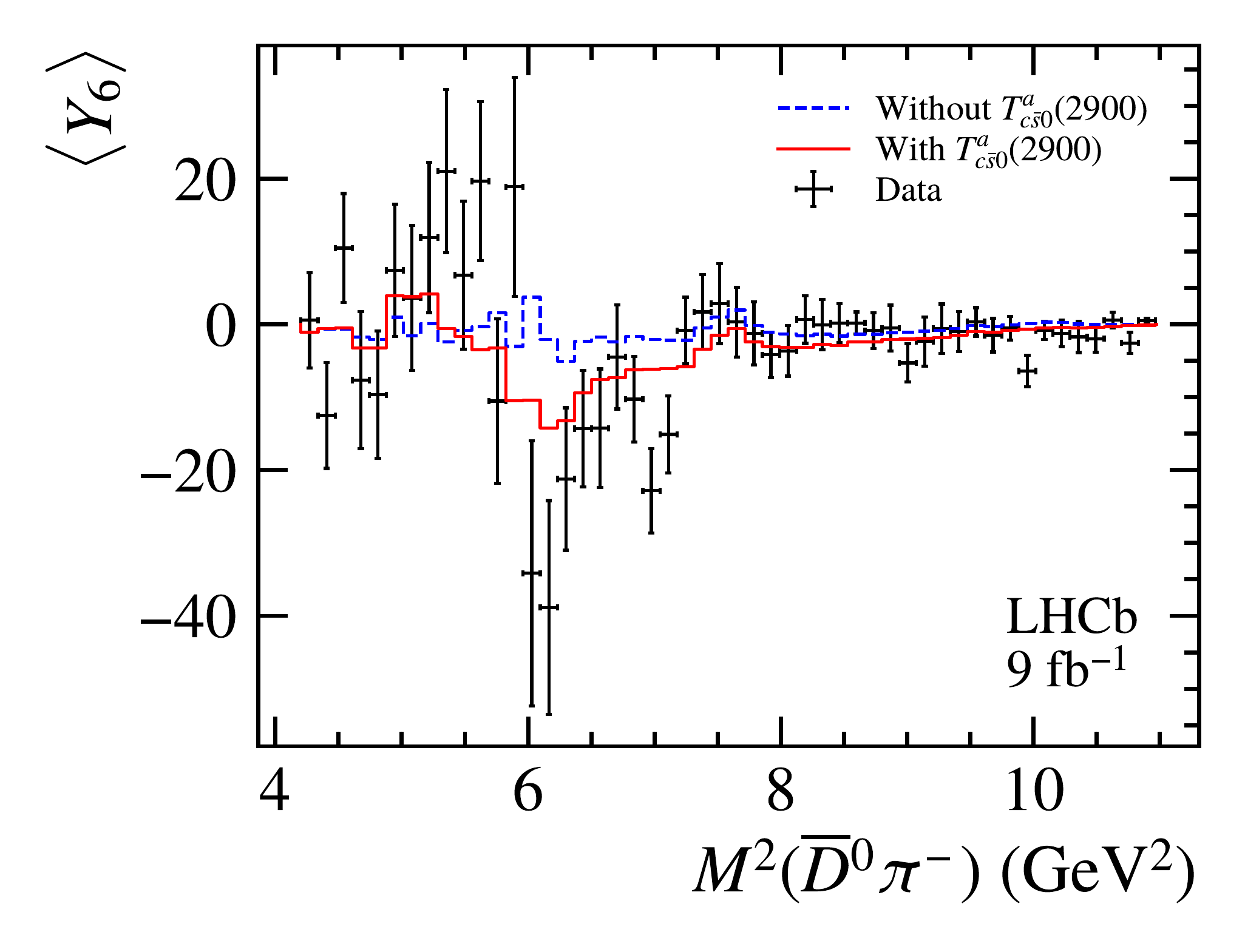}
    \includegraphics[width=0.32\linewidth]{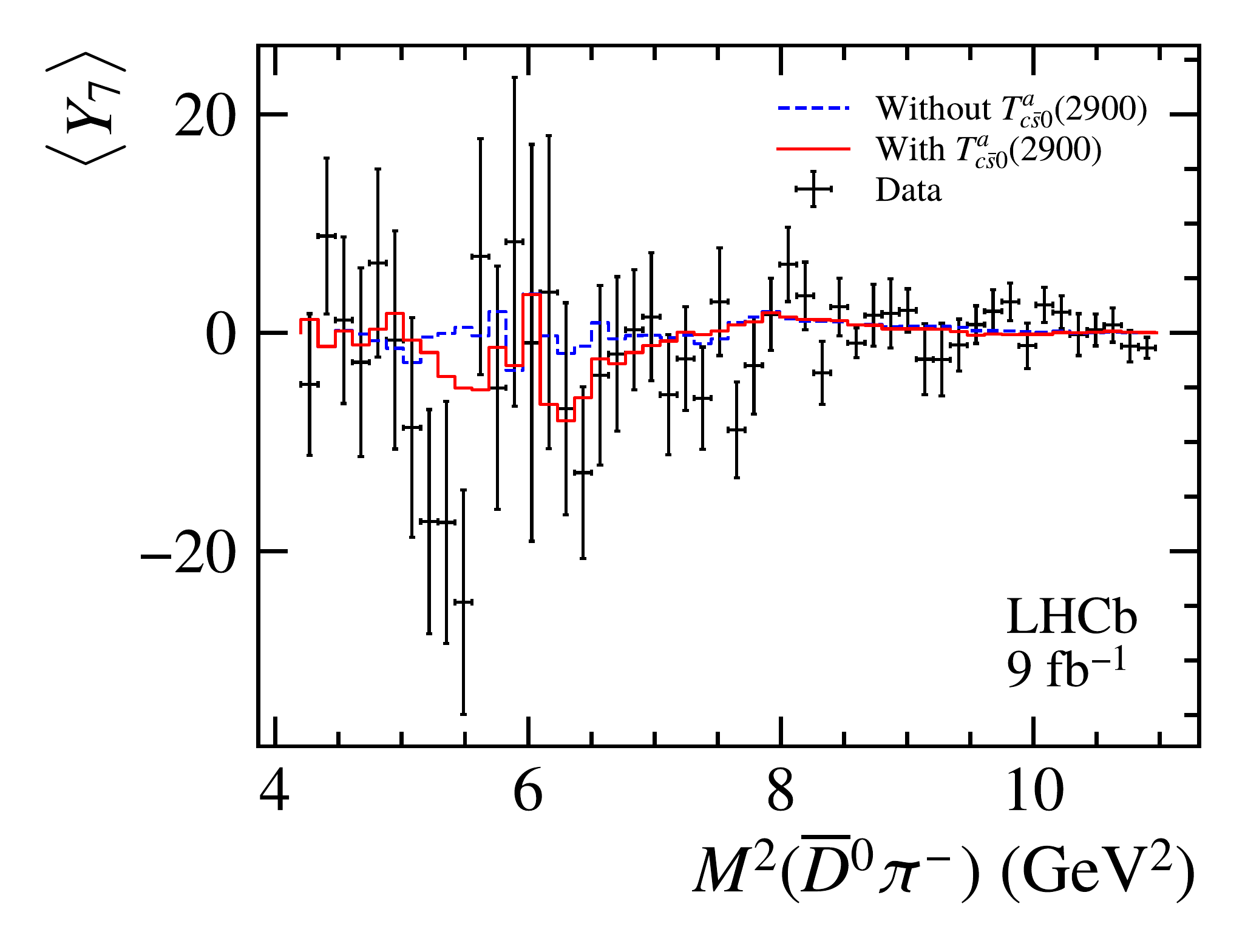}
    \includegraphics[width=0.32\linewidth]{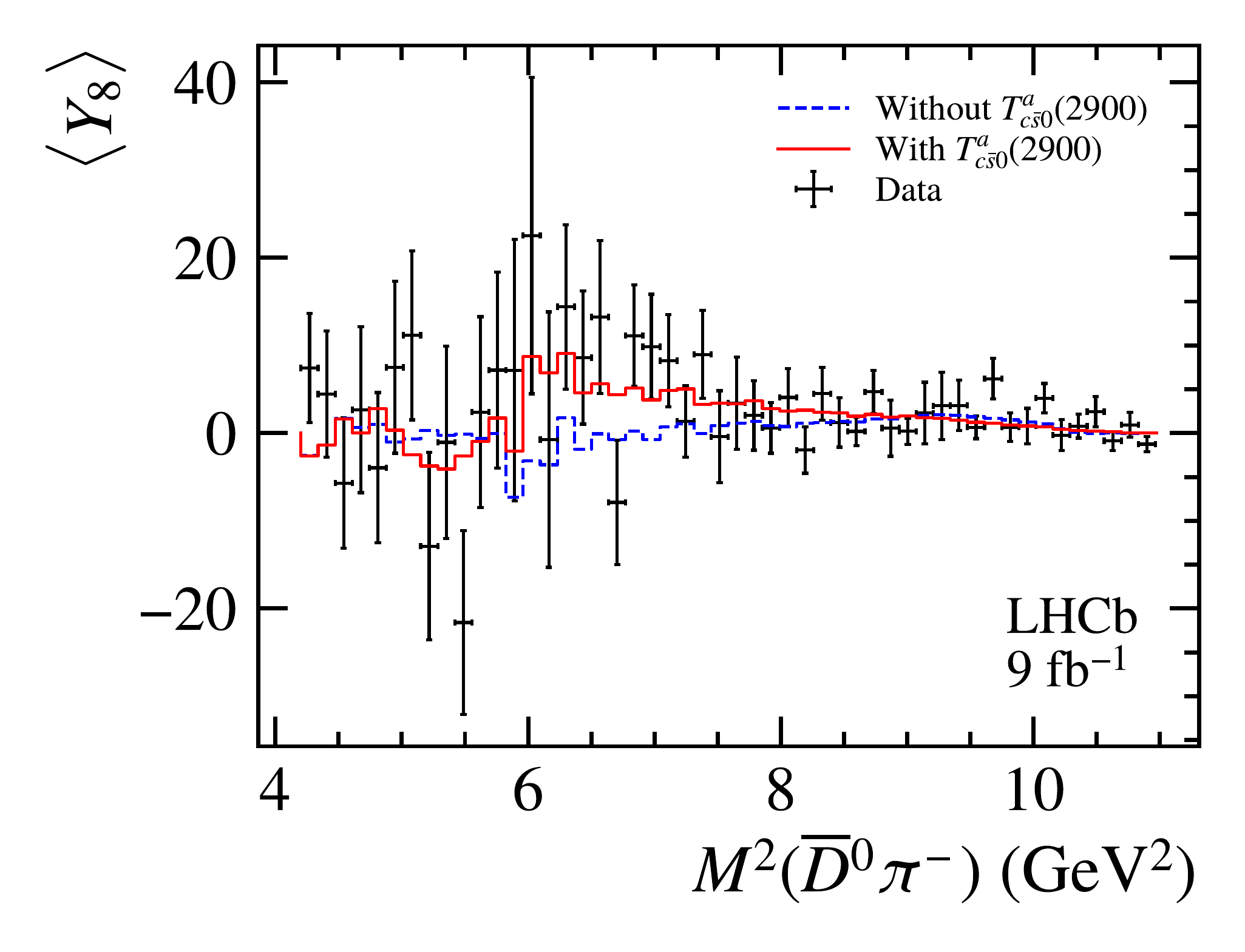}
  \end{center}
  \caption{Moments analysis of \BztoDzbarDsppim on $M^2(\Dzb\pim)$. The black points indicate the data, while the blue and red histogram indicate the fit results without and with the \Zz states separately.}
  \label{fig:mom_MDpi_Bz}
\end{figure}

\begin{figure}[htb]
  \begin{center}
    \includegraphics[width=0.32\linewidth]{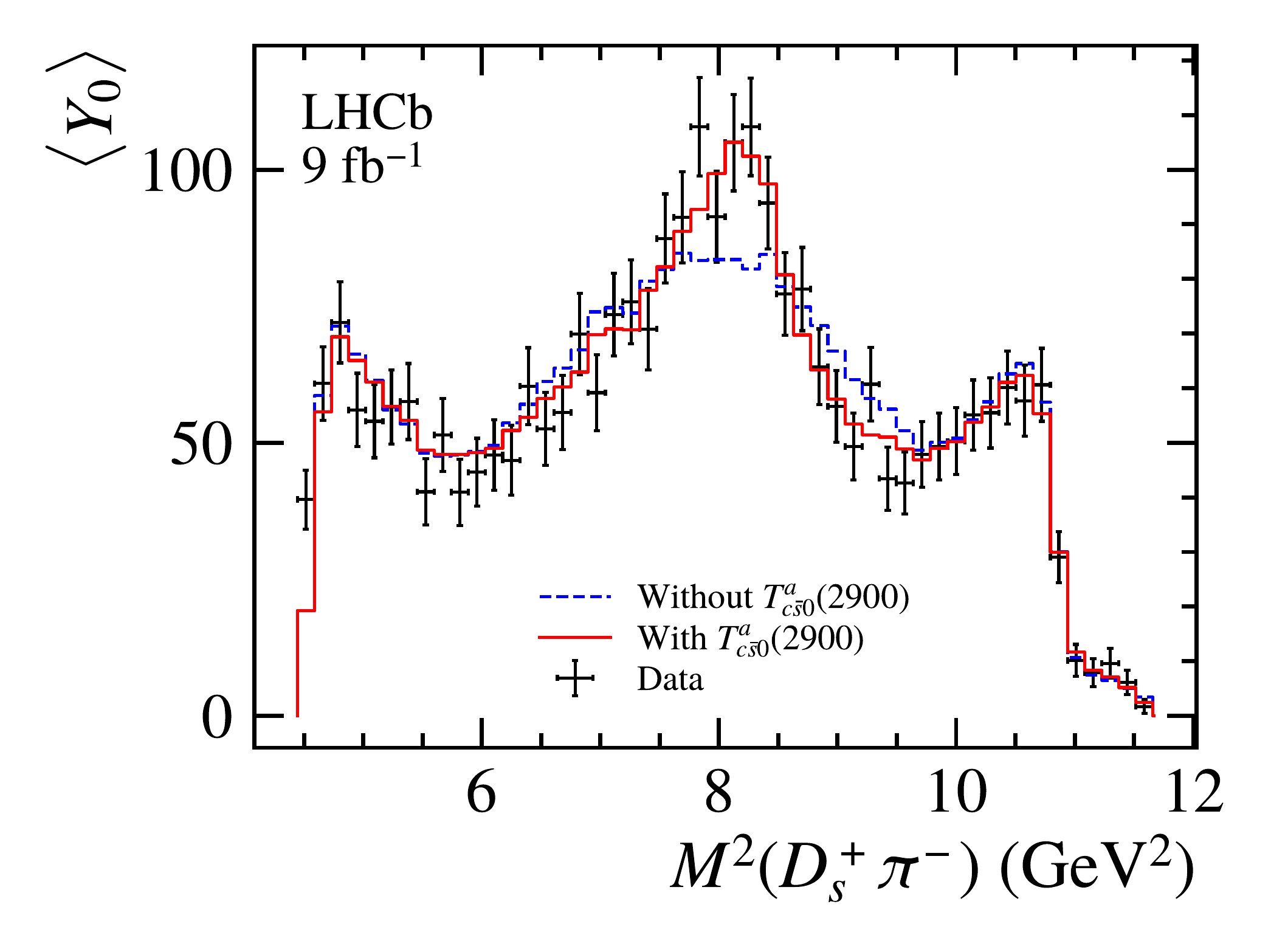}
    \includegraphics[width=0.32\linewidth]{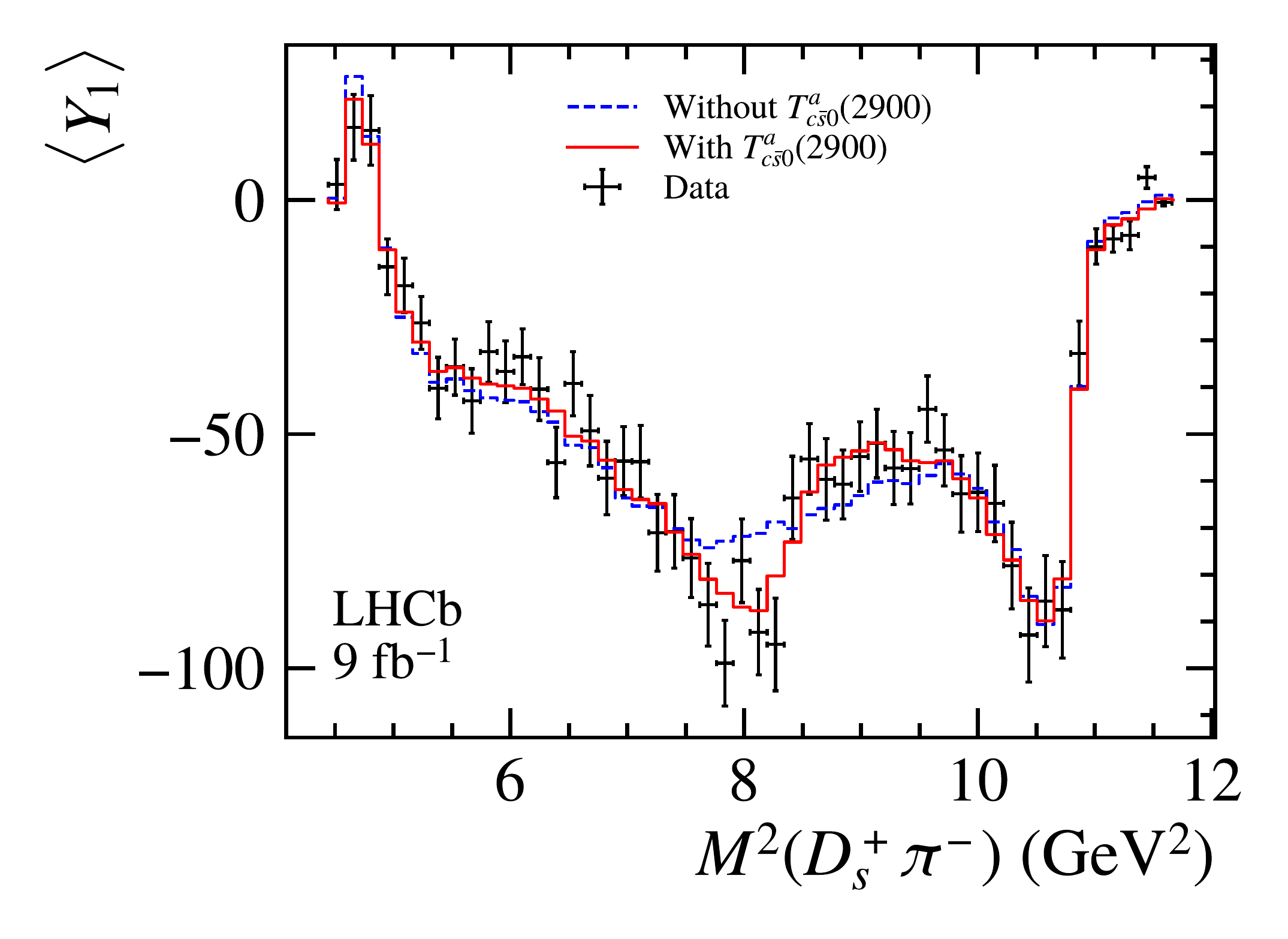}
    \includegraphics[width=0.32\linewidth]{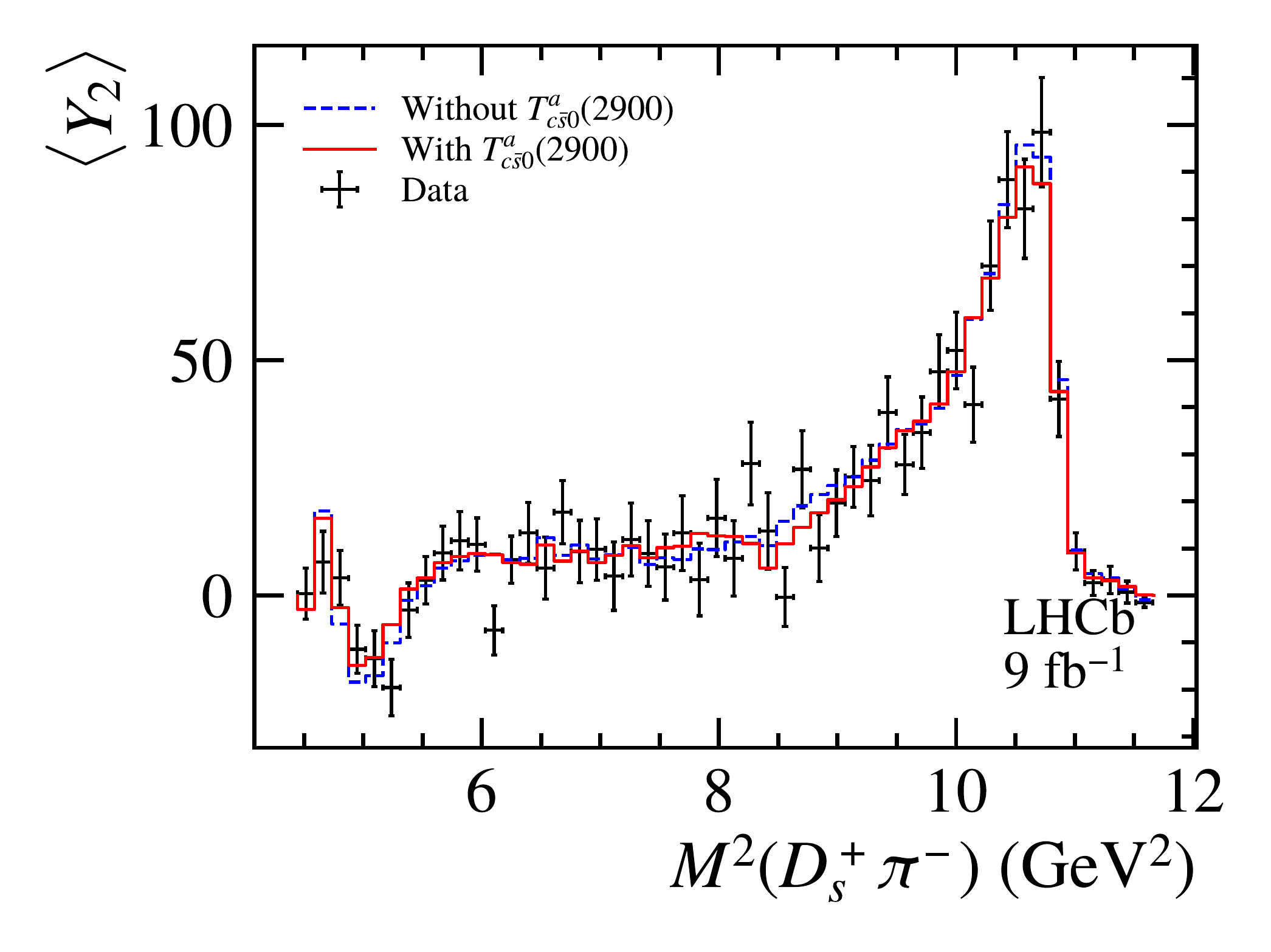}
    
    \includegraphics[width=0.32\linewidth]{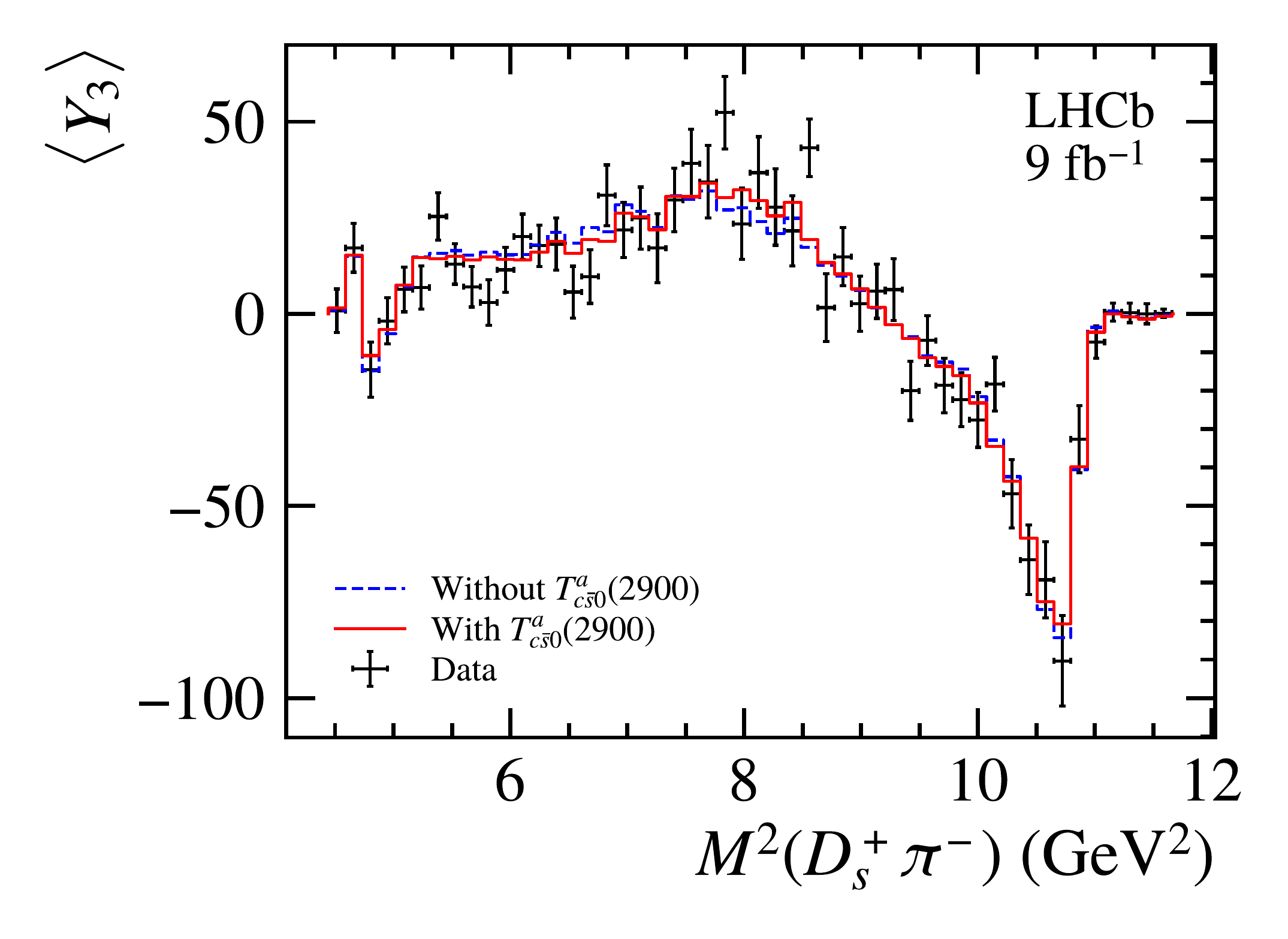}
    \includegraphics[width=0.32\linewidth]{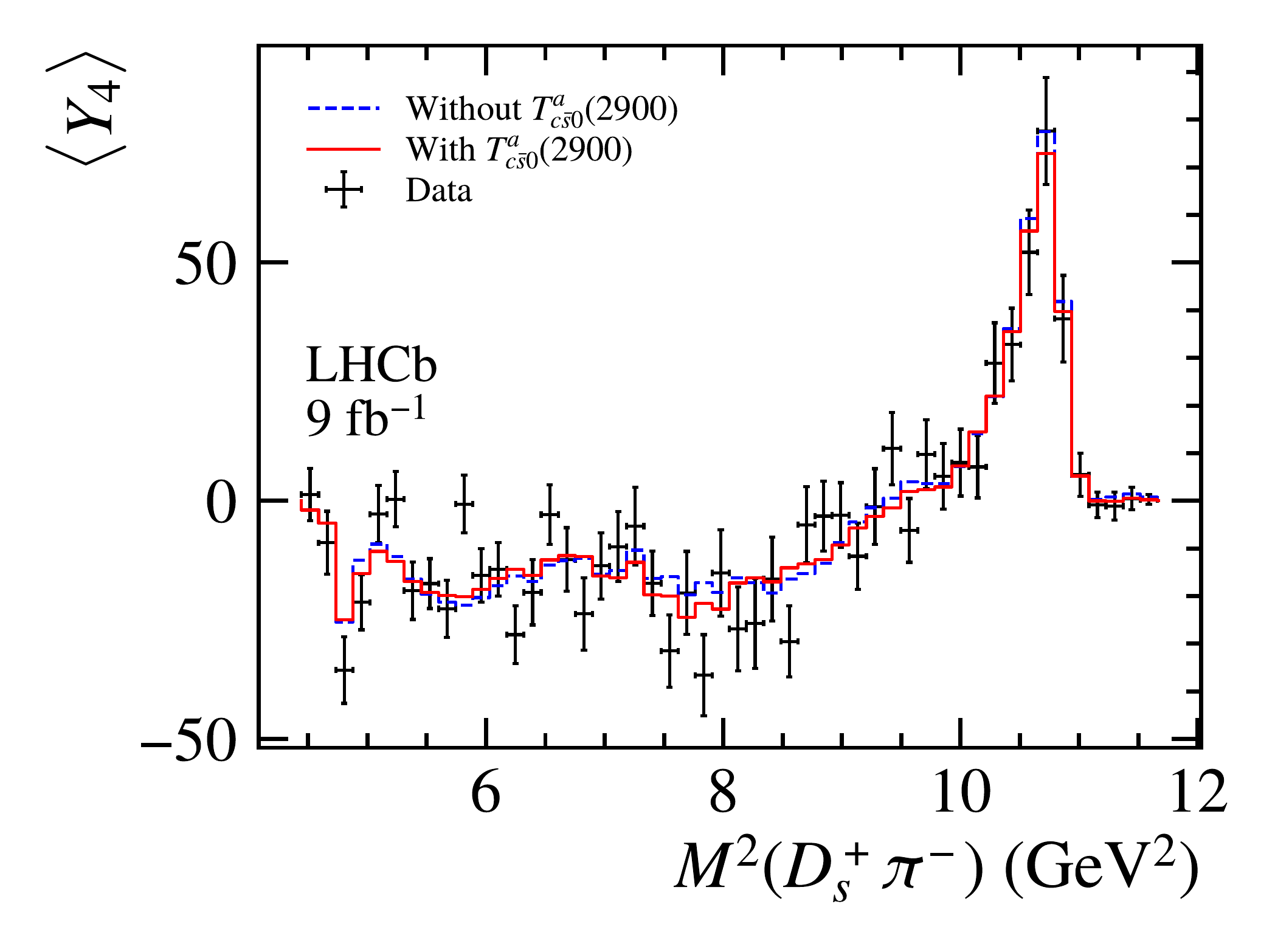}
    \includegraphics[width=0.32\linewidth]{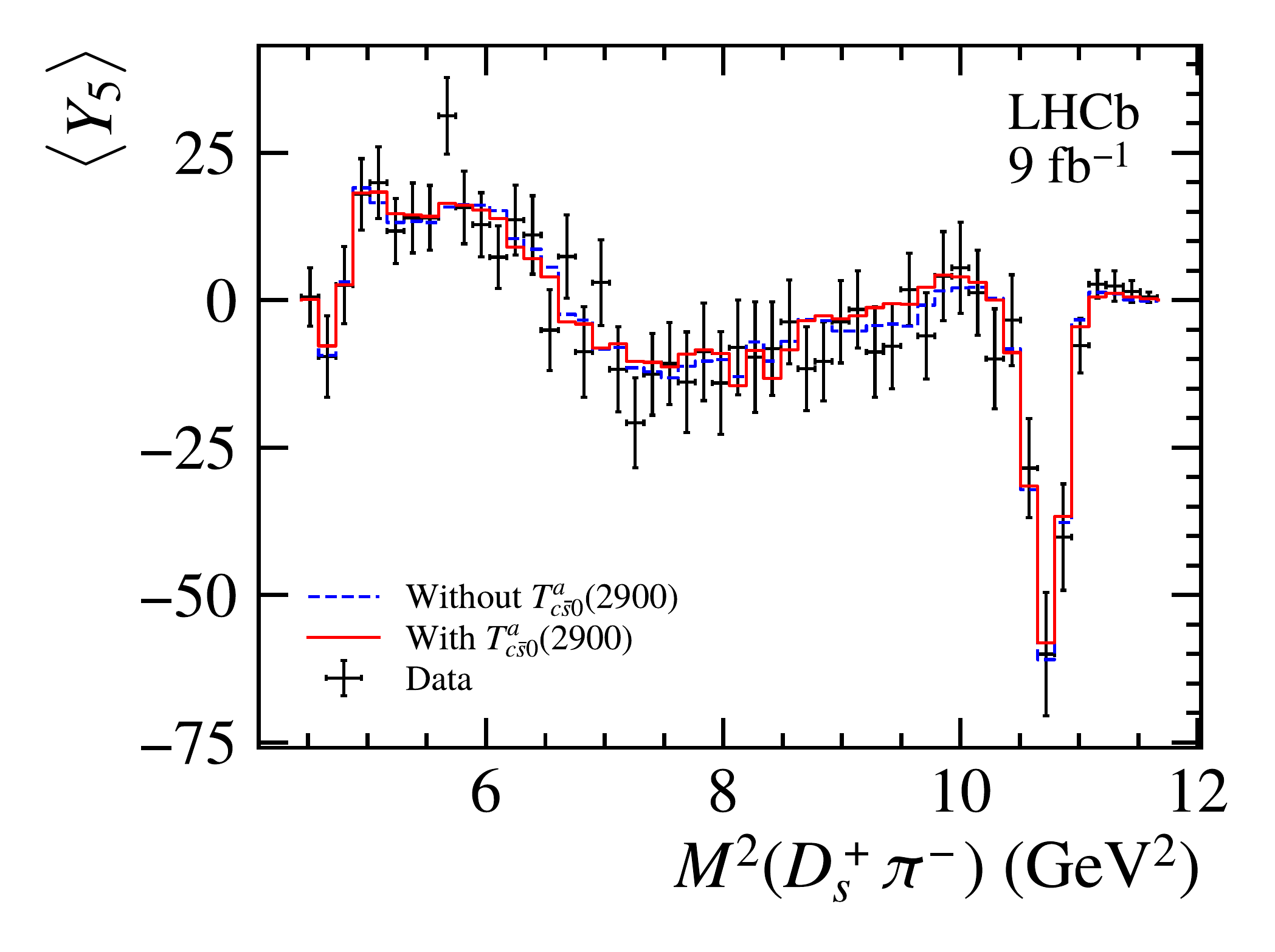}
    
    \includegraphics[width=0.32\linewidth]{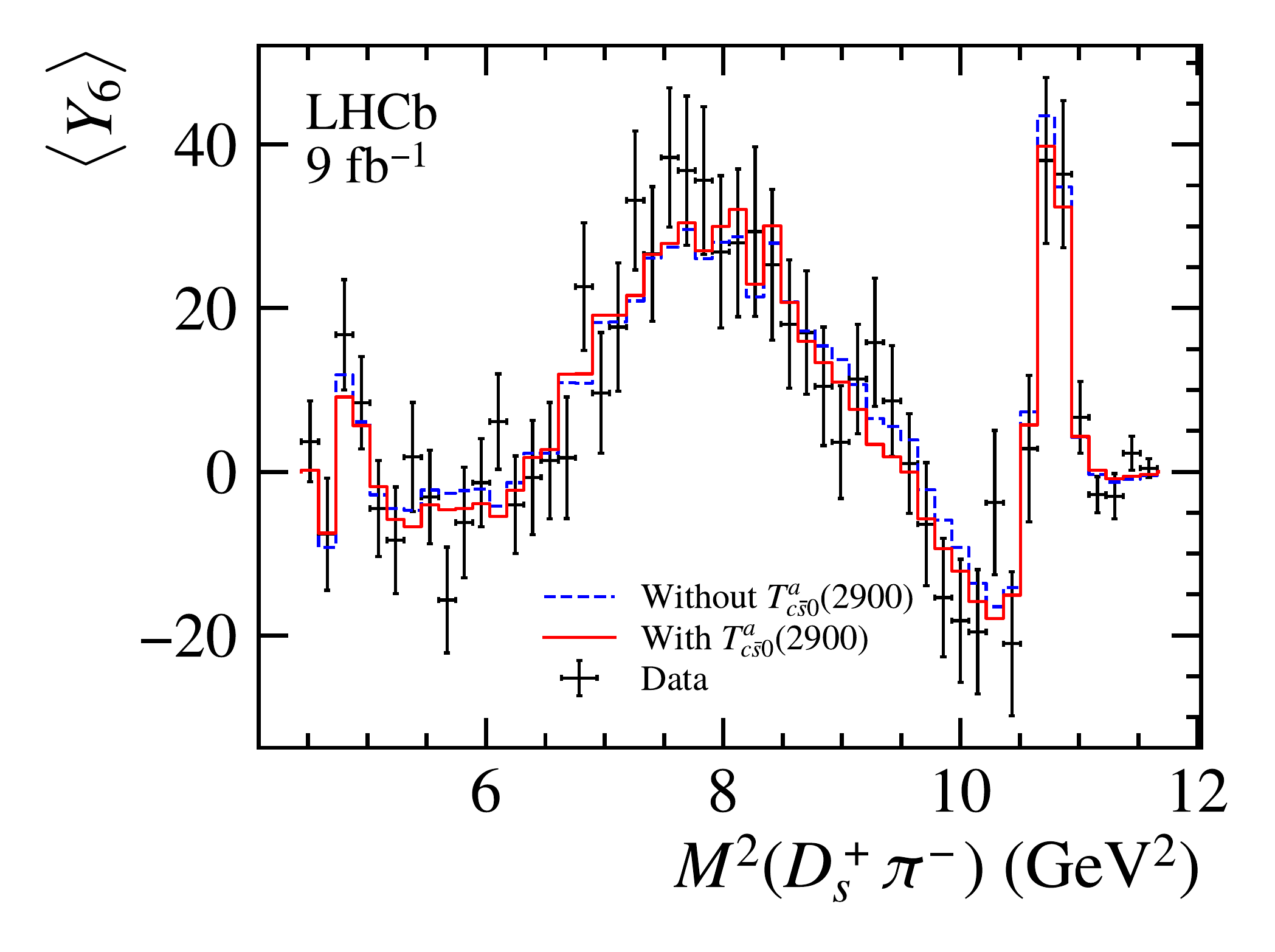}
    \includegraphics[width=0.32\linewidth]{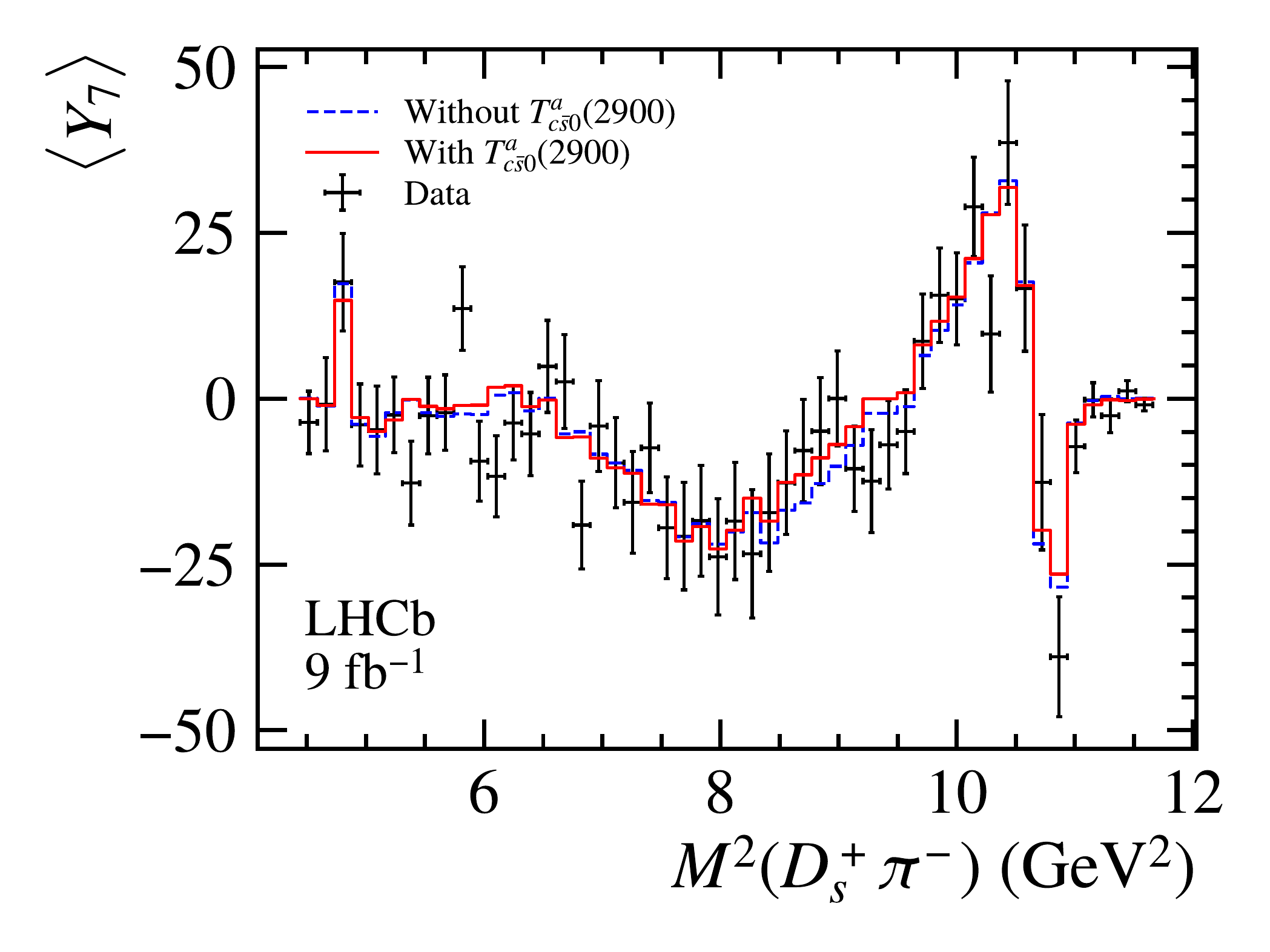}
    \includegraphics[width=0.32\linewidth]{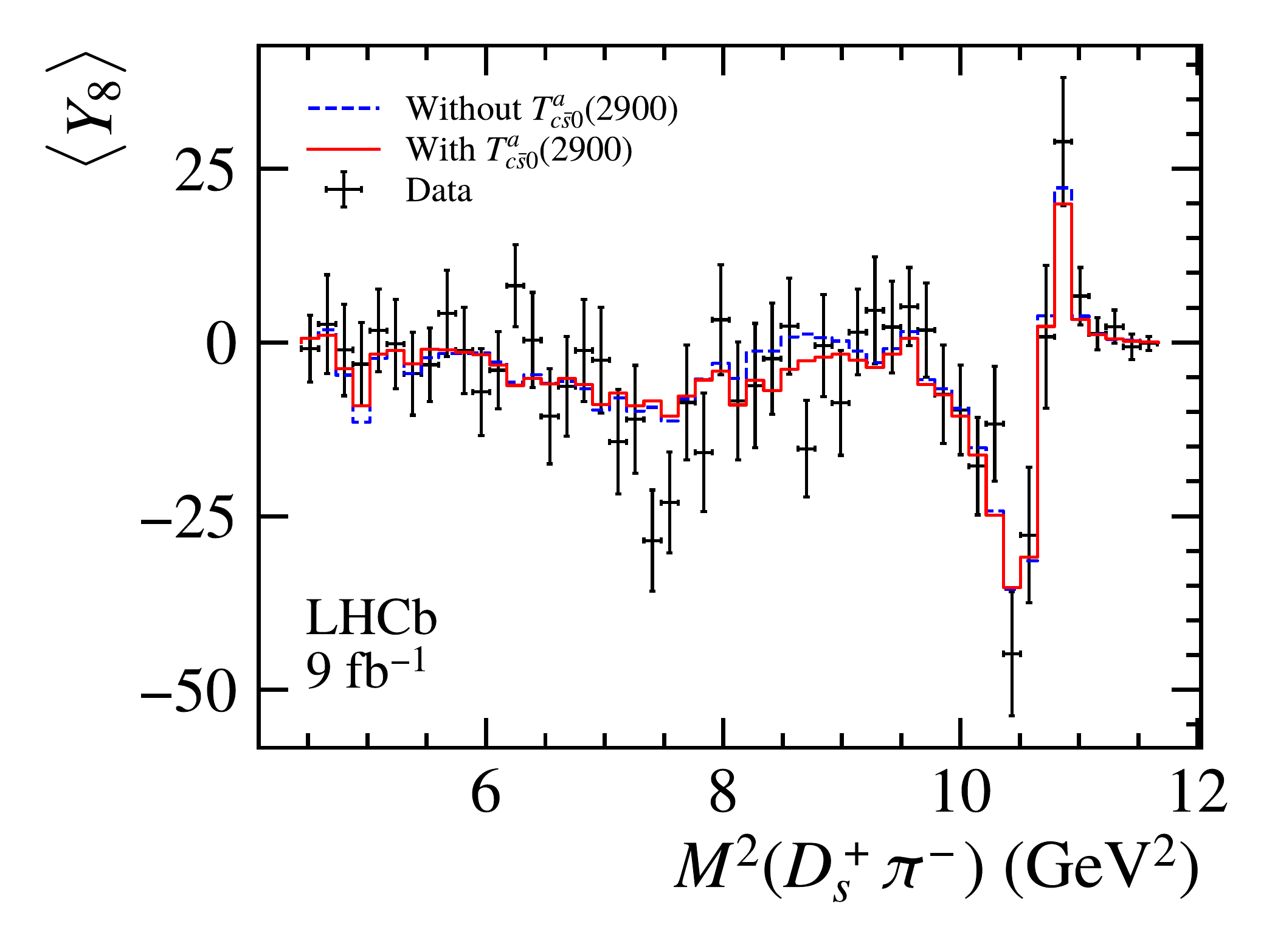}
  \end{center}
  \caption{Moments analysis of \BztoDzbarDsppim on $M^2(\Dsp\pim)$. The black points indicate the data, while the blue and red histogram indicate the fit results without and with the \Zz states separately.}
  \label{fig:mom_MDspi_Bz}
\end{figure}

\begin{figure}[htb]
  \begin{center}
    \includegraphics[width=0.32\linewidth]{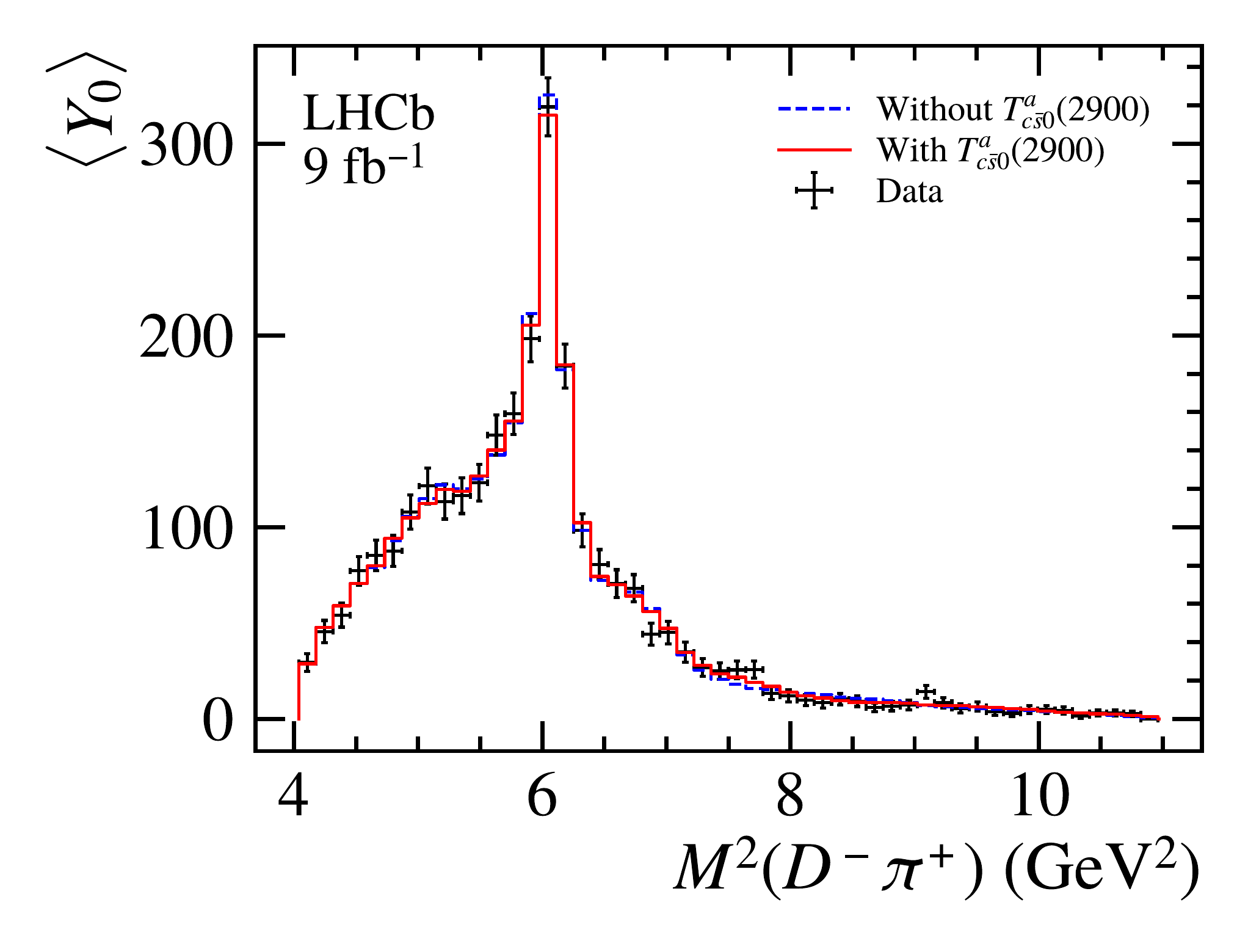}
    \includegraphics[width=0.32\linewidth]{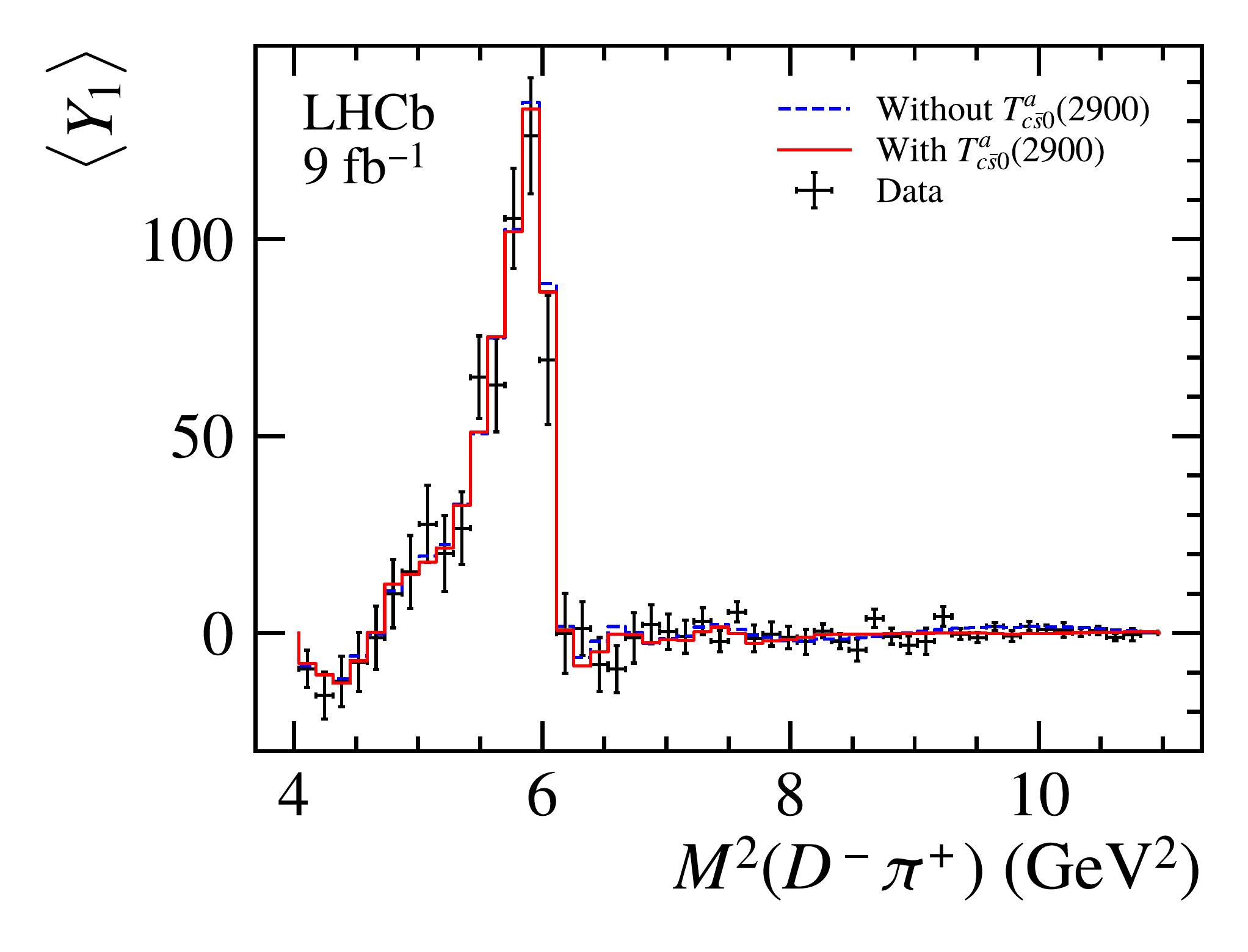}
    \includegraphics[width=0.32\linewidth]{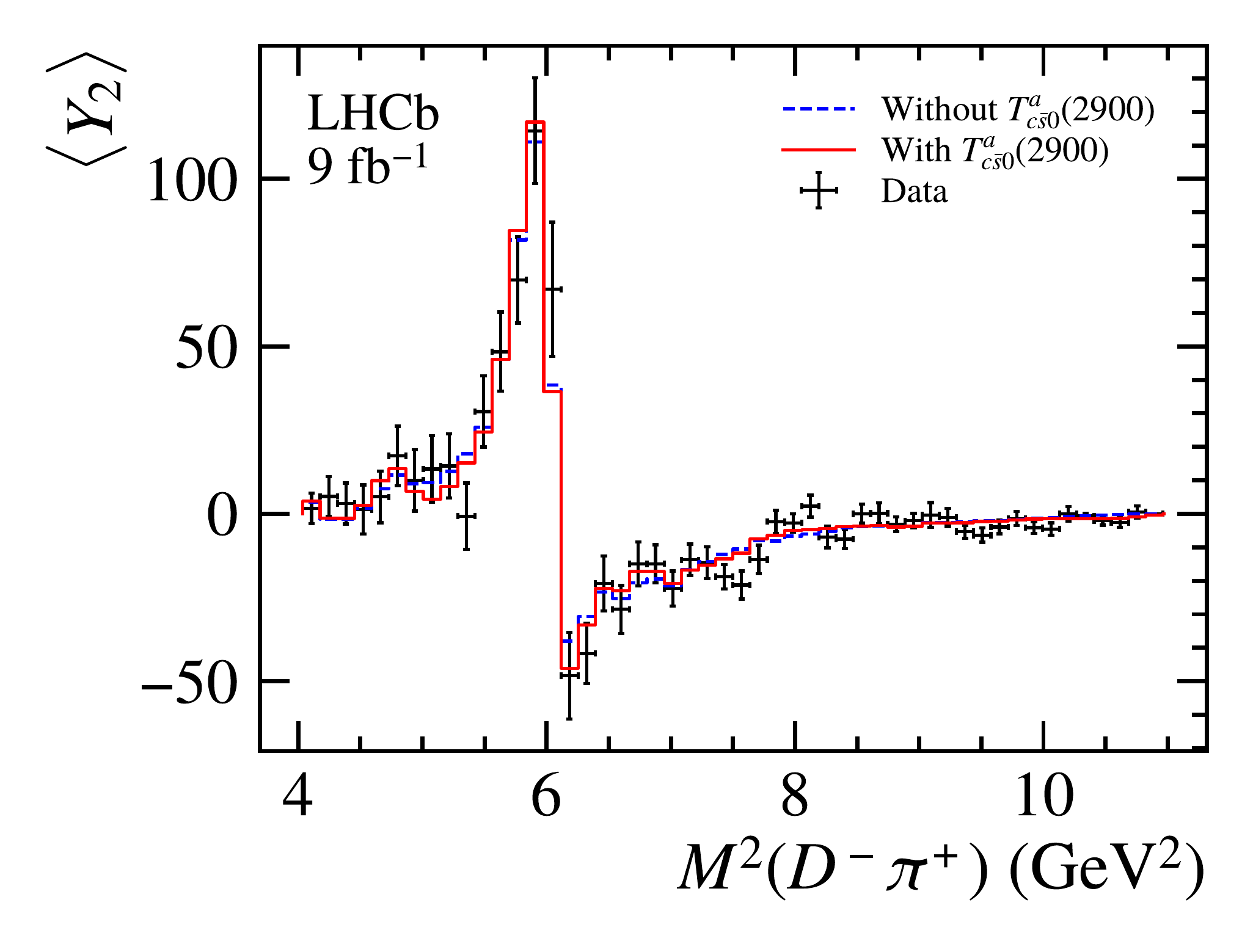}
    
    \includegraphics[width=0.32\linewidth]{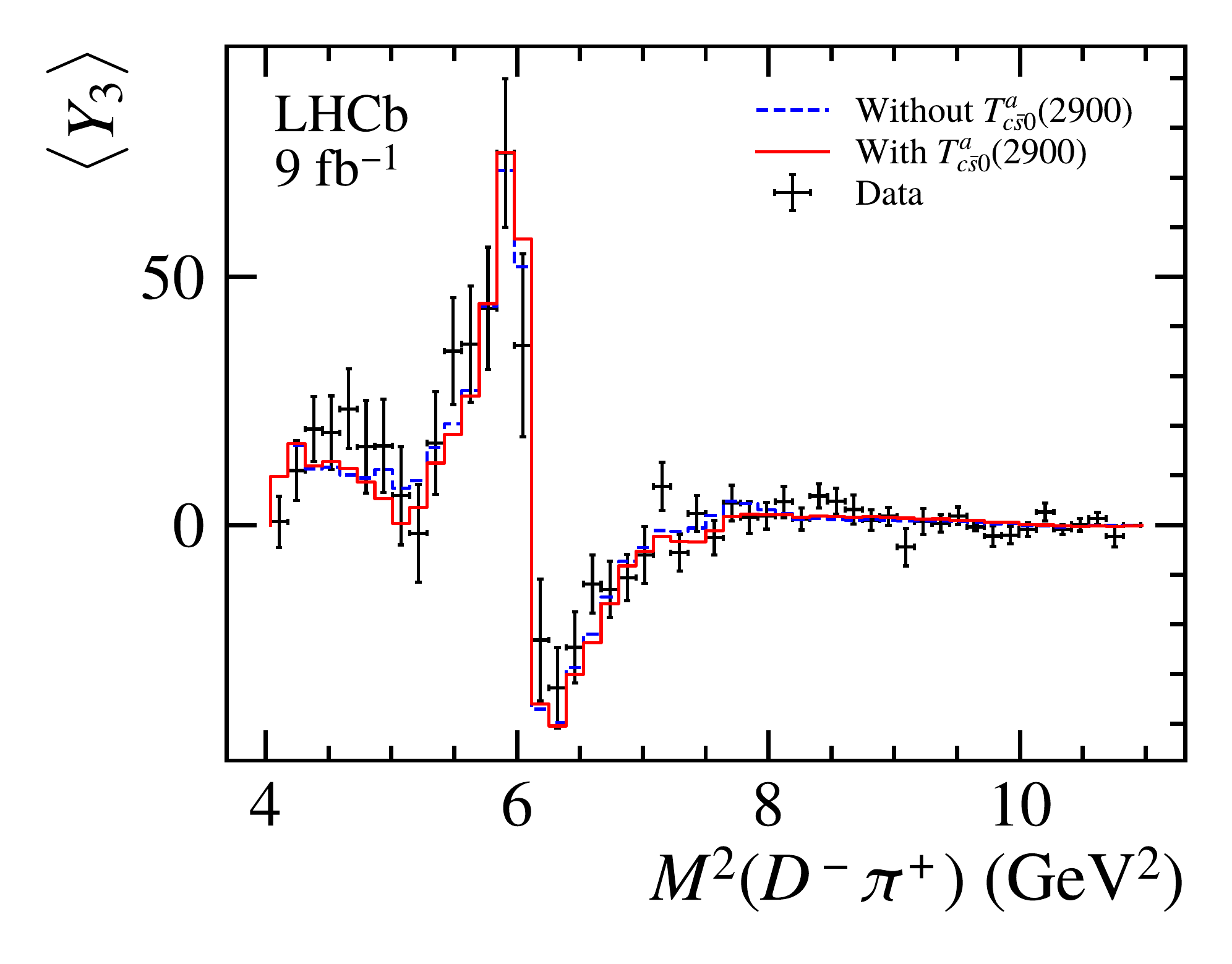}
    \includegraphics[width=0.32\linewidth]{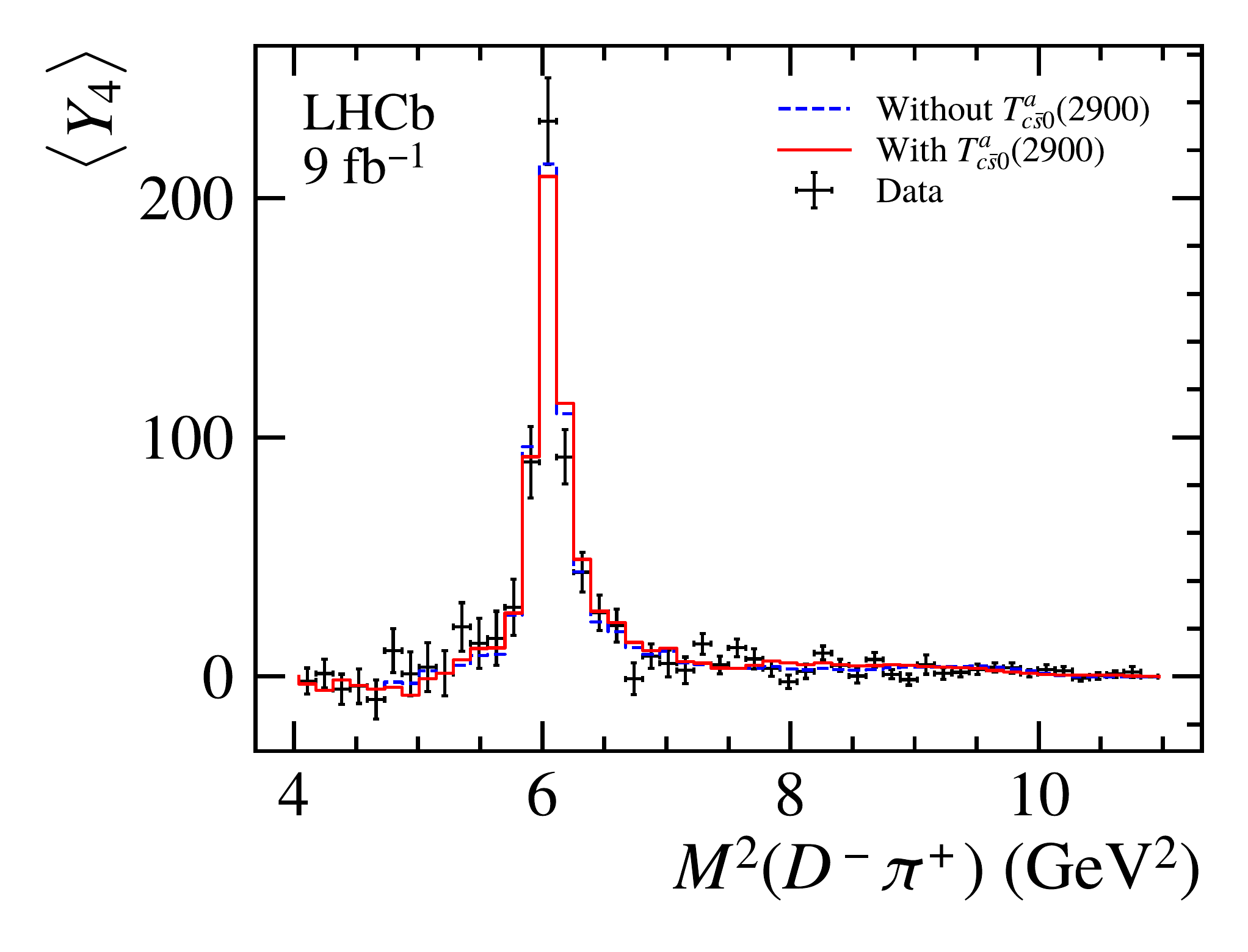}
    \includegraphics[width=0.32\linewidth]{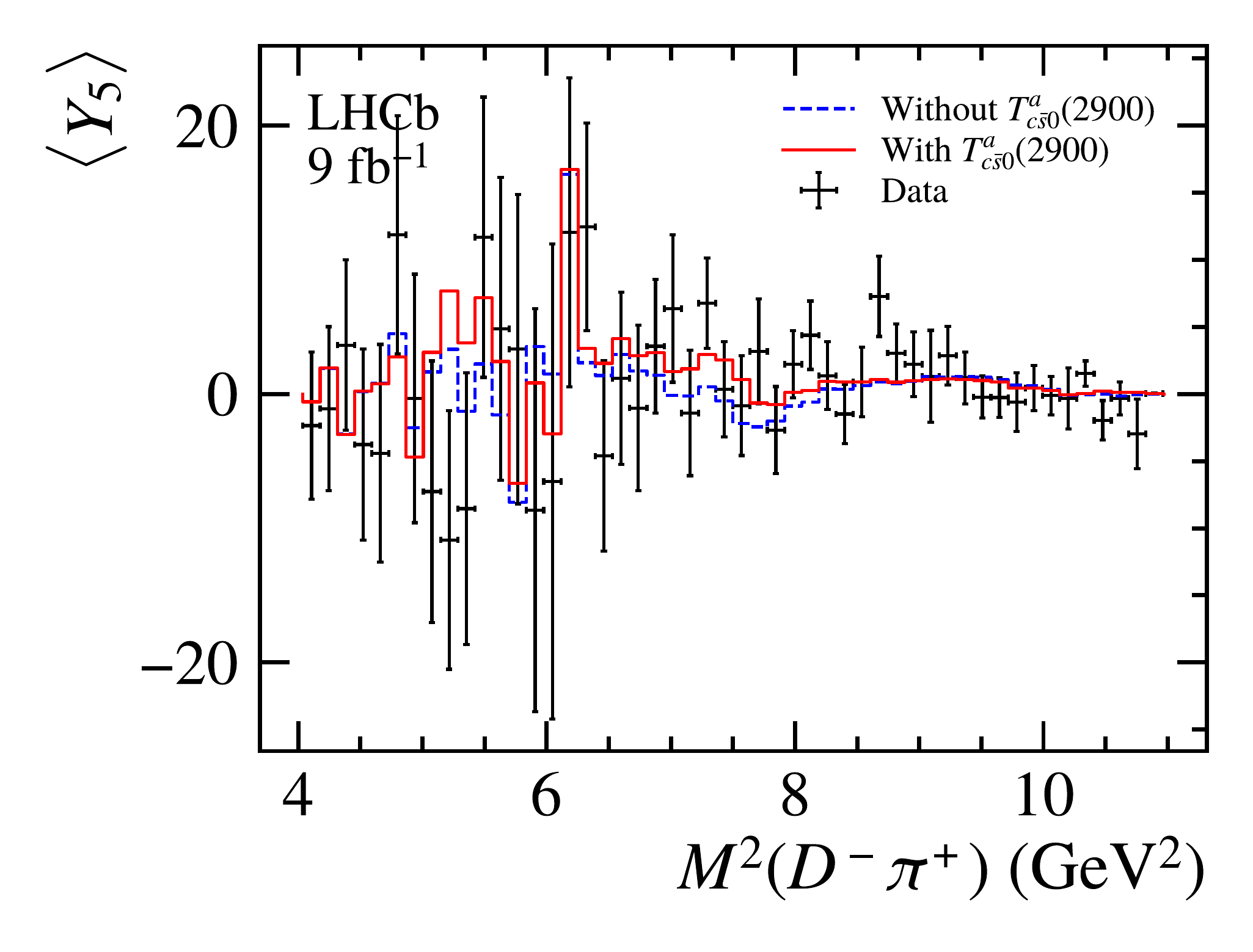}
    
    \includegraphics[width=0.32\linewidth]{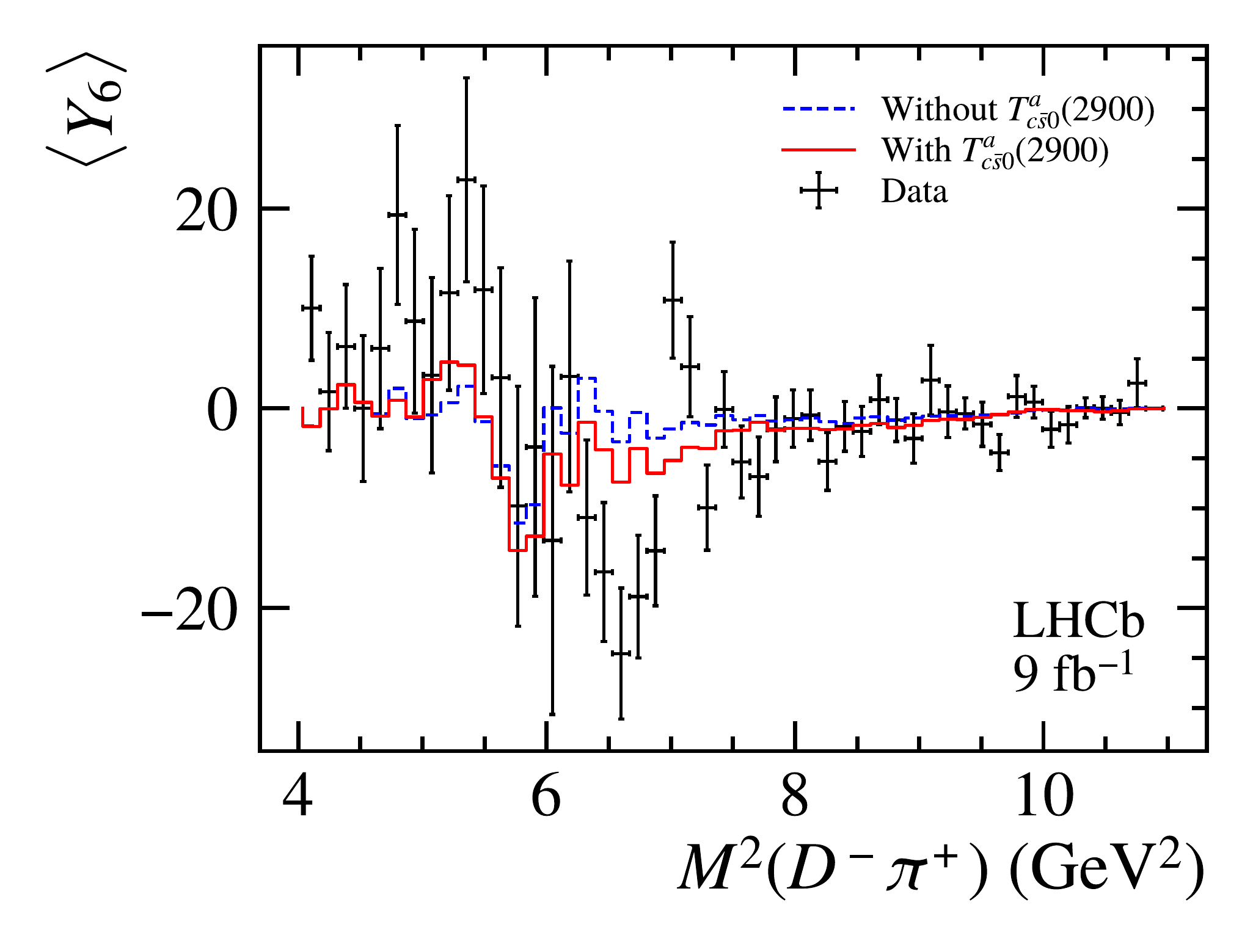}
    \includegraphics[width=0.32\linewidth]{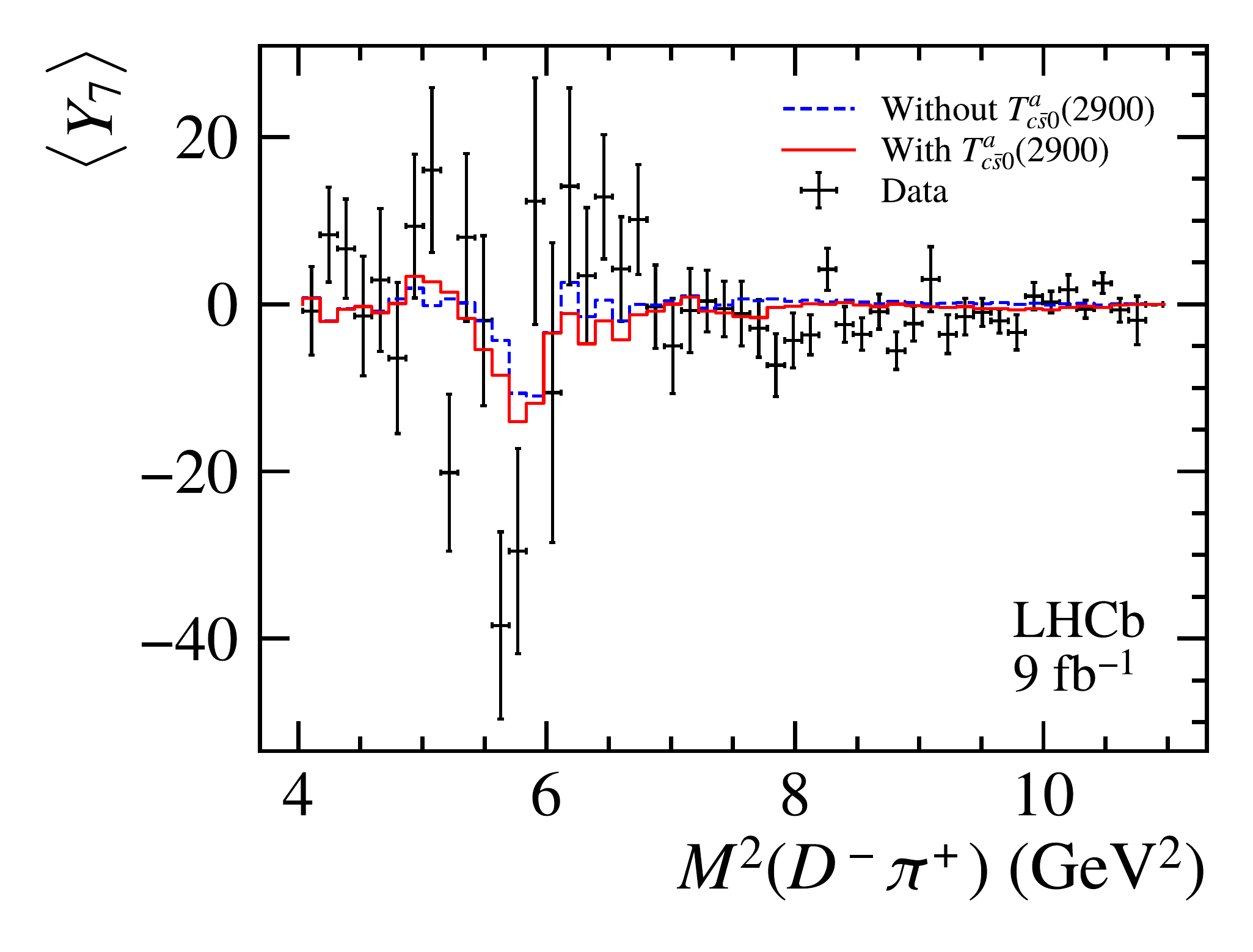}
    \includegraphics[width=0.32\linewidth]{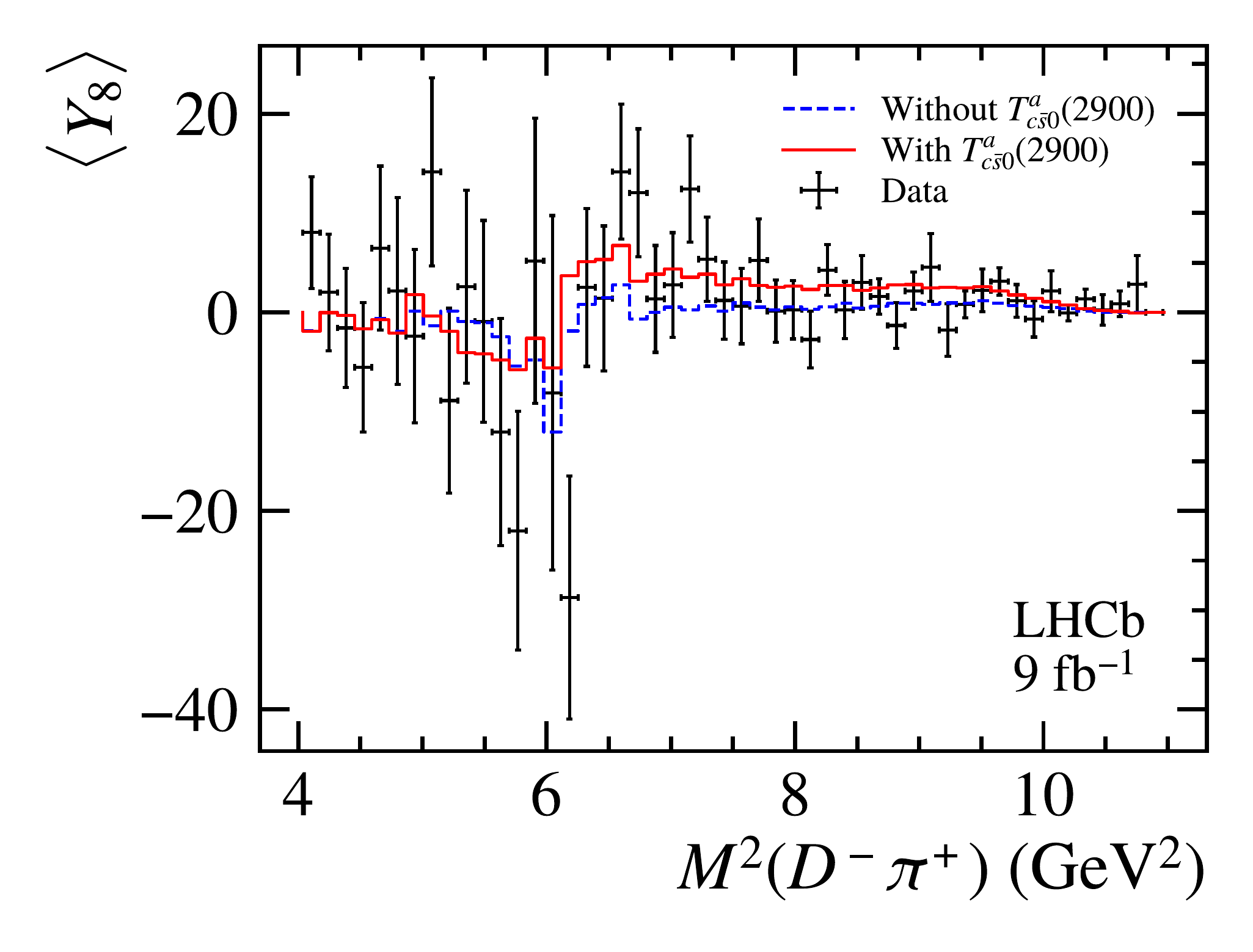}
  \end{center}
  \caption{Moments analysis of \BptoDmDsppip on $M^2(\Dm\pip)$. The black points indicate the data, while the blue and red histogram indicate the fit results without and with the \Zz states separately.}
  \label{fig:mom_MDpi_Bp}
\end{figure}

\begin{figure}[htb]
  \begin{center}
    \includegraphics[width=0.32\linewidth]{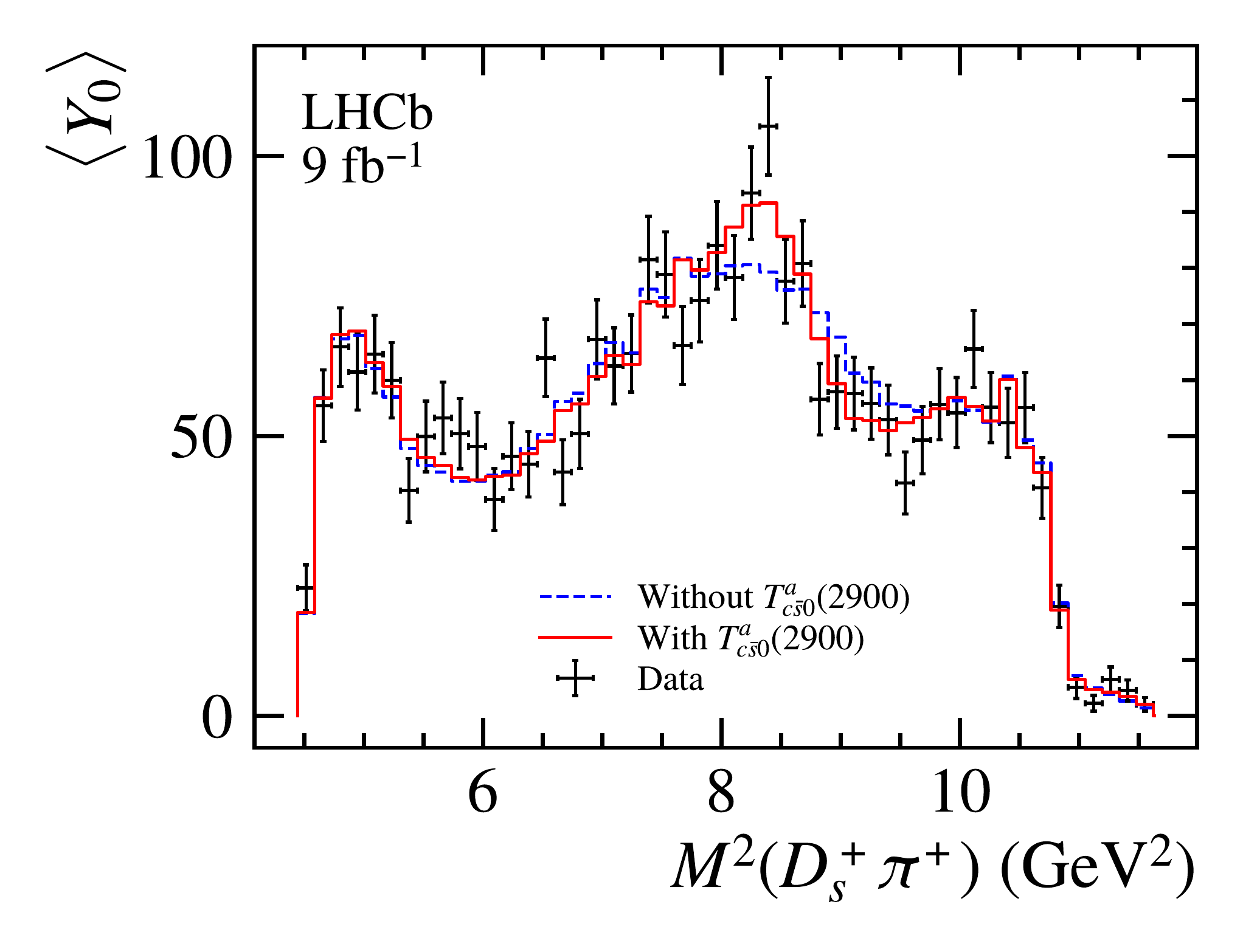}
    \includegraphics[width=0.32\linewidth]{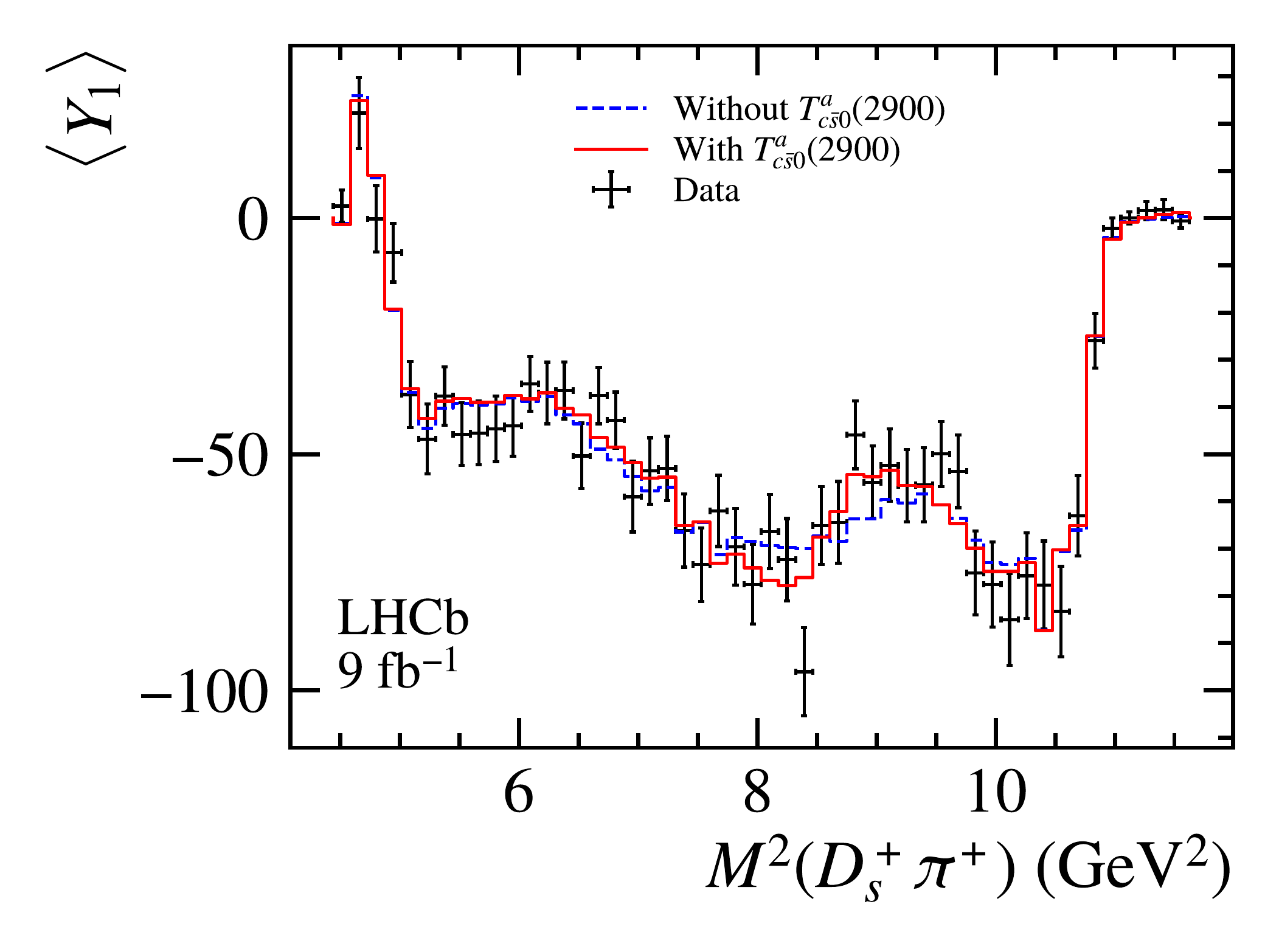}
    \includegraphics[width=0.32\linewidth]{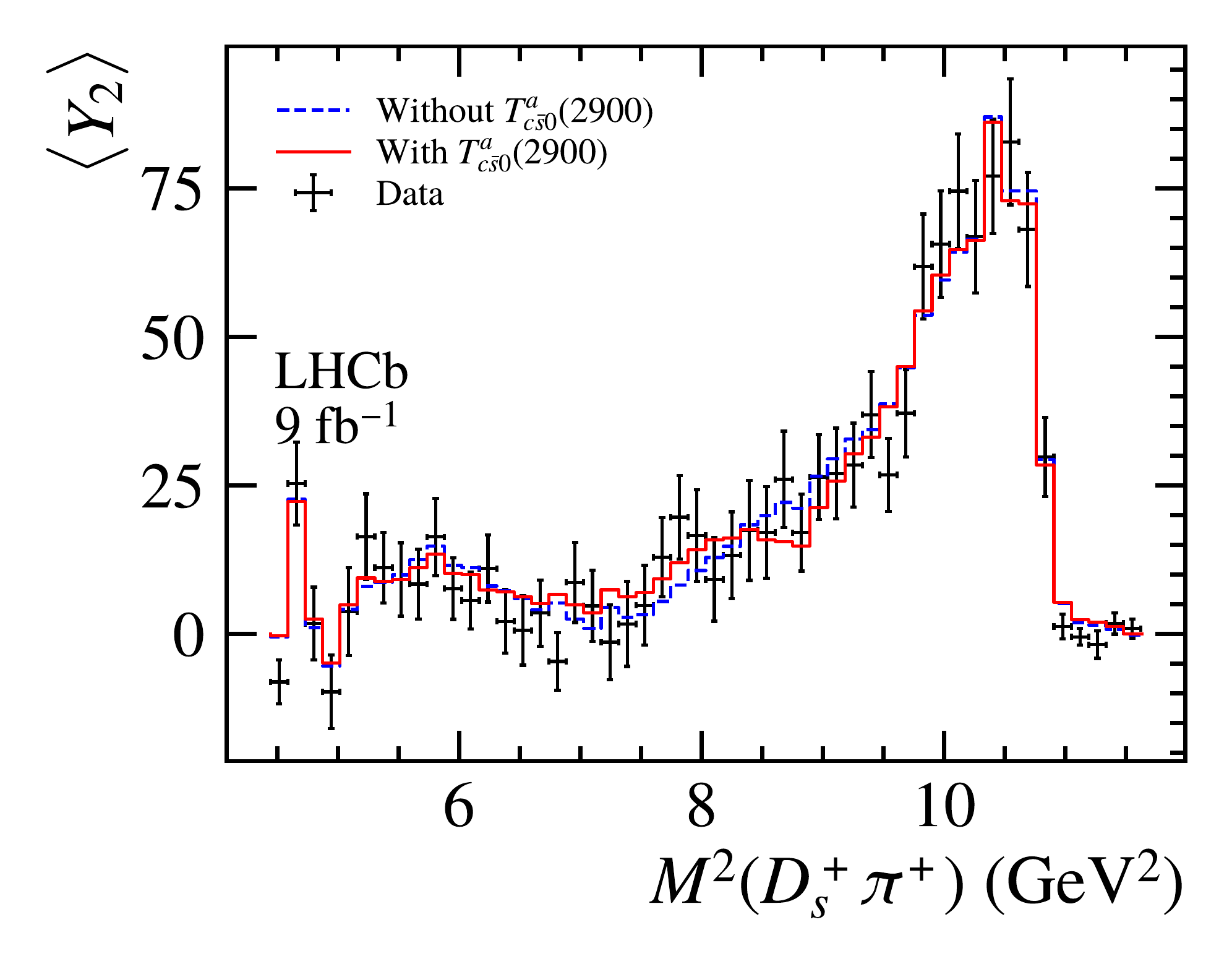}
    
    \includegraphics[width=0.32\linewidth]{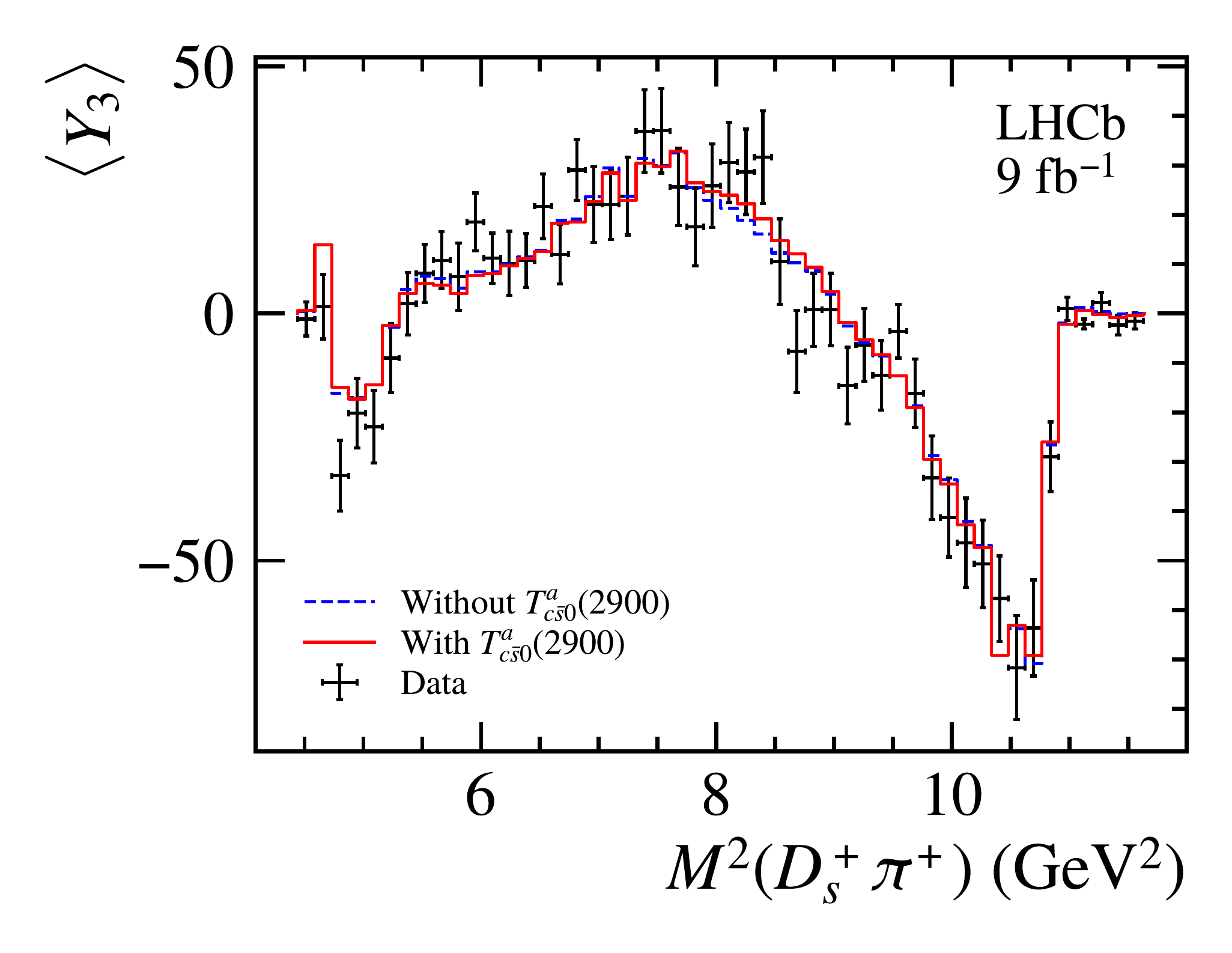}
    \includegraphics[width=0.32\linewidth]{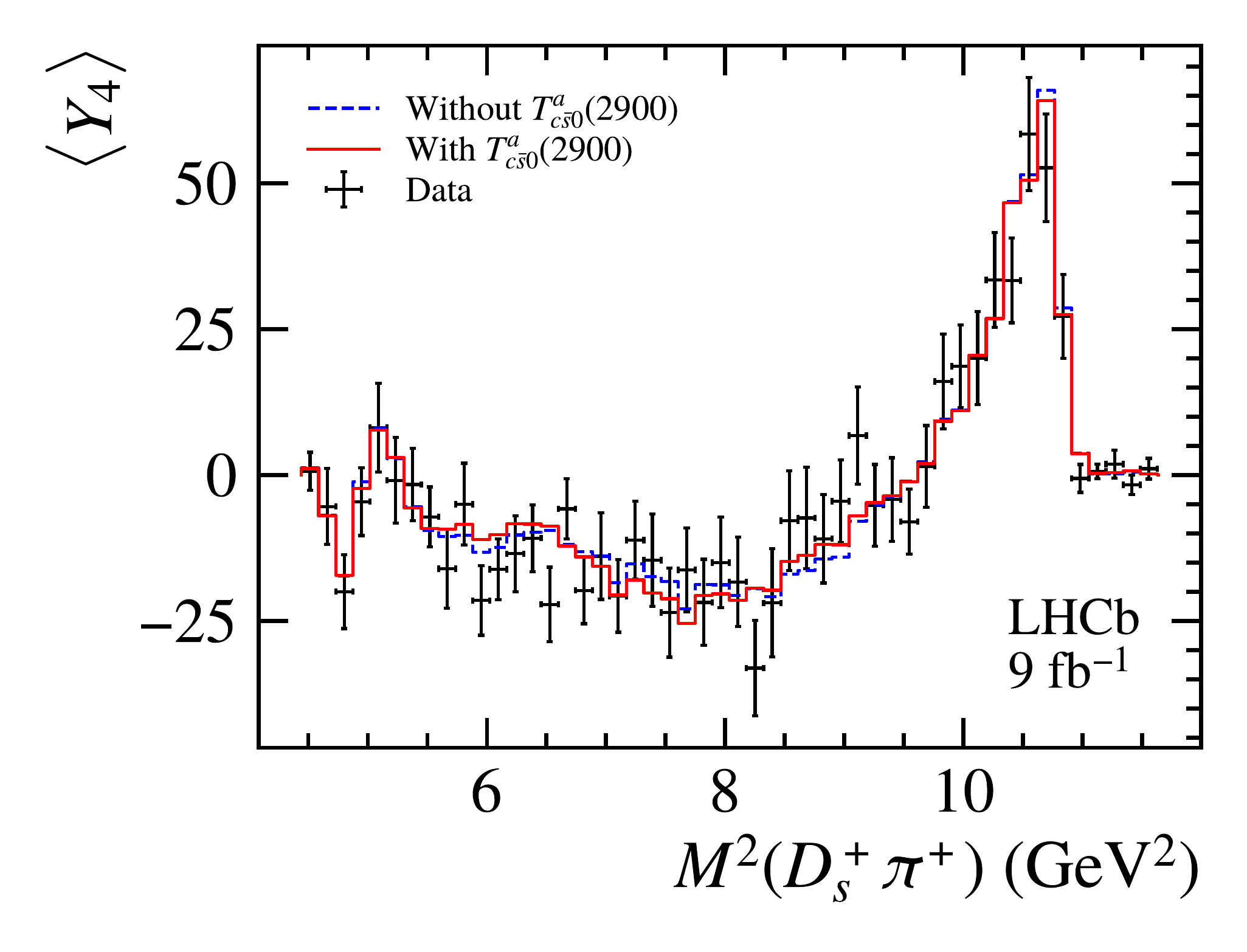}
    \includegraphics[width=0.32\linewidth]{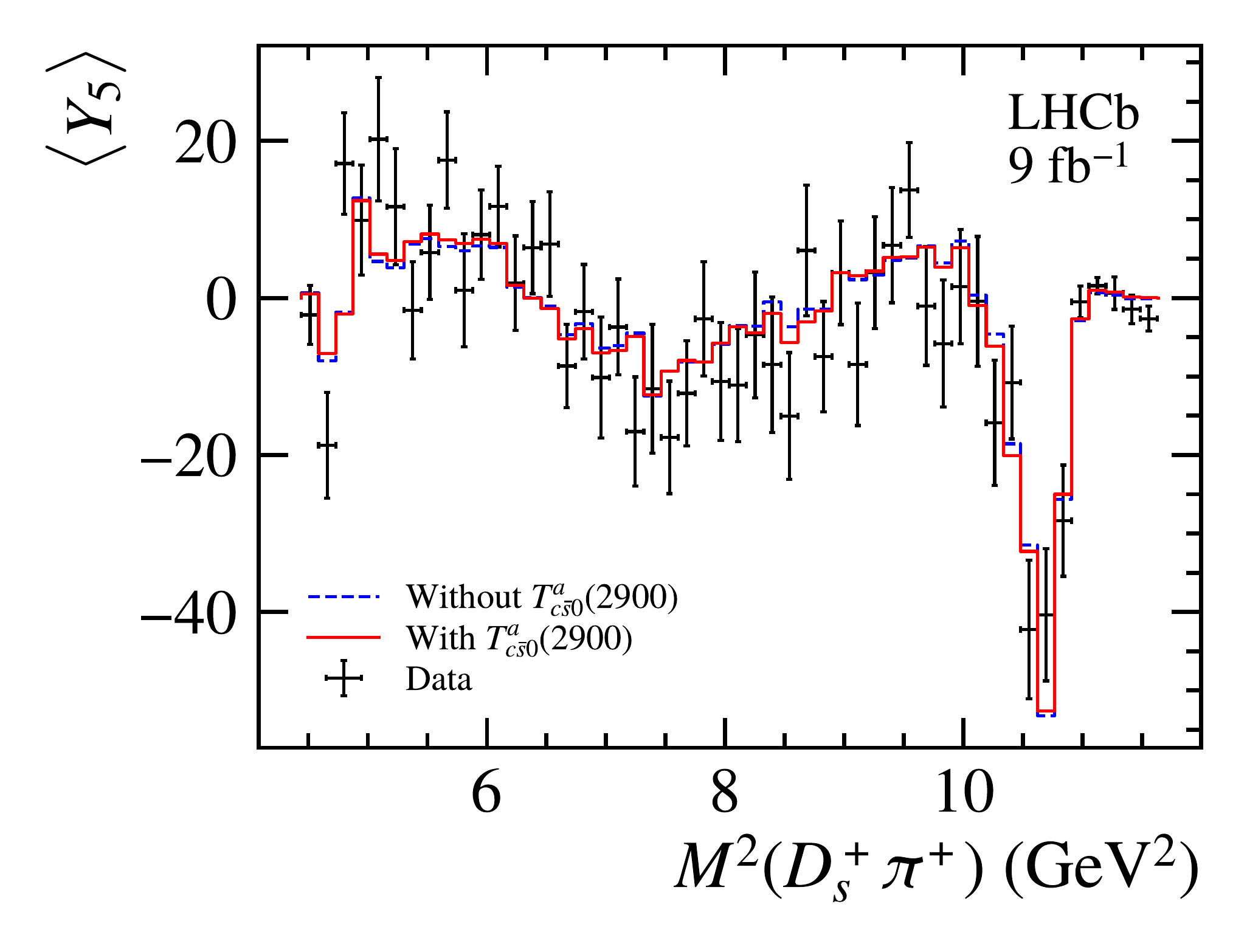}
    
    \includegraphics[width=0.32\linewidth]{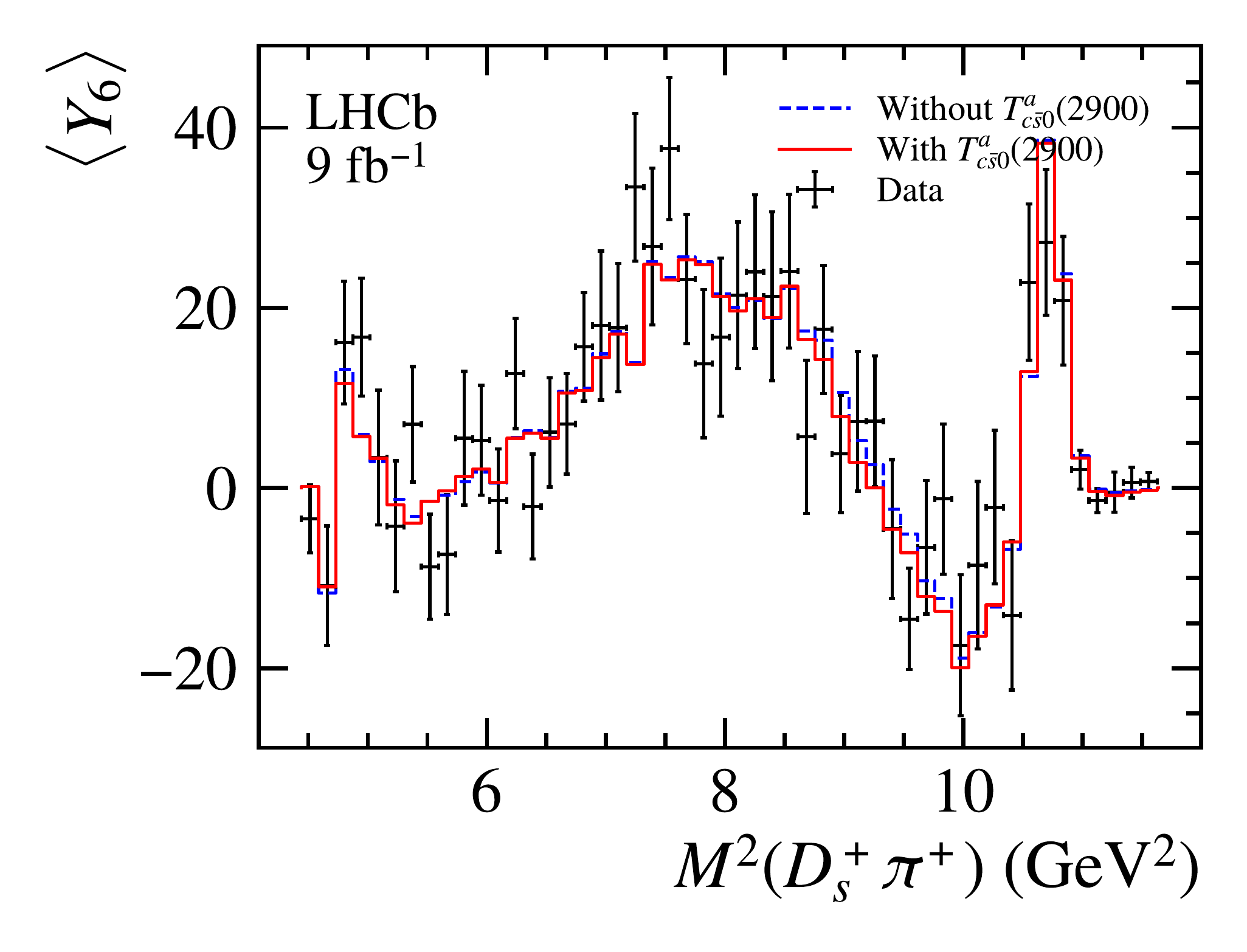}
    \includegraphics[width=0.32\linewidth]{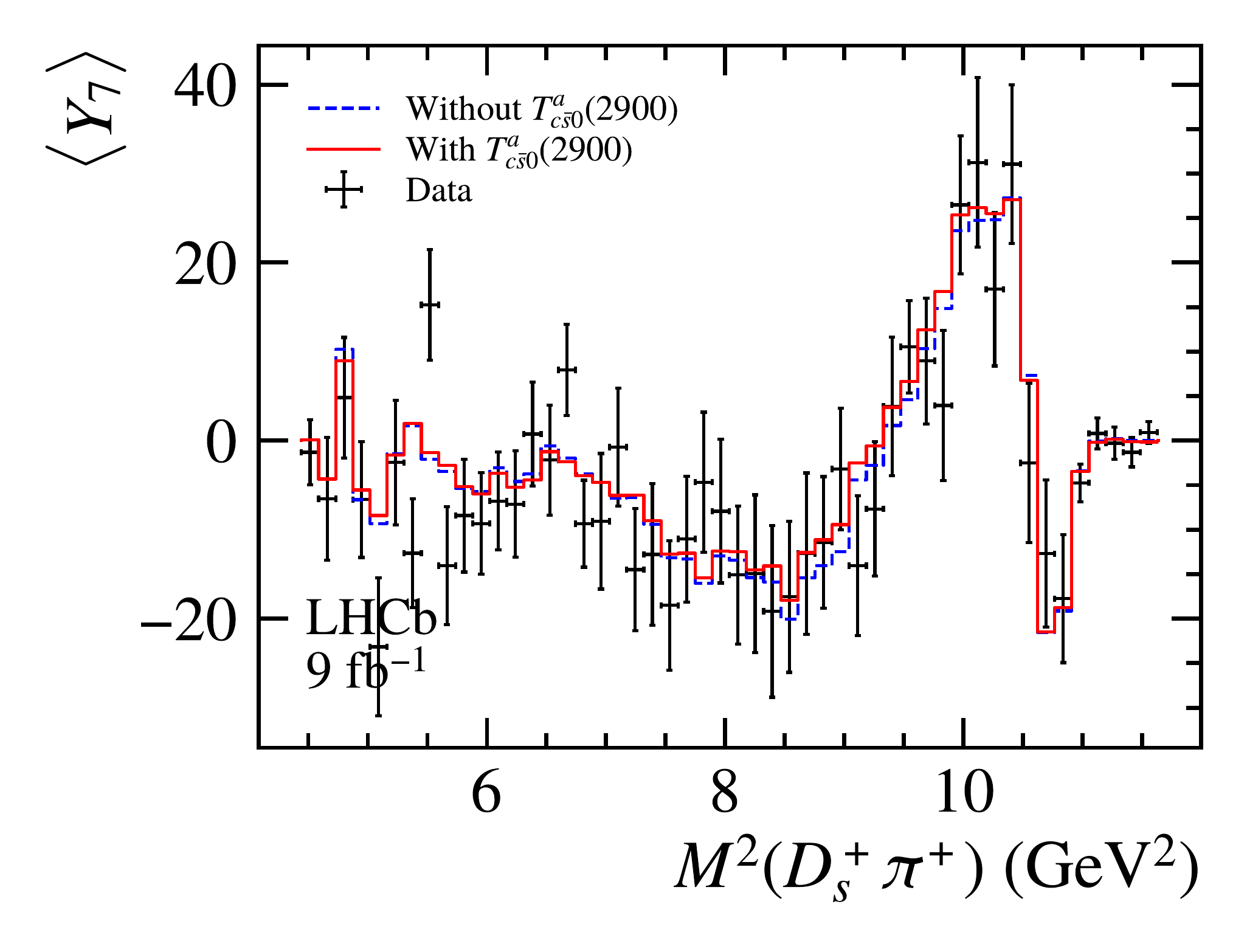}
    \includegraphics[width=0.32\linewidth]{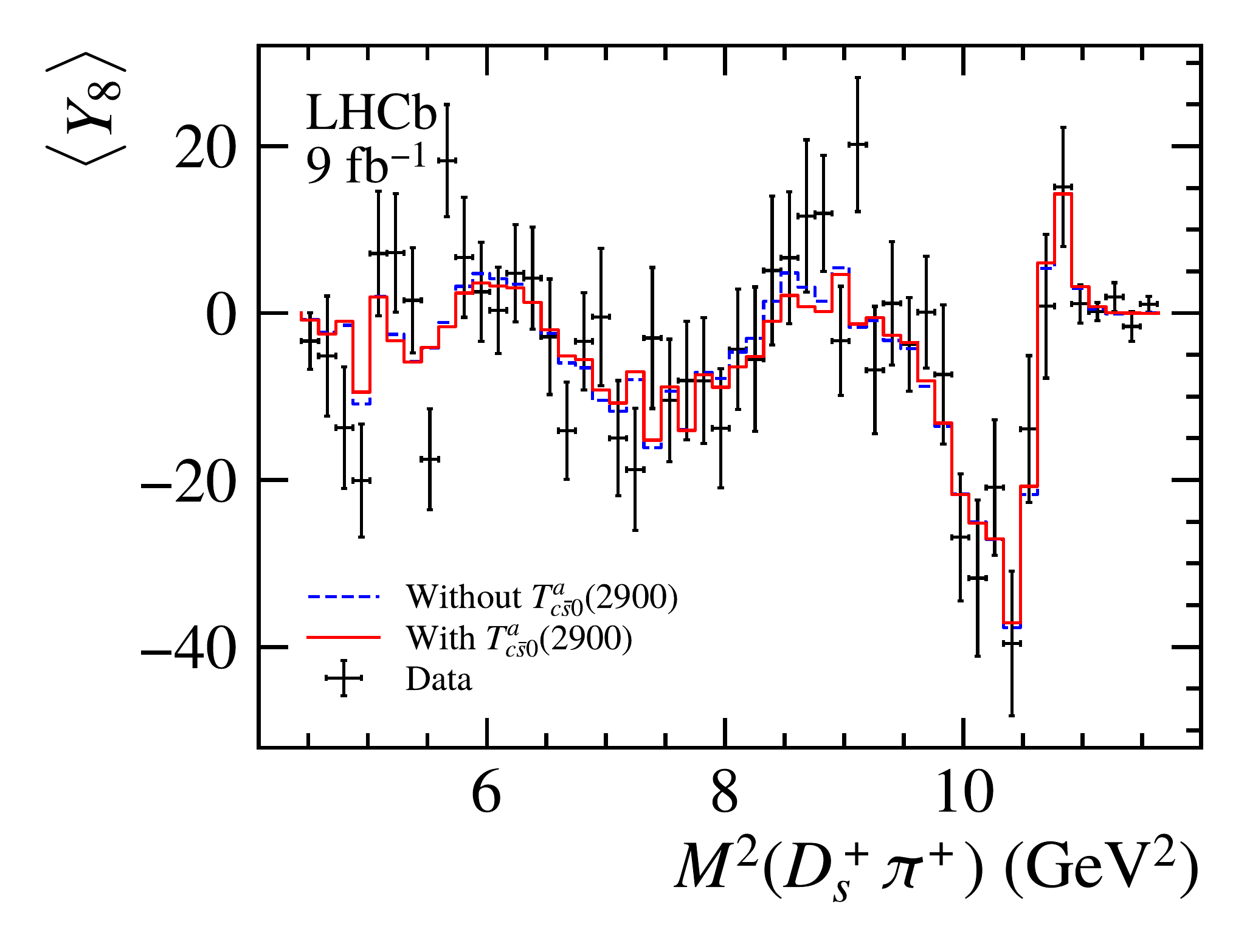}
  \end{center}
  \caption{Moments analysis of \BptoDmDsppip on $M^2(\Dsp\pip)$. The black points indicate the data, while the blue and red histogram indicate the fit results without and with the \Zz states separately.}
  \label{fig:mom_MDspi_Bp}
\end{figure}

By analytical calculation, the relationship between the Legendre-weighted total amplitude $\left<Y_k\right>$ and combination of different orders of partial waves~\cite{LHCb-PAPER-2016-026} can be bridged. Considering the existence of partial waves with the first $J$ orders of orbital momentum, only $\left<Y_k\right>$ of $k$ up to $2J$ are nonzero.
The weighted distributions can be visualized on the $M^2(D\pi)$ axis, which can be helpful to distinguish the structures from different ordered partial waves. The moments on the two axes, $M^2(D\pi)$, $M^2(D_s\pi)$, are shown up to the eighth order. The results, which are shown in Figs.~\ref{fig:mom_MDpi_Bz}, \ref{fig:mom_MDspi_Bz}, \ref{fig:mom_MDpi_Bp}, and \ref{fig:mom_MDspi_Bp}, are taken from the separate fits to \BztoDzbarDsppim and \BptoDmDsppip, where the data are background-subtracted and efficiency-corrected.

% This should be taken out in the final paper
%\input{supplementary-app}

\clearpage
\addcontentsline{toc}{section}{References}
%\setboolean{inbibliography}{true}
%\bibliographystyle{LHCb}
%\bibliography{main,standard,LHCb-PAPER,LHCb-CONF,LHCb-DP,LHCb-TDR}
%\input{main.bbl}
\ifx\mcitethebibliography\mciteundefinedmacro
\PackageError{LHCb.bst}{mciteplus.sty has not been loaded}
{This bibstyle requires the use of the mciteplus package.}\fi
\providecommand{\href}[2]{#2}

\newpage
% LHCb collaboration author list
% Data extracted on May 26th, 2023 at 3:31pm for paper reference LHCb-PAPER-2022-027
\centerline
{\large\bf LHCb collaboration}
\begin
{flushleft}
\small
R.~Aaij$^{32}$\lhcborcid{0000-0003-0533-1952},
A.S.W.~Abdelmotteleb$^{50}$\lhcborcid{0000-0001-7905-0542},
C.~Abellan~Beteta$^{44}$,
F.~Abudin{\'e}n$^{50}$\lhcborcid{0000-0002-6737-3528},
T.~Ackernley$^{54}$\lhcborcid{0000-0002-5951-3498},
B.~Adeva$^{40}$\lhcborcid{0000-0001-9756-3712},
M.~Adinolfi$^{48}$\lhcborcid{0000-0002-1326-1264},
P.~Adlarson$^{77}$\lhcborcid{0000-0001-6280-3851},
H.~Afsharnia$^{9}$,
C.~Agapopoulou$^{13}$\lhcborcid{0000-0002-2368-0147}
% ...

{\footnotesize \it

% ...
$^{9}$Universit{\'e} Clermont Auvergne, CNRS/IN2P3, LPC, Clermont-Ferrand, France\\
% ...
$^{13}$LPNHE, Sorbonne Universit{\'e}, Paris Diderot Sorbonne Paris Cit{\'e}, CNRS/IN2P3, Paris, France\\
% ...
$^{32}$Nikhef National Institute for Subatomic Physics, Amsterdam, Netherlands\\
% ...
$^{40}$Instituto Galego de F{\'\i}sica de Altas Enerx{\'\i}as (IGFAE), Universidade de Santiago de Compostela, Santiago de Compostela, Spain\\
% ...
$^{44}$Physik-Institut, Universit{\"a}t Z{\"u}rich, Z{\"u}rich, Switzerland\\
% ...
$^{48}$H.H. Wills Physics Laboratory, University of Bristol, Bristol, United Kingdom\\
% ...
$^{50}$Department of Physics, University of Warwick, Coventry, United Kingdom\\
% ...
$^{53}$School of Physics and Astronomy, University of Glasgow, Glasgow, United Kingdom\\
$^{54}$Oliver Lodge Laboratory, University of Liverpool, Liverpool, United Kingdom\\
% ...
$^{77}$Department of Physics and Astronomy, Uppsala University, Uppsala, Sweden, associated to $^{53}$\\
% ...
}
\end{flushleft}

\end{document}